\documentclass{article}

\usepackage{arxiv}

\RequirePackage{tikz}
\usetikzlibrary{decorations.pathreplacing,calligraphy}

\usepackage[utf8]{inputenc} 
\usepackage[T1]{fontenc}    
\usepackage[colorlinks]{hyperref}       
\usepackage{url}            
\usepackage{booktabs}       
\usepackage{amsfonts}       
\usepackage{nicefrac}       
\usepackage{microtype}      
\usepackage{graphicx}
\usepackage[numbers]{natbib}
\usepackage{doi}

\usepackage[normalem]{ulem}
\usepackage{amsmath}
\usepackage{amsthm}
\usepackage{amssymb}
\usepackage{amsfonts,amsbsy,amscd}
\usepackage{times}
\usepackage{latexsym}
\usepackage{subcaption}
\usepackage{xcolor}
\usepackage{siunitx}
\usepackage[nolist]{acronym}
\usepackage{comment}
\usepackage{colortbl}
\usepackage{tabularx}
\usepackage{pgfplots}
\usepgfplotslibrary{fillbetween}
\usepackage{mathtools,cuted}
\usepackage{multirow}
\usepackage{enumitem}
\usepackage{bm}

\renewcommand{\bm}[1]{\boldsymbol{#1}}
\newcommand{\DS}{\displaystyle}
\newcommand{\half}{\frac{1}{2}}

\renewcommand{\top}{T}
\definecolor{HgrauTU}{RGB}{230,230,230}
\definecolor{myBlue}{rgb}{0.0000,0.4470,0.7410}
\definecolor{darkred}{RGB}{139,0,0}
\definecolor{lightGray}{RGB}{230,230,230}
\definecolor{lightRed}{RGB}{255,153,153}
\definecolor{lightBlue}{RGB}{153,204,255}
\definecolor{darkerBlue}{RGB}{51,153,255}
\newcommand{\dif}[2]{\frac{\DS \partial #1}{\DS \partial #2}}
\newcommand{\difn}[2]{\frac{\DS \D #1}{\DS \D #2}}
\newcommand{\gat}[4]{\mathsf{D}_{#1}\,#2(#3)[#4]}
\newcommand{\GKap}{\bm{\kappa}}
\newcommand{\GBet}{\bm{\beta}}
\newcommand{\GThe}{\bm{\theta}}
\newcommand{\D}{\mathrm d}
\newcommand{\diftwo}[3]{\frac{\partial^2 #1}{\partial #2 \partial #3}}
\newcommand{\diftwon}[3]{\frac{\D^2 #1}{\D #2 \D #3}}
\newcommand{\fr}[1]{\frac{\DS 1}{\DS #1}}
\DeclareMathOperator{\grad}{grad}
\newcommand{\tr}{\mathrm{tr}\,}
\newcommand{\nums}{{n_{\textrm{s}}}}       
\newcommand{\numkappa}{{n_{\kappa}}}

\newcommand{\T}[1]{\boldsymbol{\mathsf{#1}}}               
\newcommand{\TP}[1]{\dot{\boldsymbol{\mathsf{#1}}}}           
\newcommand{\Tt}[1]{\tilde{\boldsymbol{\mathsf{#1}}}}        
\newcommand{\TMT}[1]{\boldsymbol{\mathsf{#1}}^{-T}}          
\newcommand{\TM}[1]{\boldsymbol{\mathsf{#1}}^{-1}}           
\newcommand{\TT}[1]{\boldsymbol{\mathsf{#1}}^{T}}            
\newcommand{\V}[1]{\vec{#1}}                   
\newcommand{\Vh}[1]{\hat{\vec{#1}}}            
\newcommand{\Vq}[1]{\bar{\vec{#1}}}            
\newcommand{\LV}[1]{\bm{#1}}    
\newcommand{\LVq}[1]{\overline{\bm{#1}}} 
\newcommand{\LVT}[1]{\bm{#1}^T}    
 
\newcommand{\LVp}[1]{\bm{\dot{#1}}}   
\newcommand{\LM}[1]{\bm{#1}}    
\newcommand{\elm}[1]{{\, \in \mathbb{R}}^{#1}}
\newcommand{\elmm}[2]{{\, \in \mathbb{R}}^{#1 \times #2}}
\newcommand{\GV}[1]{\bm{#1}}    
\newcommand{\GVT}[1]{\bm{#1}^T} 
\newcommand{\GVq}[1]{\skew{2}\overline{\bm{#1}}}  
\newcommand{\GVc}[1]{\check{\bm{#1}}}    
\newcommand{\GVh}[1]{\skew{2}\hat{\bm{#1}}}
\newcommand{\GVhp}[1]{\skew{2}\dot{\skew{2}\hat{\bm{#1}}}}
\newcommand{\GVt}[1]{\tilde{\bm{#1}}} 
\newcommand{\GVtT}[1]{{\tilde{\bm{#1}}}^{\;T}} 
 
\newcommand{\GVqq}[1]{\overline{\overline{\bm{#1}}}} 
\newcommand{\GVP}[1]{\bm{\dot{#1}}} 
\newcommand{\GVqc}[1]{\check{\overline{\bm{#1}}}}
\newcommand{\GM}[1]{\bm{#1}}    
\newcommand{\GMT}[1]{{\bm{#1}}^T}    
\newcommand{\GMM}[1]{{\bm{#1}}^{-1}} 
\newcommand{\GMq}[1]{\overline{\bm{#1}}} 
\newcommand{\GMqT}[1]{\overline{\bm{#1}}^T} 
\newcommand{\GMqq}[1]{\overline{\overline{\bm{#1}}}} 
\newcommand{\GMt}[1]{\tilde{\bm{#1}}}    
    
\newcommand{\U}{\GV{u}}                
\newcommand{\Uq}{\GVq{u}}
\newcommand{\Ua}{\GV{u}_{\text{a}}}
\newcommand{\Uh}{\hat{\GV{u}}}           
\newcommand{\Q}{\GV{q}}                
\newcommand{\Glam}{\boldsymbol{\lambda}}
\newcommand{\GLam}{\boldsymbol{\Lambda}}
\newcommand{\mzweiv}[2]{\left\{\begin{matrix}#1\\#2\end{matrix}\right\}}
\newcommand{\tn}{t_{n}}
\newcommand{\tnp}{t_{n+1}}
\newcommand{\dtn}{\Delta t_n}

\newcommand{\Unp}{\ensuremath{\GV{u}\INDNP} }
\newcommand{\Qn}{\ensuremath{\GV{q}\INDn} }
\newcommand{\Qnp}{\ensuremath{\GV{q}\INDNP} }

\newcommand{\nel}{{n_{\textrm{el}}}}    
\newcommand{\nnodes}{{n_{\textrm{nodes}}}} 
\newcommand{\numFE}{{n_{\textrm{F}}^{(\hat{E})}}}     
\newcommand{\numgu}{{n_{\textrm{GP}}^e}}   
\newcommand{\numue}{{n_{\textrm{u}}^e}}    
\newcommand{\numu}{{n_{\textrm{u}}}}       
\newcommand{\nump}{{n_{\textrm{p}}}}       
\newcommand{\numQ}{{n_{\textrm{Q}}}}       
\newcommand{\numq}{{n_{\textrm{q}}}}       
\newcommand{\numexp}{{n_{\textrm{exp}}}} 
\newcommand{\numexpE}{{n_{\textrm{exp}}^{(\hat{E})}}} 
\newcommand{\numNe}{{n_{\textrm{N}}^{(\hat{E})}}} 
\newcommand{\numdE}{{n_{\textrm{D}}^{(\hat{E})}}} 
\newcommand{\numuE}{{n_{\textrm{u}}^{(\hat{E})}}} 
\newcommand{\numutE}{{\tilde{n}_{\textrm{u}}^{(\hat{E})}}}    
\newcommand{\numD}{{n_{\textrm{D}}}}     
 
\newcommand{\numcol}{{n_{\text{col}}}}
\newcommand{\numcolneu}{{n_{\text{col}}^\text{N}}}
\newcommand{\numcoldir}{{n_{\text{col}}^\text{D}}}
\newcommand{\detop}{\operatorname{det}}  
\newcommand{\divop}{\operatorname{div}}  
\newcommand{\Divop}{\operatorname{Div}}  
\newcommand{\Grad}{\operatorname{Grad}}  
\newcommand{\argmax}{\operatorname{arg}\operatorname{min}}    
\newcommand{\argstat}{\operatorname{arg}\operatorname{stat}}
\def \INDA {_{\text{A}}}
\def \INDe {_{\text{e}}}
\def \INDen {_{\text{e},0}}
\def \INDp {_{\text{p}}}
\def \INDpn {_{\text{p},0}}
\def \INDpe {_{\text{p,e}}}
\def \INDa {_{\text{a}}}
\def \INDas {_{\text{as}}}
\def \INDd {_{\text{d}}}
\def \INDc {_{\text{c}}}
\def \INDR {_{\text{R}}}
\def \INDr {_{\text{r}}}
\def \INDq {_{\text{q}}}

\def \INDn {_{n}}
\def \INDNP {_{n+1}}

\def \INDO {_{\text{O}}}

\def \INDS {_{\text{S}}}
\def \INDs {_{\text{s}}}

\def \INDel {_{\text{el}}}
\def \INDup {_{\text{up}}}
\def \INDu {_{\text{u}}}
\def \INDD {_{\text{D}}}
\def \INDN{_{\text{N}}}
\def \INDVF{_{\mathcal{V}}}
\def \EXPf {^{\text{f}}}
\def \EXPp {^{\text{p}}}
\def \EXPr {^{\,\text{r}}}

\def \EXPfr {^{\text{fr}}}
\def \EXPcol {^{\text{c}}}
\def \EXPN {^{\text{N}}}
\def \EXPD {^{\text{D}}}
\newcommand{\norm}[1]{\left\lVert#1\right\rVert}
\begin{acronym}
\acro{ANN}[ANN]{artificial neural network}
\acro{FFNN}[FFNN]{feed-forward neural network}
\acro{DIC}[DIC]{digital image correlation}
\acro{FEM}[FEM]{finite element method}
\acro{PDE}[PDE]{partial differential equation}
\acro{PINN}[PINN]{physics-informed neural network}
\end{acronym}

\newtheorem{remark}{Remark}

\newcommand{\Rset}{\mathbb{R}}

\newlength{\boxwidth}
\def\btheorem{\begin{theorem}}
\def\etheorem{\end{theorem}}
\def\blemma{\begin{lemma}}
\def\elemma{\end{lemma}}
\def\bproposition{\begin{proposition}}
\def\eproposition{\end{proposition}}
\def\bcorollary{\begin{corollary}}
\def\ecorollary{\end{corollary}}
\def\bdefinition{\begin{definition}}
\def\edefinition{\end{definition}}
\def\bexample{\begin{example}}
\def\eexample{\end{example}}
\def\bremark{\begin{remark}}
\def\eremark{\end{remark}}

\DeclareMathOperator{\argmin}{{arg\,min}}

\newcommand{\be}{\begin{equation}}
\newcommand{\ee}{\end{equation}}
\newcommand{\beq}{\begin{eqnarray}}
\newcommand{\eeq}{\end{eqnarray}}
\newcommand{\bem}{\begin{multline}}
\newcommand{\eem}{\end{multline}}
\newcommand{\ba}{\begin{align}}
\newcommand{\ea}{\end{align}}

\title{Reduced and All-at-Once Approaches for Model Calibration and Discovery in Computational Solid Mechanics}

\date{}

\author{\href{https://orcid.org/0000-0002-1277-7509}{\includegraphics[scale=0.06]{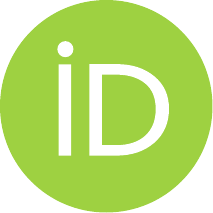}\hspace{1mm}Ulrich Römer} \\
	Institute for Acoustics and Dynamics\\
	Technische Universität Braunschweig\\
	Langer Kamp 19 \\
	38106 Braunschweig, Germany \\
	\texttt{u.roemer@tu-braunschweig.de} \\
	\And
	\href{https://orcid.org/0000-0003-1849-0784}{\includegraphics[scale=0.06]{orcid.pdf}\hspace{1mm}Stefan Hartmann} \\
	Institute of Applied Mechanics\\
	Clausthal University of Technology\\
	Adolph-Roemer-Stra{\ss}e 2a \\
	38678 Clausthal-Zellerfeld, Germany \\
	\texttt{stefan.hartmann@tu-clausthal.de} \\
	\And
	\href{https://orcid.org/0000-0002-4999-4558}{\includegraphics[scale=0.06]{orcid.pdf}\hspace{1mm}Jendrik-Alexander Tröger} \\
	Institute of Applied Mechanics\\
	Clausthal University of Technology\\
	Adolph-Roemer-Stra{\ss}e 2a \\
	38678 Clausthal-Zellerfeld, Germany \\
	\texttt{jendrik-alexander.troeger@tu-clausthal.de} \\
	\And
\href{https://orcid.org/0000-0002-0888-0220}{\includegraphics[scale=0.06]{orcid.pdf}\hspace{1mm}
        David Anton} \\
	Institute for Computational Modeling \\ in Civil Engineering\\
	Technische Universität Braunschweig\\
	Pockelsstra{\ss}e 3 \\
	38106 Braunschweig, Germany\\
	\texttt{d.anton@tu-braunschweig.de} \\
	\And
	\href{https://orcid.org/0000-0002-2542-1130}{\includegraphics[scale=0.06]{orcid.pdf}\hspace{1mm}Henning Wessels} \\
	Institute for Computational Modeling in Civil Engineering\\
	Technische Universität Braunschweig\\
	Pockelsstra{\ss}e 3 \\
	38106 Braunschweig, Germany\\
	\texttt{h.wessels@tu-braunschweig.de} \\
	\And
	\href{https://orcid.org/0000-0002-1365-9272}{\includegraphics[scale=0.06]{orcid.pdf}\hspace{1mm}Moritz Flaschel} \\
	Computational Mechanics Group \\
	Eidgen\"ossische Technische Hochschule Z\"urich\\
	Tannenstra{\ss}e 3 \\
	8092 Zürich, Switzerland\\
	\texttt{moritz.flaschel@wias-berlin.de} \\
	\And
	\href{https://orcid.org/0000-0003-2748-3287}{\includegraphics[scale=0.06]{orcid.pdf}\hspace{1mm}Laura De Lorenzis} \\
	Computational Mechanics Group \\
	Eidgen\"ossische Technische Hochschule Z\"urich\\
	Tannenstra{\ss}e 3 \\
	8092 Zürich, Switzerland\\
	\texttt{ldelorenzis@ethz.ch} \\	
}

\renewcommand{\headeright}{}
\renewcommand{\undertitle}{}
\renewcommand{\shorttitle}{Reduced and All-at-Once Approaches in Computational Solid Mechanics}

\hypersetup{
pdftitle={Reduced and All-at-Once Approaches for Model Calibration and Discovery in Computational Solid Mechanics},
pdfsubject={},
pdfauthor={Ulrich Römer, Stefan Hartmann, Jendrik-Alexander Tröger, David Anton, Henning Wessels, Moritz Flaschel, Laura de Lorenzis},
pdfkeywords={Parameter identification, all-at-once approach, virtual fields method, model discovery, parameter uncertainty},
urlcolor=black,
citecolor=blue
}

\begin{document}
\maketitle

\begin{abstract}
In the framework of solid mechanics, the task of deriving material parameters from experimental data has recently re-emerged with the progress in full-field measurement capabilities and the renewed advances of machine learning. In this context, new methods such as the virtual fields method and physics-informed neural networks  have been developed as alternatives to the already established least-squares and finite element-based approaches. Moreover, model discovery problems are starting to emerge and can also be addressed in a parameter estimation framework. 
These developments call for a new unified perspective, which is able to cover both traditional parameter estimation methods and novel approaches in which the state variables or the model structure itself are inferred as well. 
Adopting concepts discussed in the inverse problems community, we distinguish between all-at-once and reduced approaches. With this general framework, we are able to structure a large portion of the literature on parameter estimation in computational mechanics -- and we can identify combinations that have not yet been addressed, two of which are proposed in this paper. 
We also discuss statistical approaches to quantify the uncertainty related to the estimated parameters, and we propose a novel two-step procedure for identification of complex material models based on both frequentist and Bayesian principles. 
Finally, we illustrate and compare several of the aforementioned methods with mechanical benchmarks based on synthetic and real data.
\end{abstract}


%\brb{To-Do:
%\begin{itemize}
%\item Orcid for authors -- only David is missing
%\item Adaption of shorttitle in line 133 for header? $\to$ Done
%\item Adaption of figure size to one column format $\rightarrow$ Done
%\item Check text/equations $\rightarrow$ Done
%\item Check references: no URL (duplicate to DOI and not always present in .bib files)? $\rightarrow$ Done
%\item Remove unnecessary comments from template $\rightarrow$ Done
%\end{itemize}
%}

\section{Introduction}
Within the framework of continuum solid mechanics, initial boundary value problems involving partial differential equations (PDEs) are solved to determine the fields of interest, including mechanical, thermal, magnetic, and/or electric fields. To be solvable, the PDEs need to be completed by closure models known as material models (or constitutive equations) that describe the relationships between kinematic and kinetic quantities (and possibly additional variables), i.e., strain and stress, or temperature gradient and heat flux, etc., for the materials under consideration. 
Constitutive equations can be of algebraic,
ordinary differential (ODE), or differential-algebraic (DAE) form -- or they can even be defined by PDEs, whereby the space discretized form of the PDEs leads to similar mathematical structures as the aforementioned forms. Models of hyperelasticity lead to algebraic equations, whereas models of viscoelasticity or viscoplasticity imply ODEs. Rate-independent plasticity, which is based on a yield condition, delivers a DAE-system, and gradient-based damage models are described by PDEs.

Since constitutive models depend on various parameters, one of the fundamental issues in the theory of materials is the identification (or calibration) of these parameters on the basis of given experimental data.
A number of identification approaches have been proposed, among which those based on least-squares (LS) and the finite element method (FEM) have been especially prominent due to their high flexibility. Further methods have been proposed in the more recent past, including so-called full-field approaches such as the virtual fields method (VFM), stochastic and machine learning approaches, including those based on Physics-Informed Neural Networks (PINNs), for example. A detailed compilation of the related references is provided in the following sections.

Even more recently, approaches that bring identification one step further have been proposed and are drawing significant attention. They advocate a new paradigm, denoted as (material) model discovery, in which the structure  of the material model itself is estimated at the same time as its unknown parameters, again based on the given experimental data. A possible strategy consists of forming a library of candidate basis functions, which are used in a model to approximate unknown functions describing the material behavior. In this way, the model discovery problem is reformulated as a parameter estimation problem, but with a typically much higher dimension of the parameter space. It is also possible to apply VFM, stochastic methods, and machine learning to model discovery.

These developments call for a new unified perspective that is able to cover both traditional and novel parameter estimation and model discovery approaches.
The purpose of this paper is, on the one hand,
to compile an overview of the available methods along with a formalization and, on the other hand, to provide such a unified perspective. In contrast to existing attempts to unify parameter estimation methods in mechanics \cite{avriletal2008,roux2020optimal}, our perspective is based on the all-at-once approach \cite{kaltenbacher_regularization_2016}, which so far has been mainly applied in the inverse problems community.  
Adopting concepts discussed in this community, we distinguish between all-at-once and reduced approaches. The all-at-once approach employs a weighted combination of a model-based and a data-based objective function, and we demonstrate that both the reduced approach and the class of VFMs can be recovered as limit cases. With this general framework, we are able to structure a large portion of the literature on parameter estimation in computational mechanics, and we can identify combinations and settings that have not yet been addressed. Moreover, by including stochastic methods, the quality of the identified parameters can be addressed as well. Among the various available methods for parameter inference, we compare Bayesian and frequentist approaches and derive new results for two-step identification methods, which are needed to calibrate complex models. 

The remainder of this paper is structured as follows. In Section~\ref{sec:formulation_identification_problems}, we start with an overview of the basic equations of the initial boundary value problems, as well as the structure of the material models. We then discuss a few aspects of the experimental possibilities, followed by a brief review of parameter identification in elasticity, viscoelasticity, and viscoplasticity. 

Section~\ref{sec:compPI} treats the numerical approaches to identify the material parameters.
We formulate the LS method based on the FEM, the equilibrium gap, and the VFM, 
as well as more recent methods such as PINNs and the combination of model selection and parameter identification (i.e., model discovery). 

As stated earlier, one of the major aims of this paper is to formulate a unified treatment for most of the available parameter identification procedures, which is the subject of Section~\ref{sec:abstract}. Section~\ref{sec:locident} focuses on the quality of the identified material parameters -- which is addressed in a statistical setting, including the topics of identifiability and uncertainty quantification.
In Section~\ref{sec:examples}, examples are provided to compare the performance of the various schemes for selected applications.

%************************************
\section{Parameter identification in solid mechanics}
\label{sec:formulation_identification_problems}
This section addresses the basic problem of parameter identification. First, the fundamental equations of solid mechanics are summarized. 
Then, we provide an overview of the basics of experimental observations and constitutive modeling using elastic, isotropic or anisotropic, and inelastic material models. 
The notation in use is defined in the following manner: geometric vectors
are symbolized by $\V{a}$ and 
second-order tensors $\T{A}$ by bold-faced letters. Furthermore, we introduce matrices and column vectors by bold-faced italic letters $\GM{A},\GV{a}$.

\subsection{Fundamental equations}
\label{sec:fundamentaleq}
Since the procedures to identify parameters behave similarly for various PDEs, without loss of generality, we restrict ourselves to the balance of linear momentum
\begin{equation}
  \label{eq:linmomentum}
  \rho(\V{x},t) \, \V{a}(\V{x},t) = \operatorname{div} \T{\sigma}(\V{x},t) + \rho(\V{x},t) \, \V{b}(\V{x},t)
\end{equation}
occurring in the field of mechanics.
Here, $t$ stands for time, $\V{x} = \V{\chi}\INDR(\V{X},t)$ symbolizes the motion of a material point $\V{X}$ placed in the reference configuration, $\rho$ denotes the density, $\V{a}$ the acceleration, $\T{\sigma}(\V{x},t)$ the Cauchy stress tensor in spatial representation, $\V{b}$ is commonly chosen as the acceleration of gravity, and $\divop$ defines the divergence operator with respect to $\V{x}$. In the following, we assume that accelerations are small, leading to 
\begin{equation}
  \label{eq:linmomentumzero}
  \divop \T{\sigma} + \rho \, \V{b} = \V{0}.
\end{equation}
In this paper, we thus do not treat parameter estimation in dynamical systems, where properties of wave propagation in solid media are used for identification purposes, see \cite{mclaughlinyoon2004}, for example. 

An alternative formulation is given by
\begin{equation}
  \label{eq:linmomentum1stPK}
  \Divop \T{P}(\V{X},t) + \rho\INDR(\V{X},t) \, \V{b} = \V{0},
\end{equation}
where $\Divop$ defines the divergence operator with respect to
$\V{X}$. Moreover, $\rho\INDR = (\det
\T{F}) \rho$ is the density in the reference configuration, and $\T{P} = (\det \T{F}) \T{\sigma}
\TMT{F}$ symbolizes the first Piola-Kirchhoff stress tensor, where $\T{F} = \Grad
\V{\chi}\INDR(\V{X},t)$ represents the deformation gradient. Here, $\Grad$ denotes the gradient operator with respect to $\V{X}$. 

The balance of angular momentum implies the symmetry of the Cauchy stress tensor, $\T{\sigma} = \TT{\sigma}$. Thus, six independent stress components have to be determined, whereas only three scalar PDEs are available. Hence, the system of equations \eqref{eq:linmomentumzero}, or equivalently
\eqref{eq:linmomentum1stPK}, is closed by formulating 
\begin{enumerate}
    \item the kinematics, i.e.,\ a relation between the displacement vector $\V{u}(\V{X},t) =
\V{\chi}\INDR(\V{X},t) - \V{X}$ (or the motion $\V{\chi}\INDR(\V{X},t)$) and the deformation gradient $\T{F}$, and
    \item a constitutive
model describing the stress state in dependence of the deformation and possibly of additional, so-called internal variables describing the process history.
\end{enumerate}
Material objectivity requirements lead to some
restrictions on the relation between the deformation gradient $\T{F}$ and the
stress state $\T{\sigma}$, see \cite{truesdell,hauptbuch2} and the references therein for details. To comply with such restrictions, it is common to formulate the constitutive model  in terms of the second
Piola-Kirchhoff stress tensor $\T{S} = (\det \T{F}) \TM{F} \T{\sigma} \TMT{F}$,
expressed as a function of either the right Cauchy-Green tensor $\T{C} = \TT{F} \T{F}$ or the Green-Lagrange strain tensor $\T{E} = (\T{C} - \T{I})/2$. For purely elastic materials, $\T{S}$ obeys
\begin{equation}
  \label{eq:elasticlarge}
  \T{S} = \Tt{h}(\T{C})
\end{equation}
or, for the case of inelastic materials,
\begin{equation}
  \label{eq:inelasticlarge}
  \begin{split}
    \T{S} &= \Tt{h}(\T{C},\LV{q}) \\
    \LV{Y} \LVp{q} &= \LV{r}\INDq(\T{C},\TP{C},\LV{q}),
  \end{split}
\end{equation}
where $\LV{q} \elm{\numq}$ represents the vector of the scalar-, vector-, or tensor-valued internal variables. Eq.~\eqref{eq:inelasticlarge} embraces a very wide class of constitutive models, including models of viscoelasticity and viscoplasticity. Models incorporating dependencies on $\TP{C}$ (or $\TP{E}$) can be found, e.g.,\ in \cite{hauptsedlan,hartmannpom04}. In the case of rate-independent plasticity, the matrix $\LV{Y}$ has the structure
\begin{equation}
  \label{eq:Adefinition}
  \LV{Y} =
  \begin{bmatrix}
    \LV{I} & \\
           & 0
  \end{bmatrix}
\end{equation}
if the last equation defines the yield condition, see \cite{peter}. In this
case, Eq.~\eqref{eq:inelasticlarge} represents a DAE-system, see \cite{shibabuska1997,peter}. Moreover, in the field of yield-function-based, rate-independent modeling, the right-hand side
$\LV{r}\INDq(\T{C},\LV{q})$ contains case distinctions (loading-unloading or
Karush-Kuhn-Tucker conditions) to be able to reproduce a non-smooth material
behavior. 

Different approaches in constitutive modeling derive systems of the form \eqref{eq:inelasticlarge} from two scalar potentials, namely the strain energy density and the dissipation rate potential, so that (under certain conditions on the properties of such potentials) thermodynamic consistency is automatically fulfilled, see \cite{biot_thermoelasticity_1956, ziegler_thermodynamik_1957,rice_inelastic_1971,halphen_sur_1975}, for example. 
Alternative approaches evaluate only the strain energy density function within the Clausius-Duhem inequality and motivate the evolution equations \eqref{eq:inelasticlarge}$_2$ in a different manner \cite{hauptbuch2}. 

Of course, there are also other constitutive models that do not fit into the aforementioned structure -- such as models described by integral equations, fractional derivatives \cite{lion1997}, or the endochronic plasticity
formulation in \cite{Valanis1971}. Alternatively to the original formulation in
\cite{Valanis1971}, the stresses can also be given
by a rate equation using a particular kernel function of
the endochronic formulation, so that the integral formulation can be transformed
into a rate formulation \cite{hauswaldtdiss2020}.

Based on the general problem of solving Eqns.~\eqref{eq:linmomentum1stPK} with $\T{P} =
\T{F} \T{S}$ and \eqref{eq:inelasticlarge}, several special cases can be
considered. A simple example is the case of small strains, i.e.,\ the strain tensor is assumed to depend linearly on the displacement vector $\V{u}$, $\T{E} = (\grad \V{u}(\V{x},t) + \grad^T \V{u}(\V{x},t))/2$, there is no distinction between reference and current configurations, and all stress tensors become the same. In this paper, we are interested in determining the material parameters $\GKap \elm{\numkappa}$, and restricting ourselves to small strains is sufficient to understand the entire identification process. Thus, we consider the problem 
\begin{equation}
  \label{eq:smallproblem}
  \begin{split}
    \divop \T{\sigma} + \rho  \V{b} &= \V{0} \\
    \T{\sigma} &= \T{h}(\T{E},\LV{q},\GKap) \\
    \LVp{q} &= \LV{r}\INDq(\T{E},\LV{q},\GKap),
  \end{split}
\end{equation}
where, without a severe loss of generality, we assumed $\LM{Y} = \LM{I}$ and independence from $\TP{E}$. An example of such a model is provided in Appendix~\ref{ap:constModels}. Eq.~\eqref{eq:smallproblem}$_1$ has to be complemented with Neumann and Dirichlet boundary conditions 
\begin{equation}
  \label{eq:bcgeneral}
  \T{\sigma} \V{n} = \Vq{t}
  \quad \text{on} \quad \Gamma\INDN,
  \qquad
  \V{u} = \Vq{u}
  \quad \text{on} \quad \Gamma\INDD,
\end{equation}
and initial conditions
\begin{equation}
  \label{eq:ic}
  \V{u}(\V{x},0) = \Vq{u}_0(\V{x}),
  \qquad
  \LV{q}(\V{x},0) = \overline{\LV{q}}_0(\V{x}).
\end{equation}
$\Vq{t}$ is the given traction vector and $\Vq{u}$ the prescribed displacement, whereas $\Vq{u}_0(\V{x})$ and $\overline{\LV{q}}_0(\V{x})$ are the known initial conditions. 
Mixed boundary conditions, where $\V{t} = \Vh{t}(\V{u})$, are allowed as well.
Consistent initial conditions imply that Eq.~\eqref{eq:smallproblem}$_1$ is also fulfilled for the initial conditions. Formulation \eqref{eq:smallproblem} covers a large number of models.

In the following, we discuss special cases of material models and issues when determining material parameters from experimental data. We start with experimental possibilities and outline measurement techniques in Section~\ref{subsec:expobs}, while parameter identification for different classes of constitutive models is discussed in Section~\ref{subsec:constmodel}. 

\subsection{Experimental observations}\label{subsec:expobs}
The constitutive models in Eqns.~\eqref{eq:smallproblem}$_{2,3}$
contain unknown parameters $\GKap$. They have to be determined in such a manner
that they reflect experimental observations. Since the models contain stresses and strains, the question of how to deduce these quantities from experimental measurements is a crucial one.  The principal difficulty is to find experiments where the stress state under given (i.e.,\ controlled) or measured external loads is known. Thus, before addressing material parameter identification, we briefly discuss the possible options for mechanical testing and the related detectable experimental data. 

\begin{figure}[ht]
    \centering
    \begin{subfigure}[b]{0.25\linewidth}
        \centering
        \includegraphics[width=\textwidth]{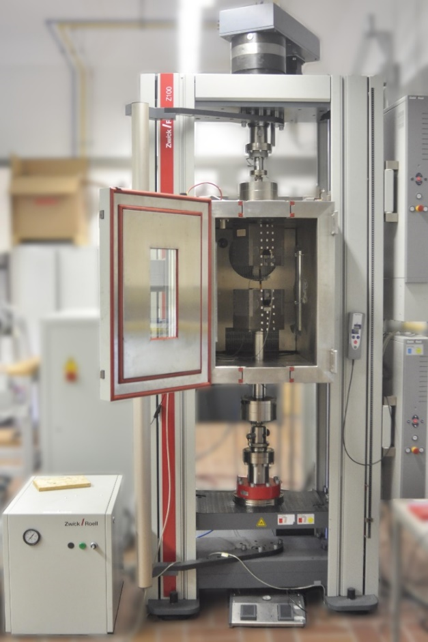}
        \caption{}
        \label{fig:tensileMachine}
    \end{subfigure}
    \hspace{0.1\linewidth}
    \begin{subfigure}[b]{0.25\linewidth}
        \centering
        \includegraphics[width=\textwidth]{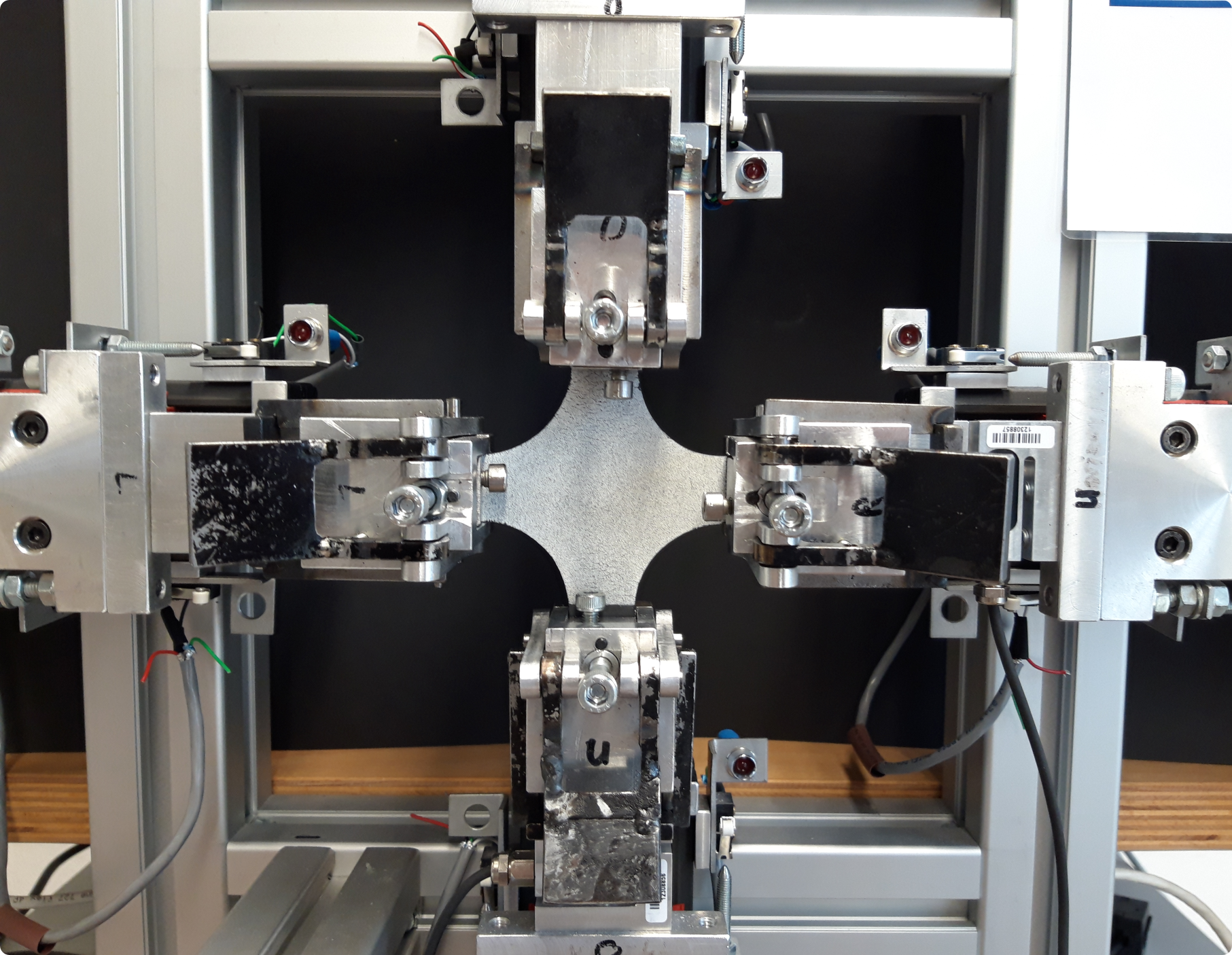}
        \caption{}
        \label{fig:biaxMachine}
    \end{subfigure}
    \caption{(a) Universal testing machine for tension, compression, and torsion tests, (b) Biaxial testing machine}
    \label{fig:expDevices}
\end{figure}

First, we consider experiments where the stress tensor
$\T{\sigma}$ is uniform in a certain area of the specimen, the simplest example being the
uniaxial tension/compression test. In this case, only Eqns.~\eqref{eq:smallproblem}$_{2,3}$ have to be solved, since Eq.~\eqref{eq:smallproblem}$_1$ is
automatically fulfilled (provided that $\rho \V{b}$ is negligible and the material properties are homogeneously distributed). In the data analysis, it is necessary to consider the specific structure of the tensorial quantities due to the assumed boundary conditions, such as the absence of stress components in the lateral direction. It should also be noted that, at any time, testing machines such as the ones depicted in Fig.~\ref{fig:expDevices}
can either control the displacement and measure the resulting force -- or vice versa. Accordingly, in order to obtain information on strains and stresses, it is necessary to know the relationship between the displacements and the strains as well as the forces and the stresses. For a uniaxial test, both relationships are known, but care must be taken to account for rigid body displacements (i.e., for the finite machine stiffness) as well as to perform a sufficiently accurate measurement of the cross-sectional area of the specimen. A further experimental choice could be to directly control the strain as a function of time rather than the displacement. This choice would require additional technical equipment and call for a synchronization of the strain measurement device with the testing machine. Of course, a temporal quasi-static process can be divided into two (or more) stages with different controls, e.g., with displacement control during loading and force control during unloading. 

Torsion tests on thin-walled tubes are an attractive alternative to tensile tests, since a uniform stress state over the wall thickness can be assumed if the wall thickness is small in comparison to the tube radius. During testing, either the torsion angle or the torsional moment is controlled. Once again, in order to compute strains and stresses, we need the relations between the torsion angle applied by the testing machine and the shear strain, as well as between the torsional moment and the shear stress. For the latter, the uniform stress assumption can be exploited.

Three-point or four-point bending tests are another simple test alternative. However, these experiments generate non-uniform stress fields, so that Eq.~\eqref{eq:smallproblem}$_1$ needs to be solved. One option is to directly solve Eq.~\eqref{eq:smallproblem}$_1$  as a local equilibrium equation (viewing the specimen as a three-dimensional solid), and another one is to recast it as a global equilibrium equation (adopting the approximations of beam theory). Respectively, the material parameters are calibrated based on full-field measurements or on discrete deflection information, see \cite{ehlerskujala2014,swainthomasselvanphilip2021}, for example.

For large deformation cases, there exist deformations -- called
universal deformations -- which fulfil the local balance of linear momentum under
the assumptions of isotropy and incompressibility, see
e.g.,\ \cite{ogdenbuch} and
\cite{hauptsedlan}. In all other cases, the entire problem \eqref{eq:smallproblem} has to be solved.

Apart from the experiments we just discussed, there are only very few other alternatives, such
as a specific type of shear test, see \cite{sguazzohartmann2018,yinetal2014} for an overview, or experiments on specific membranes, see \cite{mansouridarijanibaghani2017}, from which the stress state can be extracted, for example.
Even a biaxial tensile test, see Fig.~\ref{fig:biaxMachine}, does not provide a uniform stress state, see \cite{hartmanngilbertsguazzo2018} for a discussion on uniformity and a measure of deviation from it. In general cases, only the forces or torques of
the testing device are directly known, and assumptions on the stress state must be provided.

\begin{figure}
    \centering
    \includegraphics[width=0.375\linewidth]{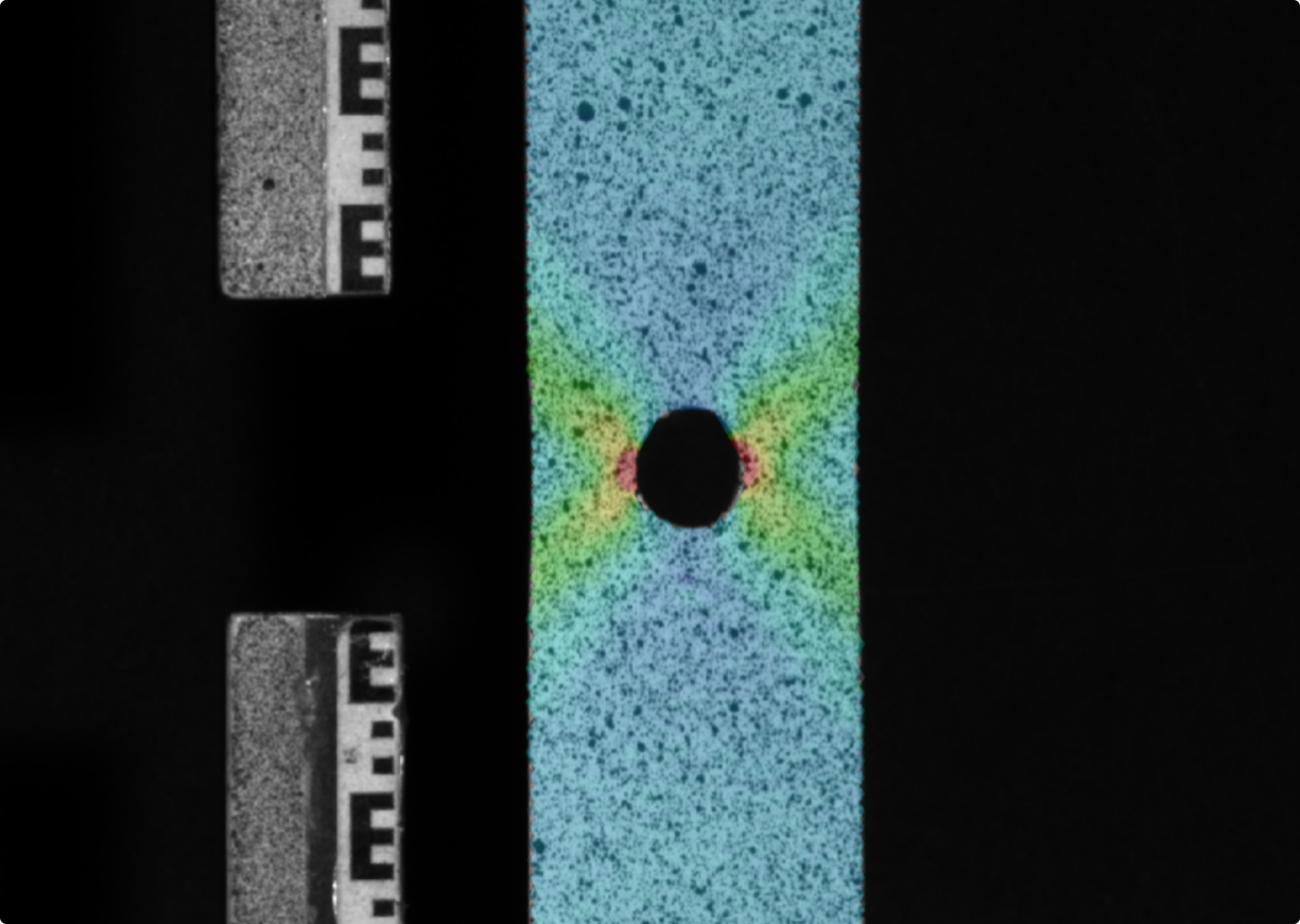}
    \caption{Full-field strain data on a plate with a hole}
    \label{fig:expAxStrain}
\end{figure}

On the other hand, if optical access to the sample surface and the ability to identify surface patterns are given, displacements  (and strains) can be determined locally on the surface using, e.g., digital image correlation (DIC, see \cite{suttonorteuschreier2009,pierron_towards_2020}), as shown in Fig.~\ref{fig:expAxStrain}. 
With this technique, out-of-plane strains cannot be obtained without introducing further assumptions.
In some situations, strain determination is only possible in an integral sense and in one direction, employing strain gauges, clip-on strain
transducers, or optical methods. Fiber-Bragg grating sensors, see
\cite{polym13050789,pereira2013,pereira2016}, are also used to
determine strains in one direction inside a component, although care must be
taken to ensure that these measurement systems do not influence the local
strain state. It is also possible to apply micro-computed tomography ($\mu$-CT) in combination with digital volume correlation (DVC) for volumetric determination of the
strains, but the recording times are currently so long that long-term changes in
the material, such as creep or relaxation effects, can only be ruled out by
estimation \cite{bay2008,boardmanmavrogordato2014}. Additional information can be obtained by the ``single-valued'' displacement or angle of the specimen holder.

In some situations, there is no optical access to the specimen, e.g.,\ in triaxial tension/compression testing machines \cite{CallochMarquis1999}, indentation tests, \cite{hubertsakmakis1999a,hubertsakmakis1999b,nakamurawangsampath2000}, or some metal-forming processes where the specimens are not visible in the forming press. In these cases, only a single displacement component and a single force component, both as functions of time, can be evaluated by interpolating discrete data. In some cases, of course, a combination of full-field information and single-valued data can be extracted and considered in the material parameter identification concept  \cite{polancoloriadaiyangrytten2012}. 

\subsection{Parameter identification for various classes of constitutive models}\label{subsec:constmodel}

The parameter identification procedure  depends on the particular constitutive model class at hand. Thus, the choice of the constitutive model class, which is based on the observed mechanical response of a specimen, e.g.,\ the presence of plastic deformations or viscous effects, directly influence on the approach to identify the sought parameters. In the following, we provide an overview of material parameter identification for a few important classes of constitutive models -- elasticity, viscoelasticity, elastoplasticity, and viscoplasticity.

\subsubsection{Linear elasticity}

For linear elasticity at small
strains, we distinguish between isotropic and anisotropic cases.

\paragraph{Isotropy}
\label{sec:isotropicelasticity}
In linear, isotropic elasticity  for
small strains, the stress state is defined by
\begin{align}
  \label{eq:linisoelast}
    \T{\sigma} &= \frac{E}{1+\nu}\left(\T{E}    + \frac{\nu}{1-2 \nu} (\tr \T{E})
    \T{I}\right) \\
    &= K (\tr \T{E}) \T{I} + 2 G \T{E}^D,
    \label{eq:linisoelastKG}
\end{align}
where $\tr \T{E} = \T{E} \cdot \T{I} = E_{kk}$ is the volumetric strain, $\T{E}^D = \T{E} - (\tr
\T{E})/3 \T{I}$ represents the deviatoric strain tensor, $E$ the Young's modulus or elasticity
modulus, $\nu$ the Poisson's ratio, $K = E/(3 (1-2 \nu))$ the compression
or bulk modulus, and $G = E/(2 (1+\nu))$ the shear modulus. The identification of this material model involves determining two independent material parameters -- such as $E$ and $\nu$ if using \eqref{eq:linisoelast}, or $K$ and $G$ if using \eqref{eq:linisoelastKG}.
Under the assumption of uniform strains
within a certain region and uniaxial tension or compression in $\V{e}_1$
direction, the strain and the stress tensors take a simplified structure, i.e., 
\begin{align}
\T{E} &= \varepsilon \V{e}_1 \otimes \V{e}_1 + \varepsilon_q (\V{e}_2 \otimes \V{e}_2 +
\V{e}_3 \otimes \V{e}_3) \, \text{and} \\ 
\T{\sigma} &= \sigma
\V{e}_1 \otimes \V{e}_1
\end{align}
so that the identification of tensorial relations reduces to that of scalar-valued functions. The axial stress is given by $\sigma := \sigma_{11} = \V{e}_1 \cdot \T{\sigma} \V{e}_1 = E
E_{11} = E \varepsilon$ (with $\varepsilon := E_{11} = \V{e}_1 \cdot
\T{E} \V{e}_1$). The lateral
strains $\varepsilon_q := E_{22} = E_{33} = \V{e}_2 \cdot \T{E} \V{e}_2$ are
given by $\varepsilon_q = -\nu \varepsilon$, i.e.,\ the
Poisson's ratio $\nu$ can be determined if the lateral strain $\varepsilon_q$ is
measured. Thus, we obtain a
decoupling of the identification of Young's modulus $E$ and Poisson's ratio
$\nu$, whereby $E$ can be determined from the stress-strain information in the axial direction and $\nu$ can be obtained by measuring the transverse strain. In both cases, furthermore, the material parameters can be determined from
linear expressions.

If one expresses the linear, isotropic elasticity relation in the form
\eqref{eq:linisoelastKG}, the material parameters $K$ and $G$ are
non-linearly included in the resulting stress-strain and lateral-axial strain relations 
\begin{equation}
\label{eq:linelastnonl}
    \sigma = \frac{9KG}{3K+G} \varepsilon,
    \qquad
    \varepsilon_q = -\frac{3K-2G}{6K+2G} \varepsilon.
\end{equation}
Then, the identification of the parameters has to be carried out, for example by a non-linear LS approach, see Subsection~\ref{sec:NLS}. Moreover, the lateral
strain data are of importance since the axial information alone is insufficient to
determine $K$ and $G$ uniquely.
A detailed discussion of the reduction of the general class of material models \eqref{eq:smallproblem}$_{2,3}$ from the 3D case to the uniaxial tensile case, which involves incorporating the 
boundary conditions into the strain and stress state, is provided in
\cite{kraemerrothehartmann2014}. 

Whether parameters, such as $K$ and $G$ in this case, can be clearly identified should be linked to a criterion. A first attempt to investigate whether material parameters can uniquely be identified is provided in \cite{hartmanngilbert2018}. There are situations where infinite combinations of parameters can yield a very accurate reproduction of the experiments, but this
raises the question of the physical meaning of the parameters. This is discussed
in the general case of optimization by \cite{beveridgeschechter70} and later on
by \cite{beckarnoldbook1977}, referred to as local identifiability. This
concept is transferred to the case of problems in solid mechanics by
\cite{hartmanngilbert2018} and a further discussion is provided by
\cite{sewerin2020}. A similar approach using the terminology of local stability
is discussed by \cite{mahnkenstein96,Vexler2004,mahnkenenzy2018}. In a more general context, a first attempt
to consider quality measures of the parameter identification in solid mechanics
can be found in \cite{KreissigBenedixGoerke2001}, where the correlation between the parameters is investigated. Here, the local identifiability concept, which can be extended to general models, is
discussed in Subsection~\ref{sec:identifiability}.

\paragraph{Anisotropy}
An extension of parameter identification for linear elastic materials to the case of anisotropy, and particularly of transverse isotropy and orthotropy (requiring five and nine material parameters, respectively, for a 3D model), is discussed in \cite{christensen2005}, unfortunately only at the theoretical level and with no supporting experiments. 
In fact, it is not possible to devise an experiment that covers one assumed boundary condition that is connected to a volumetric deformation. Thus, not all parameters can be uniquely inferred by the procedure in \cite{christensen2005}. This issue is addressed in \cite{hartmannidentTI2021} for transverse isotropy. It is shown by analytical considerations that there is one volumetric deformation process required -- apart from tensile and shear deformations -- to uniquely obtain all parameters. To implement this deformation process, a compression tool is developed \cite{hartmannidentTI2021}. However, the evaluation of the data shows that the measured lateral stresses are very sensitive to the geometric accuracy of the specimen and of the tool itself, and also to the friction conditions within the testing equipment. Consequently, the measurement results are not reliable and show large fluctuations. For the case of transverse isotropy with a specific preferred direction (perpendicular to the isotropy plane) in axisymmetric problems, the identification is discussed in \cite{talesnickleehaimson1995}. 

The case of orthotropy is addressed in
\cite{dileephartmann2022}, where $\mu$-CT data are evaluated to obtain the material parameters appearing in the analytical results obtained by assuming a homogenized material response. The difficulty is again to perform a reliable compression test to identify one of the parameters. For plane problems, the number of parameters can be reduced, see \cite{lecompteetal2007}, for example. Again, the main difficulty is the need for special compression tests and their evaluation, since these tests provide only a very small amount of measurement data, which are also uncertain. Even DIC, which delivers only the surface deformation, does not circumvent the problem of making all material parameters ``uniquely'' identifiable. The problem is partly solved by using representative volume elements with individual material properties of the constituents to numerically estimate the homogenized material behavior.  

\subsubsection{Hyperelasticity}

The identification of the material parameters in hyperelasticity (elasticity at large strains, whereby the existence of a strain energy density function is postulated) has to be discussed in more detail. First, hyperelasticity relations are chosen either as part of the models of viscoelasticity, rate-independent plasticity, and viscoplasticity, or they are chosen to model purely elastic material behavior. The latter is typically the case for rubber and of some biological materials, which are commonly modeled as weakly compressible. Thus, we discuss the case of isotropy and incompressibility first. 

Incompressibility is modelled either by an undetermined pressure, which is calculated by the geometrical constraint equation of no volume change \cite{truesdell,hauptbuch2}, or by introducing an arbitrarily large bulk modulus. In the latter case, the bulk modulus can be interpreted as a penalty factor that is defined by a large number -- and not
determined by parameter identification. This choice of the penalty factor, i.e.,\ the bulk modulus, is made by the user and can lead to lateral deformations deformations that do not precisely represent incompressibility.Further, the resulting lateral stretches might even become non-physical at very large strains \cite{ehlerseipper}. For a possible modification requiring appropriate forms of the strain-energy density function for the compressible part, see \cite{hartmannneff}. 
With the choice of incompressibility, the experimental data of the lateral stretch from a uniaxial tension/compression test are not necessary. 

In the most common hyperelastic material models -- apart from the micro-mechanically motivated approaches, e.g.,\ \cite{arrudaboyce} -- the strain energy density is expressed by a polynomial function of the first and second invariants of the right Cauchy-Green tensor \cite{rivlin,hartmannneff}, or by a polynomial in the eigenvalues of the right stretch tensor (eigenstretches) \cite{ogden}. The latter are called Ogden-type models. The main problem in identifying the parameters from uniaxial tensile tests, torsion tests, or combined tension-torsion tests is that some parameter values might lead to highly oscillatory tension-compression results or to highly sensitive results, which is discussed in
\cite{hartmannacta} for a number of sub-models of Rivlin-Saunders-type. These phenomena occur especially outside the range of the measurement data used for calibration.

One possible approach to solve this issue is to assume a priori that all polynomial coefficients are positive \cite{hartmannijss}. This leads to a linear LS identification problem with inequality constraints for uniaxial tension and torsion (universal
deformations). The inequality constraints (enforcing positivity of the parameters) can be connected to stability requirements \cite{baker}, and they raise the general question of the existence of a solution in relation to the constraints on the material parameters. This is discussed in \cite{hartmannneff} for a new class of polynomial models in the framework of polyconvex strain energy functions. Here, the assumption of non-negative polynomial coefficients (material parameters) is even necessary.

Further models permit arbitrary signs for the parameters, see \cite{yeoh} or \cite{kao}. So far, it has not been proven that negative material parameters exclude a physical behavior in all deformation stages. Thus, special attention must be given when applying such models.

As mentioned earlier, in Ogden-type models, the strain energy density function depends non-linearly on the eigenstretches. At first consideration, this has advantages for simple tests where the loading directions coincide with the eigendirections of the stretch tensor. Here, however, the material parameters occur non-linearly in the stress-strain relation, necessitating the solving of a non-linear optimization problem even for simple cases \cite{OgdenSaccomandiSgura2004,twizell,benjeddou}. It is hardly possible to guarantee the sensitivity of the result in material parameter identification (due to the choice of the initial guess of the parameters in an iterative approach) or, consequently, the uniqueness of the estimated parameters.

For a recent overview of identification in hyperelasticity, see
\cite{mahnken2022}, where appropriate weighting functions are chosen to identify the parameters. A comparison of hyperelasticity models can be found in \cite{SeibertSchoeche2000,marckmannverron2006,rickerwriggers2023}. An overview of analytical expressions for determining the material parameters for specific isotropic hyperelasticity models is given in \cite{mihaigoriely2017}. 

The treatment of material parameter identification for large strains in connection with anisotropic materials is similar to the investigations for anisotropy at small strains. The applications here are primarily designed for biological tissues, as these have preferred directions due to the collagen fibers present.
In \cite{gaolicaiberryluo2015}, the anisotropic material behavior of a soft issue is experimentally investigated and calibrated using a finite strain, anisotropic hyperelasticity relation. Even the quality of the parameters is studied, which is very difficult to be guaranteed in layered tissues as it is the case of arteries, see \cite{gilberthartmann2016}. \cite{shariff2022} considers experiments on various tissues and calibrates the material parameters based on data given in the literature. Unfortunately, the quality of the parameters is not studied. In \cite{schroederneffbalzani05}, a polyconvex model for transverse isotropy is calibrated based on virtual experiments generated with a different constitutive model. In this respect, we refer to \cite{avrilevans2017} as well. 
Alternative applications of anisotropic hyperelasticity are woven fabrics, see \cite{makhoolbalzani2024}. There, however, the simpler case of a plane stress problem is considered which essentially reduces the number of unknown parameters.

\subsubsection{Viscoelasticity}

In the case of viscoelasticity, we have to treat either integral equations, ODEs, or DAEs to consider fading memory properties in the material. For this type of model, after infinitely long holding times of the applied load, the stress state coincides with the equilibrium stress state (which has no hysteresis). One common model structure goes back to overstress-type models, i.e.,\ the stress state is decomposed into an equilibrium stress state, formulated by an elasticity relation, and an overstress part. This model type is sometimes also called hyper-viscoelasticity. The overstress part can be formulated as a sum of Maxwell models, i.e.\ ODEs of first order. Each Maxwell element contains two material parameters. Since all the parameters are highly correlated, a suitable procedure must be developed to determine the parameters in a reproducible way. Another approach takes a fractional derivative model as a surrogate model to determine the individual parameters for given relaxation spectra \cite{HauptLionBackhaus2000}. Unfortunately, the consistent reduction from the
three-dimensional modeling to the uniaxial tension is not addressed. An alternative scheme is proposed in \cite{leistnerdiss2022}, where inequality constraints are used to successively determine the parameters. For a more recent investigation and literature survey, see \cite{jalochaconstantinescuNeviere2015}. 

Since the model has a modular structure, the parameters related to the equilibrium stress and to the
overstress can be determined successively, i.e.,\ one arrives at the
identification of the parameters of an elasticity relation (discussed in the previous subsections) and of the parameters of an ODE system. In \cite{hauptsedlan}, the successive
identification is addressed under the assumption of a universal deformation
(tension-torsion). However, no quality measures for the identification are
discussed. Even the specification of the identification procedure is
missing. Because of problems in identification, singular value decompositions are employed in \cite{gerlachmatzenmiller2007}. 

Two questions arise here: First, how can the boundary conditions for homogeneous
deformations be included in three-dimensional material models
\eqref{eq:smallproblem}$_{2,3}$? Second, which methods are available to
determine the parameters in ODEs and DAEs in the context of LS
approaches? The first issue is treated in
\cite{kraemerrothehartmann2014}, where a procedure for arbitrary models is
proposed. The aspect of determining the parameters in transient problems is summarized in \cite{schittkowskibook2002}, where three
types of schemes are proposed, namely \textit{simultaneous simulation of
  sensitivities} (\cite{dunker1984,leiskramer1985,LeisKramer1988}), \textit{internal numerical differentiation} (\cite{bock1983}), and
\textit{external numerical differentiation} -- see \cite{hartmann2017} for an application in non-linear FEM. For the case of viscoelasticity
using full-field measurements, which is discussed in Section~\ref{sec:finiteelements}, the reader is referred to
\cite{KleuterMenzelSteinmann2007, hartmanngilbert2021}.

\subsubsection{Elastoplasticity and viscoplasticity}
There is a large amount of articles on parameter
identification in the context of plasticity. However, the quality of the estimated parameters is not frequently addressed. Among the first papers to investigate the correlation between parameters are
\cite{Ekh2001,JohanssonJohanEkh2006}. Models based on yield functions contain 
elastic parameters, whose identification is discussed in
Subsection~\ref{sec:isotropicelasticity}. The determination of the yield stress is much more difficult for most metals. Either there are L\"uders bands resulting in a spatial motion of dislocations, so that the resulting stress-strain response is not deterministically predictable, or the yield point is not very pronounced. Thus, the first yield stress is a very uncertain value. In the plastic region, a number of material parameters can describe the non-linear hardening behavior (isotropic and kinematic hardening, for example), and these parameters can be strongly correlated. A linear correlation between the driving  and the saturation term of the Armstrong \& Frederick
kinematic hardening model, see \cite{armstrongfrederick}, can be concluded by
analytical considerations on this model
\cite{kraemerrothehartmann2014,dileephartmann2022}. For pressure-dependent yield functions in soil mechanics, the identification using a triaxial compression testing device is discussed in \cite{fossumsensenypfeiflemellegard1995}. The quality of the estimated parameters is also examined. The contribution \cite{mahnkenjohanssonrunesson1998} discusses the identification of material parameters for a complex viscoplasticity model using a gradient-based method. A detailed explanation of the calculation of the sensitivities is provided, which fits into the \textit{internal numerical differentiation} procedure \cite{bock1983}.

The more complex the models are, the more difficult it is to determine the parameters from simple experiments. Therefore, a secondary field of material parameter
identification can serve as a means of model improvement and possibly model reduction (model identification). However, this field is often not documented, as only the final result of a model is published. Thus, when devising new material models, attention must be paid to their identifiability from experiments already during the development phase. This is discussed in detail in \cite{dileephartmann2022}. The authors of \cite{zhangvanbaelandradecamposcoppieters2022} discuss the practical identifiability of the parameters for a more complex yield function, whereas \cite{furukawayagaw1997} focuses on a non-linear kinematic hardening model that is calibrated from uniaxial tensile tests on the basis of one-dimensional considerations by applying an evolutionary algorithm. A further discussion on parameter identification using stress data from tensile experiments is provided by \cite{shutovkreissig2010}. This is extended by investigations on noisy data by \cite{shutovKaygorodtseva2019}. A review regarding the calibration of various plasticity models is also provided by \cite{rossietal2022review}. 

%************************************
\section{Computational approaches for parameter identification}
\label{sec:compPI}
As mentioned earlier, only very few experiments lead to uniform stress and strain states. If non-uniform stress/strain distributions occur within a specimen, problem \eqref{eq:smallproblem} -- consisting of (a) the balance equations (PDEs), (b) the constitutive equations, and (c) the kinematic relations -- has to be evaluated to obtain the parameters. Since the resulting systems of equations are similar for uniform deformations and spatially discretized non-uniform deformation problems, we will now discuss the latter, more challenging case. Also, the emphasis here is on the identification procedure and not on the numerical optimization schemes.

This section is structured as follows. 
First, details on the space and time discretization using finite elements are given, together with a detailed description of the three representative problems of linear elasticity for small strains, hyperelasticity, and inelasticity (these problems are later discussed from the perspective of identification). Then, a broad classification of the different identification methods is provided in Section~\ref{sec:classification}, while an in-depth comparative study is deferred to Section~\ref{sec:abstract}. 
We start our review of identification methods with the non-linear LS method using the FEM (Section~\ref{sec:nlsfemdic}) and continue with the equilibrium gap method and the VFM (Section~\ref{secmf:virtual_fields_method}). Surrogate models and PINNs are the topic of Section~\ref{sec:surrogates}. Finally, an overview of model discovery and Bayesian approaches is provided in Sections \ref{sec:discovery} and \ref{sec:bayesian}, respectively. 

\subsection{Finite element method}
\label{sec:finiteelements}
Very often, the FEM is formulated in terms of nodal displacements. In experiments, however, there is usually also a force, either measured or controlled, that can be formalized using the method of Lagrange multipliers, 
see 
\cite{hartmannquinthamkar08}.  In the following presentation, we adopt the method of vertical lines, i.e.,\ we carry out the spatial discretization first, by introducing  shape functions for real and virtual displacements, resulting in the system of DAE
\begin{equation}
  \GV{F}(t,\GV{y}(t),\GVP{y}(t))
  = \GV{0},
  \label{eq:DAELagrMult}
\end{equation}
where
\begin{equation}
    \label{eq:DAEdef}
    \GV{F}(t,\GV{y}(t),\GVP{y}(t))
  :=
      \left\{
  \begin{matrix}
    \GV{g}\INDa(t,\Ua(t),\GV{q}(t)) - \GM{M} \GV{p}(t) \\
         \GV{C}\INDc(t,\Ua(t)) \\
         \GVP{q}(t) - \GV{r}\INDq(\Ua(t),\GV{q}(t))
  \end{matrix}
  \right\}
\end{equation}
with the discretized weak form $\GV{g}\INDa$, the geometric constraints $\GV{C}\INDc$, and the evolution equations for the internal variables, %$\GVP{q}(t)$ $\GV{r}\INDq(\Ua(t),\GV{q}(t))$,
and where 
\begin{equation}
    \label{eq:state}
    \GVT{y}(t) = \{\Ua^T(t),\GVT{p}(t),\GVT{q}(t)\}, \quad \GV{y} \elm{\numu+2\nump+\numQ}.
\end{equation}
Here, $\Ua^T = \{\U^T,\Uh^T\}\elm{\numu+\nump}$ represents the vector of all nodal displacements, which can be decomposed into those with prescribed values, $\Uh\elm{\nump}$ (we initially assume that these are unknown as well), and those which are unknown, $\U\elm{\numu}$. The size $n_Q$ of $\GV{q}$, which now (with a slight abuse of notation) represents the vector of the internal variables in the spatially discrete setting, will be defined shortly later on (and we have $n_q$ of these variables in the continuum setting, see Section \ref{sec:fundamentaleq}). Moreover, 
\begin{equation}
  \label{eq:constraint}
  \GV{C}\INDc(t,\Ua) =  \Uh - \Uq(t) = \GMT{M} \Ua - \Uq(t) = \GV{0},
\end{equation}
with
\begin{equation}
    \label{eq:mdef}
    \GM{M} = 
    \begin{bmatrix}
    \GM{0}_{{\numu}  \times{\nump}}
    \\ 
    \GM{I}_{\nump}
    \end{bmatrix},
\end{equation}
represents the constraint equation, i.e.,\ the prescribed displacements $\Uq(t) \elm{\nump}$ should be identical to $\Uh$.  The incidence matrix $\GM{M} \elmm{(\numu+\nump)}{\nump}$ extracts the concerned displacements from the vector $\Ua$. The Lagrange multipliers are denoted with $\GV{p} \elm{\nump}$ and can be interpreted as the nodal reaction forces. The last equation in Eq.~\eqref{eq:DAEdef} results from an assembly procedure of all internal variables $\GV{q}\elm{\numQ}$ 
\begin{equation}
  \label{eq:qassembling}
  \GV{q}(t) = \sum_{e=1}^\nel \sum_{j=1}^\numgu {\GM{Z}\INDq^{\,e,j}}^T \LV{q}^{e,j}(t),
\end{equation}
or 
\begin{equation}
    \label{eq:qextract}
    \LV{q}^{e,j}(t) = \GM{Z}\INDq^{\,e,j} \GV{q}(t),
\end{equation}
where the matrices $\GM{Z}\INDq^{\,e,j} \elmm{\numq}{\numQ}$ represent data management matrices containing only zeros and ones (Boolean matrices). Also, $\numgu$ is the number of integration points (commonly Gauss points) within element $e$,  and $\numQ =
\big( \sum_{e=1}^{\nel} \numgu \big) \times \numq$ defines the total number of internal variables of a mesh with $\nel$ elements.

The vector $\GV{g}\INDa \elm{\numu+\nump}$ contains all equations
resulting from the spatially discretized weak formulation of the balance of linear momentum
\begin{equation}
    \GVh{g}\INDa(t,\U\INDa,\GV{p},\Q) 
    =
    \mzweiv{\GV{g}(\GV{y})}
             {\GVq{g}(\GV{y}) - \GV{p}(t)}
   =
   \mzweiv{\DS \sum_{e=1}^{\nel} {\GM{Z}^{\,e}}^T
   \LV{p}_{\textrm{int}}^e
   - \GVq{p}(t)
   }
   {\DS \sum_{e=1}^{\nel} {\GMq{Z}^{\,e}}^T
   \LV{p}_{\textrm{int}}^e
   - \GV{p}(t)
   },
\label{eq:alleqlm}
\end{equation}
with
\begin{equation}
    \label{eq:pint}
    \LV{p}_{\textrm{int}}^e
    :=
    \sum_{j=1}^\numgu
    w^{e,j}
    {\LM{B}^{e,j}}^T \;
    \LV{h} \left(
              \LV{E}^{e,j},
              \LV{q}^{e,j}
           \right)
    \det \LM{J}^{e,j}
\end{equation}
which represent the nodal internal forces, whereas $\GVq{p}(t) \elm{\numu}$ defines the given equivalent nodal force
vector. Here, $\GM{Z}^{\,e} \elmm{\numue}{\numu}$ and
$\GMq{Z}^{\,e} \elmm{\numue}{\nump}$ symbolize incidence matrices assembling all element contributions into a large system of equations (representing the assembly procedure). Moreover, $\LV{E}^{e,j} \elm{6}$ denotes the column vector representation of the element strain tensor (Voigt notation), which depends linearly on the element nodal
displacement vector $\LV{u}^{e} \elm{\numue}$, $\LV{E}^{e,j} =
\LM{B}^{e,j} \LV{u}^{e}$, with $\LV{u}^{e} = \GM{Z}^{\, e} \U + \GMq{Z}^{\; e} \Uh$. The number of element nodal displacement degrees of freedom is denoted with $\numue$, and $w^{e,j}$ denote the weighting factors of the spatial integration in an element. Furthermore, $\LM{B}^{e,j} \elmm{6}{\numue}$ is
the  strain-displacement matrix of element $e$ evaluated
at the $j$-th Gauss point, $j=1,\ldots,\numgu$. 
$\LM{J}^{e,j} \elmm{3}{3}$ symbolizes the Jacobian matrix of the coordinate
transformation between reference element coordinates and global coordinates. The symmetric stress tensor \eqref{eq:smallproblem}$_2$ is recast into vector
$\LV{T} = \LV{h}(\LV{E}^{e,j},\LV{q}^{e,j})\elm{6}$, which is
evaluated at the $j^{th}$ Gauss point. It depends, via the strain vector, on the displacements $\Ua$ and the internal variables $\Q$.

If the constraint \eqref{eq:constraint} in the DAE-system \eqref{eq:DAELagrMult}, with definition \eqref{eq:DAEdef}, is assumed to hold exactly, the reduced DAE-system
\begin{equation}
    \label{eq:DAEreduced}
    \GV{F}(t,\GVP{y}(t),\GV{y}(t))
    =
    \mzweiv{\GVh{g}\INDa(t,\U(t),\GV{p}(t),\GV{q}(t))}
    {\GVP{q}(t) - \GV{r}\INDq(t,\U(t),\GV{q}(t))}
    = \GV{0}
\end{equation}
results. 

The DAE-system \eqref{eq:DAELagrMult} is solved using the initial conditions 
\begin{equation}
  \label{eq:initcond}
  \GV{y}(t_0) 
  := 
  \left\{\begin{matrix}\Ua(t_0) \\ \GV{p}(t_0) \\\GV{q}(t_0) \end{matrix}\right\}
  = 
  \left\{\begin{matrix}{\Ua}_0 \\ \GV{p}_0 \\ \GV{q}_0 \end{matrix}\right\} 
  =: 
  \GV{y}_0.
\end{equation}
For this purpose, time discretization is needed. For example, the application of the backward Euler method yields
\begin{equation}
  \label{eq:BEforce}
  \begin{aligned}
    \GV{g}\big(\tnp,\Unp,\Uh\INDNP,\Qnp\big) &= \GV{0}, \\
    \GVq{g}\big(\Unp,\Uh\INDNP,\Qnp\big) - \GV{p}\INDNP
    &= \GV{0}, \\
    \GV{C}_c\big(\tnp,\Uh\INDNP\big) &= \GV{0}, \\
    \GV{l}\big(\Unp,\Uh\INDNP,\Qnp\big) &= \GV{0},
  \end{aligned}
\end{equation}
with
\begin{equation}
  \label{eq:ldef}
  \GV{l}\big(\Unp,\Uh\INDNP,\Qnp\big) = 
  \Qnp -
  \Qn - \dtn \GV{r}\INDq\big(\Unp,\Uh\INDNP,\Qnp\big).
\end{equation}
Here, $\tnp = \tn + \dtn$, $n=1,\ldots,N_t-1$, where $N_t$ is the number of load (time) steps.
For details on solving DAE-systems, see \cite{hairerII}. 
Independently of the non-linear solver, $\Uh \approx \Uq$ is obtained after one successful iteration and the non-linear system
\begin{equation}
  \label{eq:BEforceII}
  \begin{aligned}
    \GV{g}\big(\tnp,\Unp,\Qnp\big) &= \GV{0}, \\
    \GVq{g}\big(\tnp,\Unp,\Qnp\big) - \GV{p}\INDNP
    &= \GV{0}, \\
    \GV{l}\big(\tnp,\Unp,\Qnp\big) &= \GV{0}
  \end{aligned}
\end{equation}
has to be solved at each time step. Since Eq.~\eqref{eq:BEforceII}$_2$ is an explicit expression of the nodal reaction force vector, the non-linear system
\begin{equation}
  \label{eq:BEforceIII}
  \begin{aligned}
    \GV{g}\big(\tnp,\Unp,\Qnp\big) &= \GV{0}, \\
    \GV{l}\big(\tnp,\Unp,\Qnp\big) &= \GV{0}
  \end{aligned}
\end{equation}
is commonly solved, whereas Eq.~\eqref{eq:BEforceII}$_2$ is used to compute the nodal reaction forces during post-processing. Although it is common to state that a Newton-Raphson method is chosen to solve problem \eqref{eq:BEforceIII}, it was shown in \cite{peter,hartmannCM05} that the Multilevel Newton algorithm proposed in \cite{rabbat} is applied (if an iterative stress algorithm is chosen at the local level -- i.e.,\ at each Gauss point). This method leads to the classical structure of local iterations, often called stress algorithm (although it is the internal variable computation), and to global iterations, where the increments of the displacement vector are obtained (on the basis of the consistent linearization stemming from the usage of the implicit function theorem). 

Let us consider some special cases. In hyperelasticity (absence of Eq.~\eqref{eq:DAEdef}$_3$), simply setting $\dot{t} = 1$ 
once again yields a DAE-system, so that the same procedure can be applied. Thus, we obtain
\begin{equation}
  \label{eq:nonlelast}
  \begin{aligned}
    \GV{g}(\tnp,\GV{u}\INDNP) &= \GV{0}, \\
    \GV{p}\INDNP &= \GVq{g}(\tnp,\GV{u}\INDNP).
  \end{aligned}
\end{equation}
Frequently, the step-wise increase of the load is chosen to be close to the solution of the Newton-Raphson method applied to Eq.~\eqref{eq:nonlelast}$_1$, see \cite{hartmanngilbert2021} for details.

For the case of linear elasticity, $\LV{\sigma} = \LV{h}(\LV{E}) = \LM{C} \LV{E}$, with $\LM{C} \elmm{6}{6}$ as the elasticity matrix, the functions in equation \eqref{eq:nonlelast} read
\begin{align}
     \GV{g}(\U) 
     &=
     \GM{K} \U + \GMq{K} \Uq - \GVq{p}, 
     \\
    \GVq{g}(\U) 
    &= 
    \GMq{K}^{\,T} \U + \GMqq{K} \Uq,
\end{align}
i.e., we have
\begin{align}
   \GM{K} \U &= \GVq{p} - \GMq{K} \Uq, 
   \label{eq:linelastKu}\\
   \GV{p}
   &=
     \GMq{K}^{\,T} \U + \GMqq{K} \Uq
     \label{eq:linelastp}
 \end{align}
or the classical representation
\begin{equation}\label{eq:system_hartmann}
\begin{bmatrix}
    \GM{K} & \GMq{K} \\
    \GMq{K}^T & \GMqq{K}
\end{bmatrix}
\mzweiv{\GV{u}}{\GVq{u}}
=
\mzweiv{\GVq{p}}{\GV{p}}.
\end{equation}
Here, the explicit time (or load) dependence is omitted since proportionality is given. The stiffness matrices are defined by 
\begin{equation}
\label{eq:Kdefinitions}
  \GM{K}
  =
    \sum_{e=1}^\nel {\GM{Z}^{\,e}}^T
    \LM{k}^e
    \GM{Z}^{\,e}, \qquad
  \GMq{K}
  =
    \sum_{e=1}^\nel {\GMq{Z}^{\,e}}^T
    \LM{k}^e
    \GM{Z}^{\,e}, \qquad
  \GMqq{K}
  =
    \sum_{e=1}^\nel {\GMq{Z}^{\,e}}^T
    \LM{k}^e
    \GMq{Z}^{\,e}
\end{equation}
with the element stiffness matrix
\begin{equation}
    \label{eq:elementstiff}
    \LM{k}^e =
    \sum_{j=1}^\numgu w^{e,j} {\LM{B}^{e,j}}^T
    \LM{C}^{e,j} \LM{B}^{e,j} \det \LM{J}^{e,j}
\end{equation}
and the elasticity matrix $\LM{C}^{e,j} \elmm{6}{6}$. 

In the following, we formally restructure the equations discretized above with the aim of material parameter identification. For this purpose, we redefine $\GV{y} \elm{\nums}$ as the state vector and introduce the vector of material parameters $\GKap \elm{\numkappa}$. The structure of the state vector depends on the problem under study.

\paragraph{Problem class I: linear elasticity}
In the framework of linear elasticity, the dependence on the material parameters $\GKap$ appears in the stiffness matrices, and Eqns.~\eqref{eq:linelastKu} and \eqref{eq:linelastp} lead us to the system
\begin{equation}
    \label{eq:Felastic}
    \GV{F}(\GV{y},\GKap) 
    =
    \GV{0}
\end{equation}
with
\begin{equation}
\label{eq:Felasticdef}
    \GV{F}(\GV{y},\GKap) 
    :=
    \GM{A}\INDup(\GKap) \GV{y} - \GV{f}(\GKap)
\end{equation}
and
\begin{equation}
    \underbrace{
    \begin{bmatrix}
        \GM{K}(\GKap) & \GM{0} \\
        -\GMq{K}^T(\GKap) & \GM{I}
    \end{bmatrix}
    }_{\GM{A}\INDup(\GKap)}
    \underbrace{
    \mzweiv{\GV{u}}{\GV{p}}
    }_{\GV{y}}
    =
    \underbrace{
    \mzweiv{\GVq{p} - \GMq{K}(\GKap) \GVq{u}}{\GMqq{K}(\GKap) \GVq{u}}
    }_{\GV{f}(\GKap)}.
    \label{eq:linelastsystem}
\end{equation}
In the forward problem, $\GKap$ is given, i.e.,\ $\GKap = \bar{\GKap}$ and \ $\GV{F}(\GV{y},\bar{\GKap})=\GV{0}$ has to hold. However, in parameter identification, $\GKap$ has to be determined. Usually, no time dependence is assumed within the framework of statics.  
To indicate that different load values are used for parameter identification (and going back to the time-continuous setting for notational simplicity), Eq.~\eqref{eq:Felastic} should actually be $\GV{F}(t,\GV{y}(t),\GKap)=\GV{0}$. While $\GM{A}\INDup(\GKap)$ in Eq.~\eqref{eq:linelastsystem} is time-independent, since the stiffness matrices do not depend on time, $\GV{f}(t,\GKap)$ is time-dependent since it contains the time-dependent prescribed displacements $\Uq(t)$ and equivalent nodal force vector $\GVq{p}(t)$. 

Different specific problem settings can be defined. In a common case, one component of a resultant reaction force $\check{p}$ is available through the load cell of the testing machine. Temporarily ignoring the time dependence, the resulting equation reads 
\begin{equation}\label{eq:force_identification}
    \check{p}(\U,\Uq,\GKap) 
    = \GVT{m} \GV{p} 
    = \GVT{m} \GMqT{K}(\GKap) \U + \GVT{m} \GMqq{K}(\GKap) \Uq,
\end{equation}
where the vector $\GV{m} \elm{\nump}$ is used to sum the relevant components of the nodal reaction force vector $\GV{p}$. The dependence of the stiffness matrices on the parameters may be non-linear. In the case of a linear dependence, the following relations can be obtained (see Appendix~\ref{ap:system_matrices}) 
\begin{equation}
    \label{eq:FEMlindepGkap}
    \begin{split}
        \GM{A}\INDS(\U,\Uq) \GKap &= \GVq{p}, \\
        \GMq{A}\INDS(\U,\Uq) \GKap &= \GV{p}, \\
        \GVT{m} \GMq{A}\INDS(\U,\Uq) \GKap &= \check{p}.
    \end{split}
\end{equation}
with the third equation stemming from \eqref{eq:force_identification}.
Defining 
\begin{equation}
    \label{eq:Aas}
    \GM{A}\INDas := 
    \begin{bmatrix}
        \GM{A}\INDs \\
        \GMq{A}\INDs
    \end{bmatrix}
    \text{ and }
    \GV{p}\INDa
    :=
    \left\{
        \begin{matrix}
        \GVq{p}\\
        \GV{p}
        \end{matrix}
    \right\}
\end{equation}
Eq.~\eqref{eq:FEMlindepGkap}$_{1,2}$ can be abbreviated by
\begin{equation}
    \label{eq:Aaskappa}
    \GM{A}\INDas(\U,\Uq) \GKap = \GV{p}\INDa.
\end{equation}
Accounting for the time dependence implies $\GM{A}\INDas(t,\U(t))$ and $\GV{p}\INDa(t,\GV{p}(t))$ in Eq.~\eqref{eq:Aaskappa},
\begin{equation}
    \label{eq:AaskapP}
    \GM{A}\INDas(t,\U) \GKap = \GV{p}\INDa(t,\GV{p}),
\end{equation}
i.e.,
\begin{equation}
    \label{eq:FdefVFM}
    \GV{F}(t,\U,\GV{p},\GKap) 
    := \GM{A}\INDas(t,\U) \GKap - \GV{p}\INDa(t,\GV{p})
    = \GV{0}.
\end{equation}
In the time-discrete setting, Eq.~\eqref{eq:FdefVFM} has to be solved for each load  or time step,  yielding different parameters $\GKap$. With this approach, further considerations have to be made to finally determine a unique set of parameters (e.g.,\ by averaging, which is the simplest possibility).

\paragraph{Problem class II: hyperelasticity}
If the underlying constitutive model is hyperelastic, the problem
\begin{equation}
    \label{eq:basicproblem}
    \GV{F}(t,\GV{y}(t),\GKap) = \GV{0},
\end{equation}
with
\begin{equation}
    \label{eq:hypernonl}
    \GV{F}(t,\GV{y}(t),\GKap) =
    \mzweiv{\GV{g}(t,\U(t),\GKap)}
           {\GVq{g}(t,\U(t),\GKap) - \GV{p}(t)} 
\end{equation}
has to be solved iteratively, see also Eq.~\eqref{eq:nonlelast} in the time-discrete setting. Consideration of the time dependence has two reasons: first, to incorporate the non-linear behavior and to consider it in the identification process; and second, for very non-linear problems, an iterative procedure applied directly to the final load value may fail, necessitating an incremental procedure where each incremental displacement state is chosen as the starting estimate for the solution at the next load step. Once again, a reaction force component $\check{p}$ could be simply identified in certain parameter identification scenarios,
\begin{equation}
    \check{p}(t) = \GVT{m} \GV{p}(t) = \GVT{m} \GVq{g}(t,\U(t),\GKap).
\end{equation}
There are hyperelastic material models which are linear in the material parameters, such as the class of Rivlin-Saunders or Hartmann-Neff models, see \cite{rivlin,hartmannneff}. This leads to the same representation as in Eq.~\eqref{eq:FEMlindepGkap}. Unfortunately, this is not the case for Ogden-type models \cite{ogden,ogdenbuch}, where the fully general case \eqref{eq:basicproblem} has to be considered.

\paragraph{Problem class III: inelasticity}
For inelastic problems, where evolution equations describe the plastic, hardening, and/or viscous behavior of the material,
\begin{equation}
    \label{eq:Finelastic}
    \GV{F}(t,\GV{y}(t),\GVP{y}(t),\GKap) = \GV{0}
\end{equation}
is given, see Eq.~\eqref{eq:DAELagrMult}. If an implicit time discretization is applied, $\GVP{y}$ disappears from the equation. However, $\GV{y}$ contains the unknowns at all discrete times, and $\GV{F}$ consists of all non-linear systems to be solved at each of those time steps. Now, as introduced in  Eq. \eqref{eq:state}, $\GV{y}$ contains the unknown nodal displacements $\U$, the nodal reaction forces $\GV{p}$, and the internal variables $\GV{q}$ evaluated at the Gauss points. In this case, it is also possible to compute the reaction forces, an exercise which is left to the reader.

\paragraph{Using the implicit function theorem}
In preparation for the later formulation of the so-called reduced approaches, we introduce here the implicit function theorem, which states that the solution $\GV{y}(t)$ implicitly depends on the material parameters $\GKap$, i.e. $\GV{y}(t) = \GVh{y}(t,\GKap)$. Inserting this into the model equations yields
\begin{equation}
    \label{eq:basicproblemkappa}
    \GV{F}(t,\GVh{y}(t,\GKap),\GKap) = \GV{0},
\end{equation}
or 
\begin{equation}
    \label{eq:basicproblemDAEkappa}
    \GV{F}(t,\GVh{y}(t,\GKap),\GVhp{y}(t,\GKap),\GKap) = \GV{0},
\end{equation}
where Eq. \eqref{eq:DAEreduced} defines $\GV{F}$ in the inelastic case. Thus, once $\GVh{y}(t,\GKap)$ is known, the parameters alone have to be determined, which is denoted as  the reduced approach, see Section~\ref{sec:reduced}. 

\subsection{Classification of different approaches}
\label{sec:classification}

Calibration is used to determine model parameters with which the model can best approximate available measurement data. A criterion for the approximation accuracy is provided in terms of a closed-form mathematical expression, i.e.,\ the objective or loss function. The latter is a norm of the difference between the available measurement data and the model predictions, obtained from the solution of the governing equations via a numerical (discretization) method or a surrogate model. One important characteristic of the different numerical methods presented here is the type of spatial (and/or temporal) discretization. After formulating an objective function, the optimal model parameters that minimize this function are identified using optimization algorithms, often referred to as optimizers. Alternatively, in a statistical setting, a single best guess of the unknown material parameters is formulated as a point estimator, i.e.,\ a function of the observed data \cite{wasserman2004all}. Here, the likelihood function plays a central role, and Bayesian approaches additionally consider a prior density. With statistical models, the uncertainty of an estimate can be assessed through confidence/credible intervals, which are typically computed with sampling approaches. 

In Tab.~\ref{tab:overview}, 
\begin{table*}[h!]
\centering
\caption{Classification of different methods for parameter identification as outlined in Section~\ref{sec:classification}}
\label{tab:overview}
\begin{tabular}{p{2.5cm} p{3.5cm} p{3.5cm} p{3.5cm}}
\toprule
\textbf{method} & \textbf{discretization}  & \textbf{parametrization} & \textbf{optimization/sampling} \\
\midrule
non-linear least-squares using finite elements (Sec.~\ref{sec:nlsfemdic}) & Galerkin; local ansatz for virtual and real displacements & material parameters $\GV{\kappa}$ & gradient-based (e.g., trust region) or gradient-free (e.g., Nelder-Mead simplex) \\
\midrule
equilibrium gap method (Sec.~\ref{secmf:virtual_fields_method})  & Galerkin; local ansatz for virtual fields & material parameters $\GV{\kappa}$ & linear system for linear problems; trust region for non-linear problems \\
\midrule
virtual fields method (Sec.~\ref{secmf:virtual_fields_method}) & Galerkin; global ansatz for virtual fields & material parameters $\GV{\kappa}$ & linear system for linear problems; trust region for non-linear problems \\
\midrule
physics-informed neural networks (Sec.~\ref{sec:surrogates}) & collocation; global ansatz parametrized in $\GV{\theta}$ & material parameters $\GV{\kappa}$, parametrization of PDE solution $\GV{\theta}$ & gradient-based (e.g., ADAM, BFGS, L-BFGS-B) \\
\midrule
surrogate models (Sec.~\ref{sec:surrogates}) & collocation/regression with polynomial ansatz, neural network, Gaussian process, and many more &  material parameters $\GV{\kappa}$, surrogate parameters $\GV{\theta}$ &  any (gradient-based, gradient-free, or sampling)\\
\midrule
model discovery (Sec.~\ref{sec:discovery}) & any & material parameters $\GV{\kappa}$; non-zero $\kappa_i$ from a large library are selected via sparse regression & coordinate descent for linear problems; trust region for non-linear problems \\
\midrule
frequentist inference [\ref{sec:frequentist}] & any & any parameter combination & optimization or sampling \\
\midrule
Bayesian inference [\ref{sec:bayesian}] & any & any parameter combination & sampling or variational inference \\
\bottomrule
\end{tabular}
\end{table*}
the methods presented in the remainder of this manuscript are classified by the aforementioned features, namely
\begin{enumerate}
    \item Discretization: Galerkin or collocation methods, as well as local or global ansatz functions.
    \item Parametrization: The objective or likelihood function depends on parameters $\GV{\kappa}$ of the material model, and possibly also on the PDE solution or on a parametrization of the latter. 
    \item Optimization/sampling: While deterministic calibration approaches mostly make use of different gradient-based optimization schemes, both optimization and sampling are common choices in a statistical setting.
\end{enumerate}

\subsection{Non-linear least-squares method using finite elements}
\label{sec:nlsfemdic}

If the material parameters cannot be identified on the basis of simple experiments, where the stress and the strain are known, the entire boundary-value problem \eqref{eq:smallproblem} of the experiment must be solved. This can be done by numerical approximation methods such as finite differences, boundary elements, finite volumes, or the FEM. In the following, we assume that the latter is the method of choice. The non-linear LS (NLS) method is applied to minimize the difference between the results of the finite element simulation of the experimental boundary conditions and both point-wise and full-field measurement data.  First, we discuss proposals for parameter identification of phenomenological constitutive models (macroscale), followed by schemes addressing models with scale separation, specifically materials with a microstructure.

\subsubsection{Macroscopic constitutive models}
\label{sec:NLS}

The combination of the FEM and a LS approach to determine parameters $\GKap$ in the finite element model goes back to \cite{kavanaghclough1971}. Later, \cite{schnurzabaras1992} linked LS and FEM with the goal to detect the location and the Young's modulus of an inclusion in a matrix material. Discrete data in combination with finite elements were also used in \cite{springmannkuna2005,olberdingfrancissuh2006,hartmanngibmeierscholtes06,nakamurawangsampath2000,kleinermannponthot2003,rauchs2006}. This approach can be followed for experiments with non-uniform stress states when access is limited to, e.g., resultant forces, such as in the case of indentation tests or metal forming processes, see \cite{rauchsbardongeorges2010,rauchsbardon2011}. 

A conceptual and practical step beyond the use of pointwise data has emerged with the availability of optical methods. Here, a special focus lies on DIC methods, see e.g., \cite{suttonorteuschreier2009,grediachild2013}, where the surface displacements of the specimen are measured during loading. Approaches taking advantage of full-field displacement data, pioneered by Mahnken \cite{andresen96,mahnkenstein96,mahnkenstein97}, were further extended by \cite{schedaydiss,riegerdiss2005,kreissig98,BenedixGoerkeKreissigKretzschmar1998,KreissigBenedixGoerke2001,kraemerdiss2016,hartmanngilbertsguazzo2018,CooremanLecompteSolVantommeDebruyne2007,rosemenzel2020,rosediss2022}, see also \cite{schmaltzwillner2014} for biaxial tensile tests and \cite{dilecceetal2022,rossilattanzibarlatkim2022} for more complex deformation cases.
In \cite{avriletal2008}, an approach of this type was denoted \textit{finite element model updating} (FEMU), a terminology that has been used continuously since then. Unfortunately, this terminology can be misleading since the same name is well-established in structural dynamics, see \cite{mottersheadfriswell1993} for a review. Following \cite{mottersheadfriswell1993,mottersheadlinkfriswell2011}, the FEMU approach is applied when the mathematical structure of the problem, here a finite element equation (e.g., the entries in the stiffness or mass matrix), is changed during the system identification process, see \cite{farhathemez1993}, for example. In contrast, when calibrating constitutive models, the finite element equation is specified once -- and only the material parameters are updated. As a result, the calibration of constitutive models has to be clearly classified as parameter identification rather than FEMU. In \cite{hartmanngilbert2021}, the combination of NLS with DIC data, where the boundary-value problem is discretized using the FEM, is denoted as NLS-FEM-DIC -- a denomination that includes the objective function, the boundary-value problem solution technique, and the experimental measurement technique.

Alternative approaches to account for optical information are considered in \cite{mahnken2000}, where gratings on the specimen surface are incorporated, in \cite{kreissigbenedixgoerkelindner2007} using Moir\'{e}-patterns, or in \cite{hartmannECCMR,hartmanntschoepe} using contour data and discrete points on the surface. \cite{avriletal2008} provides an overview of other schemes, such as various ``gap methods'' mainly applied in model updating using vibrational data, and a comparison to determine the elastic parameters, see \cite{avriletal2008}. Further contributions propose the ``constitutive relation error'' and ``modified constitutive relation error'' approaches, \cite{ladevezeleguillon1983,huangfeisselvillon2016}. A study of the integrated DIC method and its references is provided by \cite{ruybalidhoefnagelssluisgeers2016}. 
Overall, the testing and identification paradigm -- which takes advantage of full-field measurement data rather than simple strain transducers or strain gauges -- in conjunction with the FEM is referred to as ``Material Testing 2.0'' in \cite{pierrongrediac2021}.

The previously mentioned approaches bear a strong relation to the mathematical literature on the least-squares method applied to the solution of ODE- or DAE-systems \cite{schittkowskibook2002}. This is explained in detail in \cite{hartmann2017} and will be briefly summarized in the following. 

In the following, $\GV{s}$ denotes the vector of simulation results, to be compared with the vector of the experimental data $\GV{d}$ (displacements or strains at single or different spatial positions and at different times, and forces -- full-field or single-valued resultant quantities). Accordingly, $\GV{y}$ contains all state quantities (unknown displacements and/or reaction forces, internal variables evaluated at all Gauss points) from all experiments and time steps. The same structure holds for $\GV{F}$ as well, see Appendix~\ref{sec:processingData}. Note that $\GV{s}$ results from the computed solution inserted into the so-called observation operator $\GV{O}(\GV{y})$. In this sense, $\GV{s}$ depends on the solution $\GV{y}$ of the problem classes  I-III introduced in Subsection~\ref{sec:finiteelements}. In solid mechanics, it is inherently assumed that the simulation results depend on the material parameters $\GKap\elm{\numkappa}$, i.e., $\GV{s} = \GV{s}(\GKap)\elm{\numD}$, with $n_D$ as the number of data. Here, the implicit function theorem is applied, leading to Eqns.~\eqref{eq:basicproblemkappa} or \eqref{eq:basicproblemDAEkappa}, i.e.\ $\GV{s}(\GKap) = \GV{O}(\GVh{y}(\GKap))$ with $\GV{y}=\GVh{y}(\GKap)$. In the LS procedure, we form the difference between the simulation results and the experimental data, called the residual 
\begin{equation}
    \label{eq:residual_nls_fem}
    \GV{r}(\GKap) = \GV{s}(\GKap) - \GV{d} = \GV{O}(\GVh{y}(\GKap)) - \GV{d}.
\end{equation}
Since the number of individual entries, their magnitude, and their physical units vary within the residual $\GV{r}$, it is common and reasonable to introduce a diagonal weighting matrix $\GM{W}\elmm{\numD}{\numD}$ and to consider the weighted residual 
$\GVt{r}(\GKap) = \GM{W} \, \GV{r}(\GKap)
                = \GM{W} \{\GV{s}(\GKap) - \GV{d}\}$.
For an approach to account for different amounts of data and different magnitudes of the physical quantities within the data, see \cite{hartmannacta}.  
The steps to pre-process the experimental full-field data as well as the data obtained from the numerical simulation are explained in detail in Appendix~\ref{sec:processingData}.

The NLS formulation requires that the square of the weighted residuals
\begin{equation}
  \phi(\GKap)
  = \half \| \GVt{r}(\GKap) \|^2
  = \frac{1}{2}\|\GM{W}\!\left(\GV{s}(\GKap) - \GV{d}\right) \|^2 
  =\half \{\GM{W} \GV{r}(\GKap)\}^{T} \GM{W} \GV{r}(\GKap) 
  =  \half \{\GM{W} \{\GV{s}(\GKap) - \GV{d}\}\}^T \{\GM{W} \{\GV{s}(\GKap) - \GV{d}\}\}   
  %\rightarrow \text{min}.
  \label{eq:objectiveFunction}
\end{equation}
is minimized, i.e.,\ we obtain a solution
\begin{equation}
    \label{eq:minproblem}
    \GKap^* = \argmin_{\GKap} \phi\left(\GV{\kappa}\right).
\end{equation}
Note that multiple minima may exist. In this case, $\GKap^*$ represents an arbitrary element of the minimization set. The necessary condition for a minimum $\GKap = \GKap^*$ is given by
\begin{equation}
  \label{eq:resNLLS}
  \difn{\phi}{\GKap}
  = \GMT{J}(\GKap) \,\GMT{W} \GM{W} \!\left\{ \GV{s}(\GKap) -
    \GV{d}\right\} = \GV{0},
\end{equation}
which represents a system of non-linear equations to determine the material parameters $\GKap$, i.e. to find the solution $\GKap^*$. Note that in some situations, it makes sense to specify equality or inequality constraints, $\GV{h}_{ec}(\GKap) = \GV{0}$, $\GV{h}_{ic}(\GKap) \le \GV{0}$.
The quantity 
\begin{equation}
  \label{eq:sensitivity}
  \GM{J}(\GKap) = \difn{\GV{r}(\GKap)}{\GKap}
  = \difn{\GV{s}(\GKap)}{\GKap}, \quad \GM{J} \elmm{\numD}{\numkappa},
\end{equation}
represents the
\textit{sensitivity matrix} (also denoted as functional matrix or Jacobian matrix). The solution of problem \eqref{eq:resNLLS} can be computed by various methods, see \cite{dennisschnabel96,nocedal_NumericalOptimization_2006,lawsonhanson95,spellucci93,bazaraasheralishetty93}. In the case of gradient-based algorithms, the derivatives \eqref{eq:sensitivity} are required to compute $\GKap^*$. Parameter identification with sensitivity assessment is also covered by \citep{johanssonetal2007}. The computation of the sensitivities is compiled in Appendix~\ref{sec:sensnlsfemdic}.

A systematization of the NLS approach with DIC data is presented in \cite{schowtjakschulteclausmeyerostwaldtekkayamenzel2022}, where various large deformation problems are treated. However, the problem of local identifiability is not addressed, see Subsection~\ref{sec:identifiability} for details.

\subsubsection{Representative volume element approaches}
\label{sec:RVE}

In some cases, materials possess a heterogeneous microstructure (an example being fiber-reinforced composites), and the goal of the identification is to determine their macroscopic properties, i.e., the parameters describing a homogenized material. For unidirectional fabrics or for orthogonal fabrics within the linear elastic regime, for example, appropriate macroscopic models are transversely isotropic elasticity (with five parameters) and orthotropic elasticity (with nine parameters), respectively. The difficulties that arise here lie in the experiments required to determine these parameters uniquely 
(see Section \ref{sec:isotropicelasticity}). One alternative is to use a representative volume element (RVE) and, assuming that the material parameters of the constituents are known, to determine the homogenized material parameters with analytical mixing rules, as discussed in \cite{halpinkardos1976,mallick2007}, or with computational homogenization. However, care must be taken to ensure that the appropriate deformation modes are excited in order for the related parameters to be determined;
see \cite{dileephartmann2022} for a specific application of an RVE to determine orthotropic material parameters. Further questions are treated in \cite{matzenmillergerlach2006,gerlachmatzenmiller2007}, discussing the issue of identifying the inelastic properties of the matrix and the interphase parameters for composite materials.

Another task might be to identify the material parameters of the microscopic constituents of a heterogeneous material. This can be done either by considering only the RVE and applying to it appropriate homogeneous loading conditions, or by relying on macroscopic experiments. The latter option is discussed by 
\cite{SchmidtMergheimSteinmann2015} in a FE$^2$ context on the basis of the LS approach. Unfortunately, local identifiability of the parameters is not discussed. Since FE$^2$ computations are very expensive, an iterative identification procedure using standard FE$^2$ codes is very time-consuming. This problem is pointed out in \cite{klingesteinmann2015}, however, no connection to an identifiability assessment is made. 
The use of integrated DIC for parameter identification within a multiscale approach is discussed in the recent contribution 
\cite{rokospeerlingshoefnaelsgeers2023}. 

An example of RVE-based identification approach for a damaging material is discussed in \cite{prahletal2002}. In order to incorporate the micro-void behavior under loading into the model, a unit cell is considered and the material parameters controlling damage evolution are identified. 

\subsection{Equilibrium gap and virtual fields method}
\label{secmf:virtual_fields_method}

As discussed in the previous section, the NLS method with DIC data seeks to minimize the square of the weighted residuals $\GVt{r}(\GKap)$ for the unknown parameters $\GKap$, thus minimizing the mismatch between FEM predictions $\GV{s}(\GKap)$ and experimental observations $\GV{d}$. 

Observing that the material parameters remain the only unknowns in a governing PDE
if the mechanical state (e.g., the displacement field) is known from experimental measurements,
another stream of research suggests minimizing the square of the residuals of the governing PDE in its discretized weak form instead of the mismatch between FEM predictions and experimental observations.
In other words, observing that forward simulations with finite elements compute the mechanical state of a system by minimizing the residuals of a PDE in its discretized weak form for a given set of material parameters, one can easily deduce an inverse problem that computes the material parameters for a given mechanical state by minimizing the same PDE residuals.
The norm of the residuals of the discretized weak form of the PDE % \brb{numerical model} 
is denoted by some authors as the equilibrium gap \cite{claire_finite_2004}. Assuming that the state $\GV{y} = \GV{u}$, i.e., the state contains only the displacements, and assuming that full-field displacement data are available, we can replace $\GV{y}$ with the data $\GV{d}$. Then, the equilibrium gap method \cite{claire_finite_2004, genet2023} seeks to determine
\begin{equation}
    \label{eq:eqgap}
    \GKap^* = \argmin_{\GKap} \half \|\GV{F}(\GV{d},\GKap)\|^2,
\end{equation}
where $\GV{F}(\GV{d},\GKap)$ are the residuals of the discretized weak form of the balance of linear momentum (after a proper treatment of boundary conditions), considering the Bubnov-Galerkin discretization for the trial and test functions as known from ordinary forward finite element problems. This leads to the necessary first-order condition
\begin{equation}
    \label{eq:VFMI}
   \left[\dif{\GV{F}(\GV{d},\GKap)}{\GV{\GKap}}\right]^T \GV{F}(\GV{d},\GKap) = \GV{0},
\end{equation}
which implies that, in general, $\GV{F}(\GV{d},\GKap) = \GV{0}$ is not fulfilled (``there is a gap to the equilibrium conditions''). Moreover, $\GV{y}$ must be known. It has to be remarked that $\U = \GV{d}\INDu$ (where $\GV{d}\INDu$ are the experimental displacement data) is very restrictive, as it only allows the use of the region of the DIC measurements (which is a subset of the domain) and limits analysis to plane problems. While the scheme is applicable for problem classes I and II, it requires additional considerations for problem class III since the internal variables are not measurable. Thus, the scheme is only applicable for very specific problems in the proposed form \eqref{eq:VFMI}. Theoretically, the scheme is extendable to force data as well. However, this may be difficult when using DIC, as the forces corresponding to the imaged regions may be unknown in this case.

In general, the equilibrium gap may be minimized for any choice of sufficiently smooth test functions (or virtual fields), which is the core idea of the VFM, \cite{grediac_principle_1989,pierron_virtual_2012,boddapati_single-test_2023}. Denoting $\mathcal{V}$ as the chosen set of virtual fields and $\GV{F}_{\mathcal{V}}(\GV{d},\GKap)$
as the corresponding residuals of the weak form of the balance of linear momentum, the objective is to find
\begin{equation}
    \label{eqmf:VFM}
    \GKap^* = \argmin_{\GKap} \half \|\GV{F}_{\mathcal{V}}(\GV{d},\GKap)\|^2.
\end{equation}
Often, the number of virtual fields in $\mathcal{V}$ is chosen such that it coincides with the number of unknown material parameters. These functions must be linearly independent.

Let us start with problem class I (linear elastic problems).
In what follows, we consider a problem for which the residual of the discretized PDE is affine with respect to the mechanical state and the material parameters. Thus, Problem \eqref{eq:system_hartmann} depends linearly on $\GKap$, see Eq.~\eqref{eq:AaskapP} (an alternative approach is provided in Appendix~\ref{ap:system_matrices}, see Eq.~\eqref{eqmf:system_virtual_fields_method}).
In the VFM, the mechanical state variables $\U$ and some resultants of nodal forces are assumed to be known from measurements, such that we can write $\GM{A}\INDas(t,\GV{d}\INDu)$ and $\GV{p}\INDa(t,\GV{d}\INDp)$, where $\GV{d}\INDp$ are the force data. In reality, of course, only very few force resultants are measured. In the following, for the sake of brevity, we discuss the entire set of equations. 
The VFM seeks to identify the material parameters by minimizing the residual of Eq.~\eqref{eq:AaskapP}, i.e.,
\begin{equation}
    \label{eqmf:VFM_linear}
    \GKap^* = \argmin_{\GKap} \half \| \GM{A}\INDas(t,\GV{d}\INDu)\GKap -  \GV{p}\INDa(t,\GV{d}\INDp) \|^2
\end{equation}
implying the necessary condition
\begin{equation}
    \label{eq:vfm}
    \GMT{A}\INDas(t,\GV{d}\INDu) \GM{A}\INDas(t,\GV{d}\INDu) \GKap 
    = \GMT{A}\INDas(t,\GV{d}\INDu) \GV{p}\INDa(t,\GV{d}\INDp),
\end{equation}
which is a linear system with $\numkappa$ unknowns. The same property also holds for problem class II (hyperelasticity) if the material parameters are linearly embedded, e.g., with Rivlin-Saunders and Hartmann-Neff models \cite{rivlin,hartmannneff}, see \cite{flaschel_unsupervised_2021}.
Thus, material parameter identification with the VFM requires only the solution of one linear system of equations, whereas the NLS-FEM approach requires to run a forward FEM simulation at each iteration of the optimization algorithm.
Note, however, that the full displacement field is here assumed to be available from experimental measurements.
A discussion on exceptions is provided in \cite{joshi_bayesian-euclid_2022}.

For problem class III (models with internal variables), the scheme explained above is not applicable. Details regarding the application of the VFM to some specific models with internal variables are provided by \cite{pierron_ExtensionVirtualFields_2010,pierron_virtual_2012}. For general considerations, we refer to Section~\ref{sec:vfm_sec4}. Please note that, given the conceptual similarities between equilibrium gap and VFM, we will only refer to the VFM in the following.

\subsection{Surrogate models and physics-informed neural networks}
\label{sec:surrogates}
In the previous sections, we formulated different minimization problems \eqref{eq:minproblem}, \eqref{eq:eqgap}, and \eqref{eqmf:VFM}, which need to be solved using iterative techniques. As follows, we consider optimization of the residual \eqref{eq:residual_nls_fem} in the generic form 
\begin{equation}
    \label{eq:surrogate_obj}
    \GV{\kappa}^* = \argmin_{\GKap} \half \|\GV{s}(\GKap) - \GV{d} \|^2.
\end{equation}
The evaluation of $\GV{s}(\GKap)$  using e.g., a FEM simulation may be costly, especially if many iterations are required and the model to be calibrated is computationally demanding. Therefore, it can be efficient to replace the simulation by a so-called surrogate or meta model. The vector of the corresponding results is denoted as  $\GV{s}^{\text{surr}}(\GKap)$. 

Surrogate models require a flexible parametric regression ansatz, such as, e.g., multivariate polynomials. Recently, machine learning approaches such as artificial neural networks (ANNs) and Gaussian processes, often referred to as Kriging, have gained increasing attention. A brief introduction to ANNs and the notation used in the manuscript is given in Appendix~\ref{sec:ann}. One of the first attempts to apply ANNs as surrogates for the identification of material parameters goes back to \cite{hubertsakmakis1999a,hubertsakmakis1999b,hubertsagrakistsakmakis2000}. Recent work in this direction can be found in \cite{meissner2022comparative,SchulteKarcaOstwaldMenzel2023}.
When using surrogate models in the context of parameter identification, the overall procedure is the following: 
\begin{enumerate}[label=\arabic*)]
    \item To train a surrogate model $\GV{s}^{\text{surr}}(\GV{\kappa}; \GV{\theta})$ as a regression function through the data 
    %$\left(\GV{\kappa}, \GV{d}\right)$
    sampled from a given number of evaluations of the physical model $\GV{s}(\GV{\kappa})$. In the context of Gaussian processes, $\GV{\theta}$ are the kernel parameters that can be fitted to the data using Bayesian model selection or by minimizing the negative log-likelihood. After fixing the kernel parameters, the GP surrogate is conditioned on the available data to obtain the regression function.  
    The training of ANNs, on the other hand, requires the identification of trainable parameters $\GV{\theta}$, i.e., the weights and biases, via optimization.
    \item To identify the physical parameters by solving the minimization problem \eqref{eq:surrogate_obj}, whereby $\GV{s}(\GV{\kappa})$ is replaced with its surrogate $\GV{s}^\text{surr}(\GV{\kappa}; \GV{\theta}=\GV{\theta}^*)$. Herein, $\GV{\theta}^*$ denotes the non-physical parameter set of the surrogate identified in step 1. Note that surrogates can also be used in the context of statistical inference in combination with sampling approaches, see sections \ref{sec:bayesian} and \ref{sec:locident}.
\end{enumerate}
Step 1 and 2 can be interconnected, which is then referred to as adaptive sampling, see \cite{fuhg2021state} for an extensive review in the context of Kriging.

\begin{figure}
    \centering
    \begin{subfigure}[b]{0.35\linewidth}
        \centering
        \begin{tikzpicture}
            \newcommand\XO{0}
            \newcommand\YO{0}
            \newcommand\MinimumSize{0.7cm}
            % Neurons
            \node[circle, draw, minimum size=\MinimumSize, inner sep=0pt, fill=lightBlue] (I_1) at (\XO,\YO+3.5){$x_{1}$};
            \node[circle, draw, minimum size=\MinimumSize, inner sep=0pt, fill=lightBlue] (I_2) at (\XO,\YO+2.5){$x_{2}$};
            \node[circle, draw, minimum size=\MinimumSize, inner sep=0pt, fill=lightBlue] (I_3) at (\XO,\YO+1.5){$x_{3}$};
            \node[circle, draw, minimum size=\MinimumSize, inner sep=0pt, fill=lightBlue] (I_4) at (\XO,\YO+0.5){$t$};
            \node[circle, draw, minimum size=\MinimumSize, inner sep=0pt, fill=lightGray] (H1_1) at (\XO+1.75,\YO+4){ };
            \node[circle, draw, minimum size=\MinimumSize, inner sep=0pt, fill=lightGray] (H1_2) at (\XO+1.75,\YO+2){ };
            \node[circle, draw, minimum size=\MinimumSize, inner sep=0pt, fill=lightGray] (H1_3) at (\XO+1.75,\YO+0){ };
            \node[circle, draw, minimum size=\MinimumSize, inner sep=0pt, fill=lightGray] (H2_1) at (\XO+3.75,\YO+4){ };
            \node[circle, draw, minimum size=\MinimumSize, inner sep=0pt, fill=lightGray] (H2_2) at (\XO+3.75,\YO+2){ };
            \node[circle, draw, minimum size=\MinimumSize, inner sep=0pt, fill=lightGray] (H2_3) at (\XO+3.75,\YO+0){ };
            \node[circle, draw, minimum size=\MinimumSize, inner sep=0pt, fill=lightRed] (O_1) at (\XO+5.5,\YO+3.5){ };
            \node[circle, draw, minimum size=\MinimumSize, inner sep=0pt, fill=lightRed] (O_2) at (\XO+5.5,\YO+2){ };
            \node[circle, draw, minimum size=\MinimumSize, inner sep=0pt, fill=lightRed] (O_3) at (\XO+5.5,\YO+0.5){ };
            % Outputs
            \draw[->] (O_1) -- ++(0:1) node[midway, above] {$u_{1}$};
            \draw[->] (O_2) -- ++(0:1) node[midway, above] {$u_{2}$};
            \draw[->] (O_3) -- ++(0:1) node[midway, above] {$u_{3}$};
            % Connections: input layer - hidden layer 1
            \draw[->] (I_1) -- (H1_1);
            \draw[->] (I_1) -- (H1_2);
            \draw[->] (I_1) -- (H1_3);
            \draw[->] (I_2) -- (H1_1);
            \draw[->] (I_2) -- (H1_2);
            \draw[->] (I_2) -- (H1_3);
            \draw[->] (I_3) -- (H1_1);
            \draw[->] (I_3) -- (H1_2);
            \draw[->] (I_3) -- (H1_3);
            \draw[->] (I_4) -- (H1_1);
            \draw[->] (I_4) -- (H1_2);
            \draw[->] (I_4) -- (H1_3);
            % Connections: hidden layer 1 - hidden layer 2
            \draw[dashed, ->] (H1_1) -- (H2_1);
            \draw[dashed, ->] (H1_1) -- (H2_2);
            \draw[dashed, ->] (H1_1) -- (H2_3);
            \draw[dashed, ->] (H1_2) -- (H2_1);
            \draw[dashed, ->] (H1_2) -- (H2_2);
            \draw[dashed, ->] (H1_2) -- (H2_3);
            \draw[dashed, ->] (H1_3) -- (H2_1);
            \draw[dashed, ->] (H1_3) -- (H2_2);
            \draw[dashed, ->] (H1_3) -- (H2_3);
            % Connections: hidden layer 1 - ouput layer
            \draw[->] (H2_1) -- (O_1);
            \draw[->] (H2_1) -- (O_2);
            \draw[->] (H2_1) -- (O_3);
            \draw[->] (H2_2) -- (O_1);
            \draw[->] (H2_2) -- (O_2);
            \draw[->] (H2_2) -- (O_3);
            \draw[->] (H2_3) -- (O_1);
            \draw[->] (H2_3) -- (O_2);
            \draw[->] (H2_3) -- (O_3);
            % Dotted lines
            \draw[dotted] (\XO+1.75,\YO+2.75) -- (\XO+1.75,\YO+3.25);
            \draw[dotted] (\XO+1.75,\YO+0.75) -- (\XO+1.75,\YO+1.25);
            \draw[dotted] (\XO+3.75,\YO+2.75) -- (\XO+3.75,\YO+3.25);
            \draw[dotted] (\XO+3.75,\YO+0.75) -- (\XO+3.75,\YO+1.25);
            % Curly brackets
            \draw [decorate, decoration = {calligraphic brace,mirror}] (\XO+1.25,\YO-0.75) --  (\XO+5.75,\YO-.75);
            \node at (\XO+3.375,\YO-1){parameterized by $\bm{\theta}$};
        \end{tikzpicture}
        \caption{}
        \label{fig:conventionalPINN}
    \end{subfigure}
    \hspace{0.1\linewidth}
    \begin{subfigure}[b]{0.35\linewidth}
        \centering
        \begin{tikzpicture}
            \newcommand\XO{0}
            \newcommand\YO{0}
            \newcommand\MinimumSize{0.7cm}
            % Neurons
            \node[circle, draw, minimum size=\MinimumSize, inner sep=0pt, fill=lightBlue] (I_1) at (\XO,\YO+4.0){$x_{1}$};
            \node[circle, draw, minimum size=\MinimumSize, inner sep=0pt, fill=lightBlue] (I_2) at (\XO,\YO+3.0){$x_{2}$};
            \node[circle, draw, minimum size=\MinimumSize, inner sep=0pt, fill=lightBlue] (I_3) at (\XO,\YO+2.0){$x_{3}$};
            \node[circle, draw, minimum size=\MinimumSize, inner sep=0pt, fill=lightBlue] (I_4) at (\XO,\YO+1.0){$t$};
            \node[circle, draw, minimum size=\MinimumSize, inner sep=0pt, fill=darkerBlue] (I_5) at (\XO,\YO+0){$\kappa$};
            \node[circle, draw, minimum size=\MinimumSize, inner sep=0pt, fill=lightGray] (H1_1) at (\XO+1.75,\YO+4){ };
            \node[circle, draw, minimum size=\MinimumSize, inner sep=0pt, fill=lightGray] (H1_2) at (\XO+1.75,\YO+2){ };
            \node[circle, draw, minimum size=\MinimumSize, inner sep=0pt, fill=lightGray] (H1_3) at (\XO+1.75,\YO+0){ };
            \node[circle, draw, minimum size=\MinimumSize, inner sep=0pt, fill=lightGray] (H2_1) at (\XO+3.75,\YO+4){ };
            \node[circle, draw, minimum size=\MinimumSize, inner sep=0pt, fill=lightGray] (H2_2) at (\XO+3.75,\YO+2){ };
            \node[circle, draw, minimum size=\MinimumSize, inner sep=0pt, fill=lightGray] (H2_3) at (\XO+3.75,\YO+0){ };
            \node[circle, draw, minimum size=\MinimumSize, inner sep=0pt, fill=lightRed] (O_1) at (\XO+5.5,\YO+3.5){ };
            \node[circle, draw, minimum size=\MinimumSize, inner sep=0pt, fill=lightRed] (O_2) at (\XO+5.5,\YO+2){ };
            \node[circle, draw, minimum size=\MinimumSize, inner sep=0pt, fill=lightRed] (O_3) at (\XO+5.5,\YO+0.5){ };
            % Outputs
            \draw[->] (O_1) -- ++(0:1) node[midway, above] {$u_{1}$};
            \draw[->] (O_2) -- ++(0:1) node[midway, above] {$u_{2}$};
            \draw[->] (O_3) -- ++(0:1) node[midway, above] {$u_{3}$};
            % Connections: input layer - hidden layer 1
            \draw[->] (I_1) -- (H1_1);
            \draw[->] (I_1) -- (H1_2);
            \draw[->] (I_1) -- (H1_3);
            \draw[->] (I_2) -- (H1_1);
            \draw[->] (I_2) -- (H1_2);
            \draw[->] (I_2) -- (H1_3);
            \draw[->] (I_3) -- (H1_1);
            \draw[->] (I_3) -- (H1_2);
            \draw[->] (I_3) -- (H1_3);
            \draw[->] (I_4) -- (H1_1);
            \draw[->] (I_4) -- (H1_2);
            \draw[->] (I_4) -- (H1_3);
            \draw[->] (I_5) -- (H1_1);
            \draw[->] (I_5) -- (H1_2);
            \draw[->] (I_5) -- (H1_3);
            % Connections: hidden layer 1 - hidden layer 2
            \draw[dashed, ->] (H1_1) -- (H2_1);
            \draw[dashed, ->] (H1_1) -- (H2_2);
            \draw[dashed, ->] (H1_1) -- (H2_3);
            \draw[dashed, ->] (H1_2) -- (H2_1);
            \draw[dashed, ->] (H1_2) -- (H2_2);
            \draw[dashed, ->] (H1_2) -- (H2_3);
            \draw[dashed, ->] (H1_3) -- (H2_1);
            \draw[dashed, ->] (H1_3) -- (H2_2);
            \draw[dashed, ->] (H1_3) -- (H2_3);
            % Connections: hidden layer 1 - ouput layer
            \draw[->] (H2_1) -- (O_1);
            \draw[->] (H2_1) -- (O_2);
            \draw[->] (H2_1) -- (O_3);
            \draw[->] (H2_2) -- (O_1);
            \draw[->] (H2_2) -- (O_2);
            \draw[->] (H2_2) -- (O_3);
            \draw[->] (H2_3) -- (O_1);
            \draw[->] (H2_3) -- (O_2);
            \draw[->] (H2_3) -- (O_3);
            % Dotted lines
            \draw[dotted] (\XO+1.75,\YO+2.75) -- (\XO+1.75,\YO+3.25);
            \draw[dotted] (\XO+1.75,\YO+0.75) -- (\XO+1.75,\YO+1.25);
            \draw[dotted] (\XO+3.75,\YO+2.75) -- (\XO+3.75,\YO+3.25);
            \draw[dotted] (\XO+3.75,\YO+0.75) -- (\XO+3.75,\YO+1.25);
            % Curly brackets
            \draw [decorate, decoration = {calligraphic brace,mirror}] (\XO+1.25,\YO-0.75) --  (\XO+5.75,\YO-.75);
            \node at (\XO+3.375,\YO-1){parameterized by $\bm{\theta}$};
        \end{tikzpicture}
        \caption{}
        \label{fig:parametricPINN}
    \end{subfigure}
    \caption{Schematic representation of (a) a conventional and (b) a parametric PINN formulation.}
    \label{fig:PINN}
\end{figure}
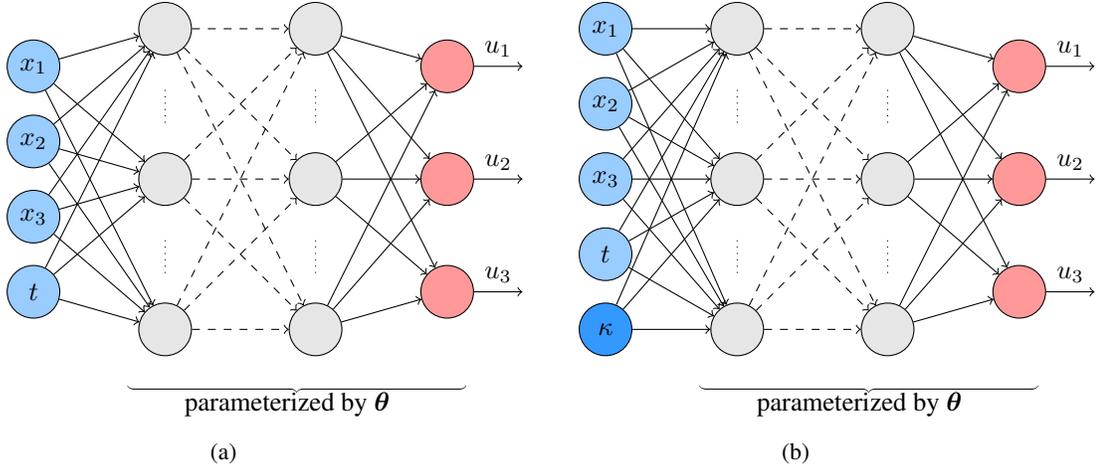

Recently, PINNs, a deep learning framework for solving forward and inverse problems involving PDEs, have gained increasing attention, also in the context of parameter identification and inverse problems. The idea behind this method was first proposed in the 1990s \cite{lagaris_ArtificialNeuralNetworks_1998, psichogios_HybridNeuralNetworkFirst_1992}. Exploiting developments in automatic differentiation \cite{baydin_AutomaticDifferentiationMachine_2018}, advanced software frameworks (such as PyTorch \cite{paszke_PyTorch_2019}, TensorFlow \cite{abadi_TensorFlowLargeScaleMachine_2015}, or JAX \cite{bradbury_JAXComposableTransformations_2018}) and hardware improvements, PINNs have recently been revived \cite{raissi_PhysicsinformedNeuralNetworks_2019}. Their key characteristic is the design of the loss function: By including the governing PDE as regularizing term, the output of the ANN is forced to satisfy the PDE at a set of chosen points, thereby incorporating physical knowledge \cite{karniadakis_PhysicsinformedMachineLearning_2021}. Thus, via the ansatz of the displacement field
\begin{equation}
\label{eq:pinn_nn}
    \LV{u}(\LV{x}, t) \approx \mathcal{U}( \LV{x}, t; \GThe),
\end{equation}
the ANN acts as a function approximator (or ansatz function) of the PDE solution. The above ansatz $\mathcal{U}$ is parameterized in the ANN weights and biases $\GV{\theta}$. Fig.~\ref{fig:conventionalPINN} shows a schematic representation of ansatz~\eqref{eq:pinn_nn}.

Spatial (and, in the case of dynamic problems, temporal) derivatives of the ansatz \eqref{eq:pinn_nn} can be calculated using automatic differentiation, see  \cite{baydin_AutomaticDifferentiationMachine_2018,abadi_TensorFlowLargeScaleMachine_2015,bradbury_JAXComposableTransformations_2018}. For dynamic problems, this approach is referred to as continuous-time PINN \cite{raissi_PhysicsinformedNeuralNetworks_2019}. Alternatively, temporal derivatives can be treated by means of discrete time-integration schemes, see \cite{raissi_PhysicsinformedNeuralNetworks_2019, wessels2020neural}. As mentioned earlier, dynamic problems are out of the scope of the representation. Instead, the time dependence in Eq.~\eqref{eq:pinn_nn} expresses that different load (time) steps within one experiment or various experiments are used to identify the set of material parameters $\GKap$. 

\paragraph{Parametric PINN as surrogate model}
PINNs may also act as surrogate models, when a parametric ansatz is chosen. A parametric PINN takes parameters (here, the material parameters $\GKap$) as additional input to the ANN, such that the ansatz \eqref{eq:pinn_nn} modifies to
\begin{equation}\label{eq:para_pinn}
    \bm{u}(\GV{x}, t, \GKap) \approx \mathcal{U}(\GV{x}, t, \GKap; \GV{\theta}).
\end{equation}
Then, with the observation operator in \eqref{eq:residual_nls_fem}, the model response is extracted from the state via
\begin{equation}
    \label{eq:state_extract_paraPINN}
     \GV{s}^{\text{surr}}\left(\GKap; \GV{\theta}\right) = \GV{O}(\mathcal{U}(\GV{x}, t, \GKap; \GV{\theta})).
\end{equation}
Fig. \ref{fig:parametricPINN} shows a schematic representation of a parametric PINN ansatz in comparison to the one in Eq. \eqref{eq:pinn_nn}. The training of parametric PINNs, i.e., the identification of the ANN parameters $\GV{\theta}$, is discussed in detail in Section~\ref{sec:PINNreduced}. Parametric PINNs have been deployed for thermal analysis \cite{hosseini2023}, magnetostatics \cite{beltran2022param}, or for the optimization of an airfoil geometry \cite{sun_PPINNAirfoilGeometry_2023}. To the best of our knowledge, they have not yet been used for the calibration of material models in solid mechanics. 

% Figure PINN 

\paragraph{Inverse or all-at-once PINNs}
An alternative to the parametric PINN approach discussed above are so-called inverse PINNs, which are related to the all-at-once formulation introduced in Section~\ref{sec:abstract}. In inverse PINNs, material and network parameters $\GKap$ and $\GThe$ are identified simultaneously, and the optimization problem \eqref{eq:surrogate_obj} does not hold anymore. Details are provided in Section~\ref{sec:aao_pinn}.

The following contributions deal with material model calibration from full-field displacement and force data using PINNs. In \cite{haghighat_PhysicsinformedDeepLearning_2021}, the authors propose a multi-network model for the identification of material parameters from displacement and stress data for linear elasticity and von Mises plasticity. 
For the analysis of internal structures and defects, the authors in 
\cite{zhang_AnalysisInternalStructures_2022} present a general framework for identifying unknown geometry and material parameters. 
In \cite{hamel_CalibratingConstitutiveModels_2022}, a framework is developed for the calibration of hyperelastic material models from full-field displacement and global force-displacement data. Unlike in the conventional PINN approach \cite{raissi_PhysicsinformedNeuralNetworks_2019}, the physical constraints are imposed by using the weak form of the PDE.
In \cite{liu_VariationalFormulationPINNs_2023}, variational forms of the residual for the identification of homogeneous and heterogeneous material properties are formulated. The method is demonstrated using the example of linear elasticity.
The study in 
\cite{zhang_PhysicsInformedNeuralNetworks_2020} considers the identification of heterogeneous, incompressible, hyperelastic materials from full-field displacement data. It uses two independent ANNs, one for the approximation of the displacement field, and another one for approximating the spatially dependent material parameters. A comparison of different data sampling strategies as well as soft and hard boundary constraints is reported in \cite{wu_EffectiveSamplingPINNs_2023}. 
In \cite{anton_PhysicsInformedNeuralNetworks_2022}, PINNs are further enhanced towards the calibration of linear elastic materials from full-field displacement and global force data in a realistic regime. The realistic regime here refers to the order of magnitude of material parameters and displacements as well as noise levels.

\subsection{Model discovery}
\label{sec:discovery}
The previously discussed methods for material characterization focus on the calibration of material parameters in an a priori known material model.
The a priori choice of the material model, i.e., the combination of mathematical operations that describes the material behavior, typically relies on the intuition and modeling experience of the user.
Inappropriate assumptions in the material model are reflected in a poor fitting accuracy of the model even after parameter calibration.
To avoid such modeling errors, strategies have been proposed recently to automatically discover a suitable symbolic representation of the material model, while simultaneously calibrating the corresponding material parameters.
Thus, the problem of material parameter calibration is generalized to the more intricate problem of material model discovery.

Algorithms for discovering symbolic models from data can be broadly classified into two types: symbolic regression methods based on genetic programming  \citep{koza_genetic_1994}, which repeatedly mutate a model until its agreement with the data is satisfactory, and sparse regression methods \citep{tibshirani_regression_1996} which select a model from a potentially large set (also called library or catalogue) of candidate models based on data.
Such methods have first been used in the physical sciences to deduce symbolic expressions of the governing equations of dynamical systems \citep{schmidt_distilling_2009,brunton_discovering_2016}.
In the field of material modeling,  symbolic regression based on genetic programming has been used since the early work of \cite{schoenauer_evolutionary_1996}, see also the more recent works in \cite{versino_data_2017,bomarito_development_2021,kabliman_application_2021,park_multiscale_2021,abdusalamov_automatic_2023}.
Sparse regression for the discovery of material models has been much less explored and has only recently gained attention \cite{flaschel_unsupervised_2021,wang_inference_2021,wang_establish_2022,meyer_thermodynamically_2023}. See also the review in \cite{georgelanguage2024}, where a novel symbolic regression approach based on formal grammars is proposed for hyperelasticity.

In \cite{flaschel_unsupervised_2021}, the authors propose EUCLID (Efficient Unsupervised Constitutive Law Identification and Discovery), a method for discovering symbolic expressions for hyperelastic strain energy density functions using sparse regression starting from displacement and force data.
The method is later extended to elastoplasticity \cite{flaschel_discovering_2022}, viscoelasticity \cite{marino_automated_2023}, and generalized standard materials \cite{flaschel_automated_2023-1}, see \cite{flaschel_automated_2023} for an overview.
A supervised version (i.e., a version based on stress data) of the EUCLID approach is experimentally validated in \cite{flaschel_automated_2023-2} using data stemming from simple mechanical tests on human brain tissue.

The idea behind EUCLID is to construct a set of candidate material models (i.e.,\ a material model library) by introducing a general parametric ansatz for the material behavior, which depends on a large number of parameters ($\GKap\in\Rset^{\numkappa}$ with $\numkappa \gg 1$) comprising all material parameters of all candidate material models.
Calibrating the parameters using one of the previously discussed calibration methods would result in a highly ill-posed inverse problem. Even assuming that such a problem is solvable, the solution would deliver a parameter vector $\GKap$ with many non-zero entries and, thus, a highly complex symbolic expression for the material model.
Therefore, a regularization term is introduced which penalizes the number of non-zero entries in $\GKap$, and, consequently, promotes simplicity of the symbolic material model expression while alleviating the ill-posedness of the inverse problem.
The so-called $\ell_p$-regularization term takes the form $\|\GKap\|_p^p=\sum_i \mid \kappa_i \mid^p$ with $0<p\leq 1$, assuming high values for dense vectors $\GKap$ and smaller values otherwise.
In the limit $p \rightarrow 0^+$, the $\ell_p$-regularization term converges to an operator that counts the number of non-zero entries in $\GKap$. For $p=1$, the approach is called Lasso-regularization. 

We first focus on problem classes I and II  and assume that the models in the material model library depend linearly on the parameters $\GKap$, see \citep{flaschel_unsupervised_2021}.
For the formulation of EUCLID for problem class III \cite{flaschel_discovering_2022,flaschel_automated_2023-1} as well as for models that do not depend linearly on the parameters, we refer to Sec.~\ref{sec:euclid_sec4}.
The regularized problem that constitutes the core of EUCLID is written as 
\begin{equation}
    \label{eqmf:EUCLID}
    \GKap^* = \argmin_{\GKap} \half \|\GM{A}(\GV{d},\Uq)\GKap   - \GVc{p}\|^2 + \lambda \|\GKap\|_p^p,
    % \bm{\GKap}^* = \argmin_{\GKap} \half \|\GM{A}\INDas(t,\GV{d})\GKap -  \GV{p}\INDa(t,\GV{d}\INDp)\|^2 + \lambda \|\GKap\|_p^p,
\end{equation}
where the first term in the objective function quantifies the mismatch between the model and the data. 
Here, similar to the VFM, see Eq.~\eqref{eqmf:VFM}, the model-data mismatch is defined indirectly by the sum of squared residuals of the discretized weak form of the balance of linear momentum. Here, however, we should note that other measures for the model-data mismatch, see Sec.~\ref{sec:nlsfemdic}, could be chosen likewise.
The matrix $\GM{A}(\GV{d},\Uq)$ and the load vector $\GVc{p}$ in Eq.~\eqref{eqmf:EUCLID} are constructed as described in Appendix \ref{ap:system_matrices} for linear elasticity at small strains.
If more sophisticated material behavior like hyperelasticity is considered, the matrix $\GM{A}(\GV{d},\Uq)$ changes accordingly, see \cite{flaschel_unsupervised_2021}.

The effect of the sparsity-promoting regularization term on the minimization problem is influenced by the weighting factor $\lambda > 0$, which can be chosen to strike a balance between the fitting accuracy and the complexity of the discovered material model, see \cite{flaschel_discovering_2022,marino_automated_2023,flaschel_automated_2023-1,flaschel_automated_2023,flaschel_automated_2023-2}.
Abandoning the deterministic setting, the EUCLID method is investigated from a Bayesian perspective in \cite{joshi_bayesian-euclid_2022} to quantify the uncertainty in the discovered models.

In the literature, two different lines of research on model discovery can be distinguished. The first aims to discover symbolic and thus interpretable expressions for the material models, as in the approaches we just described. The second is concerned with the training of non-interpretable black-box machine learning models for encoding the material response. Examples for the latter include machine learning models like splines \cite{sussman_model_2009}, Gaussian processes \cite{frankel_tensor_2020}, neural ordinary differential equations \cite{tac_data-driven_2022}, and -- the arguably most popular choice -- ANNs \cite{ghaboussi_knowledgebased_1991}.
Although purely data-driven constitutive models have shown some success \cite{ghaboussi_knowledgebased_1991,huang2021}, much effort has been addressed to constrain ANNs such that fundamental physical requirements (such as objectivity, material stability, and thermodynamical consistence) are respected, while maintaining their expressivity.
In the context of hyperelasticity, for example, \cite{asad_mechanics-informed_2022,klein_polyconvex_2022,thakolkaran_nn-euclid_2022,linkakuhl2023} choose specifically designed ANN architectures to ensure the (poly-)convexity of the strain energy density function. 
In the context of dissipative materials, \cite{masi_thermodynamics-based_2021,huang_variational_2022,masi2023evolution,meyer_thermodynamically_2023} consider ANNs that do not violate thermodynamic requirements (see \cite{rosenkranz2023} for a review).
An often disregarded drawback of ANNs is their data-hungriness.
Supervised training frameworks for ANNs require a large amount of labeled stress-strain data pairs, which are experimentally not accessible and are therefore typically acquired through computationally expensive microstructure simulations, which require the microstructural properties of the material to be known.
To counteract the data scarcity, unsupervised training frameworks that solely rely on experimentally available displacement and reaction force data like the NN-EUCLID proposed by \cite{thakolkaran_nn-euclid_2022} are indispensable.

\subsection{Bayesian approaches}
\label{sec:bayesian}
Bayesian statistical methods are gaining momentum for parameter identification in mechanics, because of their ability to incorporate prior information and because they naturally achieve a regularization of ill-posed problems. Moreover, Bayesian inference delivers a distribution over the sought parameters, which can be used for uncertainty quantification. The Bayesian approach \cite{gelman1995bayesian} formulates a posterior density as
\begin{equation}
    p(\GKap|\GV{d}) \propto p(\GV{d}|\GKap) p(\GKap),
\end{equation}
where $p(\GKap)$ refers to the parameter prior and $p(\GV{d}|\GKap)$ represents the likelihood function, which can be evaluated by solving the mechanical model. 

More precisely, writing $\GV{d} = \GV{s}(\GKap) + \GV{e}$, with $\GV{e}$ as the observation noise vector, we obtain the Likelihood as
\begin{equation}
    p(\GV{d}|\GKap) = \pi_{\GV{e}}(\GV{d} - \GV{s}(\GKap)),
\end{equation}
where $\pi_{\GV{e}}$ denotes the probability density function of the measurement noise.
Many different parameter identification approaches can be recovered as Bayesian point estimates. For instance, with a normally distributed prior $p(\GKap) \sim \mathcal{N}(\GV{\mu}_{\GKap},\GV{\Sigma}_{\GKap})$ and $\pi_{\GV{e}} \sim \mathcal{N}(\GV{0},\GV{\Sigma}_{\GV{e}})$, we can compute the maximum a posteriori estimate
\begin{equation}
    \GKap^* = \argmin_{\GKap} -\ln \left(p(\GV{d}|\GKap) p(\GKap)  \right)
    = \argmin_{\GKap} \| \GV{d} - \GV{s}(\GKap) \|_{\GV{\Sigma}_{\GV{e}}^{-1}} + \| \GKap -  \GV{\mu}_{\GKap}\|_{\GV{\Sigma}_{\GKap}^{-1}},
\end{equation}
with the weighted norm $\| \GV{x} \|_{\GM{A}}^2 = \GVT{x} \GV{A} \GV{x}$ for any positive definite matrix $\GV{A}$, and recover the NLS approach with Tikhonov regularization, see \cite{calvetti2018inverse} for details and further links between Bayesian approaches and regularization. One can even obtain information about the posterior based on this optimization-based approach, by perturbing the data $\GV{d}$ with a randomly drawn noise vector according to the density $\pi_{\GV{e}}$. This so-called randomize-then-optimize approach is investigated in a number of papers, see \cite{bardsley2014randomize} for instance. 

When dealing with model discovery in the context of a possible large library of candidate basis functions, Bayesian priors are also frequently used to enforce sparsity. For instance, maximum a posteriori estimates yield, together with the Bayesian LASSO and spike-and-slab priors, the $\ell_1$ and $\ell_0$ regularization, respectively. A unifying framework to Bayesian sparsity priors has recently been proposed in \cite{shin2022neuronized}.

Fully Bayesian approaches, however, go beyond point estimation and determine the posterior distribution $p(\GKap|\GV{d})$, most often through sampling-based Markov chain Monte Carlo analysis. Here, a major bottleneck is the large computational cost, which is often alleviated by replacing the simulation model with a surrogate. A popular choice consists in employing Gaussian process modeling with adaptive sampling, see \cite{sinsbeck2017sequential}, for example. A sampling-free approach is put forth in \cite{rosic2013parameter}, where polynomial chaos filtering is used to avoid costly sampling. An alternative consists in building a surrogate model for the likelihood function or the posterior directly, which in turn gives easy access to posterior moments and the marginal likelihood function \cite{wagner2021bayesian}. The drawback here is the concentration of Likelihood and posterior in the large data-small noise limit, which requires dedicated adaptive surrogate modeling approaches. 

Bayesian parameter estimates have been employed in the context of biological hyperelastic models \cite{romer2022surrogate}, but also for phase-field fracture modeling \cite{noii2021bayesian,wu2021parameter}, composite modeling \cite{goguetal2013}, viscoelasticity \cite{rappel2018bayesian,yueetal2023}, and plasticity \cite{ibrahimbegovic2020reduced,papadimasdodwell2021}, to name a few. See \cite{noiietal2022} for various applications in the context of computational mechanics and \cite{aguilo2013overview} for a comparison of Bayesian and deterministic parameter estimation approaches. 

Another feature that makes Bayesian parameter estimation interesting for mechanics is the ability to compare competing models based on model selection criteria, such as Bayes' scores. This is proposed, e.g., in \cite{fitt2019uncertainty} for comparing hyperelastic material models and in \cite{prudencio2015computational} for selecting continuum damage mechanics models.  

The Bayesian reasoning also underlies machine learning approaches, in particular Gaussian process models and Bayesian ANNs. A recent survey covering Bayesian methods in the context of PINNs is given in \cite{viana2021survey}. 

Bayesian methods rely on an accurate model producing the simulation data; hence, a model error term needs to be accounted for if the simulation results are corrupted by numerical errors or simplifying model assumptions. A simple way to account for model error is to add a discrepancy term $\GV{\epsilon}$ as
\begin{equation}
    \GV{d} =  \GV{s}(\GV{\beta}) + \GV{\epsilon} + \GV{e},
\end{equation}
see \cite{kennedy2001bayesian}. Recently, the role of model error has been emphasized in the statistical finite element framework \cite{girolami2021statistical}, which is mainly designed to update an uncertain state from noisy data. The reader is referred to \cite{duffin2022low} for a reaction-diffusion model and \cite{narouie2023inferring}  for the case of hyperelastic material models. The importance of accounting for errors in the state representation for parameter identification is highlighted, e.g., in \cite{calvetti2018iterative}. Therein, the authors show that a statistical representation of the discretization error leads to improved parameter estimates for several inverse problems. A similar approach is adopted for the calibration of hyperelastic materials in \cite{romer2022surrogate}. In the most basic case, one could simply inflate the covariance matrix to account for measurement and model error at the same time.

Note that there exist many other approaches to quantify uncertainties -- employing, for instance, frequentist or interval-based methods. These methods have been applied in solid mechanics to some extent, but are not in the focus of the paper. Various frequentist approaches and a comparison to the Bayesian approach are included and cited in Section~\ref{sec:locident}. Since publications in this direction typically do not employ the frequentist label, to the best of our knowledge, a systematic review of these approaches is not yet available.

%************************************
\section{Unified framework for parameter estimation}\label{sec:abstract}
This section presents an abstract and unifying view on parameter identification. In particular, we introduce a parameter estimation framework which covers most available approaches in the computational mechanics literature. In the inverse problems community, parameter estimation problems are often classified in  ``reduced'' and ``all-at-once'' approaches, see \cite{haber2001preconditioned,burger2002iterative}. The reduced approach is mainly centered around the NLS method and aims at identifying the unknown model parameters. The all-at-once approach additionally incorporates the state equation, e.g., Eq. \ref{eq:Felastic} for problem class I,  directly in the identification procedure. Thus, it allows to  simultaneously identify the mechanical state and the model parameters. This approach has also been discussed in the context of the equation error method, see \cite{ito_augmented_1990} for a combination of the NLS and equation error method. In computational mechanics, there exists a third important approach, the VFM. In the following, these three important cases are discussed and the links between them are established.

In the solid mechanics literature, the all-at-once concept is not widely discussed. As we point out later, the all-at-once setting is less restrictive regarding the model assumptions than the reduced approach and it provides a suitable framework for the recently introduced PINN-based identification approaches. Moreover, it may result in quite diverse iterative parameter estimation methods, regarding implementation and cost.
In the following, we will first briefly introduce the basic concepts of reduced and all-at-once approaches. Comparative aspects regarding first-order optimality conditions  are the topic of Section~\ref{sec:comparison}. Then, different variants of reduced (Section~\ref{sec:reduced}) and all-at-once  approaches (Section~\ref{sec:all-at-once}) are discussed. VFM and EUCLID are the subject of Section~\ref{sec:virtualfields}. 
As novel contributions, we propose both an FEM- and VFM-based all-at-once formulation in Subsection~\ref{sec:nlsfemdicallatonce}. 

\subsection{Introducing reduced and all-at-once approaches}
Recalling the notation of Section~\ref{sec:compPI}, i.e.,\ Eq.~\eqref{eq:Felastic}, and considering elastic problems for the sake of simplicity, the parameter identification problem is governed by the following set of equations 
\begin{equation}
    \label{eq:abstract_identification}
    \begin{split}
    \GV{F}(\GV{y},\GKap) &= \GV{0}, \quad \text{(state equation)}\\
    \GV{O}(\GV{y}) &= \GV{d}, \quad \text{(observation)}
    \end{split}
\end{equation}
The state and parameter vector $\GV{y}$ and $\GKap$ take values in the sets $\Omega_{\GV{y}} \subset \mathbb{R}^{n_{\GV{y}}}, \Omega_{\GKap} \subset \mathbb{R}^{n_{\GKap}}$. 
The so-called observation operator $\GV{O}$ relates the model state $\GV{y}$ to the observational data $\GV{d}$, which may be the raw data or some post-processed version of it.

In most of the approaches discussed  far, e.g. the one in Subsection~\ref{sec:nlsfemdic}, a dependence of $\GV{y}$ on $\GKap$ in Eq.~\eqref{eq:abstract_identification} is assumed, $\GV{y} = \GVh{y}(\GKap)$. In this case, we minimize 
\begin{equation}
    \label{eq:loss_reduced}
    \phi\INDr(\GKap) = d\INDO(\GV{O} (\GVh{y}(\GKap)) ,\GV{d}),
\end{equation}
with the definition $d(\GV{a},\GV{b}) := \|\GV{a} - \GV{b}\|^2/2$ of a distance function, 
see Eq.~\eqref{eq:objectiveFunction}. Approaches of this type correspond to the so-called \textit{reduced formulation}. The main advantage of this approach is that the minimization is carried out for the parameter $\GKap$ only, and Eq.~\eqref{eq:abstract_identification}$_1$ is fulfilled exactly.
It must be mentioned that the observation operator $\GV{O}(\GVh{y}(\GKap))$ can take many forms and may also account for integrated and/or indirectly measured quantities. For instance, $\GV{O}$ may be chosen to select a part of the full displacement vector in the case of DIC data. If, instead, the data $\GV{d}$ consist of strains, the observation operator computes the strains from the displacements first. Also, if measurements are available at a set of points in the computational domain, $\GV{O}$ can be defined  to interpolate the discrete solution at the sensor locations. Here, ``sensor locations'' are the experimental evaluation positions, which, in the case of optical measurement data, may be very numerous.

If the equality signs in Eqns.~\eqref{eq:abstract_identification} are assumed to be a too strong requirement because data are noisy in practice and the state equation cannot always be enforced exactly, a possible alternative is to minimize the distances
\begin{align}
    d\INDS(\GV{F}(\GV{y},\GKap),\GV{0}) &= \half \|\GV{F}(\GV{y},\GKap) \|^2, 
    \label{eq:distF}
    \\
    d\INDO(\GV{O} \left(\GV{y}\right) ,\GV{d}) &= \half \|\GV{O} \left(\GV{y}\right) - \GV{d} \|^2,
    \label{eq:distO}
\end{align}
determining at the same time both $\GV{y}$ and $\GKap$ (hence the name of \textit{all-at-once approaches}).
In the case of elastic materials and displacement-independent loads for finite element equations, the function $d\INDS(\GV{F}(\GV{y},\GKap),\GV{0})$ represents the distance between internal and external forces, which is a helpful interpretation for the later discussion.

Of course, other choices are possible; for example minimizing a weighted norm of the state equation and observations with a weighting matrix $\GM{W}$. Hence, in general an all-at-once approach employs an objective function $\phi_\text{aao}$ of the form 
\begin{equation}
\label{eq:loss_all-at-once}
    \phi_{\text{aao}}(\GV{y},\GKap) = w\INDs \underbrace{d\INDS(\GV{F}(\GV{y},\GKap),\GV{0})}_{\text{physics}} + w\INDd \underbrace{d\INDO(\GV{O} \left(\GV{y}\right) ,\GV{d})}_{\text{data}},
\end{equation}
which contains both physics- and data-based loss terms. The constants $w\INDs$ and $w\INDd$ can be used to weigh the importance of the individual contributions or to simply rescale the different loss terms. Such a conditioning of the objective function is often crucial to achieve numerical convergence, see \cite{anton_PhysicsInformedNeuralNetworks_2022}, for example. In the setting introduced so far, the residual $\GV{F}$ can be discretized with any numerical method. 

\subsection{First-order optimality conditions}
\label{sec:comparison}
In the following, we will discuss connections and differences between reduced and all-at-once approaches as well as the VFM by inspecting necessary first-order conditions (FOCs). Exemplarily, we thereby consider elastic problems, i.e., problem classes I and II.
The main lines of this section follow \cite{kaltenbacher2016regularization}, which reports a comparison of reduced and all-at-once approaches in an abstract Banach space setting. 

Minimizing the objective function $\phi\INDr$ defined in Eq.~\eqref{eq:loss_reduced} yields the following optimization problem for the reduced approach
\begin{equation}\label{eq:argmin_reduced_abstract}
\begin{aligned}
    \GKap^* &= \argmin_{\GKap} \phi\INDr(\GKap). 
\end{aligned}
\end{equation}
In general, there may be multiple parameters for which the minimum is attained, which would require to write $\GKap^* \in \argmin_{\GKap} \phi\INDr(\GKap)$ in  Eq.~\eqref{eq:argmin_reduced_abstract}. In this case, the equal sign simply chooses a single element of the set.

For the all-at-once approach and the objective function $\phi_{\text{aao}}$ defined in Eq.~\eqref{eq:loss_all-at-once}, we formulate the optimization problem
\begin{equation}
    \{\GV{y}^*, \GKap^*\} = \argmin_{\GV{y}, \GKap} \phi_{\text{aao}}(\GV{y}, \GKap).
\end{equation}
Next, we derive the FOCs for both approaches. 

We start with the \textbf{reduced approach}, where we are concerned with the minimization of 
\begin{equation}
\label{eq:phirdef}
    \phi\INDr(\GKap) = 
    \half \| \GV{s}(\GKap) - \GV{d} \|^2 
    =
    \half \| \GV{O}(\GVh{y}(\GKap)) - \GV{d}\|^2
\end{equation}
subjected to the equality constraint
\begin{equation}
    \label{eq:RF-hyperconstraint}
    \GV{F}(\GVh{y}(\GKap),\GKap) = \GV{0}.
\end{equation}

\noindent Applying the Lagrange multiplier method, we seek a stationary point of 
\begin{equation}
    \label{eq:RF-LMhyper}
    \hat{\phi}\INDr(\GKap,\GLam)
    =
    \phi\INDr(\GKap) + \GLam^T \GV{F}(\GVh{y}(\GKap),\GKap).
\end{equation}
The necessary FOCs are
\begin{align}
    \gat{\GKap}{\hat{\phi}\INDr}{\GKap,\GLam}{\Delta \GKap}
    &= 0, \label{eq:diffglambdaI}\\
    \gat{\GLam}{\hat{\phi}\INDr}{\GKap,\GLam}{\Delta \GLam}
    &= 0.
    \label{eq:difflambda}
\end{align}
where 
\begin{equation}
\gat{x}{f}{x}{h} = \difn{f(x+\lambda h)}{\lambda}\Big|_{\lambda=0}
\end{equation}
denotes the directional derivative. While the second condition (valid for arbitrary $\Delta \GLam$) implies that the constraint Eq.~\eqref{eq:RF-hyperconstraint} is satisfied, it follows from Eq.~\eqref{eq:diffglambdaI} that
\begin{equation}
    \gat{\GKap}{\hat{\phi}\INDr}{\GKap,\GLam}{\Delta \GKap}
    =
    \gat{\GKap}{\phi\INDr}{\GKap}{\Delta \GKap}
    \\
    +
    \GLam^T 
    \gat{\GKap}{\GV{F}}{\GVh{y}(\GKap),\GKap}{\Delta \GKap}
    = 0.
    \label{eq:dphidkap}
\end{equation}
Since $\GV{F}(\GVh{y}(\GKap),\GKap)=\GV{0}$ holds for arbitrary parameters $\Delta \GKap$, it is
\begin{equation}
    \gat{\GKap}{\GV{F}}{\GVh{y}(\GKap),\GKap}{\Delta \GKap}
    =
    \difn{\GV{F}}{\GKap} \Delta \GKap
    = \GV{0}
\end{equation}
for arbitrary $\Delta \GKap$, hence the total derivative $\D{\GV{F}}/\D{\GKap}$ must vanish
\begin{equation}
    \difn{\GV{F}}{\GKap}
    =
    \dif{\GV{F}}{\GV{y}}
    \difn{\GVh{y}}{\GKap}
    + 
    \dif{\GV{F}}{\GKap}
    =
    \GM{0}.
    \label{eq:dFdkappa}
\end{equation}
This also implies that the first term in Eq.~\eqref{eq:dphidkap} has to vanish, i.e.,
\begin{equation}
    \gat{\GKap}{\phi\INDr}{\GKap}{\Delta \GKap}
    =
    \Delta \GKap^T 
    \left[
        \difn{\GVh{y}}{\GKap}
    \right]^T
    \difn{\phi\INDr}{\GV{y}}
    = 0,
\end{equation}
for arbitrary $\Delta \GKap$. For $\phi\INDr$ defined in Eq.~\eqref{eq:phirdef}, we obtain
\begin{equation}
    \left[
        \difn{\GVh{y}}{\GKap}
    \right]^T
    \difn{\phi\INDr}{\GV{y}}
    =
    \left[
        \difn{\GVh{y}}{\GKap}
    \right]^T
    \left[
        \difn{\GV{O}}{\GV{y}}
    \right]^T
    \{ \GV{O}(\GVh{y}(\GKap)) - \GV{d} \}
    =
    \GMT{J}(\GKap) \{ \GV{O}(\GVh{y}(\GKap)) - \GV{d} \}
    = \GV{0}
    \label{eq:resultRFNLS}
\end{equation}
with
\begin{equation}
    \label{eq:RF-jacobian}
    \GM{J}
    =
    \difn{\GV{s}}{\GKap}
    =
    \difn{\GV{O}}{\GV{y}}
    \difn{\GVh{y}}{\GKap}.
\end{equation}
Clearly, Eq.~\eqref{eq:resultRFNLS} is equivalent to Eq.~\eqref{eq:resNLLS}, where the weighting matrix is omitted for brevity. Let us summarize the resulting equations. We have to solve Eqns.~\eqref{eq:RF-hyperconstraint} and \eqref{eq:resultRFNLS}
\begin{equation}
    \label{eq:resultRFNLS-summarized}
    \text{(reduced-1)}
    \left \{
    \begin{aligned}
        \GV{F}(\GV{y},\GKap) 
        &= 
        \GV{0} \\
        \left[
        \difn{\GV{O}}{\GV{y}} \difn{\GVh{y}}{\GKap}
        \right]^T 
        \{ \GV{O}(\GV{y}) - \GV{d} \}
        &= \GV{0}    
    \end{aligned}
    \right .
\end{equation}
with the derivative evaluated by solving \eqref{eq:dFdkappa},
\begin{equation}
\label{eq:reddydkap}
    \difn{\GVh{y}}{\GKap}
    =
    - \left[\dif{\GV{F}}{\GV{y}}\right]^{-1}
    \left[\dif{\GV{F}}{\GKap}\right].
\end{equation}

There is a second option to formulate the reduced approach, where the stationarity conditions are derived before expressing the variable $\GV{y}$ as a function of $\GKap$. This approach is discussed next because it allows for a better comparison between the reduced and all-at-once approaches. Hence, we proceed by reformulating the reduced optimization problem as
\begin{equation}
\begin{aligned}
    \{\GV{y}^*,\GKap^*\} &= \argmin_{\GV{y},\GKap}  d\INDO(\GV{O} \left(\GV{y}\right) ,\GV{d}), \\
    \ \text{subjected to }  &\GV{F}(\GV{y},\GKap)=\GV{0}.
\end{aligned}
\end{equation}
This constrained optimization problem can once again be treated with the Lagrange multiplier method, which yields
\begin{equation}
    \{\GV{y}^*, \GKap^*, \GLam^*\} 
    = \argstat_{\GV{y},\GKap,\GLam}  \Big( \frac{w\INDd}{2} \| \GV{O}(\GV{y}) - \GV{d} \|^2
    + \GLam^T  \GV{F}(\GV{y},\GKap) \Big),
\end{equation}
where $\argstat$ denotes the solution of the saddle point problem. Here, $w\INDd$ is only introduced to later point out some similarities to the all-at-once approach.
The FOCs for the second formulation of the reduced approach read
\begin{equation}
\label{eq:reduced-first-order-necessary}
\text{(reduced-2)}
\left \{
\begin{aligned}
    &\left[\dif{\GV{F}}{\GKap} \right]^T \GV{\Lambda} 
    = \GV{0}, \\ 
    &\left[\dif{\GV{F}}{\GV{y}}\right]^T \GV{\Lambda} = - w\INDd \left[\difn{\GV{O}}{\GV{y}}\right]^T \{\GV{O}(\GV{y}) - \GV{d}\}, \\
    &\bm{F}(\GV{y},\GKap) = \GV{0}. 
\end{aligned}
\right. 
\end{equation}
Next, we would like to point out the equivalence of the FOCs for the  reduced-1 and reduced-2 approaches. To show this, we start from Eq.~\eqref{eq:reduced-first-order-necessary} and replace $\GV{y}$ with $\GVh{y}(\GKap)$. Multiplying Eq.~\eqref{eq:reduced-first-order-necessary}$_2$ with $[\D \GVh{y}/\D \GKap]^T$ and adding the resulting equation to Eq.~\eqref{eq:reduced-first-order-necessary}$_1$, we obtain 
\begin{equation}    
\left[\dif{\GV{F}}{\GV{y}}
    \difn{\GVh{y}}{\GKap}
    + 
    \dif{\GV{F}}{\GKap} \right]^T \GV{\Lambda}
    = - w\INDd \left[\difn{\GV{O}}{\GV{y}} \difn{\GVh{y}}{\GKap}\right]^T \{\GV{O}(\GV{y}) - \GV{d}\}.
\end{equation}
The left-hand side vanishes because of condition \eqref{eq:dFdkappa}, and Eq.~\eqref{eq:RF-jacobian} leads us to Eq.~\eqref{eq:resultRFNLS-summarized}$_2$. Reversely, starting from Eq.~\eqref{eq:resultRFNLS-summarized}, we define $\GLam$ as the solution of Eq.~\eqref{eq:reduced-first-order-necessary}$_2$, compactly written as 
\begin{equation}
    \left[\dif{\GV{F}}{\GV{y}}\right]^T \GV{\Lambda} = - w\INDd  \difn{\phi\INDr}{\GV{y}}.
\end{equation}
Multiplying both sides with $\left[ \D{\GVh{y}}/\D{\GKap} \right]^T$ and using Eq.~\eqref{eq:resultRFNLS} yields 
\begin{equation}
    \left[ \difn{\GVh{y}}{\GKap} \right]^T \left[\dif{\GV{F}}{\GV{y}}\right]^T \GV{\Lambda} = \GV{0}.
\end{equation}
With Eq.~\eqref{eq:dFdkappa}, 
we arrive at Eq.~\eqref{eq:reduced-first-order-necessary}$_1$. Hence, we have shown the equivalence of Eq.~\eqref{eq:reduced-first-order-necessary} and Eq.~\eqref{eq:resultRFNLS-summarized}.

For the \textbf{all-at-once formulation}, with 
\begin{equation}
    \{\GV{y}^*, \GKap^*\} 
    = \argmin_{\GV{y},\GKap}  \half \Big( w\INDs \| \GV{F}\left(\GV{y}, \GKap\right) \|^2
    +   w\INDd \| \GV{O}(\GV{y}) - \GV{d} \|^2  \Big),
\end{equation}
we obtain the FOCs 
\begin{equation}
    \label{eq:FOC_aoo_naive}
    \text{(AAO)}
\left \{
\begin{aligned}
    w\INDs \left[\dif{\GV{F}}{\GV{y}}\right]^T \GV{F}(\GV{y},\GKap) &= - w\INDd\left[\difn{\GV{O}}{\GV{y}}\right]^T \{\GV{O}(\GV{y}) - \GV{d}\}, \\
    \left[\dif{\GV{F}}{\GKap}\right]^T \GV{F}(\GV{y},\GKap) &= \GV{0}.
\end{aligned}
\right .
\end{equation}
At first sight, Eqns.~\eqref{eq:FOC_aoo_naive} do not resemble the FOCs \eqref{eq:reduced-first-order-necessary} of the reduced approach. However, we can reveal some strong similarities. To this end, we introduce $\GLam$ as the solution of Eq.~\eqref{eq:reduced-first-order-necessary}$_2$. This allows us to compare Eq.~\eqref{eq:reduced-first-order-necessary}$_2$ with Eq.~\eqref{eq:FOC_aoo_naive}, which results in
\begin{equation*}
    w\INDs \GV{F}(\GV{y},\GKap) = \GV{\Lambda} 
    \quad \text{and} \quad
    \left[\dif{\GV{F}}{\GKap}\right]^T \GV{\Lambda} = \GV{0}.
\end{equation*}
Hence, we can recast the all-at-once FOCs as 
\begin{equation}
    \label{eq:allatonce-first-order-necessary}
\begin{aligned}
    \left[\dif{\GV{F}}{\GKap} \right]^T \GV{\Lambda} 
    &= \GV{0}, \\ 
    \left[\dif{\GV{F}}{\GV{y}}\right]^T \GV{\Lambda} &= - w\INDd\left[\difn{\GV{O}}{\GV{y}}\right]^T \{\GV{O}(\GV{y}) - \GV{d}\}, \\
    w\INDs \bm{F}(\bm{y},\bm{\kappa}) &= \GV{\Lambda},
\end{aligned}
\end{equation}
where a strong connection to the reduced FOCs \eqref{eq:reduced-first-order-necessary} becomes apparent. In the limit $w\INDs \rightarrow \infty$, we recover the reduced optimality conditions from those of the all-at-once formulation. 

An interesting observation arises in connection with the \textbf{VFM}. Even the VFM can be recovered as a special case of the all-at-once approach. If we divide Eq.~\eqref{eq:FOC_aoo_naive}$_1$ by $w\INDd$ and pass to the limit of $w\INDd \rightarrow \infty$, we obtain
\begin{equation}
    \label{eq:aoo-to-virtualfields-1}
  \left[\difn{\GV{O}}{\GV{y}}\right]^T \{\GV{O}(\GV{y}) - \GV{d}\} = \GV{0}.  
\end{equation}
In case of the VFM, $\GV{O}(\GV{y}) = \GV{y}$ holds, and hence, Eq.~\eqref{eq:aoo-to-virtualfields-1} reduces to $\GV{y} = \GV{d}$. Then, Eq.~\eqref{eq:FOC_aoo_naive}$_2$ can be recast as 
\begin{equation}
    \left[\dif{\GV{F}}{\GKap}\right]^T \GV{F}(\GV{d},\GKap) = \GV{0},
    \label{eq:FOCVFMI}
\end{equation}
which is precisely the FOC of the VFM. 

\subsection{The reduced formulation}
\label{sec:reduced}
The main idea of the reduced formulation is to eliminate the state and to predict the data solely based on the parameters. This is typically achieved by expressing the state directly as a function of the parameters, $\GV{y} = \GVh{y}(\GKap)$, a function which is referred to as the solution map. In other words, the implicit function theorem is applied as discussed in \cite{krantzparks2003}, which holds for
\begin{equation}
    \label{eq:conditions-implicit-function-theorem}
   \left\| \left[ \dif{\GV{F}(\GV{y},\GKap)}{\GV{y}}  \right]^{-1} \right\| \leq C, \quad \forall \GV{y} \in \Omega_{\GV{y}}, \GKap \in \Omega_{\GKap}.
\end{equation}
Assuming that condition \eqref{eq:conditions-implicit-function-theorem} holds, the implicit function theorem implies 
\begin{align}
    \GV{F}(\GVh{y}(\GKap),\GKap) &= \GV{0}, \label{eq:reduced_state}\\
    \GV{O} \left(\GVh{y}(\GKap)\right) &= \GV{d} \label{eq:reduced_observation}.
\end{align}
Following this approach, the parameter-to-observable map $\GV{s}(\GKap) = \GV{O}(\hat{\GV{y}}(\GKap))$ predicts the data $\GV{d}$ for each $\GKap$, 
see also Table~\ref{tab:reduced_maps}.

The reduced approach has already been anticipated in Eqns.~\eqref{eq:basicproblemkappa} or \eqref{eq:basicproblemDAEkappa}, and is in particular associated with the NLS-FEM approach, cf. the minimum problem \eqref{eq:minproblem} and Section~\ref{sec:NLS}. The reduced approach for the calibration of mechanical models is reviewed, e.g., in \cite{mahnkenstein96} with a focus on inelastic models. 

Based on Eq.~\eqref{eq:reduced_observation}, the parameter vector can now be tuned to minimize the prediction-data mismatch $d\INDO$. The main advantage of this approach is that only an optimization problem in the parameter domain needs to be solved. Moreover, the solver may be employed as a black box, providing a data prediction for each parameter value. The drawback is the restriction \eqref{eq:conditions-implicit-function-theorem}, which may not hold in practice. In the following, we proceed by discussing several specific problems.

\begin{table}
\caption{Solution and parameter-to-observable maps that enable the reduced formulation.}
\label{tab:reduced_maps}
\centering
\begin{tabular}{|c|c|}
\hline
     $\GVh{y}\left(\GKap\right)$ & solution map \\
     \hline 
     $\GV{s}\left(\GKap\right)= \GV{O}\left(\GVh{y}\left(\GKap\right)\right)$ & parameter-to-observable map \\
     \hline
\end{tabular}
\end{table}

\subsubsection{NLS-FEM-based reduced approach}
\label{sec:RF-nlsfemdic}

In the following, the reduced approach using the NLS-FEM method is discussed. 

\paragraph{Problem classes I and II: elasticity}

The NLS-FEM approach of Section~\ref{sec:nlsfemdic} can be related to the reduced formulation. For models in which the material parameters or the state quantities are included linearly, see, for example, Eqns.~\eqref{eq:Felasticdef} or \eqref{eq:FdefVFM}, there is no significant advantage with regard to the optimization problem. Thus, the iterative computation of equations \eqref{eq:resultRFNLS-summarized} has to be carried out.

\paragraph{Problem class III: inelasticity}
We can formulate problems with internal variables either in the time-continuous or in the time-discrete setting. We start with the time-continuous formulation, i.e.,\ we consider the more general case of DAEs, see Eq.~\eqref{eq:basicproblemDAEkappa}. Thus, we have to solve the minimum problem \eqref{eq:argmin_reduced_abstract} with $\phi\INDr$ defined in Eq.~\eqref{eq:phirdef} under the equality constraint (eq. \eqref{eq:basicproblemDAEkappa}, here repeated for convenience)
\begin{equation}
    \label{eq:RF-DAEconstraint}
    \GV{F}(t,\GVhp{y}(t,\GKap),\GVh{y}(t,\GKap),\GKap) = \GV{0}.
\end{equation}
To solve the problem we apply a procedure similar to that in Section \ref{sec:comparison}. The method of Lagrange multipliers requires solving the DAE-system \eqref{eq:RF-DAEconstraint} and the non-linear system of equations \eqref{eq:resultRFNLS} with the Jacobian~\eqref{eq:RF-jacobian}. Furthermore, Eq.~\eqref{eq:dFdkappa} now reads
\begin{equation}
\label{eq:dFdkappaDAE}
\begin{aligned}
    %&\brb{\text{D}_{\GKap} \GV{F} \left(\GVh{y}(\GKap),\GKap\right) ??? =}
    \difn{\GV{F}}{\GKap}
     = \dif{\GV{F}}{\GVP{y}} \dif{\GVhp{y}}{\GKap}
    +
    \dif{\GV{F}}{\GV{y}} \dif{\GVh{y}}{\GKap}
    +
    \dif{\GV{F}}{\GKap}
    =
    \GM{0}.
\end{aligned}
\end{equation}
For the semi-discrete inelastic model $\GV{F}$ defined in Eq.~\eqref{eq:DAEreduced} with $\GVh{g}\INDa$ introduced in Eq.~\eqref{eq:alleqlm}, we obtain
\begin{equation}
\label{eq:DAEvonkappa}
\GV{F}(t,\GVP{y}(t,\GKap),\GV{y}(t,\GKap),\GKap)
    =
    \mzweiv{\GVh{g}\INDa(t,\U(t,\GKap),\GV{p}(t,\GKap),\GV{q}(t,\GKap))}
    {\GVP{q}(t,\GKap) - \GV{r}\INDq(t,\U(t,\GKap),\GV{q}(t,\GKap),\GKap)}
    = \GV{0},
\end{equation}
Eq.~\eqref{eq:dFdkappaDAE} implies the linear matrix DAE-system:
\begin{equation}
    \label{eq:DAEsensitivity}
    \begin{split}
    \GM{0}
    &=
    \dif{\GV{g}}{\U} \dif{\Uh}{\GKap}
    +
    \dif{\GV{g}}{\Q} \dif{\GVh{q}}{\GKap}
    +
    \dif{\GV{g}}{\GKap} 
    \\
    \GM{0}
    &=
    \dif{\GVq{g}}{\U} \dif{\Uh}{\GKap}
    +
    \dif{\GVq{g}}{\Q} \dif{\GVh{q}}{\GKap}
    +
    \dif{\GVq{g}}{\GKap}
    -
    \dif{\GV{p}}{\GKap}\\
        \frac{\D}{\D t} \dif{\GVh{q}}{\GKap}
    &=
    \dif{\GV{r}\INDq}{\U} \dif{\Uh}{\GKap}
    +
    \dif{\GV{r}\INDq}{\Q} \dif{\GVh{q}}{\GKap}
    +
    \dif{\GV{r}\INDq}{\GKap}
    \\
    \end{split}
\end{equation}
Here, the last equation results from Eq.~\eqref{eq:DAEreduced}$_2$. 
The solution of this DAE-system yields the sensitivities $\partial \GVh{q} / \partial \GKap$, $\partial \Uh / \partial \GKap$, $\partial \GV{p} /\partial \GKap$. In \cite{dunker1984,leiskramer1985,schittkowskibook2002}, the system \eqref{eq:DAEsensitivity} is denoted as ``simultaneous sensitivity equations'', see also \cite{hartmann2017} in the context of the FEM.

An alternative approach can be derived if a time discretization scheme is first applied to the DAE-system \eqref{eq:DAEvonkappa}.
In this case, the sensitivities are computed at each individual time $\tnp$; this is termed ``internal numerical differentiation'' since it is based on analytical derivatives provided upfront, see \cite{schittkowskibook2002}. The details within the FEM context are discussed in \cite{hartmann2017} and form the basis of most implementations. In the time-discrete formulation with the backward-Euler method, see Eq.~\eqref{eq:BEforceII} with the abbreviation \eqref{eq:ldef} as well as \eqref{eq:alleqlm}, we have at each time
\begin{equation}
    \label{eq:FFEred}
    \GV{F}(\tnp,\GVh{y}\INDNP(\GKap),\GVh{y}\INDn(\GKap),\GKap)
    =
    \GV{0},
\end{equation}
i.e.,\ a large system 
\begin{equation}
    \label{eq:FAFEred}
    \GV{F}\INDA(\GVh{y}\INDA(\GKap),\GKap)
    =
    \GV{0},
\end{equation}
with
\begin{align}
    \GV{F}^T\INDA
    &=
    \{
      \GVT{F}(t_1,\ldots),\ldots,\GVT{F}(t_N,\ldots)
    \} 
    \nonumber \\
    \GVh{y}^T\INDA
    &=
    \{
      \GVh{y}_1^T(\GKap),\ldots,\GVh{y}_N^T(\GKap)
    \} 
    \label{eq:abbrevFA}\\
    \GVh{y}^T\INDNP
    &=
    \{
      \GVh{u}\INDNP^T(\GKap),\GVh{p}\INDNP^T(\GKap),\GVh{q}^T\INDNP(\GKap)
    \} 
    \nonumber 
\end{align}
(where the initial conditions $\GV{y}_0$ are assumed to be independent of $\GKap$). In a concrete setting, Eq.~\eqref{eq:FFEred} reads
\begin{equation}
    \GV{F}\big(\tnp,\GVh{y}\INDNP(\GKap),\GVh{y}\INDn(\GKap),\GKap\big)
    =
    \left\{
    \begin{matrix}
        \GV{g}\big(\tnp,\GVh{u}\INDNP(\GKap),\GVh{q}\INDNP(\GKap),\GKap\big) \\
        \GVq{g}\big(\tnp,\GVh{u}\INDNP(\GKap),\GVh{q}\INDNP(\GKap),\GKap\big) - \GVh{p}\INDNP(\GKap) \\
        \GV{l}\big(\tnp,\GVh{u}\INDNP(\GKap),\GVh{q}\INDNP(\GKap),\GVh{q}\INDn(\GKap),\GKap\big) 
    \end{matrix}
    \right\}
\end{equation}
with
\begin{equation}
    \GV{l}\big(\tnp,\GVh{u}\INDNP(\GKap),\GVh{q}\INDNP(\GKap),\GVh{q}\INDn(\GKap),\GKap\big)
    =
    \GVh{q}\INDNP(\GKap) - \GVh{q}\INDn(\GKap)
    - \dtn \GV{r}\INDq\big(\tnp,\GVh{u}\INDNP(\GKap),\GVh{q}\INDNP(\GKap),\GKap\big).
\end{equation}
Eq.~\eqref{eq:FAFEred} represents the first equation in the FOC \eqref{eq:resultRFNLS-summarized}. The linear system in Eq.~\eqref{eq:reddydkap} 
\begin{equation}
    \left[
    \dif{\GV{F}\INDA}{\GV{y}\INDA}
    \right]
    \difn{\GVh{y}\INDA}{\GKap}
    =
    -
    \dif{\GV{F}\INDA}{\GKap}
    \label{eq:senslinsys}
\end{equation}
leads to linear systems at each time $\tnp$ of the form
{
\begin{align}
    \dif{\GV{g}}{\Unp} \difn{\GVh{u}\INDNP}{\GKap}
    +
    \dif{\GV{g}}{\GV{q}\INDNP} \difn{\GVh{q}\INDNP}{\GKap}
    &=
    - \dif{\GV{g}}{\GKap} 
    \nonumber\\
    \dif{\GVq{g}}{\Unp} \difn{\GVh{u}\INDNP}{\GKap}
    +
    \dif{\GVq{g}}{\GV{p}\INDNP} \difn{\GVh{p}\INDNP}{\GKap}
    +
    \dif{\GVq{g}}{\GV{q}\INDNP} \difn{\GVh{q}\INDNP}{\GKap}
    &=
    - \dif{\GVq{g}}{\GKap} 
    \label{eq:FEsens}\\
    \dif{\GV{l}}{\Unp} \difn{\GVh{u}\INDNP}{\GKap}
    +
    \dif{\GV{l}}{\GV{q}\INDNP} \difn{\GVh{q}\INDNP}{\GKap}
    +
    \dif{\GV{l}}{\GV{q}\INDn} \difn{\GVh{q}\INDn}{\GKap}
    &=
    - \dif{\GV{l}}{\GKap}
    \nonumber
\end{align}}
with
\begin{equation}
    \dif{\GVq{g}}{\GV{p}\INDNP}
    =
    - \GV{I},
    \qquad 
    \dif{\GV{l}}{\GV{q}\INDNP}
    =
    \GV{I} - \dtn \dif{\GV{r}\INDq}{\GV{q}\INDNP},
    \qquad
    \dif{\GV{l}}{\GV{q}\INDn}
    =
    -\GV{I},
    \qquad
    \dif{\GV{l}}{\GKap}
    =
    -\dtn \dif{\GV{r}\INDq}{\GKap}.
\end{equation}
Eqns.~\eqref{eq:FEsens}$_1$ and \eqref{eq:FEsens}$_3$ can be combined leading to several matrices which are already available within classical finite element implementations, where a Multilevel-Newton algorithm is applied. This leads to the functional matrices $\D \GVh{u}\INDNP/\D \GKap$ and $\D \GVh{q}\INDNP/\D \GKap$. Particularly, the evaluation of Eq.~\eqref{eq:FEsens}$_3$ can be carried out at Gauss-point level, since $\GV{l}$ are only formally assembled into a large vector. See \cite{hartmann2017} for details, and \cite{hartmanngilbert2021} for a concrete implementation.

Since direct access to the finite element code is required in the internal numerical differentiation version, an alternative for the case of black box finite element programs has to considered. This is called ``external numerical differentiation'', see \cite{schittkowskibook2002,hartmann2017}, where finite differences are chosen to obtain the derivatives in Eq.~\eqref{eq:senslinsys}.

\subsubsection{Parametric PINN-based reduced approach}
\label{sec:PINNreduced}

In the following, the general procedure for parameter identification with parametric PINNs is presented for the different problem classes. We thereby consider PINN formulations based on the strong form of the balance of linear momentum \eqref{eq:smallproblem}, which can be regarded as a meshfree collocation method. Note that it is equally possible to approximate other physical principles using ANNs, e.g., the minimum of potential energy. This approach is known as the deep energy method, see \cite{samaniego_EnergyApproachSolution_2020} and \cite{fuhg2022mixed} for a mixed formulation. 

From here on, the superscript $\bullet\EXPcol$ denotes quantities defined at collocation points $\GV{x}\EXPcol$. For an introduction to the notation of ANNs, the reader is referred to Appendix~\ref{sec:ann}.

\paragraph{Problem class I \& II: elasticity}

In elasticity, the parametric ansatz below is used for the displacements
\begin{equation}\label{eq:pinn_ansatz_para}
    \bm{u}(\GV{x}, t, \GKap) \approx \mathcal{U}(\GV{x}, t, \GKap; \GV{\theta})
\end{equation}
to approximate the local equilibrium conditions $\GV{g}\EXPcol$. The latter must be satisfied at all $\numcol$ spatial collocation points $\GV{x}\EXPcol$,
\begin{equation}
    \begin{aligned}
    \label{eq:balance_lin_pinn}
        \GV{g}\EXPcol 
        &= 
        \left\{ 
           \begin{matrix} 
            \divop \LV{h}(\LV{E}(\mathcal{U}(\LV{x}\EXPcol_1, t, \GKap; \GV{\theta})),\GKap) + \rho \, \LV{b} \\
            \vdots \\
            \divop \LV{h}(\LV{E}(\mathcal{U}(\GV{x}\EXPcol_{\numcol}, t, \GKap; \GV{\theta})),\GKap) + \rho \, \LV{b}
            \end{matrix} 
        \right\}
        = \GV{0}.
    \end{aligned}
\end{equation}
Neumann boundary conditions are defined at $\numcolneu$ collocation points $\LV{x}\EXPN\subset\LV{x}\EXPcol$ on the Neumann boundary
\begin{equation}
\label{eq:neumann_lin_pinn}
    \LV{C}\INDN :=
    \left\{
    \begin{matrix}
        \LV{h}(\LV{E}(\mathcal{U}(\LV{x}_1\EXPN, t, \GKap; \GV{\theta})),\GKap) \GV{n} - \LVq{t}(\GV{x}_1\EXPN, t)\\
        \vdots \\
        \LV{h}(\LV{E}(\mathcal{U}(\LV{x}_{\numcolneu}\EXPN, t, \GKap; \GV{\theta})),\GKap) \GV{n} - \LVq{t}(\LV{x}_{\numcolneu}\EXPN, t) \\
    \end{matrix}
    \right\} = \GV{0}
\end{equation}
and Dirichlet boundary conditions at $\numcoldir$ collocation points on the Dirichlet boundary $\LV{x}\EXPD\subset\LV{x}\EXPcol$
\begin{equation}
\label{eq:dirichlet_soft}
    \begin{aligned}
        \LV{C}\INDD := 
        \left\{
        \begin{matrix}
            \mathcal{U}(\LV{x}_1\EXPD, t, \GKap; \GV{\theta}) - \LVq{u}(\LV{x}_1\EXPD, t) \\
            \vdots \\
\mathcal{U}(\LV{x}_{\numcoldir}\EXPD, t, \GKap; \GV{\theta}) - \LVq{u}(\LV{x}_{\numcoldir}\EXPD, t)
        \end{matrix}
        \right\} = \GV{0}
        .
    \end{aligned}
\end{equation}
The semi-discrete model summarizes Eqns.~\eqref{eq:balance_lin_pinn}-\eqref{eq:dirichlet_soft} and results in
\begin{equation}\label{eq:F_elastic_pinn}
        \GV{F}\EXPcol(t, \mathcal{U}(\GV{x}\EXPcol, t, \GKap; \GV{\theta}),\GKap)
        =
        \left\{\begin{matrix}
            \GV{g}\EXPcol(\mathcal{U}(\GV{x}\EXPcol, t, \GKap; \GV{\theta}),\GKap) \\
            \LV{C}\INDN(t, \mathcal{U}(\GV{x}\EXPN, t, \GKap; \GV{\theta}),\GKap) \\
            \LV{C}\INDD(t, \mathcal{U}(\GV{x}\EXPD, t, \GKap; \GV{\theta}))
        \end{matrix} \right\}
        = \GV{0}.
\end{equation}
Herein, reaction forces at Dirichlet boundaries have been neglected. Since these are relevant in the context of displacement controlled-experiments, they could be accounted for by an additional loss term or by using the method of Lagrange multipliers, see Section~\ref{sec:finiteelements} in the context of the FEM.

Spatial derivatives occurring in the semi-discrete model \eqref{eq:F_elastic_pinn} are computed using automatic differentiation.
Alternatively, the automatic differentiation can be enriched or even replaced by incorporating discrete derivative operators known from other numerical methods, e.g., peridynamics \cite{haghighat2021nonlocal}.

For the sake of a unified notation, we define the state vector $\GVh{y}\INDn\left(\GKap;  \GV{\theta}\right)$ at one specific time (load) step $t_n$ as 
\begin{equation}\label{eq:pinn_state_para}
    \hat{\GV{y}}_n(\GKap; \GV{\theta}) = 
    \left\{ 
           \begin{matrix} 
           \mathcal{U}(\LV{x}\EXPcol_1, t_n, \GKap; \GV{\theta}) \\
            \vdots \\
            \mathcal{U}(\GV{x}\EXPcol_{\numcol}, t_n, \GKap; \GV{\theta})
            \end{matrix} 
        \right\}, \quad \ n=1,\ldots,t\INDN
\end{equation}  
and assemble the full state vector $\GVh{y}\INDa(\GKap; \GThe)$ as 
\begin{equation}
\label{eq:assembly_states}
    \GVh{y}\INDa(\GKap; \GThe) = \{ \GVh{y}^T_1 \ldots \GVh{y}^T\INDN\}^T.
\end{equation}
Of course, the state vector \eqref{eq:pinn_state_para} can be generalized to incorporate different experiments as well, following the notation in Appendix~\ref{sec:processingData}. In this sense, the full discrete model is assembled to
\begin{equation}\label{eq:Ffull_param}
    \GV{F}\EXPcol\INDa(\hat{\GV{y}}\INDa(\GKap; \GThe), \GKap) = 
    \left\{
    \begin{matrix}
    \GV{F}\EXPcol(t_1, \GVh{y}_1(\GKap; \GThe), \GKap) \\
    \vdots \\
    \GV{F}\EXPcol(t\INDN, \GVh{y}\INDN(\GKap; \GThe), \GKap)
    \end{matrix}
    \right\}= \GV{0}.
\end{equation}
The network parameters $\GV{\theta}$ are identified by solving an optimization problem, where the sum of the norm of $\GV{F}\INDa\EXPcol$, evaluated at a set of Monte Carlo or other evaluation points $\GKap^{(i)}$ in the parameter domain, is minimized as
\begin{equation}    
\label{eq:min_para_elas_1}
\begin{aligned}
    \GV{\theta}^* = \argmin_{\GV{\theta}} \sum_{i=1}^n \half \left\| \GV{F}\EXPcol\INDa\big(\hat{\GV{y}}\INDa(\GKap^{(i)}; \GV{\theta}), \GKap^{(i)} \big)\right\|^2.
\end{aligned}
\end{equation}
Note that the mean squared error is often minimized in machine learning, see \cite{henkes_PhysicsInformedNeural_2022}. Here, we keep the squared error for the sake of consistency with the rest of the paper. After training, the material parameter vector $\GKap$ can be identified using a reduced approach. 

Dirichlet conditions \eqref{eq:dirichlet_soft} enforced via the loss function are referred to as soft boundary conditions. Alternatively, the ansatz function \eqref{eq:pinn_ansatz_para} can be modified such that it fulfills boundary conditions by construction; these are referred to as hard boundary conditions \cite{berg_UnifiedDeepArtificial_2018, wessels2020neural, henkes_PhysicsInformedNeural_2022}. Hard boundary conditions can be formulated using a boundary extension $\GV{G}$ and  distance function $\GV{D}$ (here for the non-parametric case),
\begin{equation}
    \GV{u}\left(\GV{x},t\right) \approx \GV{G}\left(\GV{x}, t\right) + \GV{D}\left(\GV{x}, t\right) \circ \mathcal{U}\left(\GV{x}, t; \GV{\theta}\right).
\end{equation}
Note that hard Neumann boundary conditions can also be imposed if the ansatz  is extended by the stress tensor $\GV{\sigma}$, Eq.~\eqref{eq:smallproblem}, see  \citep{haghighat_PhysicsinformedDeepLearning_2021}:
\begin{equation}
    \{\GV{u}(\GV{x}, t), \GV{\sigma}(\GV{x}, t) \} 
    = \mathcal{U}^{\GV{u}, \GV{\sigma}}(\GV{x}, t; \GThe)
\end{equation}
Once trained in an offline stage according to the minimization problem \eqref{eq:min_para_elas_1}, the parametric PINN can, in a second step, be used for online parameter identification. This second step is related to the reduced approach, with the loss, i.e.,\ distance function
\begin{equation}\label{eq:loss_red_pinn}
    \phi\INDr \left(\GKap\right)
    = \half \left\| \GV{O}(\hat{\GV{y}}\INDa\left(\GKap; \GThe\right)) - \GV{d} \right\|^2
\end{equation}
and the optimization problem 
\begin{equation}\label{eq:argmin_red}
    \GKap^* = \argmin_{\GKap} \phi\INDr\left(\GKap\right).
\end{equation}
The necessary condition is given by
\begin{equation}
\label{eq:optimality_red_pinn}
    \gat{\GKap}{{\phi}\INDr}{\GKap}{\Delta \GKap}
    = 
    \Delta \GKap^T 
    \left[
        \dif{\GVh{y}\INDa}{\GKap}
    \right]^T
    \difn{\phi\INDr}{\GV{y}\INDa}
    = \GV{0}
\end{equation}
and, with $\phi\INDr$ given by the loss function \eqref{eq:loss_red_pinn},
\begin{equation}\label{eq:resultRF_redpinn}
    \begin{aligned}
    \left[
        \dif{\GVh{y}\INDa}{\GKap}
    \right]^T
    \difn{\phi\INDr}{\GV{y}\INDa}
    &=
    \left[
        \difn{\GV{O}}{\GV{y}\INDa}
        \difn{\GVh{y}\INDa}{\GKap}
    \right]^T
    \{ \GV{O}(\hat{\GV{y}}\INDa) - \GV{d} \}
    \\
    &=
    \GMT{J} \{ \GV{O}(\hat{\GV{y}}\INDa) - \GV{d} \}
    = \GV{0}.
    \end{aligned}
\end{equation}
The Jacobian $\GV{J}$ is defined by
\begin{equation}
    \label{eq:jacobian_pinn}
    \GM{J}
    =\dif{\GV{s}^{\text{surr}}}{\GKap}
    =
    \difn{\GV{O}}{\GV{y}\INDa}
    \difn{\GVh{y}\INDa}{\GKap}
\end{equation}
where $\GV{s}^{\text{surr}}$ denotes the parametric PINN surrogate model, see also Section~\ref{sec:surrogates}.
It becomes apparent that Eq.~\eqref{eq:resultRF_redpinn} is analogous to the reduced formulation of the FEM-based reduced approach \eqref{eq:resultRFNLS}. The partial derivatives ${\partial \GVh{y}\INDa}/{\partial \GKap}$ can be computed using automatic differentiation, see also Appendix~\ref{sec:ann}.

\paragraph{Problem class III: inelasticity}

For inelastic problems, the constitutive model becomes a function of the internal variables $\GV{q}$, see Eq.~\eqref{eq:smallproblem}. Consequently, a parametric ansatz is required for both the displacements and the internal variables
\begin{equation}
\label{eq:para_pinn_inelas}
    \begin{aligned}
    \mzweiv{\GV{u}\left(\GV{x}, t, \GKap \right)}{\GV{q}\left(\GV{x}, t, \GKap \right)} 
        \approx 
        \left\{ \begin{matrix}
            \mathcal{U}\left(\GV{x}, t, \GKap; \GV{\theta}_{\GV{u}}\right) \\
            \mathcal{Q}\left(\GV{x}, t, \GKap ; \GV{\theta}_{\GV{q}}\right)
        \end{matrix}\right\}.
    \end{aligned}
\end{equation}
Here, separate networks with trainable parameters $\GV{\theta}_{\GV{u}}$, $\GV{\theta}_{\GV{q}}$ are used as an ansatz to simplify the notation, and we define $\GThe = \left\{\GThe_{\GV{u}}^T \ \GThe_{\GV{q}}^T\right\}^T$.
Note that both networks depend on the same set of material parameters $\GKap$. In analogy to its FEM counterpart \eqref{eq:Finelastic}, the semi-discrete model operator is completed by evolution equations for the internal variables $\GV{q}$ and reads
\begin{equation}
    \begin{aligned}\label{eq:F_inelastic_pinn}
        \GV{F}\EXPcol\Big(t, \mathcal{U}\left(\GV{x}\EXPcol, t, \GKap; \GThe_{\GV{u}}\right)&, \mathcal{Q}\left(\GV{x}\EXPcol, t, \GKap; \GThe_{\GV{q}}\right),\dot{\mathcal{Q}}\left(\GV{x}\EXPcol, t, \GKap; \GThe_{\GV{q}}\right), \GKap \Big) = \\
                &= 
        \footnotesize{\left\{\begin{matrix}
            \GV{g}\EXPcol\left(\mathcal{U}\left(\GV{x}\EXPcol, t, \GKap; \GThe_{\GV{u}} \right), \mathcal{Q}\left(\GV{x}\EXPcol, t, \GKap; \GThe_{\GV{q}} \right),\GKap \right) \\
            \LV{C}\INDN\left(t, \,  \mathcal{U}\left(\GV{x}\EXPN, t, \GKap; \GThe_{\GV{u}} \right), \mathcal{Q}\left(\GV{x}\EXPN, t, \GKap; \GThe_{\GV{q}}\right),\GKap \right) \\
            \LV{C}\INDD\left(t, \, \mathcal{U}\left(\GV{x}\EXPD, t, \GKap; \GThe_{\GV{u}}\right)\right) \\
            \dot{\mathcal{Q}}(\GV{x}\EXPcol, t, \GKap; \GThe_{\GV{q}})  - \GV{r}\INDq\left(\mathcal{U}(\GV{x}\EXPcol, t, \GKap;   \GThe_{\GV{u}}), \mathcal{Q}(\GV{x}\EXPcol, t, \GKap; \GThe_{\GV{q}}),\GKap\right)
        \end{matrix} \right\}}
    \end{aligned}
\end{equation}
There are two options to handle the time- (history-) dependence in the inelastic model above. First, the ansatz \eqref{eq:para_pinn_inelas} is referred to as a continuous time PINN \cite{raissi_PhysicsinformedNeuralNetworks_2019}. Alternatively, a time discretization scheme such as the backward Euler method can be applied, leading to a so-called discrete-time PINN \cite{raissi_PhysicsinformedNeuralNetworks_2019, wessels2020neural}. In the latter case, a separate discrete-time PINN needs to be trained at each time (load) step. In the following, we adopt the continuous time formulation, which is trained with discrete samples in space and time.  The inelastic state vector at one time (load) step $t_n$ is defined by 
\begin{equation}\label{eq:state_inelas_tn}
    \GV{y}_n\left(\GKap; \GV{\theta}\right) = 
    \left\{ 
           \begin{matrix} \mathcal{U}\left(\LV{x}\EXPcol_1, t_n, \GKap; \GV{\theta}_{\GV{u}}\right) \\
            \vdots \\\mathcal{U}\left(\GV{x}\EXPcol_{\numcol}, t_n, \GKap; \GV{\theta}_{\GV{u}}\right)\\
            \mathcal{Q}\left(\LV{x}\EXPcol_1, t_n, \GKap; \GV{\theta}_{\GV{q}}\right) \\
            \vdots \\\mathcal{Q}\left(\GV{x}\EXPcol_{\numcol}, t_n, \GKap; \GV{\theta}_{\GV{q}}\right)
            \end{matrix} 
        \right\}, \quad n=1,\ldots,t\INDN
\end{equation}  
and its rate by
\begin{equation}\label{eq:state_rate_inelas_tn}
    \dot{\GV{y}}_n\left(\GKap; \GThe\right) =  
    \left\{ 
           \begin{matrix} \dot{\mathcal{U}}\left(\LV{x}\EXPcol_1, t_n, \GKap; \GV{\theta}_{\GV{u}}\right) \\
            \vdots \\\dot{\mathcal{U}}\left(\GV{x}\EXPcol_{\numcol}, t_n, \GKap; \GV{\theta}_{\GV{u}}\right)\\
            \dot{\mathcal{Q}}\left(\LV{x}\EXPcol_1, t_n, \GKap; \GV{\theta}_{\GV{q}}\right) \\
            \vdots \\\dot{\mathcal{Q}}\left(\GV{x}\EXPcol_{\numcol}, t_n, \GKap; \GV{\theta}_{\GV{q}}\right)
            \end{matrix} 
        \right\}, \quad n=1,\ldots,t\INDN,
\end{equation}
whereby the full state vector $\GV{y}\INDa\left(\GKap; \GV{\theta}\right)$ and its rate $\dot{\GV{y}}\INDa\left(\GKap; \GV{\theta}\right)$ are obtained through the assembling step \eqref{eq:assembly_states}. 

Next, the  full inelastic discrete model $\GV{F}\EXPcol\INDa\left(\GV{y}\INDa\left(\GKap; \GV{\theta}\right), \dot{\GV{y}}\INDa\left(\GKap; \GV{\theta}\right), \GKap\right)$ is assembled as in Eq.~\eqref{eq:Ffull_param} and the training follows in analogy to Eq.~\eqref{eq:min_para_elas_1}. Parameter identification once again requires again minimizing \eqref{eq:loss_red_pinn} by solving the minimum problem \eqref{eq:argmin_red}. Here, we restrict ourselves to cases where the observation operator depends on the state $\GV{y}$ only, and not on its rate $\dot{\GV{y}}$. As a consequence, the same necessary FOC \eqref{eq:optimality_red_pinn} as in the elastic case can be applied.

\subsection{Virtual fields method and EUCLID}
\label{sec:virtualfields}

There are methods that do not fit exactly into either the reduced formulation or the all-at-once approach. The VFM, for instance, is based on an optimization problem over the parameter domain only, while replacing the unknown state with experimental data directly. Hence, it shares similarities with the reduced approach. In the following, we discuss the material parameter identification process for the VFM in Section~\ref{sec:vfm_sec4}. A generalization of the VFM to model discovery is EUCLID, presented in Section~\ref{sec:euclid_sec4}.  

\subsubsection{Virtual fields method}
\label{sec:vfm_sec4}
In the case of the VFM -- and for the equilibrium gap method as a particular case -- the minimization problem is given by 
\begin{equation}
\label{eq:argmin_VFM_elas}
    \GKap^{*} = \argmin_{\GKap} = \half \|\GV{F}\INDVF(t,\GV{d},\GKap) \|^2
\end{equation}
where the unknown state $\GV{y}$ is fully replaced by the experimental data vector $\GV{d}$. As mentioned earlier, the equilibrium gap method can be interpreted as finite element equations $\GV{F}(t,\GV{d},\GKap)=\bf{0}$, whereas the VFM adopts special ansatz functions for the virtual displacements, leading to modified equations $\GV{F}\INDVF(t,\GV{d},\GKap)=\bf{0}$. However, the general considerations are very similar. Thus, we omit the index $\mathcal{V}$ in the following.
The minimum problem~\eqref{eq:argmin_VFM_elas} implies the FOC
\begin{equation}
\label{eq:FOCVFMred}
    \left[\dif{\GV{F}(t,\GV{d},\GKap)}{\GKap}\right]^T \GV{F}(t,\GV{d},\GKap) = \GV{0},
\end{equation}
see Eq.~\eqref{eq:FOCVFMI} as well.

\paragraph{Problem class I: linear elasticity}
For the case of linear elasticity, the equilibrium conditions are defined as
\begin{equation}
\label{eq:FVFM}
    \GV{F}(t,\U,\GV{p},\GKap) := \GM{A}\INDas(t,\U) \GKap - \GV{p}\INDa(t,\GV{p}),
\end{equation}
see Eq.~\eqref{eq:FdefVFM}, or Eq.~\eqref{eqmf:system_virtual_fields_method} for a specific implementation. Here, however, the experimental data are inserted, including displacements $\U = \GV{d}\INDu$ and reaction forces $\GV{p} = \GV{d}\INDp$,
\begin{equation}
    \GV{F}(t,\GV{d},\GKap) = \GM{A}\INDas(t,\GV{d}\INDu) \GKap - \GV{p}\INDa(t,\GV{d}\INDp).
\end{equation}
In this case, the FOC~\eqref{eq:FOCVFMred} reads
\begin{equation}
\label{eq:redVFMresES}
    \GMT{A}\INDas(t,\GV{d}\INDu) \GM{A}\INDas(t,\GV{d}\INDu) \GKap 
    = 
    \GMT{A}\INDas(t,\GV{d}\INDu) \GV{p}\INDa(t,\GV{d}\INDp). 
\end{equation}
In other words, a very small system of linear equations has to be solved, which makes the method attractive for the problems under consideration.

To further reduce the effort of the matrix-matrix multiplication in Eq.~\eqref{eq:redVFMresES}, a different implementation can be considered, see App.~\ref{ap:system_matrices}, Eq.~\eqref{eqmf:system_virtual_fields_method}, where only one resulting force is extracted.

We further note that the VFM is closely related to the FEM-based reduced approach for problem class I. This is shown in \cite{roux_optimal_2020}, where it is observed that the objective functions of the minimization problems of the VFM and the FEM-based reduced approach differ only by the choice of the norm that measures the difference between the measured and the simulated displacements. 

\paragraph{Problem class II: hyperelasticity}

If the constitutive equations of hyperelasticity are linear in the parameters, Eq.~\eqref{eq:redVFMresES} holds as well. If the constitutive model depends non-linearly on $\GKap$, such as, e.g., for Ogden-type hyperelasticity models, the FOC~\eqref{eq:FOCVFMred} reads
\begin{equation}
    \left[
        \difn{\GV{g}\INDa(\GV{d},\GKap)}{\GKap}
    \right]^T
    \GV{g}\INDa(\GV{d},\GKap)
    =
    \left[
        \difn{\GV{g}}{\GKap}
    \right]^T \GV{g}(\GV{d}\INDu,\GKap)
    +
    \left[
        \difn{\GVq{g}}{\GKap}
    \right]^T \{\GVq{g}(\GV{d}\INDu,\GKap) - \GV{d}\INDp\}
    = \GV{0}
\end{equation}
with 
\begin{equation}
    \GV{F}(\GV{d},\GKap) = \GV{g}\INDa(\GV{d},\GKap) =
    \mzweiv{\GV{g}(\GV{d}\INDu,\GKap)}
           {\GVq{g}(\GV{d}\INDu,\GKap) - \GV{d}\INDp}.
\end{equation}
Hence, a system of $\numkappa$ non-linear equations must be solved to determine $\GKap$.

\paragraph{Problem class III: inelasticity}

For the case of inelasticity, various approaches are possible. We start by writing
\begin{equation*}
    \GKap^* = \argmin_{\GKap,\GV{q}\INDA} \half \|\GV{F}\INDA(\GV{d}\INDA,\GV{q}\INDA,\GKap)\|^2
\end{equation*}
with $\GV{F}\INDA$ defined in Eq.~\eqref{eq:abbrevFA} for the time-discretized DAE-system, where the displacements and the reaction forces are substituted by the experimental displacements and forces, whereas the internal variables are unknown. In this case, the FOCs read
\begin{equation}
    \begin{aligned}
        \left[
        \dif{\GV{F}\INDA}{\GV{q}\INDA}
        \right]^T
        \GV{F}\INDA &= \GV{0}\\
        \left[
        \dif{\GV{F}\INDA}{\GKap}
        \right]^T
        \GV{F}\INDA &= \GV{0}
    \end{aligned}
\end{equation}
and we obtain a coupled system of non-linear equations. For stiff problems or unstable evolution equations, as discussed in \cite{leistnerhartmannablizziegmann2020}, the fact that the integration step for the evolution equations are not exactly satisfied may lead to problems. Thus, a different approach may be needed, which is discussed in the following. 

First of all, as with the reduced formulation using the FEM, a time-continuous and a time-discrete formulation can be considered. In the time-continuous case, we can write 
\begin{align}
\label{eq:argmin_VFM_plas}
    \GKap^{*} &= \argmin_{\GKap}  \half \|\GV{g}\INDa(t,\GV{d},\GV{q}(t,\GKap),\GKap) \|^2 \\
    \text{subjected to } &\GVP{q}(t,\GKap) - \GV{r}\INDq(t,\GV{d},\GV{q}(t,\GKap),\GKap) = \GV{0},
\end{align}
see Eq.~\eqref{eq:DAEreduced} with $\GV{g}\INDa$ defined in Eq.~\eqref{eq:alleqlm}. In this case, the differential part of the DAE-system is treated in such a manner that it is fulfilled exactly. As before, the dependence on $\GV{d}$ only indicates that
the state $\GV{y}$ is replaced by the experimental data $\GV{d}$. Using Lagrange multipliers, we thus write 
\begin{equation}
    \phi(t,\GLam,\GVP{q}(t,\GKap),\GV{q}(t,\GKap),\GKap) 
    = \half \|\GV{g}\INDa(t,\GV{d},\GV{q}(t,\GKap),\GKap) \|^2 
    + \GLam^T \{\GVP{q}(t,\GKap) - \GV{r}\INDq(t,\GV{d},\GV{q}(t,\GKap),\GKap)\} \rightarrow \text{stat.}
    \label{eq:phiVFMDAE}
\end{equation}
The differential of this function with respect to $\GLam$, equated to zero, yields
\begin{equation}
\label{eq:ODEVFM}
    \GVP{q}(t,\GKap) - \GV{r}\INDq(t,\GV{d},\GV{q}(t,\GKap),\GKap) = \GV{0},
\end{equation}
implying
\begin{equation}
\label{eq:VFMmatrixODE}
    \dif{}{\GKap} \left\{\GVP{q}(t,\GKap) - \GV{r}\INDq(t,\GV{d},\GV{q}(t,\GKap),\GKap)\right\} = \GM{0}.
\end{equation}
Thus, differentiation of Eq.~\eqref{eq:phiVFMDAE} with respect to $\GKap$ leads to
\begin{equation}
\label{eq:VFMalgebraic}
    \left[
    \dif{\GV{g}\INDa}{\GV{q}} \dif{\GVh{q}}{\GKap} + \dif{\GV{g}\INDa}{\GKap}
    \right]^T
    \GV{g}\INDa(t,\GV{d},\GV{q}(t,\GKap),\GKap)
    =
    \GV{0}.
\end{equation}
In other words, an algebraic equation is required to obtain the parameters $\GKap$, and Eq.~\eqref{eq:ODEVFM} is necessary to compute the remaining unknowns. The term $\partial \GVh{q}/\partial \GKap$ in Eq.~\eqref{eq:VFMalgebraic} can be obtained by the matrix ODE-system \eqref{eq:VFMmatrixODE}
\begin{equation}
    \difn{}{t} \dif{\GV{q}}{\GKap} = \dif{\GV{r}\INDq}{\GV{q}} \dif{\GVh{q}}{\GKap} + \dif{\GV{r}\INDq}{\GKap}.
\end{equation}
This can be interpreted as the simultaneous sensitivity equation for the VFM. 

Alternatively, we can discretize Eq.~\eqref{eq:ODEVFM} by a time discretization scheme first, and determine the derivatives, such as for the internal numerical differentiation approach \cite{schittkowskibook2002}. 
In this case, we solve the minimum problem
\begin{equation}
\begin{aligned}
    \GKap^* &= \argmin_{\GKap} \half \|\GV{g}\INDA(\GV{d}\INDA,\GVh{q}\INDA(\GKap),\GKap\|^2 \\
    \text{subjected to }
    &\GV{l}\INDA(\GV{d}\INDA,\GVh{q}\INDA(\GKap),\GKap) = \GV{0}.
    \label{eq:minVFMreddiscr}
\end{aligned}
\end{equation}
Here, $\GVT{d}\INDA = \{\GVT{d}_1, \ldots, \GVT{d}_N\}$ contains the measured displacements and reaction forces, with $\GVT{d}\INDNP = \{{\GVT{d}\INDu}\INDNP, {\GVT{d}\INDp}\INDNP\}$. Further, $\GV{g}\INDA$ and $\GV{l}\INDA$ contain the discretized weak form and the integration step for the internal variables at all times $\tn$, $n=1,\ldots,t_N$. Similarly, we have
{\footnotesize
\begin{equation}
    \begin{aligned}
        \GVT{g}\INDA 
        &=
        \{
        \GVT{g}\INDa(t_1,{\GV{d}\INDu}_1,{\GV{d}\INDp}_1,\GV{q}_1,\GKap),
        \ldots,
        \GVT{g}\INDa(t_N,{\GV{d}\INDu}_N,{\GV{d}\INDp}_N,\GV{q}_N,\GKap)
        \} \\
        \GVT{l}\INDA 
        &=
        \{
        \GVT{l}(t_1,{\GV{d}\INDu}_1,\GV{q}_1,\GV{q}_0,\GKap),
        \ldots,
        \GVT{l}(t_N,{\GV{d}\INDu}_N,\GV{q}_N,\GV{q}_{N-1},\GKap)
        \}.
    \end{aligned}
\end{equation}}
Applying the Lagrange multiplier method to the problem  \eqref{eq:minVFMreddiscr} yields the objective function
\begin{equation}
    \phi(\GKap,\GVh{q}\INDA(\GKap),\GLam)
    =
    \half \| \GV{g}\INDA(\GV{d}\INDA,\GVh{q}\INDA(\GKap),\GKap) \|^2
    +
    \GLam^T \GV{l}\INDA(\GV{d}\INDA,\GVh{q}\INDA(\GKap),\GKap)
\end{equation}
leading to the necessary FOCs
\begin{equation}
\label{eq:gaTga}
    \begin{aligned}
        \left[
        \dif{\GV{g}\INDA}{\GV{q}\INDA}
        \dif{\GVh{q}\INDA}{\GKap}
        +
        \dif{\GV{g}\INDA}{\GKap}
        \right]^T
        \GV{g}\INDA
        &= \GV{0}, \\
        \GV{l}\INDA(\GV{d}\INDA,\GVh{q}\INDA,\GKap) 
        &= \GV{0},
    \end{aligned}
\end{equation}
where the matrix 
\begin{equation}
    \dif{\GV{l}\INDA}{\GV{q}\INDA}
    \dif{\GVh{q}\INDA}{\GKap}
    +
    \dif{\GV{l}\INDA}{\GKap}
    = \GV{0}
\end{equation}
vanishes due to condition \eqref{eq:gaTga}$_2$. In other words, Eq.~\eqref{eq:gaTga}$_1$ represents a small system of $\numkappa$ non-linear equations, which is coupled to the integration step of the internal variables \eqref{eq:gaTga}$_2$. Various special cases can be treated -- such as linearity in the material parameters, pre-calculation of the internal variables, and others, depending on the structure of the constitutive equations. This is not discussed in detail here.

\subsubsection{EUCLID}\label{sec:euclid_sec4}

In EUCLID, instead of a specific material model, a so-called model library with a large number of material parameters (e.g., in the order of hundreds \cite{marino_automated_2023} or thousands \cite{flaschel_automated_2023-2} parameters) is employed in the formulation of the model $\GV{F}$.
Applying the previously discussed identification methods to such a model library would result in a highly ill-posed minimization problem and in uninterpretable and impractical models with a large number of non-zero parameters.
In order to obtain interpretable models, the number of non-zero parameters contained in the identified $\GKap$ must be as small as possible, which is ensured using Lasso regularization by modifying the objective functions \eqref{eq:argmin_VFM_elas} and \eqref{eq:argmin_VFM_plas}. 

\paragraph{Problem class I: linear elasticity}
We note that, for the linear elastic case there is not much freedom in how to choose the material model.
The material characterization problem essentially boils down to identifying the elasticity tensor.
Thus, formulating a model library for linear elasticity is not meaningful, and applying EUCLID is unnecessary.

\paragraph{Problem class II: hyperelasticity}
For nonlinear elastic materials, the choice of the specific material model is not trivial, and EUCLID can be leveraged to select a suitable material model from a predefined model library.
The minimization problem that constitutes the core of EUCLID is formulated as a regularized version of \eqref{eq:argmin_VFM_elas}
\begin{equation}
    \GKap^{*} = \argmin_{\GKap} \half \| \GV{F}(t,\GV{d},\GKap) \|^2 + \lambda \|\GKap\|_p^p,
\end{equation}
with $\lambda > 0$ and $0 < p \leq 1$. 
A possible choice for a hyperelastic material model library is to choose a parametric ansatz for the strain energy density function that is linear in the material parameters $\GKap$ (see \cite{flaschel_unsupervised_2021} for details). 
In this way, the minimization problem simplifies to
\begin{equation}
\label{eqmf:EUCLID_linear}
    \GKap^{*} = \argmin_{\GKap} \half \|\GM{A}\INDas(t,\U) \GKap - \GV{p}\INDa(t,\GV{p})\|^2 + \lambda \|\GKap\|_p^p.
\end{equation}
Due to the non-smooth regularization term, deriving a linear system of equations via the FOC as for the VFM is not possible and the solver has to be chosen thoughtfully. 
Efficient solvers for problems of this type have been proposed, for instance, by \cite{tibshirani_regression_1996,efron_least_2004}.
The assumption that the strain energy density depends linearly on the material parameters covers well-known hyperelastic material models like Rivlin-Saunders or Hartmann-Neff-type models. Other models, e.g., those of Ogden type, do not fall into this category. In \cite{flaschel_automated_2023-2,marino_automated_2023}, a discretization of the parameter space is used to recast the latter models in the form \eqref{eqmf:EUCLID_linear}. So far, EUCLID has been proposed in conjunction with the VFM, which is also the basis for the following presentation. Nevertheless, the concept can be transferred to other identification approaches as well.

\paragraph{Problem class III: inelasticity}
For inelastic problems, EUCLID is formulated as a regularized version of \eqref{eq:argmin_VFM_plas}, i.e.,
\begin{equation}
\begin{aligned}
    \GKap^{*} &= \argmin_{\GKap}  \half \|\GV{g}\INDa(t,\GV{d},\GV{q}(t,\GKap),\GKap) \|^2 + \lambda \|\GKap\|_p^p \\
    \text{subjected to } &\GVP{q}(t,\GKap) - \GV{r}\INDq(t,\GV{d},\GV{q}(t,\GKap),\GKap) = \GV{0}.
\end{aligned}
\end{equation}
The choice of the model library, which determines the characteristics of $\GV{g}\INDa(t,\GV{d},\GV{q}(t,\GKap),\GKap)$ and the ODE constraint, is not as straightforward as for elastic problems. In \cite{flaschel_discovering_2022}, a Fourier expansion of the yield surface is proposed to formulate a general library for plasticity models, and in \cite{flaschel_automated_2023-1}, the concept of generalized standard materials \cite{halphen_sur_1975} is leveraged to construct a general model library containing elastic, viscoelastic, plastic, and viscoplastic material models.

\subsection{All-at-once approach}
\label{sec:all-at-once}

The all-at-once approach, also referred to as the "piggyback" or one-shot approach to inverse problems, see \cite{guth2020ensemble}, directly employs the parameter identification problem \eqref{eq:abstract_identification} and the objective function \eqref{eq:loss_all-at-once}. Hence, the implicit function theorem does not need to be applied, and assumption \eqref{eq:conditions-implicit-function-theorem} as well as the necessity of defining a solution operator are avoided. The approach can be seen as a parameter estimation problem for the joint vector $\GBet = \{\GVT{y} \; \GKap^T\}^T$, which is, in the case of elasticity, related to the data by 
\begin{equation}
    \label{eq:allAtOnce_abstract}
    \left\{ 
    \begin{array}{c}
        \GV{F}(\GV{y},\GKap) \\
        \GV{O}(\GV{y})
    \end{array}
    \right\}
    =
    \left\{ 
    \begin{matrix}
        \GV{0} \\
        \GV{d}
    \end{matrix}
    \right\}.
\end{equation}
Then, the generalized minimization problem of the loss function \eqref{eq:loss_all-at-once} reads
\begin{equation}
    \GBet^* = \argmin_{\GBet} \phi_{\text{aao}} \left(\GBet\right)
\end{equation}
In the all-at-once approach, we jointly estimate the state and the parameter vector over $\Omega_{\GV{y}} \times \Omega_{\GKap}$, which is a much larger space than in the reduced formulation. Inverse problems involving the estimation of the state vector are typically ill-posed, and, therefore, the aspect of regularization becomes particularly important, see \cite{benning2018modern} for a recent overview. In general, a regularization may contain both the state and the parameter vector, allowing for a robust solution of the inverse problem. The minimization problem with regularization reads 
\begin{equation}
    \GBet^* = \argmin_{\GBet} \phi_{\text{aao}} \left(\GBet\right) + r\left(\GBet\right)
\end{equation}
The Tikhonov regularization, for instance, reads
\begin{equation}
    \label{eq:general_regularization}
    r(\GV{y},\GKap) = \frac{\gamma_{\text{S}}}{2}\|\GV{y} - \GV{y}_0\|^2 + \frac{\gamma_{\text{P}}}{2}\|\GKap - \GKap_0\|^2.
\end{equation}
Here, $\GV{y}_0$ and $\GKap_0$ denote nominal values, which are either zero or express prior knowledge about the expected state/parameter values, and $\gamma_{\text{S}},\gamma_{\text{P}}>0$ are penalty parameters. However, we can also consider cases where $r(\GV{y},\GKap)$ only contains a penalty regularization on the state, as discussed in \cite{kaltenbacher2016regularization}, because the high-dimensional state vector is the main source of ill-posedness in the inverse problem. An alternative approach to solve ill-posed inverse problems without explicit regularization is the Landweber iteration, a particular form of gradient descent. Interestingly, the Landweber iteration results in quite different implementations for the reduced and all-at-once approaches, as noted in \cite{kaltenbacher_regularization_2016}. In particular, the reduced Landweber iteration requires the repetitive solution of systems of equations, whereas the all-at-once Landweber iteration only requires matrix-vector multiplications. A short description of the method can be found in Appendix~\ref{app:Landweber}. 

Despite the challenges of a high-dimensional parameter space, the all-at-once approach is appealing because of its flexibility, but also because it may result in efficient iterative methods. A Bayesian all-at-once approach is outlined, for instance, in \cite{schlintl2021all}. In \cite{guth2020ensemble}, the all-at-once approach is presented in combination with an ensemble Kalman inversion method. Therein, a similarity to the PINN setting is indicated \cite{guth2020ensemble}, and in Section~\ref{sec:aao_pinn}, we elaborate on this connection in detail.

\subsubsection{FE- and VFM-based all-at-once approaches} %(NLS-FEM-DIC)
\label{sec:nlsfemdicallatonce}
In this section, we propose a new FEM-based all-at-once approach.
For simplicity, we drop the regularization term \eqref{eq:general_regularization} for the following derivations. Starting from the objective function
\begin{equation}
    \phi_{\text{aao}}(\GV{y},\GKap) 
    =
    \frac{\sigma\INDs}{2}
    \|\GV{F}(\GV{y},\GKap)\|^2
    +
    \frac{\sigma\INDd}{2}
    \|\GV{O}(\GV{y}) - \GV{d}\|^2
\end{equation}
with $\sigma\INDs,\sigma\INDd>0$, the necessary FOCs yield the two coupled systems of non-linear equations
\begin{equation}
    \label{eq:resultnlsfemdicAAO}
    \begin{split}
    \sigma\INDs
    \left[
        \dif{\GV{F}}{\GV{y}}
    \right]^T
    \GV{F}(\GV{y},\GKap)
    +
    \sigma\INDd
    \left[
        \difn{\GV{O}}{\GV{y}}
    \right]^T
    \{\GV{O}(\GV{y}) - \GV{d}\}
    &= \GV{0},
    \\
    \left[
        \dif{\GV{F}}{\GKap}
    \right]^T
    \GV{F}(\GV{y},\GKap)  
    &=
    \LV{0},
    \end{split}
\end{equation}
consisting of $\nums$ and $\numkappa$ equations, respectively. Note that the discretized equilibrium conditions $\GV{F}(\GV{y},\GKap)=\GV{0}$ are no longer exactly fulfilled, see Eq.~\eqref{eq:resultnlsfemdicAAO}$_2$. If a Newton-like scheme is applied, the second derivatives of $\GV{F}$ with respect to $\GV{y}$ and $\GKap$ are required, making the scheme difficult from an implementation point of view. Thus, a Gauss-Newton scheme that circumvents this drawback is of higher interest, or the Landweber iteration may be applied, see Appendix~\ref{app:Landweber}.

\paragraph{Problem class I: linear elasticity}

For linear elastic problems, the function $\GV{F}(\GV{y},\GKap)$ is given by Eq.~\eqref{eq:Felasticdef}. In this case, Eqns.~\eqref{eq:resultnlsfemdicAAO} can be written as
\begin{equation}
    \label{eq:AAOfemlin}
    \begin{split}
        \sigma\INDs \GMT{A}\INDup \big\{\GM{A}\INDup \GV{y} - \GV{f}\big\}
        +\sigma\INDd \left[\difn{\GV{O}}{\GV{y}}\right]^T \{\GV{O}(\GV{y}) - \GV{d}\}
        &= \GV{0}, \\
        \GMT{A}\INDas \big\{\GM{A}\INDas \GKap - \GV{p}\INDa\big\}
        &= \GV{0}
    \end{split}
\end{equation}
where we also used Eq.~\eqref{eq:Aaskappa}. Since $\GM{A}\INDup(\GKap)$ depends on $\GKap$ and $\GM{A}\INDas(\U,\Uq)$ depends on $\U$, the equations are certainly non-linear (in the displacements $\U$). For example, an iterative Block-Gauss Seidel method could be considered to obtain an efficient procedure, where step-wise systems of linear equations are solved to determine $\U$, $\GV{p}$, and $\GKap$. An alternative is provided in Appendix~\ref{app:Landweber}. An open question is the influence of uncertain measurement data in Eq.~\eqref{eq:AAOfemlin}$_1$, which affect the violation of the equilibrium conditions due to the coupling to Eq.~\eqref{eq:AAOfemlin}$_2$.

\paragraph{Problem class II: hyperelasticity}
In the case of hyperelasticity with constitutive models that depend linearly on the material parameters, we have $\GV{F}(\GV{y},\GKap)$ defined in Eq.~\eqref{eq:hypernonl}, which is coupled with Eq.~\eqref{eq:AAOfemlin}$_2$ (of course, with a matrix $\GM{A}\INDas(\U,\Uq)$ written for large strains). Thus, we obtain a very similar structure of equations. If this is not the case, the more general form \eqref{eq:resultnlsfemdicAAO} has to be evaluated. 

\paragraph{Problem class III: inelasticity}

In a consistent approach, we would have to consider the DAE-system~\eqref{eq:DAELagrMult} with the functions \eqref{eq:DAEdef}, so that even the geometric boundary conditions would not be exactly fulfilled later. In the following, however, we use version \eqref{eq:DAEreduced}, where Dirichlet boundary conditions are fulfilled exactly. We define the objective function 
\begin{equation}
\label{eq:AAOFEMclassIII}
\begin{split}
    \phi(\GVP{y},\GV{y},\GKap) 
    &=
    \frac{\sigma\INDs}{2}
    d\INDS(\GV{F}(\GVP{y},\GV{y},\GKap),\GV{0}) 
    + \frac{\sigma\INDd}{2} d\INDO(\GV{O}(\GV{y}),\GV{d}) \\
    &=
    \frac{\sigma\INDs}{2}
    \|\GV{F}(\GVP{y},\GV{y},\GKap)\|^2
    +
    \frac{\sigma\INDd}{2}
    \|\GV{O}(\GV{y}) - \GV{d}\|^2,
\end{split}
\end{equation}
which must be minimized. The FOCs are the three coupled equations 
\begin{equation}
    \label{eq:resultDAEAAO}
    \begin{split}
    \left[
        \dif{\GV{F}}{\GVP{y}}
    \right]^T
    \GV{F}(\GVP{y},\GV{y},\GKap)  
    &=
    \LV{0} \\
    \sigma\INDs
    \left[
        \dif{\GV{F}}{\GV{y}}
    \right]^T
    \GV{F}(\GVP{y},\GV{y},\GKap)
    +
    \sigma\INDd
    \left[
        \difn{\GV{O}}{\GV{y}}
    \right]^T
    \{\GV{O}(\GV{y}) - \GV{d}\}
    &= \GV{0},
    \\
    \left[
        \dif{\GV{F}}{\GKap}
    \right]^T
    \GV{F}(\GVP{y},\GV{y},\GKap)  
    &=
    \LV{0}.
    \end{split}
\end{equation}
With $\GV{F}$ given by Eq.~\eqref{eq:DAEreduced}, Eq.~\eqref{eq:resultDAEAAO}$_1$ reads 
\begin{equation}
\label{eq:qpminr}
    \left[\GM{0} \; \GM{I}\right] \mzweiv{\GVh{g}\INDa(t,\U,\GV{p},\GV{q},\GKap)}{\GVP{q}(t) - \GV{r}\INDq(t,\U,\GV{q},\GKap)}
    =
    \GVP{q}(t) - \GV{r}\INDq(t,\U,\GV{q},\GKap) = \GV{0},
\end{equation}
implying the fulfillment of the differential part of the DAE-system. In Appendix~\ref{sec:derivAAOFEM} the functional matrices of the decomposed system are derived, leading to the DAE-system
\begin{equation}
\begin{aligned}
    \sigma\INDs 
    \left\{
        \Big[\dif{\GV{g}}{\U}\Big]^T \{\GV{g}-\GVq{p}\}
        +
        \Big[\dif{\GVq{g}}{\U}\Big]^T \{\GVq{g}-\GV{p}\}
    \right\}
    +
    \sigma\INDd
    \Big[\difn{\GV{O}\INDu}{\U}\Big]^T \{\GV{O}\INDu-\GV{d}\INDu\}
    &= \GV{0}, 
    \\
    -\sigma\INDs 
    \{\GVq{g}-\GV{p}\}
    +
    \sigma\INDd
    \Big[\difn{\GV{O}\INDp}{\GV{p}}\Big]^T \{\GV{O}\INDp-\GV{d}\INDp\}
    &= \GV{0}, 
    \\
    \Big[\dif{\GV{g}}{\GKap}\Big]^T \{\GV{g}-\GVq{p}\}
    +
    \Big[\dif{\GVq{g}}{\GKap}\Big]^T \{\GVq{g}-\GV{p}\}
    &= \GV{0},
    \\
    \GVP{q}(t) - \GV{r}\INDq(t,\U,\GV{q},\GKap) 
    &= \GV{0}
\end{aligned}
\label{eq:concreteFEAAO}
\end{equation}
where Eq.~\eqref{eq:qpminr} represents the differential part of the DAE-system.

An alternative approach might be to apply the implicit function theorem $\GVP{y}(\GV{y})$ inserted into Eq.~\eqref{eq:AAOFEMclassIII}. Then, a reduction of the number of equations is obtained. Alternatively, the time discretization of the DAE-system can be performed first, resulting in a similar approach as for problems class II. 

\paragraph{Further considerations}

Regarding the all-at-once approach of the case of linear elasticity, two aspects are considered. First, an efficiency treatment reducing the matrix-matrix multiplication in Eq.~\eqref{eq:AAOfemlin} can be carried out, and only displacement data are taken into account 
\begin{equation}
    \label{eq:aao_Euclidean_norm}
    \begin{aligned}
    \{\U^*,\GKap^*\} 
    &= \argmin_{\U,\GKap} \frac{\sigma\INDs}{2} \Big(\| \GM{A}(\U,\Uq)\GKap - \GVc{d}\INDp \|^2\Big) + \frac{\sigma\INDd}{2} \Big( \| \U - \GV{d}\INDu \|^2\Big),\\
    &= \argmin_{\U,\GKap} \frac{\sigma\INDs}{2}\Big(\| \GM{K}\EXPfr(\GKap) \U + \GMq{K}\EXPfr(\GKap) \Uq - \GVc{d}\INDp\|^2\Big) + \frac{\sigma\INDd}{2} \Big( \| \U - \GV{d}\INDu \|^2\Big).
    \end{aligned}
\end{equation}
Here, $\GVc{d}\INDp$ describes the measurement data of the reaction forces. In the following, the above method is referred to as AAO-FEM. The matrices, e.g.,\ $\GM{K}\EXPfr$, are defined in Eq.~\eqref{eq:Kfr}.

Second, it should be noted that we have only considered the Euclidean norm in the objective function above. Inspired by the considerations by \cite{roux_optimal_2020} -- who show that different calibration methods based on full-field data can be interpreted as similar minimization problems which differ only in the choice of norm -- we also explore the possibility of changing the norm to the semi-norm $\left\| \GV{v} \right\|_{\GM{K}\EXPfr} = \left\| \GM{K}\EXPfr \GV{v} \right\|$, here applied to the distance function of the observation 
\begin{equation}
    \label{eq:aao_K_norm}
    \begin{aligned}
    \{\U^*,\GKap^*\}
    &= \argmin_{\U,\GKap} \frac{\sigma\INDs}{2} \Big(\| \GM{K}\EXPfr(\GKap) \U + \GMq{K}\EXPfr(\GKap) \Uq - \GVc{p} \|^2\Big) + \frac{\sigma\INDd}{2} \Big( \| \U - \GV{d}\INDu \|^2_{\GM{K}\EXPfr(\GKap)}\Big),\\
    &= \argmin_{\U,\GKap} \frac{\sigma\INDs}{2} \Big(\left\| \GM{A}(\U,\Uq)\GKap  - \GVc{p}\right\|^2 \Big) + \frac{\sigma\INDd}{2} \Big( \left\| \left[\GM{A}(\U,\Uq)-\GM{A}(\GV{d}\INDu,\Uq)\right]\GKap \right\|^2\Big).   
    \end{aligned}
\end{equation}
Since in \cite{roux_optimal_2020} the Euclidean norm is related to the FEM-based reduced approach and the $\GM{K}\EXPfr$-semi-norm is related to the VFM, we denote the method \eqref{eq:aao_K_norm} as AAO-VFM.
As changing the norm to a semi-norm has an influence on the solution uniqueness, a proper regularization might be necessary to treat the problem, which is beyond the scope of this work. 

\subsubsection{PINN-based all-at-once approach}
\label{sec:aao_pinn}
The training of an inverse \ac{PINN} can be classified as an all-at-once approach: Both network parameters $\GV{\theta}$ and material parameters $\GKap$ are identified simultaneously, expressed by the vector of trainable parameters $\GBet = \left \{\GKap^T \ \GVT{\theta}\right \}^T$, see also Table~\ref{tab:variables_unified}. In comparison to the AAO-FEM approach, the state vector $\GV{y}$ is not identified directly, but through its parametrization in network parameters $\GV{\theta}$, as outlined below for the different problem classes.

\begin{table}
\caption{Notation of state and parameter vectors used in the unified framework.}
\label{tab:variables_unified}
\centering
\begin{tabular}{|c|c|}
\hline
     $\GV{\kappa}$ & vector of material parameters \\
     \hline 
     $\GV{\theta}$ & neural network parameter \\
     & (state parametrization) \\
     \hline 
     $\GV{\beta}$ & joint parameter vector \\
    & $\GV{\beta}=\{\GKap^T \GVT{y}\}^T$ for AAO-FEM and AAO-VFM \\
    & $\GV{\beta}=\{\GKap^T \GVT{\theta}\}^T$ for AAO-PINN \\
     \hline
\end{tabular}
\end{table}

\paragraph{Problem classes I \& II: elasticity}

The ansatz function for AAO-PINNs in elasticity is given by
\begin{equation}\label{eq:ansatz_forward_pinn}
    \GV{u}(\GV{x}, t; \GV{\theta}) \approx \mathcal{U}(\GV{x}, t; \GV{\theta})
\end{equation}
and the states at one time (load) step $t_n$ are defined as
\begin{equation}\label{eq:pinn_state_aao}
    \GV{y}_n\left(\GV{\theta}\right) = 
    \left\{ 
           \begin{matrix} \mathcal{U}\left(\LV{x}\EXPcol_1, t_n; \GV{\theta}\right) \\
            \vdots \\\mathcal{U}\left(\GV{x}\EXPcol_{\numcol}, t_n; \GV{\theta}\right)
            \end{matrix} 
        \right\}, \quad n=1,\ldots,t\INDN.
\end{equation}  
As for parametric PINNs, the full state vector $\GV{y}\INDa\left(\GV{\theta}\right)$ and the full elastic discrete model $\GV{F}\EXPcol\INDa\left(\GV{y}\left(\GThe\right)\right)$  are assembled in analogy to \eqref{eq:assembly_states} and \eqref{eq:Ffull_param}, respectively.

For parameter identification, the PINN needs to map some data $\GV{d}$, which can be either strain or displacement fields, as reflected by the observation operator $\GV{O}(\GVh{y}(\GV{\theta}))$. This constraint arises from the discretization of the parameter identification problem \eqref{eq:abstract_identification} with $\GVh{y}\INDa(\GThe)$
\begin{equation}
\label{eq:loss_inverse}
    \begin{aligned}
        {\{\GV{\theta}^*, \GV{\kappa}^*\}} &= \argmin_{\GV{\theta}, \GKap} \phi_{\text{aao}}(\GV{y}\INDa(\GThe), \GKap)\\
        &= \argmin_{\GV{\theta}, \GKap} \half \big( \sigma\INDs \| \GV{F}\EXPcol\INDa(\GV{y}\INDa(\GV{\theta}), \GV{\kappa}) \|^2 +\ \sigma\INDd\|\GV{O}(\GV{y}\INDa(\GV{\theta}))- \GV{d} \|^2\big).
    \end{aligned}
\end{equation}
To avoid trivial solutions and ensure the identifiability of material parameters $\GV{\kappa}$, it is important to incorporate force information into the loss function. This can be either achieved via (i) volumetric forces, (ii) Neumann boundary conditions, or (iii) the balance of internal and external work, see \cite{henkes_PhysicsInformedNeural_2022, anton_IdentificationMaterialParameters_2022}, for example. For the aspect of identifiability, see also Sec.~\ref{sec:identifiability}.

Since network and material parameters $\GThe$ and $\GKap$ differ considerably in their respective order of magnitude, the formulation of the loss function  is crucial for the convergence of \eqref{eq:loss_inverse}, as discussed in detail in \cite{anton_IdentificationMaterialParameters_2022}. This might be achieved, for example, by adaptive weighting of the loss terms \cite{wang_UnderstandingMitigatingGradient_2021, mcclenny2023self}. 

The necessary FOCs related to \eqref{eq:loss_inverse} are given by
\begin{equation}
\label{eq:foc_pinn_aao}
\begin{aligned}
    \gat{\GThe}{\phi_{\text{aao}}}{\GBet}{\Delta \GThe} &= \GV{0}, \\ 
    \gat{\GKap}{\phi_{\text{aao}}}{\GBet}{\Delta \GKap} &= \GV{0}, \\
\end{aligned}
\end{equation}
which yields
\begin{equation}
\begin{aligned}
    \sigma\INDs \left[ \dif{\GV{F}\INDa\EXPcol}{\GV{y}\INDa} \, \difn{{\GVh{y}\INDa}}{\GThe} \right]^T \GV{F}\INDa\EXPcol(\GVh{y}\INDa(\GThe), \GKap) + \sigma\INDd \left[\difn{\GV{O}}{\GVh{y}\INDa} \difn{{\GVh{y}\INDa}}{\GThe} \right] \{ \GV{O}(\GVh{y}\INDa(\GThe)) - \GV{d} \} &= \GV{0} \\
    \left[ \dif{\GV{F}\INDa\EXPcol}{\GKap}\right]^T \GV{F}\INDa\EXPcol(\GVh{y}\INDa(\GThe), \GKap) &= \GV{0}
\end{aligned}
\end{equation}

\paragraph{Problem class III: inelasticity}

For inelastic problems, the ansatz~\eqref{eq:ansatz_forward_pinn} needs to be extended by the internal variables $\GV{q}$
\begin{equation}\label{eq:ansatz_inelastic}
    \begin{aligned}
    \mzweiv{\GV{u}\left(\GV{x}, t\right)}{\GV{q}\left(\GV{x}, t\right)} 
        \approx 
        \left\{ \begin{matrix}
            \mathcal{U}\left(\GV{x}, t; \GV{\theta}_{\GV{u}}\right) \\
            \mathcal{Q}\left(\GV{x}, t; \GV{\theta}_{\GV{q}}\right)
        \end{matrix}\right\}.
    \end{aligned}
\end{equation}
As for inelastic parametric PINNs, two different networks  parametrized in $\GV{\theta}_{\GV{u}}$ and $\GV{\theta}_{\GV{q}}$, respectively, are employed for ease of notation. We again define the state vector for each loading step by
\begin{equation}
    \GV{y}_n\left(\GV{\theta}\right) = 
    \left\{ 
           \begin{matrix} \mathcal{U}(\LV{x}\EXPcol_1, t_n; \GV{\theta}) \\
            \vdots \\
            \mathcal{U}(\GV{x}\EXPcol_{\numcol}, t_n; \GV{\theta})\\
            \mathcal{Q}(\LV{x}\EXPcol_1, t_n; \GV{\theta}) \\
            \vdots \\
            \mathcal{Q}(\GV{x}\EXPcol_{\numcol}, t_n; \GV{\theta})
            \end{matrix} 
        \right\}, \quad n=1,\ldots,t\INDN
\end{equation}  
and assemble the full state vector $\GV{y}\INDa(\GThe)$ and its rate $\GVP{y}\INDa\left(\GThe\right)$ in analogy to Eqs.~\eqref{eq:assembly_states} and \eqref{eq:state_rate_inelas_tn}, respectively. The full inelastic discrete model  $\GV{F}\EXPcol\INDa(\GVP{y}\INDa(\GThe),\GV{y}\INDa(\GThe)),\GKap)$ follows in analogy to Eqns.~\eqref{eq:F_inelastic_pinn} and \eqref{eq:Ffull_param}. Finally, the material parameters are identified by determining
\begin{equation}\label{eq:argmin_inverse_inelas}
    \begin{aligned}
        \{\GV{\theta}^*, \GKap^*\} &= \argmin_{\GV{\theta}, \GKap} \phi_{\text{aao}}(\GVP{y}\INDa(\GThe),\GV{y}\INDa(\GThe), \GKap)\\
        &= \argmin_{\GV{\theta}, \GKap} 
        \half \Big( \sigma\INDs 
        \| \GV{F}\EXPcol\INDa(\GVP{y}\INDa(\GV{\theta}),\GV{y}\INDa(\GV{\theta}), \GV{\kappa}) \|^2 + \ \sigma\INDd \|\GV{O}(\GV{y}\INDa(\GV{\theta}))- \GV{d} \|^2\Big).
    \end{aligned}
\end{equation}
The FOCs \eqref{eq:foc_pinn_aao} of the minimum problem \eqref{eq:argmin_inverse_inelas} read
\begin{align}
\begin{split}
    \sigma\INDs 
    \left[ 
        \dif{\GV{F}\INDa\EXPcol}{\GVP{y}\INDa} \, 
        \difn{ \dot{\GVh{y}}\INDa}{ \GThe} 
    \right]^T 
    \GV{F}\INDa\EXPcol(\GVhp{y}\INDa(\GThe),\GVh{y}\INDa(\GThe),\GKap) 
    + \sigma\INDs 
    \left[ 
        \dif{\GV{F}\INDa\EXPcol}{\GV{y}\INDa} \, 
        \difn{ \GVh{y}\INDa}{ \GThe} 
    \right]^T 
    \GV{F}\INDa\EXPcol(\GVhp{y}\INDa\left(\GThe\right),\GVh{y}\INDa(\GThe),\GKap) \\
    + \sigma\INDd 
    \left[
        \difn{\GV{O}}{\GV{y}\INDa} \difn{\GVh{y}\INDa}{\GThe} 
    \right] 
    \left\{ \GV{O}(\GVh{y}\INDa(\GThe)) - \GV{d} \right\} &= \GV{0},
    \end{split}\\
    \left[ 
        \dif{ \GV{F}\INDa\EXPcol}{ \GKap}
    \right]^T 
    \GV{F}\INDa\EXPcol(\GVhp{y}\INDa(\GThe),\GVh{y}\INDa(\GThe),\GKap) &= \GV{0}.    
\end{align}

%*************************************
\section{Statistical parameter inference}
\label{sec:locident}

In this section, we address the parameter identification problem in a statistical setting and discuss issues of identifiability and uncertainty quantification. The section covers both frequentist and Bayesian approaches. The former assume a true, but unknown, deterministic material parameter vector, whereas the latter treat the unknown material parameter vector as a random variable with a prior distribution that is conditioned on data. 
To illustrate the concepts, we first consider problem classes I and II (elasticity) in the following. The statistical counterpart of the deterministic parameter identification problem defined in Eq.~\eqref{eq:abstract_identification}  reads
\begin{equation}
\label{eq:model_with_noise}
\begin{aligned}
    \GV{F}(\GV{y},\GKap) &= \GV{\epsilon}, \\
    \GV{O} \left(\GV{y}\right) &= \GV{d} + \GV{e},
\end{aligned}    
\end{equation}
which now additionally contains the observation noise vector $\GV{e}$ and the model error $\GV{\epsilon}$. In the following, we assume that $\GV{e}$ is normally distributed with mean value $\GV{0}$ and positive definite covariance matrix $\GV{\Sigma}_{\GV{e}}$, i.e., $\GV{e} \sim \mathcal{N}(\GV{0},\GV{\Sigma}_{\GV{e}})$. More generally, $\GM{\Sigma}_{\GV{a}}$ will from now on refer to the covariance matrix of a random vector $\GV{a}$, defined by 
\begin{equation}
    \GM{\Sigma}_{\GV{a}} = \mathbb{E}[\left\{\GV{a} -\mathbb{E}\left\{ \GV{a} \right\}\right\} \left\{\GV{a} -\mathbb{E}\left\{ \GV{a} \right\}\right\}^T].
\end{equation} 
The quantity $\GV{\epsilon}$ can represent some missing part of the physics that is difficult to resolve, or simply numerical errors, see also Section~\ref{sec:bayesian}. In the following, we restrict ourselves to problems where $\GV{\epsilon}= \GV{0}$. 

With $\GV{s}(\GKap)=\GV{O}(\GVh{y}(\GKap))$, the statistical parameter identification problem \eqref{eq:model_with_noise} in the \textit{\textbf{reduced approach}} gives rise to the  conditional probability density
\begin{equation}\label{eq:statistical_model}
\begin{aligned}
    p(\GV{d} \vert \GKap) &= \mathcal{N}(\GV{s}(\GKap),\GV{\Sigma}_{\GV{e}}) \\
    &=  \frac{1}{(2 \pi)^{\numD/2} \det(\GV{\Sigma}_{\GV{e}})^{1/2}} \:\mathrm{exp}\left(-\frac{1}{2} \left(\GV{s}(\GKap) - \GV{d} \right)^\top\GV{\Sigma}_{\GV{e}}^{-1} \left( \GV{s}(\GKap) - \GV{d}\right)\right),
\end{aligned}
\end{equation}
i.e., we model the probability density of the data conditional on a specific value of the parameters. Here $\numD$ represents the number of data variables. The conditional density can be linked to the likelihood function as $p(\GV{d}\vert\GKap) = L_{\GV{d}}(\GKap)$. The likelihood function plays a crucial role for both frequentist and Bayesian approaches. 

Regarding the \textit{\textbf{all-at-once formulation}}, cf. \eqref{eq:allAtOnce_abstract}, we consider the statistical model 
\begin{equation}\label{eq:aao_statistic}
\begin{aligned}
    \tilde{\GV{s}}\left(\GBet\right) =
    \left\{ \begin{matrix}
        \GV{F}(\GV{y},\GKap) \\
        \GV{O}\left(\GV{y}\right)
    \end{matrix}
    \right\}
    =
    \left\{
    \begin{matrix}
        \GV{0} \\
        \GV{d} + \GV{e},
    \end{matrix}
    \right\} = \tilde{\GV{d}} + \tilde{\GV{e}}
\end{aligned}    
\end{equation}
where again $\GVt{e} \sim \mathcal{N}(\GV{0},\GV{\Sigma}_{\GVt{e}})$. Then, we obtain the likelihood function $L_{\GVt{d}}(\GBet)$ in complete analogy to \eqref{eq:statistical_model} as
\begin{equation}
\begin{aligned}
\label{eq:statistical_model_aao}
    p(\GVt{d} \vert \GBet) 
    &=  \frac{1}{(2 \pi)^{(\numD+n_s)/2} \det(\GV{\Sigma}_{\GVt{e}})^{1/2}} \:\mathrm{exp}\left(-\frac{1}{2} \left(\GVt{s}(\GBet) - \GVt{d} \right)^\top\GV{\Sigma}_{\GVt{e}}^{-1} \left( \GVt{s}(\GBet) - \GVt{d}\right)\right).
\end{aligned}
\end{equation}
Here, $n_s$ represents the number of state variables. It should be noted that the high-dimensional parameter vector $\GBet$ poses severe problems for the inference approach. Bayesian approaches are particularly appealing in this case because the prior enables regularization, which is needed when inferring a high-dimensional state variable. A Bayesian all-at-once approach was outlined in \cite{schlintl2021all}.

The remaining part of this section covers frequentist approaches in Section \ref{sec:frequentist}, while Bayesian approaches are discussed in Section \ref{sec:bayesian-inference}. A two-step approach is covered in Section \ref{sec:two-step-inference}, which is important to calibrate complex models. 

\subsection{Frequentist approach}
\label{sec:frequentist}

Here, we mainly cover aspects of estimation and asymptotic uncertainty analysis, whereas hypothesis testing is not  addressed. Our goal is to define a point estimator $\GKap^*$, which represents a single best guess for the parameters given the data. We thereby focus on the reduced approach. Various methods, such as the method of moments and the maximum likelihood method are available to this end, see \cite{wasserman2004all}. With the maximum likelihood method -- the most common method for parametric frequentist inference -- we recover the ordinary LS formulations
\begin{equation}\label{eq:frequentist_point_estimate}
    \begin{aligned}
    \GKap^* = \text{arg max}_{\GKap} \, L_{\GV{d}}(\GKap)&= \argmin_{\GKap} \Big(- \ln L_{\GV{d}}(\GKap)\Big) = \argmin_{\GKap} \half \|\GV{s}(\GKap) - \GV{d} \|^2
    %&= \argmin_{\GKap} \half \|\GV{r}(\GKap) \|^2
    \end{aligned}
\end{equation}
in the \textit{\textbf{reduced formulation}}, $\GM{\Sigma}_{\GV{e}}=\GM{I}$, see also Eq.\eqref{eq:objectiveFunction}. The minimum of the LS objective \eqref{eq:frequentist_point_estimate} can be characterized by the normal equation 
\begin{equation}
    \GMT{J}(\GKap) \{\GV{s}(\GKap) - \GV{d}\} = \GV{0}, 
\end{equation}
which, in the linear case $\GV{s}(\GKap) = \GV{J} \GKap$, simplifies to 
\begin{equation}
    \GMT{J} \GM{J} \GKap = \GMT{J} \GV{d}.
\end{equation}
Provided that some regularity conditions on the parameter domain and the map $\GKap \mapsto \GV{s}(\GKap)$ are satisfied, the NLS estimator asymptotically follows a normal distribution, i.e., 
\begin{equation}
    \label{eq:normality-NLS}
     \sqrt{\numD} \{\GKap^* - \GKap_0\} \rightarrow \mathcal{N}\big(\GV{0},\GMM{Q}(\GKap_0) \GM{Z}(\GKap_0) \GMM{Q}(\GKap_0)\big)
\end{equation}
in the infinite-data limit $\numD \rightarrow \infty$, where $\GKap_0$ denotes the true but unknown parameter value and where the matrices $\GM{Q}$ and $\GM{Z}$ are detailed in Section~\ref{sec:uncertainty-inference} and \ref{sec:twoStep_frequentist}, in the context of uncertainty quantification. Before that, the important topic of parameter identifiability is discussed in Section~\ref{sec:identifiability}.

Note that the \textit{\textbf{all-at-once approach}} can be handled along the same lines by using a $\tilde{\bullet}$ for the quantities $\GV{s}$, $\GV{d}$, and $\GV{e}$ and by replacing $\GKap$ with $\GBet$, see also Eq.~\eqref{eq:aao_statistic}.

%************************************************
\subsubsection{Identifiability}
\label{sec:identifiability}
Another important aspect concerns the question whether the parameters can in theory be recovered from the data. In the context of this paper, this mainly concerns the material parameters $\GKap$, which are identifiable, if for any $\GKap,\GKap'$,
\begin{equation}
    \label{eq:identifiability}
    L_{\GV{d}}(\GKap) = L_{\GV{d}}(\GKap') \ \Longrightarrow \GKap = \GKap', \quad  \forall \GV{d} \in \Omega_D,
\end{equation}
see \cite{mccullagh2002statistical}. Next, we study identifiability in the context of the \textit{\textbf{reduced approach}}. Noting that a maximizer of $L_{\GV{d}}(\GKap)$ corresponds to a minimizer of $\phi\INDr(\GKap)$, we proceed in analogy to \cite{beveridgeschechter70,beckarnoldbook1977} by performing a Taylor expansion,
\begin{equation}
  \label{eq:quadapprox}
  \hat{\phi}\INDr(\GKap)
  = \phi\INDr(\GKap^*) + \left.\left\{\difn{\phi\INDr(\GKap)}{\GKap}\right\}^T\right\rvert_{\GKap =
    \GKap^*} \Delta \GKap 
  + \fr{2} \Delta \GKap^T \left.\left[\diftwon{\phi\INDr(\GKap)}{\GKap}{\GKap}\right]\right\rvert_{\GKap =
    \GKap^*} \Delta \GKap,
\end{equation}
with $\Delta \GKap = \GKap - \GKap^*$. Here, 
\begin{equation}
  \label{eq:hessian}
  {H_{ij}}(\GKap) =
    \diftwo{\phi\INDr(\GKap)}{\kappa_i}{\kappa_j}=
  \left[
    \sum_{k=1}^{\numD}W_{kk}^2
    \left(
      \diftwo{s_k(\GKap)}{\kappa_i}{\kappa_j} (s_k(\GKap) - d_k)
        +
      \dif{s_k(\GKap)}{\kappa_i} \dif{s_k(\GKap)}{\kappa_j}
    \right)
  \right]
\end{equation}
are the components of the Hessian matrix. Remember that $\D \phi\INDr(\GKap)/\D \GKap$ vanishes in the local minimum $\GKap^*$ according to Eq.~\eqref{eq:resNLLS} and $W_{kk}$ are the diagonal entries in the weighting matrix $\GM{W}$. Since at $\GKap^*$, $s_k(\GKap) - d_k \approx 0$ holds, the Hessian is usually approximated by
\begin{equation}
  \label{eq:hessianapprox}
  \GM{H} \approx \GMT{J}\GMT{W} \GM{W} \GM{J} \quad \text{with} \quad
  H_{ij}\approx
  \left[\sum_{k=1}^{\numD}
     W_{kk}^2
    \dif{s_k(\GKap)}{\kappa_i}
    \dif{s_k(\GKap)}{\kappa_j}
  \right].
\end{equation}
The evaluation of the Hessian matrix $\GM{H}$ provides information on whether a unique solution exists locally, at the obtained solution $\GKap^*$. If the determinant of the Hessian matrix or any sub-determinant vanishes, $\detop \GM{H} = 0$, there is no unique solution, see \cite{beveridgeschechter70,beckarnoldbook1977}. This concept is called \textit{local identifiability} and has been investigated for common material models of solid mechanics in \cite{hartmanngilbert2018}. This approach can be adopted if there are no constraints in the optimization problem (or if none of the constraints are active in the solution). Moreover, it is also difficult to decide whether the determinant is really zero, especially since very small values can occur because of large numbers of data and/or the assigned units. To circumvent this issue, \cite{beckarnoldbook1977} describe a measure which is based on the relation between the eigenvalues of the Hessian matrix. In \cite{Vexler2004} and \cite{mahnkenenzy2018} this is linked to stability investigations, where also the eigenvalues are evaluated. Here, stability implies that small perturbations of the experimental data do not lead to any significant changes in the parameters \cite{mahnken2022}. It is noteworthy that the evaluation of the local identifiability is useful for investigations of different parameter identification procedures. Thus, the studies are usually performed in re-identifications of given parameters with synthetic data.

The concept of local identifiability has been applied in several works, see \cite{hartmanngilbert2018,sewerin2020,hartmannidentTI2021}, for example. The case of finite strain viscoelasticity using DIC-data is addressed in \cite{hartmanngilbert2021} and finite strain viscoplasticity by \cite{dileephartmann2022}. With regard to the identifiability of parameters, see also \cite{cobellidistefano1980}.

For parameter identification using parametric PINNs, the same conditions on identifiability apply as for the FEM-based reduced approach. In the context of \textit{\textbf{all-at-once}} PINNs, the Hessian of the loss function with respect to the trainable parameters takes the form 
\begin{equation}
    \frac{\D^2 \phi_{\text{aao}}}{\D \GBet^2} = 
    \begin{bmatrix} 
        \frac{\DS \partial^2 \phi_{\text{aao}}}{\DS \partial \GV{\theta}^2} & \ \frac{\DS \partial^2 \phi_{\text{aao}}}{\DS \partial \GV{\theta} \partial \GKap} \\[1ex]
        \frac{\DS \partial^2 \phi_{\text{aao}}}{\DS \partial \GKap \partial \GV{\theta} } & \ \frac{\DS \partial^2 \phi_{\text{aao}}}{\DS \partial \GKap^2}
    \end{bmatrix}.
\end{equation}
Again, the same conditions on identifiability apply to ${\partial^2 \phi_{\text{aao}}}/{\partial \GKap^2}$. Note that in the context of ANNs, identifiability of $\partial^2 \phi_{\text{aao}}/{\partial \GThe^2}$ is usually not considered.

\subsubsection{Uncertainty analysis}\label{sec:uncertainty-inference}

The following considerations are presented for the \textit{\textbf{reduced approach}}, but can be transferred to the all-at-once formulation by using a $\tilde{\bullet}$ for the quantities $\GV{s}$, $\GV{d}$, and $\GV{e}$ and by replacing $\GKap$ with $\GBet$, see Eq.~\eqref{eq:aao_statistic}. Note, however, that the numerical realization of the all-at-once approach may be more complex and may require additional steps to achieve robustness. In the following, we illustrate the quantification of parameter uncertainty  in the context of a  weighted LS problem. The derivation holds for any surrogate, e.g., the parametric PINNs of Section~\ref{sec:PINNreduced}, provided that some moderate regularity conditions are satisfied. The development covers the misspecified case, where the data is not generated by the model used for inference, which is important to cover, for instance, surrogate approximation errors. Consider the relation
\begin{equation}
    \GKap^* = \argmin_{\GKap} \big(-\log L_{\GV{d}}(\GKap)\big) = \argmin_{\GKap} \frac{1}{2} \| \GVt{r}(\GKap)\|_{\GV{\Sigma}_{\GV{e}}^{-1}}^2. 
\end{equation}
The weighting matrix in the LS formulation can be related to the diagonal noise covariance by
\begin{equation}\label{eq:weightMat_LS}
    \GV{W} = \GV{\Sigma}_{\GV{e}}^{-1/2},
\end{equation}
where 
\begin{equation}
    \GM{\Sigma}_{\GV{e}} = \mathbb{E}[\left\{\GV{e} -\mathbb{E}\left\{ \GV{e} \right\}\right\} \left\{\GV{e} -\mathbb{E}\left\{ \GV{e} \right\}\right\}^T] = \mathrm{diag}((\mathrm{Var}(e_i))_i)
\end{equation}
and $\mathrm{Var}$ denotes the variance. If we assume that $\mathrm{Var}(e_i)=\sigma_{\GV{e}}^2$ for all measurements, the covariance of the estimator $\bm{C}$ can be written and approximated as
\begin{equation}
    \bm{C} =  \frac{1}{\numD} \GV{Q}_0^{-1} \GV{Z}_0 \GV{Q}_0^{-1} \approx \sigma_{\GV{e}}^2 [\bm{J}_0^\top \bm{J}_0]^{-1}.
\end{equation}
Here, we have used the Hessian approximation derived in the preceding subsection and the subscript $_0$ refers to evaluation at the true value $\GKap_0$. Note that $\sigma_{\GV{e}}^2 [\GMT{J}_0 \bm{J}_0]^{-1}$ corresponds to the inverse Fisher information matrix, whereas $\GV{Q}_0,\GV{Z}_0$ represent the factors appearing in the Huber sandwich \cite{huber1967behavior}. 

A consistent estimator is obtained by replacing $\GKap_0$ with $\GKap^*$. Hence, we can quantify the uncertainty as
\begin{equation}
    \label{eq:covarMat_full}
    \GM{C}  \approx  \sigma_{\GV{e}}^2 [\GMT{J}(\GKap^*)  \GM{J}(\GKap^*)]^{-1} \approx s^2 [\GMT{J} \GM{J}]^{-1}, 
\end{equation}
where $e_i = d_i - s_i(\GKap^*) = r_i(\GKap^*)$ and the assumption of independent and identically distributed (i.i.d.) measurement noise leads to the unbiased estimate
\begin{equation}
    \label{eq:stddev}
    s^2 = \fr{\numD - 1} \, {\GV{r}}^T(\GKap^*) \GV{r}(\GKap^*),
\end{equation}
see, for example, \cite{bjoerck96}. It is worth mentioning that the unweighted residuals have to be used in Eq.~\eqref{eq:stddev} and the estimate holds only for constant measurement noise. Moreover, it is noteworthy that the fraction in Eq.~\eqref{eq:stddev} is sometimes applied as $1/\left(\numD - \numkappa\right)$, see \cite{beckarnoldbook1977}, leading to comparable results since $\numD$ is usually large. The diagonal entries of $\bm{C}$ can be used as a variance estimate for each parameter from which a confidence interval can be derived, e.g.,
\begin{equation}
    \label{eq:confidenceInt}
      \GKap_{\text{conf}} = \GKap^* \pm 1.96 \Delta \GKap 
\end{equation}
for a confidence level of $95\%$, with the uncertainty
\begin{equation}
    \label{eq:uncertaintyFrequentist}
    \Delta \kappa_i = \sqrt{C_{ii}},
    \quad
    i = 1,\ldots, \numkappa.
\end{equation}
If the sought parameter is high-dimensional, a regularization is required. Although this can be achieved in a frequentist setting, e.g., via Tikhonov regularization, see Section~\ref{sec:all-at-once}, we will consider regularization in the form of a prior model in the next section.

\subsection{Bayesian approach}\label{sec:bayesian-inference}

The following considerations are derived for the \textit{\textbf{reduced approach}}. In analogy to Section~\ref{sec:frequentist}, the \textit{\textbf{all-at-once approach}} can be handled along the same lines by using a $\tilde{\bullet}$ for the quantities $\GV{s}$, $\GV{d}$, and $\GV{e}$ and by replacing $\GKap$ with $\GBet$, see also Eq.~\eqref{eq:aao_statistic}.

In the Bayesian approach, a prior density $p(\GKap)$ is formulated in addition to the likelihood function. Hence, a Bayesian procedure posits a model for the data (likelihood) and a distribution over the parameter space (prior), which expresses plausible ranges or other physical constraints without any recurrence to an observation. The result is a distribution over the parameter domain, conditional on the observations (posterior), which expresses our uncertainty about the unknown parameters and which allows to extract an estimate of the parameters themselves. Bayesian inference can be carried out by treating the model as a black box, similar to LS-based approaches, however, the model structure can also be used explicitly to improve the numerical efficiency. 

In a Bayesian approach, the unknown parameter $\GKap$ is treated as a random variable with a prior distribution, which is updated using Bayes' law as
\begin{equation*}
    p(\bm{\kappa}\vert\bm{d}) \propto L_{\bm{d}}(\bm{\kappa}) p(\bm{\kappa}),
\end{equation*}
where $p(\bm{\kappa}\vert\bm{d})$ represents the posterior density. A fully Bayesian approach would seek to compute the entire density, providing simultaneously a point estimate and a quantification of uncertainty. For instance, we can employ the mean $\GKap^* = \mathbb{E}_{\GKap \vert \GV{d}}[\GKap]$ or maximum a posteriori estimate $\GKap^* = \mathrm{argmax}_{\GKap} \, p(\GKap\vert\GV{d})$ and derive some credible intervals from $p(\GKap\vert\GV{d})$. Then, if we consider a Gaussian prior 
\begin{equation}
\label{eq:prior-Gauss}
p(\bm{\kappa}) \propto \exp \left( -\frac{1}{2} \| \bm{\kappa}\|^2/\sigma_\kappa^2  \right),
\end{equation}
the maximum a posteriori estimate is given by
\begin{equation}
\GKap^* = \argmax_{\GKap} -\log p(\GKap\vert\bm{d}) = \argmax_{\GKap} -\Big( \log L_{\GV{d}}(\GKap) + \log p(\bm{\kappa})\Big),
\end{equation}
which, together with Eqns.~\eqref{eq:statistical_model} and \eqref{eq:prior-Gauss},  leads to 
\begin{equation}
\GKap^* = \argmin_{\GKap} \frac{1}{2} \Big(\| \bm{r}(\bm{\kappa})\|^2/\sigma_{\GV{e}}^2 +  \| \bm{\kappa} \|^2/\sigma_\kappa^2 \Big).
\end{equation}
Hence, the prior naturally leads to a regularized LS problem, and the frequentist estimate (\ref{eq:frequentist_point_estimate}) is recovered for a uniform (non-informative) prior.

But even in the case of normally distributed data and a normally distributed prior, the posterior is not normally distributed for finite sample sizes. This is because of the nonlinear dependence between the parameter vector and simulated data $\GV{s}(\GKap)$. The posterior density therefore needs to be approximated numerically, where a large variety of algorithms are now in use, see, e.g., \cite{brooks2011handbook} for an overview of Markov chain Monte Carlo approaches and \cite{jaakkola2000bayesian} for variational approaches to Bayesian parameter estimation. 

Once approximated, the posterior distribution provides insights on the parameter dependence structure and on the parameter uncertainty. For instance, we can extract the posterior covariance matrix, which is appealing if the posterior is unimodal and close to a normal distribution. In the case of a multimodal non-Gaussian posterior, empirical credible intervals can be derived directly from the Markov chain.  

\paragraph{Linkage to frequentist approach}
Here, we highlight the connection of the Bayesian to the frequentist approach. The frequentist and Bayesian approaches can be connected in the so-called large data~/~small noise regime, where the influence of the prior vanishes. Precisely, there holds 
\begin{equation}
\label{eq:BvM-Theorem}
    d_{\mathrm{TV}}\left( p(\GKap|\GV{d}), \mathcal{N}\big(\GKap^*, \sigma_{\GV{e}}^2 \big[\GMT{J}_0 \GM{J}_0\big]^{-1}\big) \right) \rightarrow 0
\end{equation}
for $\numD \rightarrow \infty$ or $\sigma_{\GV{e}} \rightarrow 0$ in the distribution, where $d_{\mathrm{TV}}$ denotes the total variation distance -- a distance measure for probability distributions. The limit statement~\eqref{eq:BvM-Theorem} is known as the Bernstein-von-Mises theorem and implies that, asymptotically, the posterior distribution contracts to the true value and that frequentist confidence intervals are asymptotically equivalent to their Bayesian counterparts derived from $p(\GKap|\GV{d})$. Note that the theorem requires the data to be generated from the model with the true parameter $\GKap_0$, see \cite{lu2017bernstein}, and it is known that Bayesian methods perform sub-optimally under model misspecification. Bernstein-von-Mises theorems for the case of a model misspecification are reviewed in \cite{bochkina2019bernstein}. 

\paragraph{Identifiablity in the Bayesian approach}

Regarding the concept of solution uniqueness, the recently published article \cite{latz2023bayesian} states (relatively mild) conditions under which a unique posterior measure exists, which depends continuously on the observed data in appropriate distances between probability measures. This underlines the global viewpoint of Bayesian methods, contrary to the local identifiability analysis outlined for frequentist approaches in the previous subsection. Please note that a precise definition of identifiability in a Bayesian context is still a topic of discussion \cite{san2010bayesian}.

\subsection{Inference for complex materials with two-step approach}
\label{sec:two-step-inference}

When parameter identification is carried out for inelastic materials, it is common to identify the elastic parameters first before estimating the remaining parameters characterizing the inelastic behavior.  The approach can be formalized in a statistical way with the concept of two-step inference methods \cite{wooldridge2010econometric}. A Bayesian counterpart is obtained by hierarchical modeling, as we show in this section. In the following, we focus on the \textit{\textbf{reduced approach}}. 

First, we partition the parameter vector as $\GKap = \{\GKap\INDe^\top,\GKap\INDp^\top\}^\top \in \mathbb{R}^{n\INDe+n\INDp}$, where the subscripts $\INDe,\INDp$ refer to elasticity and plasticity, respectively. We also partition the data and simulated results as 
\begin{equation}\label{eq:two_step_1}
\GV{d} = \{\GV{d}\INDe^\top,\GV{d}\INDp^\top\}^\top, \quad \GV{s} = \{\GV{s}\INDe^\top,\GV{s}\INDp^\top\}^\top.
\end{equation}
Multiple load steps are particularly necessary for calibration in the plasticity regime, where the vectorial responses are all stacked into a single response and data vector, see also the notation in Appendix~\ref{sec:processingData}.

The underlying assumption is that $\GV{s}\INDe(\GKap) = \check{\GV{s}}\INDe(\GKap\INDe)$, i.e., that the elastic data can be explained with the elastic material parameters only. Hence, $\GKap\INDe$ can be estimated first, independently of the remaining plastic parameters. The uncertainty analysis is more involved in this case because the uncertainty of the initial elasticity inference step needs to be taken into account when inferring the plasticity parameters. 

In the following, a two-step frequentist and the corresponding hierarchical Bayesian approach are presented in Sections~\ref{sec:twoStep_frequentist} and~\ref{sec:twoStep_bayes}, respectively.

\subsubsection{Two-step frequentist approach}
\label{sec:twoStep_frequentist}

This section covers a derivation of uncertainty estimates through asymptotic normality considerations and Gaussian error propagation, which, to the best of our knowledge, is new in the context of a two-stage parameter identification framework in solid mechanics. 

\paragraph{Uncertainty via asymptotic normality}

A detailed derivation of the results presented in this section is provided in Appendix \ref{sec:two-step-appendix}. We denote with $m\INDe$ and $m\INDp$ the number of loading steps, i.e., the number of data vectors, in the elastic and plastic regime, respectively. The FOCs associated to the two-step NLS problem read 
\begin{align*}
    \difn{\phi\INDe(\GKap\INDe)}{\GKap\INDe}  &= \frac{1}{m\INDe} \GMT{J}\INDe(\GKap\INDe) \left\{\check{\GV{s}}\INDe(\GKap\INDe) - \GV{d}\INDe \right\} = \GV{0}, \\ 
    \dif{\phi\INDp(\GKap\INDe^*,\GKap\INDp)}{\GKap\INDp}  &= \frac{1}{m\INDp} \GMT{J}\INDp(\GKap\INDe^*,\GKap\INDp) \left\{ \GV{s}\INDp(\GKap\INDe^*,\GKap\INDp) - \GV{d}\INDp \right\} = \GV{0},
\end{align*}
and their solutions are denoted as $(\GKap\INDe^*,\GKap\INDp^*)$. Please note the introduction of scaling factors $m\INDe,m\INDp$ to simplify the derivations. 

For the identification of elasticity parameters, similar to Section~\ref{sec:frequentist},
\begin{equation}
    \GKap\INDe^* \sim \mathcal{N}\left(\GKap\INDen,\Big[\GMT{J}\INDe(\GKap\INDen) \GV{\Sigma}_{\GV{r}\INDe} \GM{J}\INDe(\GKap\INDen) \Big]^{-1} \right)
\end{equation}
holds, where $\GKap\INDen$ denotes the unknown true value of the elasticity parameter vector. It is, however, more difficult to obtain an accurate uncertainty estimate for the plasticity parameters because the additional uncertainty in $\GKap\INDe^*$ needs to be considered as well.

First, observe that the structure of the asymptotic covariance once again reads
\begin{equation}
    \label{eq:variance_plasticity}
    \GV{C}_{\GKap\INDp}
    = 
    \frac{1}{m\INDp}\GMM{Q}\INDp(\GKap_0) \GM{Z}\INDp(\GKap_0) \GMM{Q}\INDp(\GKap_0).
\end{equation}
Then, based on mean-value expansions, we can derive 
\begin{equation}
    \sqrt{m\INDp} \{\GKap\INDp^* - \GKap\INDpn\}  
    = - \GMM{Q}\INDp(\GKap_0) \sqrt{m\INDp} \left\{ \GV{v}_1 + \GV{v}_2 \right\},
\end{equation}
where
\begin{align}
    \GV{v}_1 &= \dif{\phi\INDp(\GKap\INDen,\GKap\INDpn)}{\GKap\INDp}, \\
    \GV{v}_2 &= \left[\dif{}{\GKap\INDe} \Big\{\dif{\phi\INDp(\GKap\INDe^+,\GKap\INDpn)}{\GKap\INDp}  \Big\}\right] \{\GKap\INDe^* - \GKap\INDen\}.
\end{align}
Note that $\GKap\INDe^+$ refers to the expansion point in a mean-value expansion, see Appendix~\ref{sec:two-step-appendix}. Note that the vector $\GV{v}_2$ represents the effect of uncertainty in $\GKap\INDe$ on the inferred $\GKap\INDp$. The matrix $\GV{Q}\INDp(\GKap_0)$ can be estimated as 
\begin{equation}
    \GV{Q}\INDp(\GKap_0) \approx \frac{2}{m\INDp} \GMT{J}\INDp(\GKap^*) 
%\GV{\Sigma}_{\GV{r}\INDp} 
    \GM{J}\INDp(\GKap^*).
\end{equation}
Finally, following the steps reported in the Appendix~\ref{sec:two-step-appendix}, we obtain
\begin{equation}
    \GM{Z}\INDp(\GKap\INDen,\GKap\INDpn) 
    = \frac{4}{m\INDp} \Big[ \sigma\INDp^2\GM{J}\INDp^{0\, T} \GM{J}\INDp^0
    + \GM{J}\INDp^{0\, T} \GM{J}\INDpe^0  \GM{\Sigma}_{\GKap\INDe^*}\GM{J}\INDpe^{0\,T} \GM{J}\INDp^0
    + \sigma\INDp^2 \GM{G}\INDpe^0 \GM{\Sigma}_{\GKap\INDe^*}  \Big],
    \label{eq:zMatrix}
\end{equation}
where the quantities $\GV{J}\INDpe$ and $\GV{G}\INDpe^0$, which are defined in the Appendix, reflect sensitivities of the plasticity parameter estimate to changes in the elasticity parameters. A computable estimate is obtained by replacing all true parameters in the previous equations with their parameter estimates. Hence, all terms appearing in the definition of the covariance matrix \eqref{eq:variance_plasticity} have been specified.

\paragraph{Uncertainty via Gaussian error propagation}

A different approach to account for the elasticity parameter uncertainty in the second identification step is put forth \cite{troegerhartmann2022,dileephartmann2022}, based on the concept of Gaussian error propagation, i.e.\ the first-order second-moment method or delta method in statistics. The uncertainty $\delta F$ of a quantity $F(\hat{\GKap})$ is estimated by
\begin{equation}
    \label{eq:uncertainty}
      \delta F = \sqrt{\sum_{k=1}^\numkappa \left(\dif{F}{\kappa_k} \Delta\kappa_k\right)^2},
\end{equation}
i.e.,\ the confidence interval reads $F \pm \delta F$. In \cite{hartmannmuellerlohsetroeger2023}, this is interpreted as the norm of the Gateaux derivative in the direction of the uncertainties of the individual parameters, where mixed partial derivatives are neglected (i.e., the parameters are assumed to be uncorrelated). Moreover, the same concept is chosen in \cite{dileephartmann2022,troegerhartmann2022,troegerhartmann2023} to also estimate the uncertainty in FEM simulations caused by uncertainties in the material parameters. The covariance matrix \eqref{eq:covarMat_full} and confidence interval \eqref{eq:confidenceInt} are often denoted as \textit{quality measures} in the frequentist approach.

\subsubsection{Hierachical Bayesian approach}
\label{sec:twoStep_bayes}
With the same notations as in the previous section, we are now going to summarize the Bayesian approach to inference in the two-step case. Once again starting with the elasticity identification step, the posterior distribution reads 
\begin{equation}
    \label{eq:independence_elasticity_Bayes}
    p(\GKap\INDe | \GV{d}) = p(\GKap\INDe | \GV{d}\INDe),
\end{equation}
because of the assumption that the data for the elasticity regime is independent of $\GKap\INDp$. The full posterior can be rewritten as 
\begin{align*}
    p(\GKap| \GV{d}) &= p((\GKap\INDp,\GKap\INDe)| \GV{d}) = p(\GKap\INDp| \GKap\INDe, \GV{d}) p(\GKap\INDe | \GV{d}),
\end{align*}
where we used the definition of a conditional density function and where $p(\GKap\INDp| \GKap\INDe, \GV{d})$ represents the density of $\GKap\INDp$, conditional on both $\GKap\INDe$ and $\GV{d}$. Assuming that $\GKap\INDp$ only depends on $\GV{d}\INDp$ and with \eqref{eq:independence_elasticity_Bayes}, we obtain 
\begin{equation}
    \label{eq:posterior_two_step_Bayes}
    p(\GKap| \GV{d}) = p(\GKap\INDp|\GKap\INDe,\GV{d}\INDp) p(\GKap\INDe|\GV{d}\INDe).
\end{equation}
Eq.~\eqref{eq:posterior_two_step_Bayes} represents a hierarchical model for the posterior of the two-step problem. The posterior distribution of the plasticity calibration step $p(\GKap\INDp|\GKap\INDe,\GV{d}\INDp)$ contains $\GKap\INDe$ as a hyperparameter, which can be inferred at the next level. The final level, i.e., the prior densities, is not explicitly re-stated here. 

%*************************************
\section{Examples}
\label{sec:examples}
This section presents illustrative examples for the theoretical considerations of Sections~\ref{sec:abstract}
and~\ref{sec:locident}. The numerical setups for the different reduced approaches, the VFM and the all-at-once approaches, outlined in Section~\ref{sec:abstract}, are the topic of Section~\ref{sec:linElas}. For the reduced approaches, results are presented both in a deterministic and in a stochastic setting based on Bayesian inference.

In Section~\ref{sec:smallStrainPlasticityExp}, we present results for the novel two-step inference procedure with inelastic materials, as discussed in Section~\ref{sec:two-step-inference}. In this context, the focus lies on the uncertainty quantification of a small-strain elasto-plasticity model, where full-field data are not considered. Instead, stress-strain data obtained from real-world experiments are used. 

Note that the main aim of this section is to illustrate how the various methods reviewed and introduced in this paper can be used in practice for parameter identification. We thereby identify the main capabilities and requirements of each approach. It is out of the scope of the paper to investigate the detailed cost-accuracy trade-off for the individual methods.

\subsection{Plate with a hole}
\label{sec:linElas}

%\brb{@subsection-schreiber: In each subsection, specify values of $\sigma\INDs$ and $\sigma\INDd$}
%\brb{MF:done / JT, DA ?}
%\bbb{JT:Wir haben $\sigma\INDs$ und $\sigma\INDd$ nur beim all-at-once Ansatz in der Form eingeführt, für AAO-PINNs ist die Angabe vorhanden. Beim reduced-1 Ansatz aus 4.1 ist die Angabe in 6.1.1 und 6.1.2 aus meiner Sicht nicht sinnvoll.} \brb{HW: Stimme zu, scheint vollständig zu sein. @UR: Kommentar gerne löschen.}

As a first example, we consider a plate with a hole under tensile load and assume linear elastic isotropic material behavior. The corresponding material parameters under consideration are $\bm{\kappa} = \{E,\nu\}^T$, %$ = (K,G)$
where we choose the true values $\bm{\kappa^*} = \{\SI{210000}{\N\per\mm\squared},0.3\}^T$. To compare the results obtained by the different calibration methods, we draw on so-called re-identifications, as follows. The boundary-value problem with prescribed material parameters is first solved with the FEM. The computed displacement data are subsequently polluted with artificially generated noise and applied as generated full-field data for the model calibration in order to re-identify the previously prescribed material parameters.

\begin{figure}[ht]
    \centering
    \includegraphics[width=0.4\linewidth]{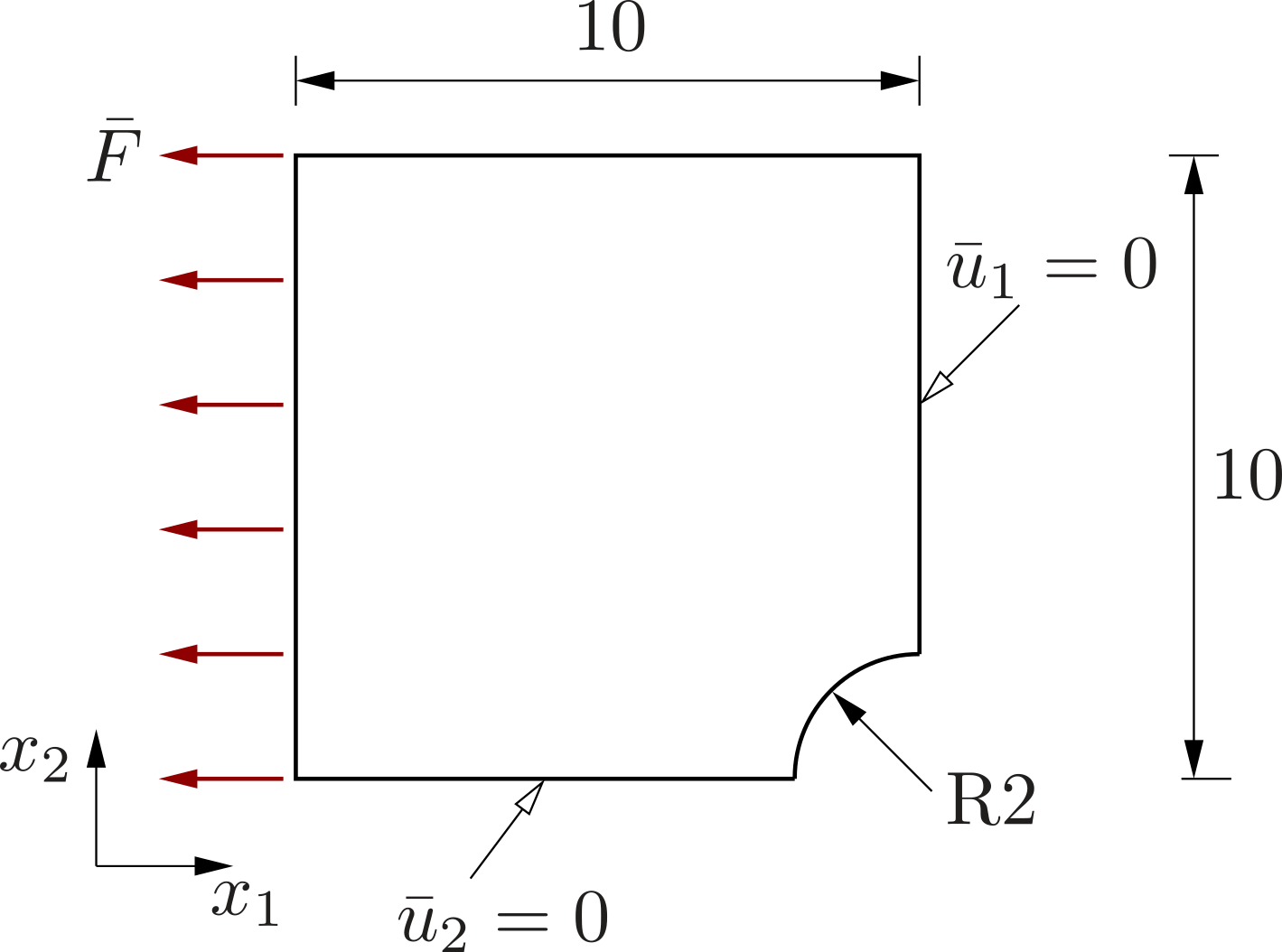}
    \caption{Geometry and loading conditions of plate with a hole}
    \label{fig:geometry_plate}
\end{figure}

\begin{figure}
    \centering
    \begin{subfigure}[b]{0.45\linewidth}
    \centering
    \includegraphics[width=\linewidth]{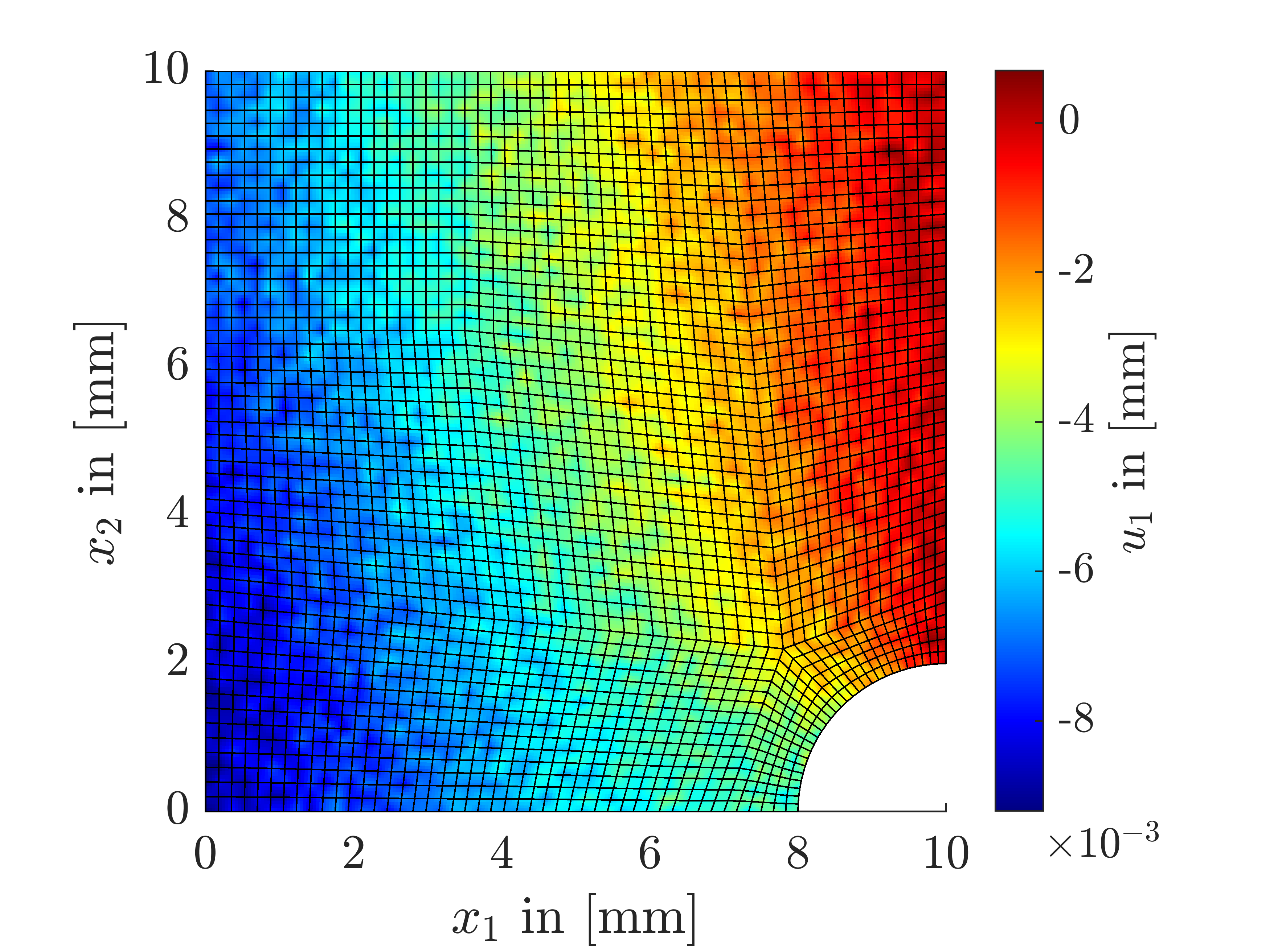}
    \caption{Generated axial displacement data with artificial noise}
    \label{fig:gen_axDisp}
    \end{subfigure}
    \hspace{0.05\linewidth}
    \begin{subfigure}[b]{0.45\linewidth}
    \centering
    \includegraphics[width=\linewidth]{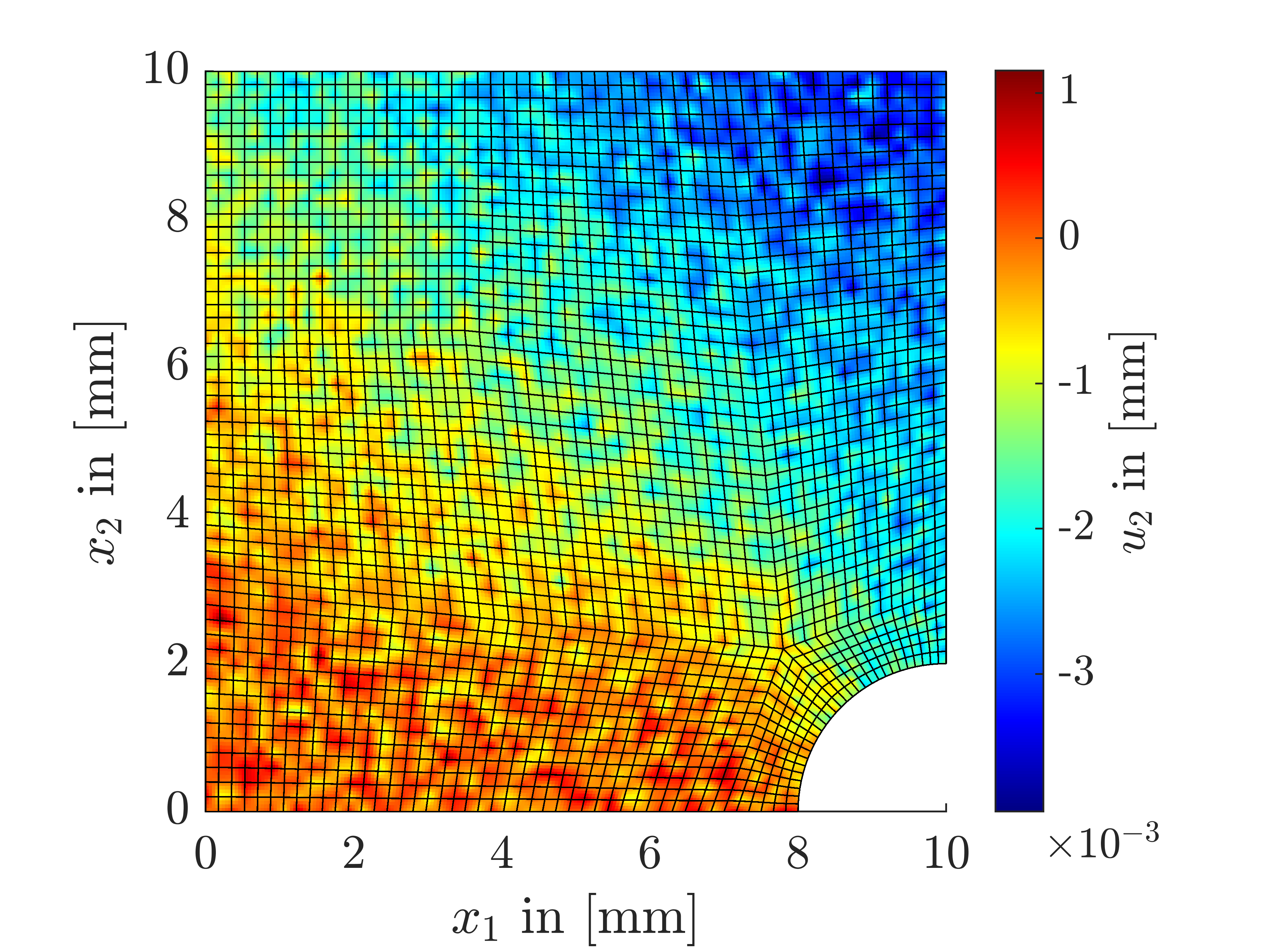}
    \caption{Generated lateral displacement data with artificial noise}
    \label{fig:gen_latDisp}
    \end{subfigure}
    \caption{Generated displacement data with artificial noise $\mathcal{N}(0,(\SI{4e-4}{\mm})^2)$ for re-identification of elasticity parameters}
    \label{fig:genDisp}
\end{figure}

The geometry of the domain is shown in Fig.~\ref{fig:geometry_plate}.
The plate is subjected to a tensile load $\bar{F} = \SI{1500}{\N}$ on the left edge, which is applied as equivalent nodal force on the corresponding nodes in the FEM. Here, symmetry is employed, i.e.,\ only a quarter of the geometry is used for the two-dimensional spatial discretization with 4-noded quadrilateral elements (bilinear Q4-elements). Further, a plane stress state is assumed. To minimize the influence of the spatial discretization error, a high-fidelity solution is computed with $\nel = 251500$ elements. To mimic DIC data, a linear interpolation of the high-fidelity displacement data to a coarser spatial discretization ($\nnodes = 3097$, $\nel = 2980$) is performed. The distance of the finite element nodes in this coarse discretization is approximately $\SI{0.2}{\mm}$ -- a spatial resolution which can be obtained by real DIC measurements. These artificial full-field displacements are treated as synthetic experimental data (with $\numD = 6194$ experimental data points, see App.~\ref{sec:processingData}) for calibration, and they are studied as clean data and with different levels of Gaussian noise $\mathcal{N}(0,\sigma^2)$. Following \cite{pierron_ExtensionVirtualFields_2010,hartmannmuellerlohsetroeger2023}, the standard deviation of the displacements obtained from DIC measurements is $\sigma = \SI{4e-4}{\mm}$. To investigate the effect of the applied noise on the calibration results, we choose a second noise level $\sigma = \SI{2e-4}{\mm}$ as well. The generated displacement data with artificial additive Gaussian noise $\mathcal{N}(0,(\SI{4e-4}{\mm})^2)$ is shown in Fig.~\ref{fig:genDisp}. \\

\subsubsection{FEM-based reduced approach}\label{sec:res_red_fem}

In the following, we first discuss the deterministic parameter identification, followed by a stochastic approach based on Bayesian inference with the FEM.

\begin{figure}
    \centering
    \begin{subfigure}[b]{0.45\linewidth}
    \centering
    \includegraphics[width=\linewidth]{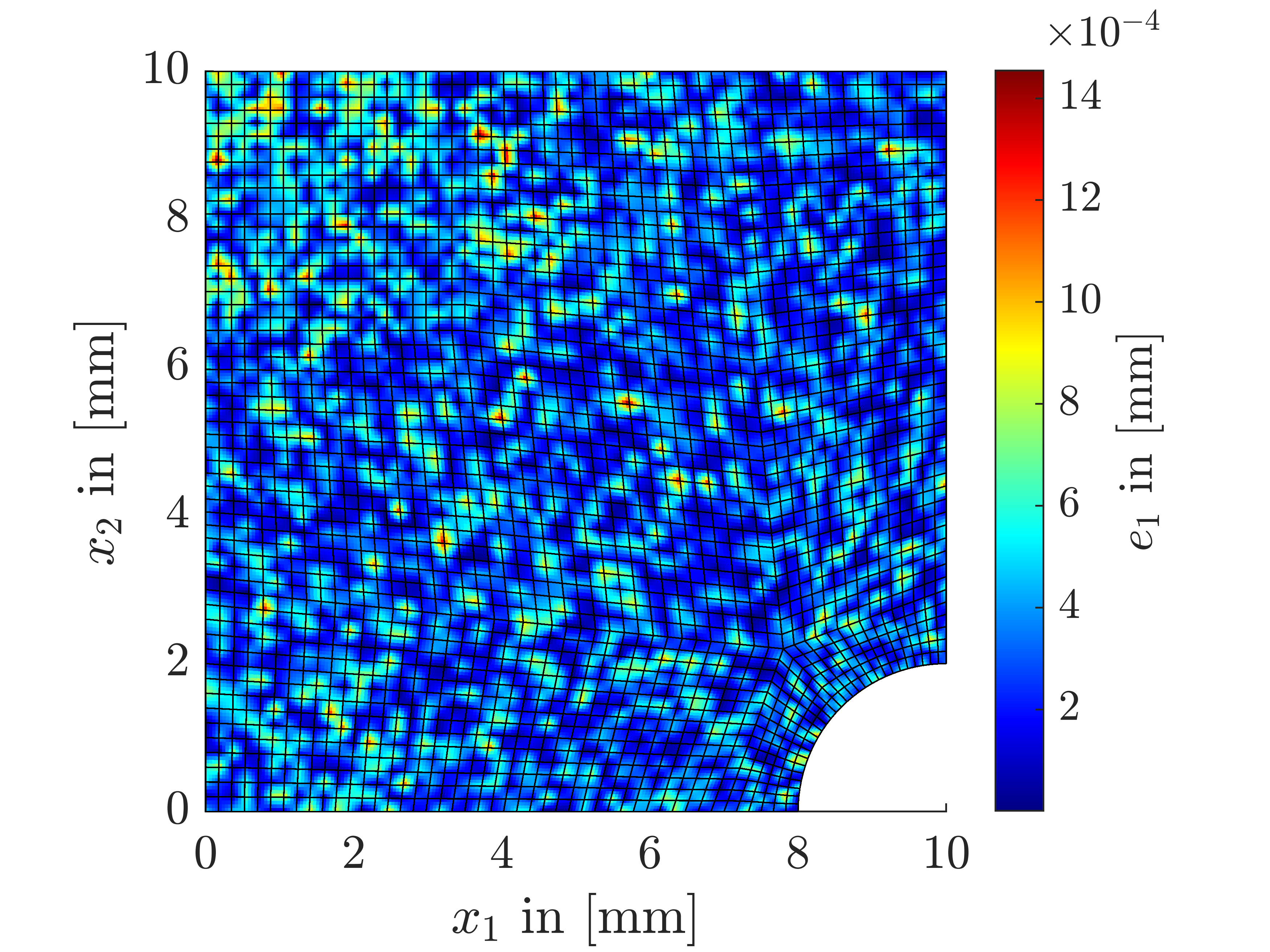}
    \caption{Absolute error in the axial displacements with artificial noise}
    \label{fig:absErr_axDisp}
    \end{subfigure}
    \hspace{0.05\linewidth}
    \begin{subfigure}[b]{0.45\linewidth}
    \centering
    \includegraphics[width=\linewidth]{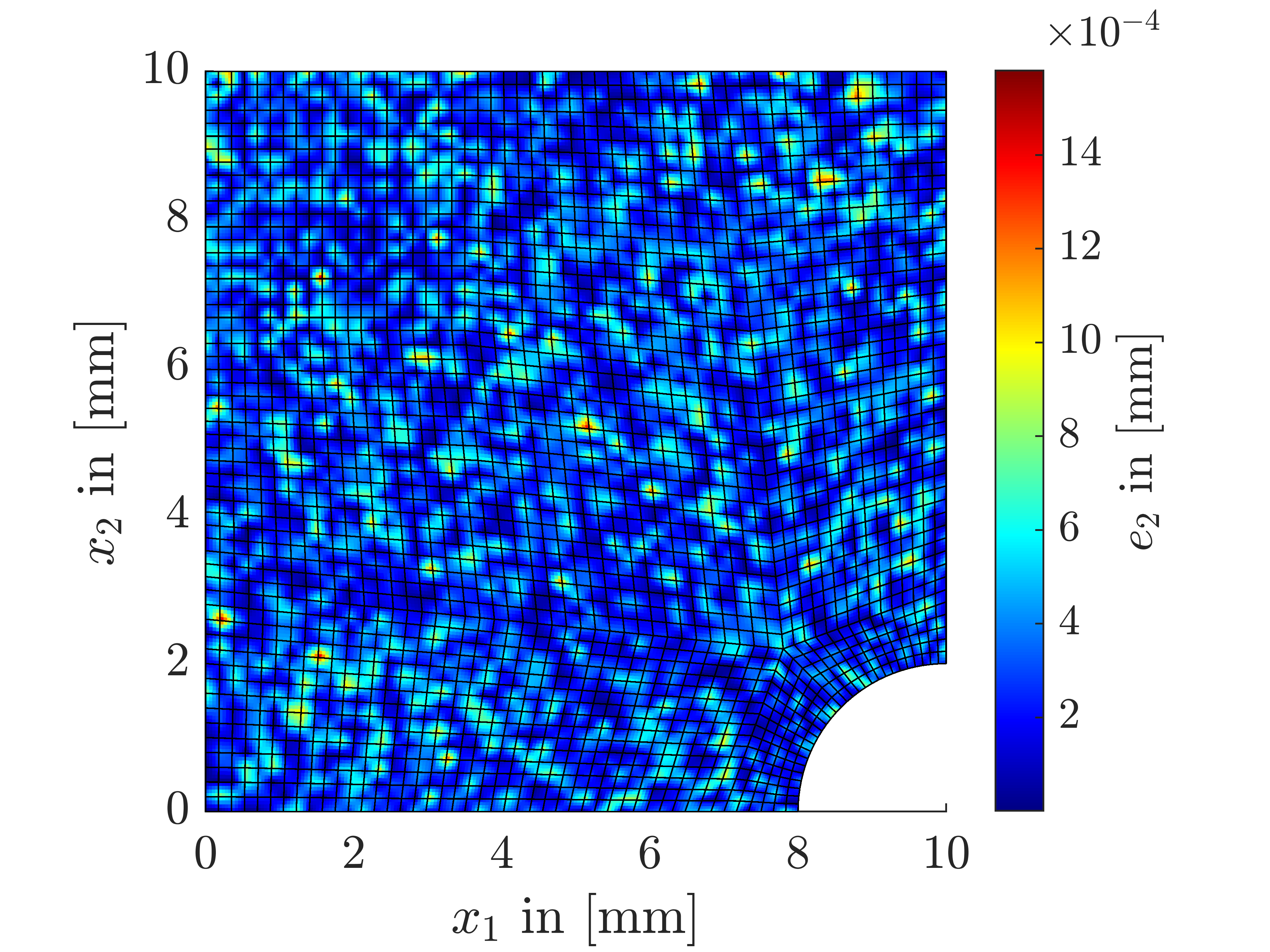}
    \caption{Absolute error in the lateral displacements with artificial noise}
    \label{fig:absErr_latDisp}
    \end{subfigure}
    \caption{Absolute error between generated displacements with $\mathcal{N}(0,(\SI{4e-4}{\mm})^2)$ and computed displacements with the identified parameters in the NLS scheme with FEM}
    \label{fig:errorNLSfit}
\end{figure}

\paragraph{Deterministic identification}
A weighted NLS scheme according to Sections~\ref{sec:nlsfemdic} or \ref{sec:RF-nlsfemdic} is applied to re-identify both parameters based on the artificially generated displacement data. The corresponding weighting factors are chosen as the maximum displacement values in each direction to account for the order of magnitude of the displacements. A trust-region reflective algorithm, implemented in the Matlab routine \texttt{lsqnonlin.m}, is applied to find the solution $\GKap^*$. The particular termination criteria for the optimizer, the change in the function value of the objective function, and the change in the arguments, are set to $10^{-8}$. The re-identified parameters in Tab.~\ref{tab:reIdent_FEdata} show good agreement with the true values, with slight deviations stemming from the interpolation of the high-fidelity data to the coarser grid. The node-wise relative error $e = \vert\vert \bf{u}^{\textrm{fit}} - \bf{u}^{\textrm{exp}} \vert\vert$ between the computed displacements $\bf{u}^{\textrm{fit}}$ in the solution $\GKap^*$ and the generated displacements $\bf{u}^{\textrm{exp}}$ are given in Fig.~\ref{fig:errorNLSfit}. % 
It is noteworthy that the applied noise level does not significantly influence the identification results. Although the determinant of the approximated Hessian \eqref{eq:hessianapprox} is small $\left(\detop \bm{H} = \SI{e-4}{\mm\tothe{4}\per\N\tothe{2}} \right)$, the parameters are uniquely identifiable from different initial values. 
% Here, the unweighted Hessian, i.e\ $W_{kk} = 1$, is evaluated to prevent from influences of the applied weighting factors. 
Further, the trust-region algorithm requires the computation of the Jacobian \eqref{eq:sensitivity}, which is done here by means of numerical differentiation employing a forward difference quotient. As a result, 12 calls of the FEM program are required as the optimizer stops after 4 iterations (including the initial evaluation of the start values). 

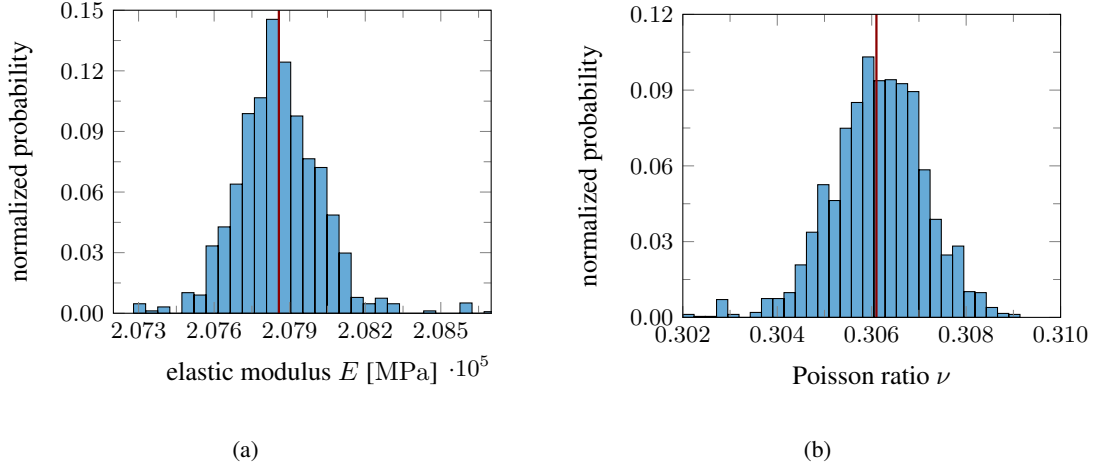
\begin{figure}
	\centering
	\begin{subfigure}[b]{0.4\linewidth}
	\centering
	\begin{tikzpicture}
	\tikzstyle{meanLine} = [draw = darkred, line width = 0.3mm]
	\begin{axis}[
	ylabel = normalized probability, 
	xlabel = elastic modulus $E$ {[\si{\MPa}]}, 
	xticklabel style={
	/pgf/number format/fixed,
	/pgf/number format/precision=3,
	/pgf/number format/fixed zerofill
	},
        yticklabel style={
	/pgf/number format/fixed,
	/pgf/number format/precision=2,
	/pgf/number format/fixed zerofill
	},
	scaled y ticks=false,
	xmin = 2.072e5, xmax = 2.087e5,
	ymin = 0.0, ymax = 0.15,
	ytick distance = 0.03,
        xtick distance = 0.003e5,
        minor tick num = 1,
	tick label style = {font=\small},
	width = \linewidth,
	height = 0.85\linewidth,
	area style
	]
	\addplot[ybar interval, fill = myBlue, fill opacity = 0.6, draw = black] table {pbU_chainE_plate.dat};
	\draw[meanLine] (axis cs:207857,0.0) -- (axis cs:207857,0.15);
	\end{axis}
	\end{tikzpicture}
	\label{fig:plate_pdfE}
	\caption{}
	\end{subfigure}
        \hspace{0.05\linewidth}
	\begin{subfigure}[b]{0.4\linewidth}
	\centering
	\begin{tikzpicture}
	\tikzstyle{meanLine} = [draw = darkred, line width = 0.3mm]
	\begin{axis}[
	ylabel = normalized probability,
	xlabel = Poisson ratio $\nu$,
	yticklabel style={
	/pgf/number format/fixed,
	/pgf/number format/precision=2,
	/pgf/number format/fixed zerofill
	},
	xticklabel style={
	/pgf/number format/fixed,
	/pgf/number format/precision=3,
	/pgf/number format/fixed zerofill
	},
	scaled y ticks=false,
	xmin = 3.02e-1, xmax = 3.1e-1,
	ymin = 0.0, ymax = 0.12,
	ytick distance = 0.03,
        xtick distance = 0.002,
        minor tick num = 1,
	tick label style = {font=\small},
	width = \linewidth,
	height = 0.85\linewidth,
	area style
	]
	\addplot[ybar interval, fill = myBlue, fill opacity = 0.6, draw = black] table {pbU_chainNu_plate.dat};
	\draw[meanLine] (axis cs:0.3061,0.0) -- (axis cs:0.3061,0.14);
	\end{axis}
	\end{tikzpicture}
	\label{fig:plate_pdfnu}
	\caption{}
	\end{subfigure}
	\caption{Posterior densities for Bayesian inference using FEM and noise level $\sigma = \SI{2e-4}{\mm}$} 
	\label{fig:plate_pdfs}
\end{figure}

\paragraph{Bayesian inference with FEM}
Next, we apply a sampling-based approach according to Section~\ref{sec:bayesian-inference} to identify the material parameters. In particular, we employ an in-house variant of the affine-invariant ensemble sampler \cite{goodman2010ensemble}, which features only one free parameter, the step size, and shows very robust performance for a wide range of densities. Compared to the LS approach, sampling approaches to Bayesian inference require a large number of model evaluations, particularly for high-dimensional problems. Hence, tractable approaches need to include surrogate modeling. Here, however, we sample the original FEM model directly, since the simulation times are manageable, and the posterior density is free of surrogate modeling errors in this case. 
We employ uniform priors, covering 10 \% variation around the nominal parameter values. The Bayesian approach permits to estimate the measurement noise size together with the unknown parameters. In our experiments, however, we prescribe this value, as is done for all other methods discussed herein. 
For all test cases, we employ an ensemble consisting of 50 chains, each of size 100. The step size is varied to achieve sound acceptance rates. In all cases, $\SI{50}{\percent}$ of all ensemble chains are removed to account for burn-in, i.e. the phase where the Markov chain explores the parameter space and the samples are not representative for the posterior distribution. All chains are merged afterwards to form a sample from the posterior density. Exemplarily, we depict the results for noise level $\sigma=\SI{2e-4}{\mm}$ and step size 5, consisting of the posterior histograms in Fig.~\ref{fig:plate_pdfs}. %
We observe a good concentration of the posterior with a small uncertainty. The expected values (red lines) are slightly off the true parameter values; however, the performance is comparable to the other methods and the slight offset can again be attributed to the coarse grid interpolation of the data. The trace plots further underline the good stationary behavior of the chains. The determined parameters are reported in Tab.~\ref{tab:reIdent_FEdata}. Note that for clean data, a step size of 4 was found to give better results. 

\subsubsection{Parametric PINN-based reduced approach}

In this section, we use a parametric \ac{PINN} surrogate for  calibration, and the material parameters $E$ and $\nu$ become additional inputs to the ANN alongside the coordinates $\GV{x}$. The displacements in both $x_1$- and $x_2$-direction are the outputs of the \ac{ANN} (see Sec.~\ref{sec:surrogates} for more details). For this example, we choose an \ac{ANN} with $4$ hidden layers with $32$ neurons each and the hyperbolic tangent as activation. The imposed displacements are enforced by hard boundary conditions \cite{berg_UnifiedDeepArtificial_2018}. In an offline phase, the parametric \ac{PINN} is first trained to learn a parameterized solution of the displacement field for a range of $[\SI{180000}{\N\per\mm\squared}, \SI{240000}{\N\per\mm\squared}]$ and $[0.2, 0.4]$ for $E$ and $\nu$, respectively. For the training, we sampled $32$ uniformly distributed $E$ and $\nu$ in the specified range, resulting in $1024$ different combinations. For each combination of material parameters, in turn, we sampled $64$ collocation points in the domain and $32$ points on each of the five boundaries.

In a subsequent online phase, we use the parametric \ac{PINN} for the re-identification of the material parameters with both the NLS method (similar to Secs.~\ref{sec:nlsfemdic} and~\ref{sec:RF-nlsfemdic}) and Bayesian inference (Secs.~\ref{sec:bayesian-inference}), where the parametric \ac{PINN} acts as a surrogate of the solution map. Both for the optimization problems resulting from \ac{PINN} training and for solving the NLS problem, we use the L-BFGS algorithm \cite{nocedal_NumericalOptimization_2006}. For Bayesian inference, we employ the same algorithm and hyperparameters as for Bayesian inference with FEM in Section~\ref{sec:res_red_fem}. The results of the re-identification are reported in Tab.~\ref{tab:reIdent_FEdata}. The results of both the NLS method and the Bayesian inference are in good agreement with the true material parameters. For the Bayesian inference, we achieve similar small uncertainties as for the Bayesian inference with  {FEM}.

\subsubsection{PINN-based all-at-once approach}
Next, an inverse \ac{PINN} is used to calibrate the linear-elastic material model from the artificially generated displacement data following the all-at once approach. The \ac{PINN} formulation used here is identical to that in \cite{anton_PhysicsInformedNeuralNetworks_2022} and is described in an abstract manner in Section \ref{sec:aao_pinn}. Equivalent to the mean squared error, for the weights $\sigma\INDs$ and $\sigma\INDd$ in \eqref{eq:loss_inverse}, we choose $\sigma\INDs=\frac{2}{3097}$ and $\sigma\INDd=\frac{2}{3097} \times 10^3$, where $3097$ is the number of observation points. To take the force information into account, the balance between internal and external energy is used as a further loss term. While the integral of the strain energy is approximated via the observation points, the external energy is approximated at $64$ points distributed uniformly on the Neumann boundary. As an initial guess for $E$ and $\nu$, we choose $\SI{210000}{\N\per\mm\squared}$ and $0.3$, respectively. Please note that a sensible choice of the initial guess is important for solving the optimization problem.
The displacement field in both $x_1$- and $x_2$-direction is approximated by two independent, fully connected feed-forward \acp{ANN} with $2$ hidden layers with $16$ neurons each, and we choose the hyperbolic tangent as activation. The resulting optimization problem is solved using the BFGS \cite{broyden_ConvergenceClassDoublerank_1970, fletcher_NewApproachVariable_1970, goldfarb_FamilyVariablemetricMethods_1970, shanno_ConditioningQuasiNewtonMethods_1970} optimizer. The re-identification yields the material parameters given in Tab.~\ref{tab:reIdent_FEdata}. For the clean displacement data, %and noise level $\sigma = \SI{2e-4}{\mm}$, 
the re-identified parameters are in good agreement with the true values. As already mentioned for the other approaches, at least for the clean data, a significant cause for the slight deviation in the re-identified Young's modulus $E$ lies in the linear interpolation of the high-fidelity data to the coarser grid. For increasing noise levels,
%$\sigma=\SI{4e-4}{\mm}$, 
however, increasingly significant deviations of the re-identified material parameters can be observed, especially for the Poisson's ratio. One possible reason for the large deviation is the overfitting of the \ac{ANN} to the noisy displacement data, see also the discussion in \cite{anton_PhysicsInformedNeuralNetworks_2022}. Although the PDE acts as regularization, additional regularization might be necessary for further improvement. Further investigation of this method for  noisy data is the subject of current research.

\subsubsection{Virtual fields method}
We now apply the VFM to identify the elastic parameters from the given displacement and force data.
We interpolate the given displacement data with local finite element ansatz functions and choose the same local finite element ansatz functions for the virtual fields to assemble the overdetermined system of equations \eqref{eqmf:system_virtual_fields_method} (see Appendix \ref{ap:system_matrices} for details).
As more information is provided in the interior of the domain than at the boundary, we choose the weighting factor $\sigma\INDr=10^4$ (see Appendix \ref{ap:system_matrices}) to increase the influence of the boundary data in the regression problem.
The overdetermined system of equations is solved in a LS sense to arrive at the desired material parameters, which are reported in Tab.~\ref{tab:reIdent_FEdata}.

The computation of the system of equations \eqref{eqmf:system_virtual_fields_method} requires computing the strain field from the given displacement data.
As the displacement data are interpolated with local finite element ansatz functions, adding artificial noise to the displacement data has a significant effect on the computed strain field.
Therefore, the results of the VFM become unreliable for the noisy data cases. 
This could be counteracted by denoising the displacement data before using them as input to the VFM.
For the sake of brevity, and because the VFM has been validated for noisy data in previous studies \cite{avriletal2008}, we omit this denoising step here and present only results of the VFM for the noiseless case.

\subsubsection{FEM- and VFM-based all-at-once approaches}
Finally, we apply the all-at-once approach, see Section~\ref{sec:nlsfemdicallatonce}. 
As for the reduced VFM, we interpolate the given displacement data with local finite element ansatz functions, and we choose the same local finite element ansatz functions for the virtual fields.
Then, the system matrices are assembled as described in Appendix \ref{ap:system_matrices}.
We choose the weights of the optimization problem as $\sigma\INDr=10^4$ (see Appendix \ref{ap:system_matrices}), $\sigma\INDs=1$, $\sigma\INDd=10^{-5}$ for the AAO-FEM and $\sigma\INDr=10^4$, $\sigma\INDs=1$, $\sigma\INDd=10^{-10}$ for the AAO-VFM.
Further, we choose the initial elasticity parameters $C_{11}=\SI{225000}{\N\per\mm\squared}, C_{12}=\SI{65000}{\N\per\mm\squared}$ (corresponding to $E = \SI{206220}{\N\per\mm\squared}$ and $\nu = 0.2889$) and the noisy artificial data for the initial displacement field.
To solve the optimization problem (\eqref{eq:aao_Euclidean_norm}, \eqref{eq:aao_K_norm}), we choose the reflective trust-region method implemented in the Matlab function \texttt{lsqnonlin.m}.
% The all-at-once objective function is minimized with the Matlab build-in trust-region optimizer \texttt{lsqnonlin.m}.
The all-at-once approach simultaneously identifies the displacement field and the material parameters.
Hence, an advantage of the all-at-once approach over, for example, the VFM is that a smooth displacement field is obtained even if the measurement data are noisy, while at the same time identifying the material parameters.
The results of the identified parameters are reported for different noise levels in Tab.~\ref{tab:reIdent_FEdata}.

\subsubsection{Discussion}
\label{sec:discuss_lin_elas}

The  material parameters $E$ and $\nu$ identified with the different schemes are provided in Tab.~\ref{tab:reIdent_FEdata}. All deterministic and stochastic reduced approaches yield satisfactory results, independently of the noise level. While a noisy case is not considered for the VFM here, the VFM yields satisfactory results for the noiseless case. The AAO-PINN exhibits quite a strong dependence on the noise level, see also \cite{anton_IdentificationMaterialParameters_2022}. Also, the accuracy of the results obtained with AAO-FEM and AAO-VFM is more affected by noise compared to their reduced counterparts. 

Further, during the numerical studies on the all-at-once approaches (AAO-PINN, AAO-FEM, AAO-VFM), a relatively high sensitivity of the identified parameters with respect to the initial guesses was observed (see also the related discussion for AAO-PINNs in  \cite{anton_PhysicsInformedNeuralNetworks_2022}). In addition to the numerical examples shown in this section, a proper mathematical analysis of the well-posedness and in particular the sensitivity with respect to the initial guess of the methods would be highly interesting -- but this goes beyond the scope of this article. As outlined in Section~\ref{sec:nlsfemdicallatonce}, regularization, e.g., using a  prior, could help to alleviate this issue.

\begin{table}[ht]
    \caption{Results of re-identification of material parameters for linear elasticity from clean and noisy data}
    \label{tab:reIdent_FEdata}
    \centering
    \begin{tabular}{m{0.2\linewidth} c c c}
        \toprule
        method & {$\sigma$ [\si{\mm}]} & {$E^*$ [\si{\N\per\mm\squared}]} & {$\nu^*$ [-]} \\
        \midrule
        true values &  & 210000 & 0.3000 \\
        \midrule
        \multicolumn{3}{l}{\textbf{reduced approaches} [\ref{sec:reduced}]} \\
        \midrule
        \multirow{3}{*}{\parbox{3.0cm}{LS (FEM) [\ref{sec:nlsfemdic}, \ref{sec:RF-nlsfemdic}]}} 
        & $0$ & $205410$ & $0.3007$ \\
        & $\num{2e-4}$ & $205963$ & $0.3025$ \\
        & $\num{4e-4}$ & $205855$ & $0.3034$ \\
        \midrule
        \multirow{3}{*}{\parbox{3.0cm}{Bayesian inference (FEM) [\ref{sec:bayesian-inference}, \ref{sec:RF-nlsfemdic}]}}
        & $0$ & $207757$ & $0.3051$ \\
        & $\num{2e-4}$ & $207857$ & $0.3061$ \\
        & $\num{4e-4}$ & $207415$ & $0.3064$ \\
        \midrule
        \multirow{3}{*}{\parbox{3.0cm}{LS (parametric PINN) [\ref{sec:PINNreduced}]}}
        & $0$ & $210121$ & $0.2999$ \\
        & $\num{2e-4}$ & $210199$ & $0.3008$ \\
        & $\num{4e-4}$ & $209795$ & $0.3012$ \\
        \midrule        
        \multirow{3}{*}{\parbox{3.0cm}{Bayesian inference (parametric PINN) [\ref{sec:bayesian-inference}, \ref{sec:PINNreduced}]}}
        & $0$ & $210688$ & $0.3011$  \\
        & $\num{2e-4}$ & $210794$ & $0.3019$ \\
        & $\num{4e-4}$ & $210373$ & $0.3021$ \\
        \midrule
        \multicolumn{3}{l}{\textbf{VFM} [\ref{sec:virtualfields}]} \\
        \midrule
        \parbox{3.0cm}{VFM [\ref{sec:vfm_sec4}]}
        & $0$ & $209294$ & $0.2993$ \\
        \midrule
        \multicolumn{3}{l}{\textbf{all-at-once approaches} [\ref{sec:all-at-once}]} \\
        \midrule
        \multirow{3}{*}{\parbox{3.0cm}{AAO-FEM [\ref{sec:nlsfemdicallatonce}]}}
        & $0$ & $210009$ & $0.2581$ \\
        & $\num{2e-4}$ & $212698$ & $0.2338$ \\
        & $\num{4e-4}$ & $206222$ & $0.2889$ \\
        \midrule
        \multirow{3}{*}{\parbox{3.0cm}{AAO-VFM [\ref{sec:nlsfemdicallatonce}]}}
        & $0$ & $210018$ & $0.2581$ \\
        & $\num{2e-4}$ & $212728$ & $0.2336$ \\
        & $\num{4e-4}$ & $206242$ & $0.2857$ \\
        \midrule
        \multirow{3}{*}{\parbox{3.0cm}{AAO-\ac{PINN} [\ref{sec:aao_pinn}]}}
        & $0$ & $203911$ & $0.2999$ \\
        & $\num{2e-4}$ & $201715$ & $0.3733$ \\
        & $\num{4e-4}$ & $195740$ & $0.4278$ \\
        \bottomrule
    \end{tabular}
\end{table}

\subsection{Two-step inference of small-strain von Mises plasticity}
\label{sec:smallStrainPlasticityExp}

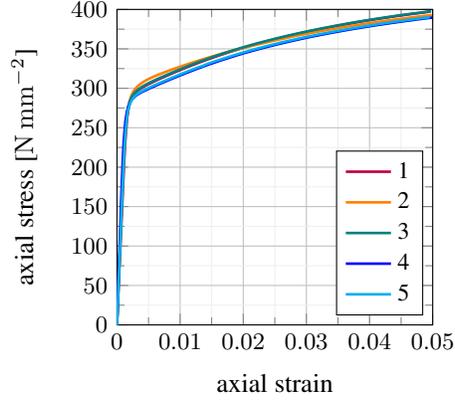
\begin{figure}[ht]
\centering
\begin{tikzpicture}
\tikzstyle{datalines} = [draw, line width = 0.35mm]
\begin{axis}[
xlabel = axial strain,
ylabel = axial stress {[\si{\N\per\mm\squared}]},
xticklabel style={
/pgf/number format/fixed,
/pgf/number format/precision=2,
%/pgf/number format/fixed zerofill
},
scaled x ticks=false,
xmin = 0, xmax = 0.05,
ymin = 0, ymax = 400,
xtick distance = 0.01,
ytick distance = 50,
tick label style = {font=\small},
grid = both,
minor tick num = 1,
major grid style = {lightgray},
minor grid style = {lightgray!25},
width = 0.35\linewidth,
height = 0.35\linewidth,
legend cell align = {left},
legend pos = south east,
legend style={font=\small}
]
\addplot[datalines, purple] table [x = {axstrain}, y = {stress}]{ssS_TS275_0001.dat};
\addplot[datalines, orange] table [x = {axstrain}, y = {stress}]{ssS_TS275_0002.dat};
\addplot[datalines, teal] table [x = {axstrain}, y = {stress}]{ssS_TS275_0003.dat};
\addplot[datalines, blue] table [x = {axstrain}, y = {stress}]{ssS_TS275_0004.dat};
\addplot[datalines, cyan] table [x = {axstrain}, y = {stress}]{ssS_TS275_0005.dat};
\legend{1,2,3,4,5}
\end{axis}
\end{tikzpicture}
\caption{Stress-strain data for calibration} 
\label{fig:expData_plasticity_stress}
\end{figure}

In this example, two-step inference is investigated for a small-strain plasticity model with von Mises yield function and non-linear kinematic hardening of Armstrong and Frederick type. The numerical treatment of the constitutive model was originally developed in \cite{HartmannHaupt1993,HartmannLuehrsHaupt1997}, and the material model is summarized in Appendix~\ref{ap:constModels}. The experimental data are obtained from tensile tests on specimens according to DIN EN ISO 6892-1 and simultaneous measurements with a DIC system. The information from the DIC system allows to determination the axial and lateral strains in the parallel region of the specimen as explained in \cite{hartmannidentTI2021}, whereas an additional compensation of rigid body movements is applied here. The specimens are made of the steel alloy TS275 and are loaded in a displacement-controlled process with $\dot{u} = \SI{0.01}{\mm\per\s}$. The experimental data of five specimens are employed up to a maximum axial strain of $5\%$. The experimental stress-strain data are shown in Fig.~\ref{fig:expData_plasticity_stress}.

The material parameters of the constitutive model can be identified in a sequential manner. First, the Young's modulus $E$ is determined from the stress-strain response of the specimens in the elastic region, which here comprises the data until a maximum axial strain of $0.1\%$. Then, the Poisson's ratio $\nu$ is identified from the lateral strain data in the elastic region. The constitutive model, see App.~\ref{ap:constModels}, requires a different set of elastic parameters $\GKap\INDe = \lbrace K, G \rbrace^T$, namely the bulk modulus $K$ and shear modulus $G$, instead of the elastic parameters $\tilde{\GKap}\INDe = \lbrace E, \nu \rbrace^T$. This has to be considered during the model calibration, as will be explained later. Finally, the plastic parameters $\GKap\INDp$, i.e., the yield stress $k$ and the kinematic hardening parameters $b$ and $c$ of the Armstrong and Frederick ansatz of kinematic hardening, are identified, i.e.\ $\GKap\INDp = \lbrace k, b, c \rbrace^T$. The experimental data obtained from the five tensile tests 
are reduced to one dataset by computing the mean values of stresses and lateral strains. To account for uncertainties, the covariance matrix $\bm{\Sigma}$ of the observations $\GV{d}$ is considered during calibration. In this example, the calibration is carried out with a frequentist approach and Bayesian inference using a similar setup to obtain comparable results, especially regarding the uncertainties of the parameters. For the theoretical considerations, see also Section~\ref{sec:two-step-inference}.

\subsubsection{Two-step frequentist approach}
Herein, a formulation corresponding to Sections~\ref{sec:NLS} and \ref{sec:reduced} is applied, where the covariance matrix $\bm{\Sigma}$ of the observations is chosen in the weighted LS approach, i.e.,\ $\bm{\Sigma}^{-1} = \GMT{W}\GM{W}$ in Eq.~\eqref{eq:objectiveFunction}. Further, the settings of the optimization algorithm are the same as in the previous example. All confidence intervals in the following are provided for a confidence level of $\SI{68}{\percent}$. \\

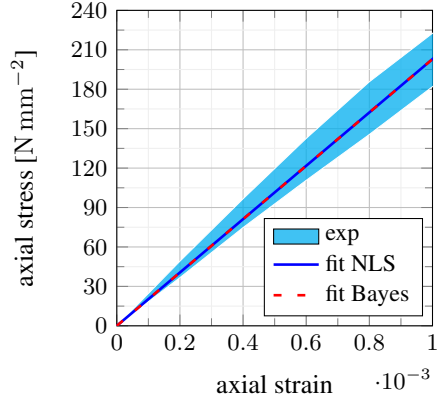
\begin{figure}[ht]
\centering
\begin{tikzpicture}
\tikzstyle{datalines} = [draw, line width = 0.35mm]
\begin{axis}[
xlabel = axial strain,
ylabel = axial stress {[\si{\N\per\mm\squared}]},
xticklabel style={
/pgf/number format/fixed,
/pgf/number format/precision=2,
%/pgf/number format/fixed zerofill
},
tick label style = {font=\small},
xmin = 0, xmax = 0.001,
ymin = 0, ymax = 240,
xtick distance = 0.0002,
ytick distance = 30,
grid = both,
minor tick num = 1,
major grid style = {lightgray},
minor grid style = {lightgray!25},
width = 0.35\linewidth,
height = 0.35\linewidth,
legend cell align = {left},
legend pos = south east,
legend style={font=\small}
]
\addplot[datalines,line width = 0.0mm, white, name path = f] table [x = {axstrain}, y = {minstress}]{ssS_fit_steel_smallStrain.dat};
\addplot[datalines,line width = 0.0mm, white, name path = g] table [x = {axstrain}, y = {maxstress}]{ssS_fit_steel_smallStrain.dat};
\addplot[fill=cyan,fill opacity=0.75] fill between [of = f and g];
\addplot[datalines,blue] table [x = {axstrain}, y = {stress}]{ssS_fit_steel_smallStrain.dat};
\addplot[datalines,loosely dashed,red] table [x = {axstrain}, y = {stress}]{sbU_fit_elastic_Bayes.dat};
\legend{,,exp,fit NLS,fit Bayes}
\end{axis}
\end{tikzpicture}
\caption{Experimental and calibrated stress-strain data in the elastic domain} 
\label{fig:fit_plasticity_stressElastic}
\end{figure}

\noindent\textit{Elasticity parameters:} The Young modulus $E$ is identified from the axial stress-strain data in the elastic region utilizing a weighted linear LS scheme as $E^* = \num[round-mode=places,round-precision=0]{202465} \pm \SI[round-mode=places,round-precision=0]{1468}{\N\per\mm\squared}$. The Hessian indicates local uniqueness of the solution $\left(\detop \bm{H} = 10^{-4}\right)$. The result of the calibration is visualized in Fig.~\ref{fig:fit_plasticity_stressElastic}, where the shaded area corresponds to the experimental data. Five data points, i.e.,\ $\numD = 5$, are used for the identification.

Moreover, the Poisson ratio $\nu$ is estimated from the lateral strain data in the elastic domain, where a weighted linear LS scheme is applied again. 

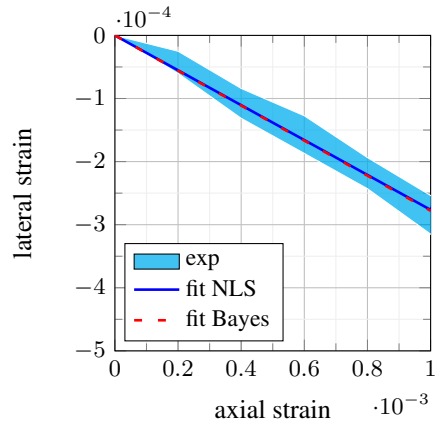
\begin{figure}[ht]
\centering
\begin{tikzpicture}
\tikzstyle{datalines} = [draw, line width = 0.35mm]
\begin{axis}[
xlabel = axial strain,
ylabel = lateral strain,
xmin = 0, xmax = 0.001,
ymin = -0.0005, ymax = 0,
xtick distance = 0.0002,
ytick distance = 0.0001,
tick label style = {font=\small},
grid = both,
minor tick num = 1,
major grid style = {lightgray},
minor grid style = {lightgray!25},
width = 0.35\linewidth,
height = 0.35\linewidth,
legend cell align = {left},
legend pos = south west,
legend style={font=\small}
]
\addplot[datalines,line width = 0.0mm, white, name path = f] table [x = {axstrain}, y = {minlatstr}]{ssS_fit_steel_smallStrain.dat};
\addplot[datalines,line width = 0.0mm, white, name path = g] table [x = {axstrain}, y = {maxlatstr}]{ssS_fit_steel_smallStrain.dat};
\addplot[fill=cyan,fill opacity=0.75] fill between [of = f and g];
\addplot[datalines,blue] table [x = {axstrain}, y = {latstrain}]{ssS_fit_steel_smallStrain.dat};
\addplot[datalines,loosely dashed,red] table [x = {axstrain}, y = {latstrain}]{sbU_fit_elastic_Bayes.dat};
\legend{,,exp,fit NLS,fit Bayes}
\end{axis}
\end{tikzpicture}
\caption{Experimental and calibrated lateral strain data in the elastic domain} 
\label{fig:fit_plasticity_latStrain}
\end{figure}

The LS method yields $\nu^* = \num[round-mode=places,round-precision=4]{0.2764} \pm \num[round-mode=places,round-precision=4]{0.0041}$. As before, the obtained solution is locally unique, although the determinant of the Hessian is small $\left(\detop \bm{H} = 10^{-11}\right)$. The experimental and calibrated lateral strain data are shown in Fig.~\ref{fig:fit_plasticity_latStrain}.

Since the elasticity relation of the constitutive model is formulated with the bulk modulus $K$ and the shear modulus $G$, see App.~\ref{ap:constModels}, both material parameters are computed by drawing on the well-known relations 
\begin{equation}
    \label{eq:ElasticityRelations}
    K = \frac{E}{3(1-2\nu)}, \quad 
    G = \frac{E}{2(1+\nu)},
\end{equation}
using a Monte Carlo approach to account for the corresponding uncertainties. To this end, we assume normally distributed $E$ and $\nu$ and randomly pick 4000 samples, which are used to evaluate Eq.~\eqref{eq:ElasticityRelations}. The computed distributions for $K$ and $G$ yield the mean value as parameter estimation $\GKap^*$ and the standard deviation for the uncertainty $\delta\GKap^*$ under consideration of error propagation effects. As a result, the elasticity parameters $K^* = \num{150991} \pm \SI{2951}{\N\per\mm\squared}$ and $G^* = \num{79321} \pm \SI{628}{\N\per\mm\squared}$ are obtained. It is worth mentioning that the Gaussian error propagation \eqref{eq:uncertainty} as an alternative approach for the uncertainty quantification yields values very similar to the Monte Carlo approach, namely  $K^* = \num{150937} \pm \SI{2984}{\N\per\mm\squared}$ and $G^* = \num{79309} \pm \SI{629}{\N\per\mm\squared}$. \\

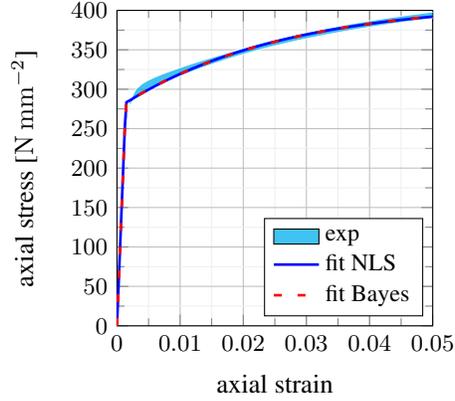
\begin{figure}[ht]
\centering
\begin{tikzpicture}
\tikzstyle{datalines} = [draw, line width = 0.35mm]
\begin{axis}[
xlabel = axial strain,
ylabel = axial stress {[\si{\N\per\mm\squared}]},
xticklabel style={
/pgf/number format/fixed,
/pgf/number format/precision=2,
%/pgf/number format/fixed zerofill
},
scaled x ticks=false,
tick label style = {font=\small},
xmin = 0, xmax = 0.05,
ymin = 0, ymax = 400,
xtick distance = 0.01,
ytick distance = 50,
grid = both,
minor tick num = 1,
major grid style = {lightgray},
minor grid style = {lightgray!25},
width = 0.35\linewidth,
height = 0.35\linewidth,
legend cell align = {left},
legend pos = south east,
legend style={font=\small}
]
\addplot[datalines,line width = 0.0mm, white, name path = f] table [x = {axstrain}, y = {minstress}]{ssS_fit_steel_smallStrain.dat};
\addplot[datalines,line width = 0.0mm, white, name path = g] table [x = {axstrain}, y = {maxstress}]{ssS_fit_steel_smallStrain.dat};
\addplot[fill=cyan,fill opacity=0.75] fill between [of = f and g];
\addplot[datalines,blue] table [x = {axstrain}, y = {stress}]{ssS_fit_steel_smallStrain.dat};
\addplot[datalines,loosely dashed,red] table [x = {axstrain}, y = {stress}]{sbU_fit_plastic_Bayes.dat};
\legend{,,exp,fit NLS,fit Bayes}
\end{axis}
\end{tikzpicture}
\caption{Experimental and calibrated stress-strain data}
\label{fig:fit_plasticity_stress}
\end{figure}

\noindent\textit{Plasticity parameters:} As previously mentioned, the plasticity parameters $\GKap\INDp$ can be identified from the stress-strain response of the specimens. For this purpose, we only apply data that have not already been considered in the calibration of the elasticity parameters. The weighted NLS method employs $\numD = 50$ data points and yields $k^* = \num[round-mode=places,round-precision=1]{282.6304} \pm \SI[round-mode=places,round-precision=2]{1.1161}{\N\per\mm\squared}$, $b^* = \num[round-mode=places,round-precision=2]{41.0409} \pm \num[round-mode=places,round-precision=2]{1.7603}$, and $c^* = \num[round-mode=places,round-precision=1]{3499.7771} \pm \SI[round-mode=places,round-precision=1]{116.7866}{\N\per\mm\squared}$. The Hessian indicates local uniqueness $\left(\detop \bm{H} = 10^{2}\right)$ for the minimum found. 
% However, different initial values lead to different solutions $\GKap^*$ indicating that multiple local minima exist. Thus, a grid-search is performed and the reported values are taken from the solution with the smallest value of the objective function. 
The calibrated stress-strain curve including elastic and plastic stages is shown in Fig.~\ref{fig:fit_plasticity_stress} together with the experimental data. \\
The aforementioned uncertainties of the plasticity parameters stem from the mismatch between experimental data and model response in the obtained solution $\GKap\INDp^*$. However, it also has to be considered that the plasticity parameters are identified with uncertain elasticity parameters $\GKap\INDe$. To account for this, we proceed according to Section~\ref{sec:twoStep_frequentist}. Then, the uncertainties $\delta k^* = \SI[round-mode=places,round-precision=2]{1.0683}{\N\per\square{\mm}}$, $\delta b^* = \num[round-mode=places,round-precision=2]{1.6829}$, and $\delta c^* = \SI[round-mode=places,round-precision=1]{111.6543}{\N\per\square{\mm}}$ of the plasticity parameters are obtained.

\subsubsection{Hierarchical Bayesian inference}
In the following, we compare the frequentist approach of Section~\ref{sec:frequentist} with a sampling-based Bayesian approach to parameter estimation and uncertainty quantification. In particular, we employ a nested Markov chain Monte Carlo approach for the hierarchical Bayesian model Eq.~\eqref{eq:posterior_two_step_Bayes}. The same experimental data as in the NLS calibration are employed, and a hierarchical model is used for the two-step calibration approach, following Section~\ref{sec:twoStep_bayes}. We directly sample the finite element model with the affine-invariant ensemble sampler \cite{goodman2010ensemble}. \\

\noindent\textit{Elasticity parameters:}
The elasticity parameters $\tilde{\GKap}\INDe = \lbrace E, \nu \rbrace^T$ are determined using uniform priors with 20\% variation around the estimates $E = \SI{200000}{\N\per\mm\squared}$ and $\nu = \num{0.275}$ with step size 12 and employing 100 chains of chain length 100. Afterwards, the chains are merged to represent a sample of the posterior density. The calibration yields $E^* = \num{202820} \pm \SI{8634}{\N\per\mm\squared}$ and $\nu^* = \num{0.2770} \pm \num{0.0143}$, which is in good agreement with the experimental data, see Figs.~\ref{fig:fit_plasticity_stressElastic} and \ref{fig:fit_plasticity_latStrain}, and with the previously reported NLS results. The posterior histograms are depicted in Fig.~\ref{fig:steel_pdfs_elastic}, and the trace plots are shown in Fig.~\ref{fig:steel_trace}, where the stationary behavior of the chains is evident. 
\begin{figure}[ht]
	\centering
	\begin{subfigure}[b]{0.4\linewidth}
	\centering
	\begin{tikzpicture}
	\tikzstyle{meanLine} = [draw = darkred, line width = 0.3mm]
	\begin{axis}[
	ylabel = normalized probability, 
	xlabel = Young's modulus $E$, 
	yticklabel style={
	/pgf/number format/fixed,
	/pgf/number format/precision=2,
	/pgf/number format/fixed zerofill
	},
	xticklabel style={
	/pgf/number format/fixed,
	/pgf/number format/precision=2,
	/pgf/number format/fixed zerofill
	},
	scaled y ticks=false,
	xmin = 0.2, xmax = 0.9,
	ymin = 0.0, ymax = 0.10,
	ytick distance = 0.02,
	xtick distance = 0.15,
        minor tick num = 1,
	tick label style = {font=\small},
	width = \linewidth,
	height = 0.85\linewidth,
	area style
	]
	\addplot[ybar interval, fill = myBlue, fill opacity = 0.6, draw = black] table {sbU_histE.dat};
	\draw[meanLine] (axis cs:0.5353,0.0) -- (axis cs:0.5353,0.12);
	\end{axis}
	\end{tikzpicture}
	\label{fig:steel_pdfE}
	\caption{}
	\end{subfigure}
        \hspace{0.05\linewidth}
	\begin{subfigure}[b]{0.4\linewidth}
	\centering
	\begin{tikzpicture}
	\tikzstyle{meanLine} = [draw = darkred, line width = 0.3mm]
	\begin{axis}[
	ylabel = normalized probability,
	xlabel = Poisson's ratio $\nu$,
	yticklabel style={
	/pgf/number format/fixed,
	/pgf/number format/precision=2,
	/pgf/number format/fixed zerofill
	},
	xticklabel style={
	/pgf/number format/fixed,
	/pgf/number format/precision=2,
	/pgf/number format/fixed zerofill
	},
	scaled y ticks=false,
	xmin = 0.08, xmax = 0.94,
	ymin = 0.0, ymax = 0.10,
	ytick distance = 0.02,
        xtick distance = 0.15,
        minor tick num = 1,
	tick label style = {font=\small},
	width = \linewidth,
	height = 0.85\linewidth,
	area style
	]
	\addplot[ybar interval, fill = myBlue, fill opacity = 0.6, draw = black] table {sbU_histNu.dat};
	\draw[meanLine] (axis cs:0.5183,0.0) -- (axis cs:0.5183,0.12);
	\end{axis}
	\end{tikzpicture}
	\label{fig:steel_pdfnu}
	\caption{}
	\end{subfigure}
	\caption{Posterior densities for elasticity parameters $E$ and $\nu$} 
	\label{fig:steel_pdfs_elastic}
\end{figure}
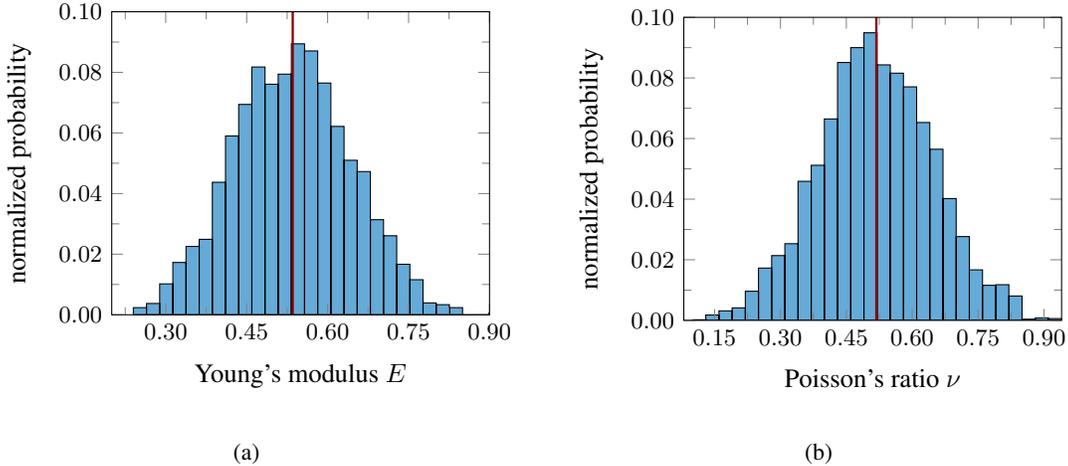
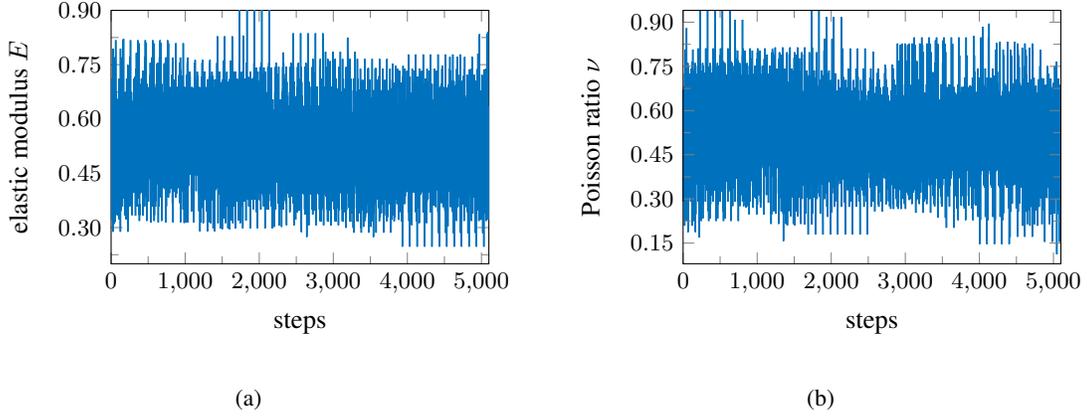
\begin{figure}
	\centering
	\begin{subfigure}{0.4\linewidth}
	\centering
	\begin{tikzpicture}
	\tikzstyle{line} = [draw = myBlue, line width = 0.2mm]
	\begin{axis}[
	ylabel = elastic modulus $E$, 
	xlabel = steps, 
	yticklabel style={
	/pgf/number format/fixed,
	/pgf/number format/precision=2,
	/pgf/number format/fixed zerofill
	},
	xticklabel style={
	/pgf/number format/fixed,
	/pgf/number format/precision=2,
	},
	scaled y ticks=false,
	xmin = 0, xmax = 5100,
	ymin = 0.2, ymax = 0.9, 
	xtick distance = 1000,
    ytick distance = 0.15,
	tick label style = {font=\small},
	minor tick num = 1,
	width = \linewidth,
	height = 0.75\linewidth,
	]
	\addplot[line] table [x expr=\coordindex, y index = 0] {sbU_chainE.dat};
	\end{axis}
	\end{tikzpicture}
	\label{fig:steel_traceE}
	\caption{}
	\end{subfigure}
        \hspace{0.05\linewidth}
	\begin{subfigure}{0.4\linewidth}
	\centering
	\begin{tikzpicture}
	\tikzstyle{line} = [draw = myBlue, line width = 0.2mm]
	\begin{axis}[
	ylabel = Poisson ratio $\nu$,
	xlabel = steps,
	yticklabel style={
	/pgf/number format/fixed,
	/pgf/number format/precision=2,
	/pgf/number format/fixed zerofill
	},
	xticklabel style={
	/pgf/number format/fixed,
	/pgf/number format/precision=2,
	},
	scaled y ticks=false,
	xmin = 0, xmax = 5100,
	ymin = 0.08, ymax = 0.94,
	xtick distance = 1000,
	ytick distance = 0.15,
	minor tick num = 1,
	tick label style = {font=\small},
	width = \linewidth,
	height = 0.75\linewidth,
	area style
	]
	\addplot[line] table [x expr=\coordindex, y index = 0] {sbU_chainNu.dat};
	\end{axis}
	\end{tikzpicture}
	\label{fig:steel_traceNu}
	\caption{}
	\end{subfigure}
	\caption{Trace plots of Markov chains for elasticity parameters $E$ and $\nu$ after burn-in} 
	\label{fig:steel_trace}
\end{figure}
The obtained Markov chain can be directly used for the computation of the elasticity parameters $\GKap\INDe = \lbrace K, G \rbrace^T$, hence, contrary to the NLS approach, no assumption regarding the distribution of the elasticity parameters $E$ and $\nu$ is required. In this way, we obtain $K^* = \num{152306} \pm \SI{11803}{\N\per\mm\squared}$ and $G^*= \num{79392} \pm \SI{3526}{\N\per\mm\squared}$. Note that reporting the mean value and standard deviation here does not imply a Gaussian distribution for the parameters. Indeed, the Bayesian sampling approach allows to estimate complex distributions from the posterior sample. \\

\begin{figure*}[ht]
    \centering
    \includegraphics[width=0.8\textwidth]{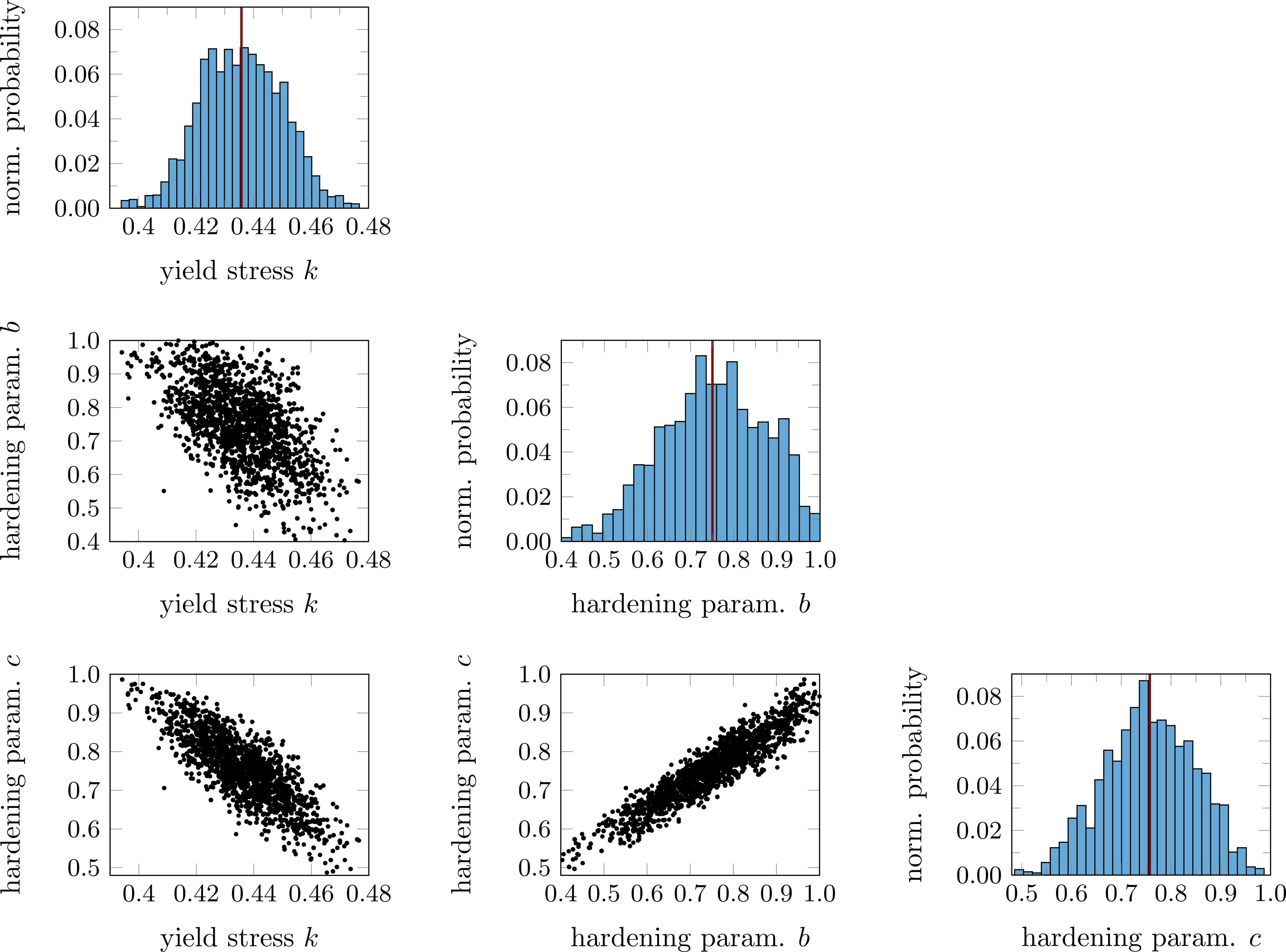}
    \caption{Corner plot for calibration of plasticity parameters $\GKap\INDp$ using Bayesian inference with elasticity parameters $\GKap\INDe = \lbrace{163299,82661 \rbrace}^T \si{\N\per\mm\squared}$}
    \label{fig:steel_plasticBayes_cornerPlot}
\end{figure*}

\begin{figure*}[ht]
    \centering
    \includegraphics[width=0.8\textwidth]{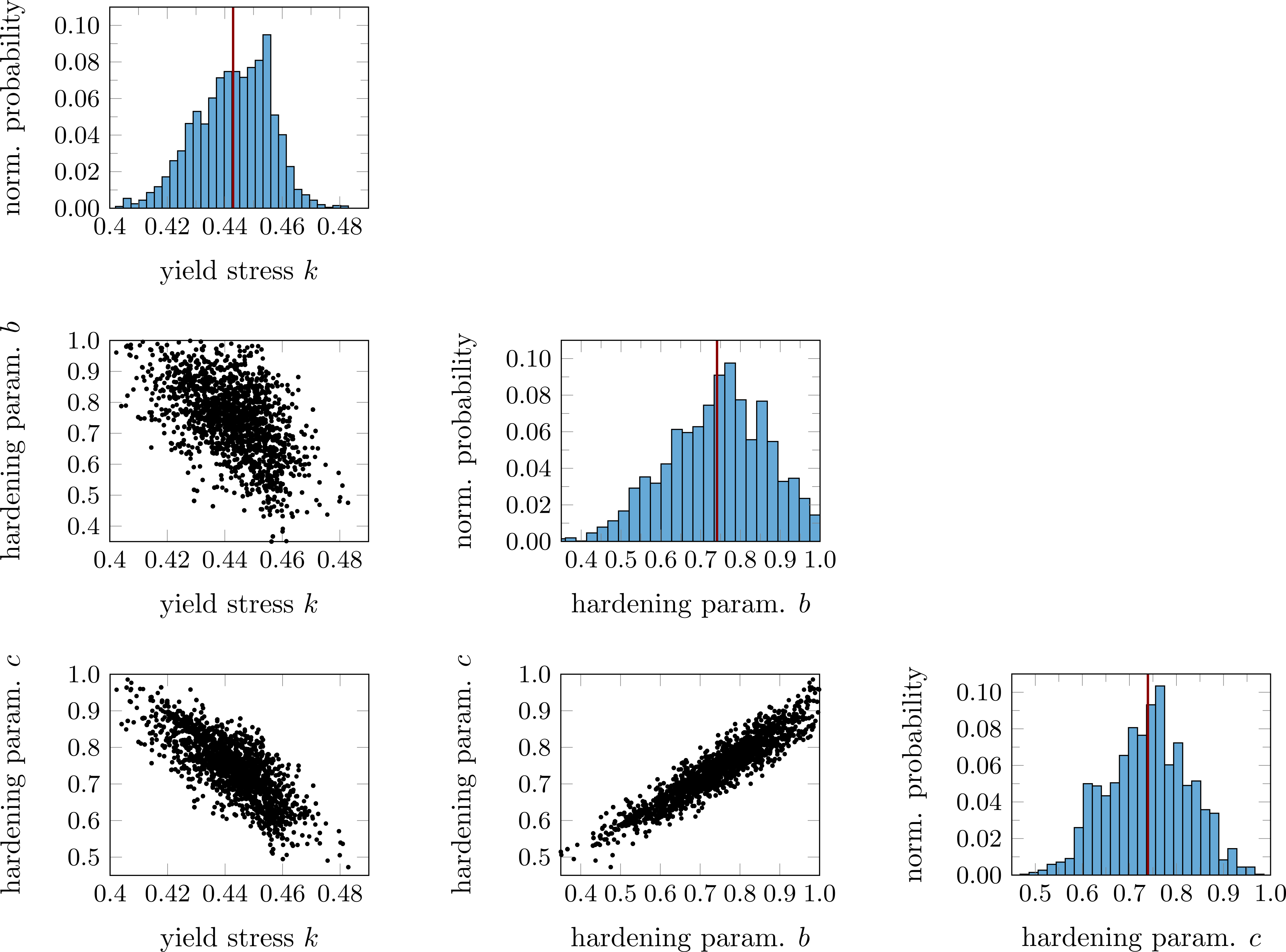}
    \caption{Corner plot for calibration of plasticity parameters $\GKap\INDp$ using Bayesian inference with elasticity parameters $\GKap\INDe = \{K, G\}^T= \lbrace{155136,75105 \rbrace}^T \si{\N\per\mm\squared}$}
    \label{fig:steel_plasticBayes_cornerPlot2}
\end{figure*}

\noindent\textit{Plasticity parameters:}
Analogously to the NLS approach, the plastic material parameters $\GKap\INDp$ are inferred from the elasto-plastic stress response of the material. Again, uniform priors are used, where 20\% variation around $\hat{k} = \SI{290}{\N\per\mm\squared}$ is chosen for the yield stress $k$ and 30\% variation around $\hat{b} = \num{35}$ and $\hat{c} = \SI{3000}{\N\per\mm\squared}$ is applied for the hardening parameters. Step sizes of 4 and 80 chains of chain length 100 are selected to obtain sufficient acceptance rates. Exemplarily, we show the calibration results for two different sets of elasticity parameters in Figs.~\ref{fig:steel_plasticBayes_cornerPlot} and \ref{fig:steel_plasticBayes_cornerPlot2}.

The parameter estimation and uncertainty quantification is done according to Section~\ref{sec:twoStep_bayes} using 1000 samples of the elasticity parameters $\GKap\INDe$. Depending on the elasticity parameters, different plasticity parameters $\GKap\INDp^*$ and uncertainties $\Delta\GKap\INDp^*$ are determined. The distributions are shown in Figs.~\ref{fig:hist_plastic_Bayes} and \ref{fig:hist_plastic_Bayes_std}, where the calibrated parameters $\GKap\INDp^*$ and uncertainties $\delta\GKap^*$ under consideration of elasticity parameter uncertainty are indicated with red lines. The model response with the calibrated parameters compared to the experimental data is shown in Fig.~\ref{fig:fit_plasticity_stress} as well.

\begin{figure*}
\centering
\begin{subfigure}[b]{0.3\textwidth}
\centering
\begin{tikzpicture}
\tikzstyle{meanLine} = [draw = darkred, line width = 0.3mm]
\begin{axis}[
	ylabel = norm. probability, 
	xlabel = yield stress $k^*$, 
	yticklabel style={
	/pgf/number format/fixed,
	/pgf/number format/precision=2,
	/pgf/number format/fixed zerofill
	},
	xticklabel style={
	/pgf/number format/fixed,
	/pgf/number format/precision=3,
	/pgf/number format/fixed zerofill
	},
	scaled y ticks=false,
	xmin = 0.43, xmax = 0.448,
	ymin = 0.0, ymax = 0.13,
	ytick distance = 0.04,
	xtick distance = 0.005,
        minor tick num = 1,
	tick label style = {font=\small},
	width = \linewidth,
	height = 0.85\linewidth,
	area style
	]
	\addplot[ybar interval, fill = myBlue, fill opacity = 0.6, draw = black] table {sbU_hist_k_UQ.dat};
	\draw[meanLine] (axis cs:0.4394,0.0) -- (axis cs:0.4394,0.15);
	\end{axis}
\end{tikzpicture}
\caption{}
\label{fig:hist_k_Bayes}
\end{subfigure}
\hspace{0.04\textwidth}
\begin{subfigure}[b]{0.3\textwidth}
\centering
\begin{tikzpicture}
\tikzstyle{meanLine} = [draw = darkred, line width = 0.3mm]
\begin{axis}[
	ylabel = norm. probability, 
	xlabel = hardening param. $b^*$, 
	yticklabel style={
	/pgf/number format/fixed,
	/pgf/number format/precision=2,
	/pgf/number format/fixed zerofill
	},
	xticklabel style={
	/pgf/number format/fixed,
	/pgf/number format/precision=2,
	/pgf/number format/fixed zerofill
	},
	scaled y ticks=false,
	xmin = 0.722, xmax = 0.78,
	ymin = 0.0, ymax = 0.09,
	ytick distance = 0.02,
	xtick distance = 0.014,
        minor tick num = 1,
	tick label style = {font=\small},
	width = \linewidth,
	height = 0.85\linewidth,
	area style
	]
	\addplot[ybar interval, fill = myBlue, fill opacity = 0.6, draw = black] table {sbU_hist_b_UQ.dat};
	\draw[meanLine] (axis cs:0.7528,0.0) -- (axis cs:0.7528,0.10);
	\end{axis}
\end{tikzpicture}
\caption{}
\label{fig:hist_b_Bayes}
\end{subfigure}
\hspace{0.04\textwidth}
\begin{subfigure}[b]{0.3\textwidth}
\centering
\begin{tikzpicture}
\tikzstyle{meanLine} = [draw = darkred, line width = 0.3mm]
\begin{axis}[
	ylabel = norm. probability, 
	xlabel = hardening param. $c^*$, 
	yticklabel style={
	/pgf/number format/fixed,
	/pgf/number format/precision=2,
	/pgf/number format/fixed zerofill
	},
	xticklabel style={
	/pgf/number format/fixed,
	/pgf/number format/precision=2,
	/pgf/number format/fixed zerofill
	},
	scaled y ticks=false,
	xmin = 0.73, xmax = 0.772,
	ymin = 0.0, ymax = 0.09,
	ytick distance = 0.02,
	xtick distance = 0.01,
        minor tick num = 1,
	tick label style = {font=\small},
	width = \linewidth,
	height = 0.85\linewidth,
	area style
	]
	\addplot[ybar interval, fill = myBlue, fill opacity = 0.6, draw = black] table {sbU_hist_c_UQ.dat};
	\draw[meanLine] (axis cs:0.7521,0.0) -- (axis cs:0.7521,0.10);
	\end{axis}
\end{tikzpicture}
\caption{}
\label{fig:hist_c_Bayes}
\end{subfigure}
\caption{Distributions of calibrated plasticity parameters $\GKap\INDp^*$, when using different sets of elasticity parameters $\GKap\INDe$}
\label{fig:hist_plastic_Bayes}
\end{figure*}
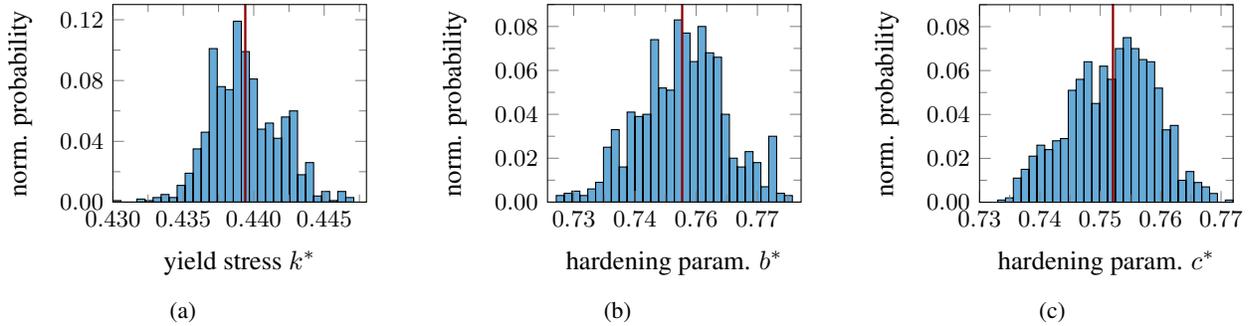
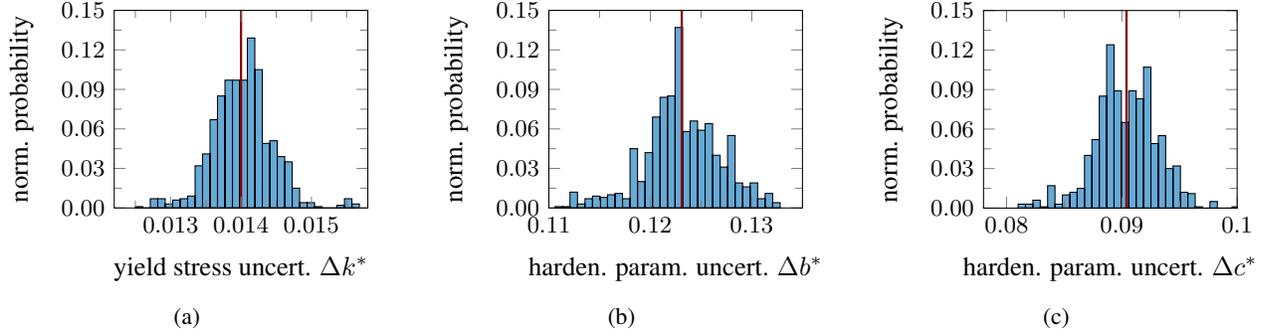
\begin{figure*}
\centering
\begin{subfigure}[b]{0.3\textwidth}
\centering
\begin{tikzpicture}
\tikzstyle{meanLine} = [draw = darkred, line width = 0.3mm]
\begin{axis}[
	ylabel = norm. probability, 
	xlabel = yield stress uncert. $\Delta k^*$, 
	yticklabel style={
	/pgf/number format/fixed,
	/pgf/number format/precision=2,
	/pgf/number format/fixed zerofill
	},
	xticklabel style={
	/pgf/number format/fixed,
	/pgf/number format/precision=4,
%	/pgf/number format/fixed zerofill
	},
	scaled x ticks = false,
	scaled y ticks = false,
	xmin = 0.0122, xmax = 0.0158,
	ymin = 0.0, ymax = 0.15,
	ytick distance = 0.03,
	xtick distance = 0.001,
        minor tick num = 1,
	tick label style = {font=\small},
	width = \linewidth,
	height = 0.85\linewidth,
	area style
	]
	\addplot[ybar interval, fill = myBlue, fill opacity = 0.6, draw = black] table {sbU_hist_k_UQ_std.dat};
	\draw[meanLine] (axis cs:0.0140,0.0) -- (axis cs:0.0140,0.15);
	\end{axis}
\end{tikzpicture}
\caption{}
\label{fig:hist_k_Bayes_std}
\end{subfigure}
\hspace{0.04\textwidth}
\begin{subfigure}[b]{0.3\textwidth}
\centering
\begin{tikzpicture}
\tikzstyle{meanLine} = [draw = darkred, line width = 0.3mm]
\begin{axis}[
	ylabel = norm. probability, 
	xlabel = harden. param. uncert. $\Delta b^*$, 
	yticklabel style={
	/pgf/number format/fixed,
	/pgf/number format/precision=2,
	/pgf/number format/fixed zerofill
	},
	xticklabel style={
	/pgf/number format/fixed,
	/pgf/number format/precision=3,
%	/pgf/number format/fixed zerofill
	},
	scaled x ticks=false,
	xmin = 0.11, xmax = 0.135,
	ymin = 0.0, ymax = 0.15,
	ytick distance = 0.03,
        xtick distance = 0.01,
        minor tick num = 1,
	tick label style = {font=\small},
	width = \linewidth,
	height = 0.85\linewidth,
	area style
	]
	\addplot[ybar interval, fill = myBlue, fill opacity = 0.6, draw = black] table {sbU_hist_b_UQ_std.dat};
	\draw[meanLine] (axis cs:0.1231,0.0) -- (axis cs:0.1231,0.15);
	\end{axis}
\end{tikzpicture}
\caption{}
\label{fig:hist_b_Bayes_std}
\end{subfigure}
\hspace{0.04\textwidth}
\begin{subfigure}[b]{0.3\textwidth}
\centering
\begin{tikzpicture}
\tikzstyle{meanLine} = [draw = darkred, line width = 0.3mm]
\begin{axis}[
	ylabel = norm. probability, 
	xlabel = harden. param. uncert. $\Delta c^*$, 
	yticklabel style={
	/pgf/number format/fixed,
	/pgf/number format/precision=2,
	/pgf/number format/fixed zerofill
	},
	xticklabel style={
	/pgf/number format/fixed,
	/pgf/number format/precision=3,
%	/pgf/number format/fixed zerofill
	},
	scaled x ticks = false,
	scaled y ticks = false,
	xmin = 0.078, xmax = 0.1,
	ymin = 0.0, ymax = 0.15,
	ytick distance = 0.03,
	xtick distance = 0.01,
        minor tick num = 1,
	tick label style = {font=\small},
	width = \linewidth,
	height = 0.85\linewidth,
	area style
	]
	\addplot[ybar interval, fill = myBlue, fill opacity = 0.6, draw = black] table {sbU_hist_c_UQ_std.dat};
	\draw[meanLine] (axis cs:0.0904,0.0) -- (axis cs:0.0904,0.15);
	\end{axis}
\end{tikzpicture}
\caption{}
\label{fig:hist_c_Bayes_std}
\end{subfigure}
\caption{Distributions of calibrated parameter uncertainties $\Delta\GKap^*\INDp$ when using different sets of elasticity parameters $\GKap\INDe$}
\label{fig:hist_plastic_Bayes_std}
\end{figure*}

\subsubsection{Discussion}
The identified material parameters are compiled in Tab.~\ref{tab:Ident_plasticity}, where the notation $\Delta\GKap^*$ is used for uncertainties computed by the model calibration scheme and $\delta\GKap^*$ for uncertainties when considering uncertain elasticity parameters.
\begin{table}[ht]
    \caption{Results from calibration of small strain elasto-plasticity constitutive model from experimental data}
    \label{tab:Ident_plasticity}
    \centering
%    \resizebox{0.475\textwidth}{!}{
    \begin{tabular}{>{\centering\arraybackslash}m{0.6cm} c c c c c}
        \toprule
        Para- & $K$ & $G$ & $k$ & $b$ & $c$ \\
        meter & [\si{\N\per\mm\squared}] & [\si{\N\per\mm\squared}] & [\si{\N\per\mm\squared}] & [-] & [\si{\N\per\mm\squared}] \\
        \midrule
        \multicolumn{6}{l}{NLS} \\
        \midrule
        $\GKap^*$ & $\num[round-mode=places,round-precision=0]{150991}$ & $\num[round-mode=places,round-precision=0]{79321}$ & $\num[round-mode=places,round-precision=1]{282.6304}$ & $\num[round-mode=places,round-precision=2]{41.0409}$ & $\num[round-mode=places,round-precision=1]{3499.7771}$ \\
        $\Delta\GKap^*$ & - & - & $\num[round-mode=places,round-precision=2]{1.1161}$ & $\num[round-mode=places,round-precision=2]{1.7603}$ & $\num[round-mode=places,round-precision=1]{116.7866}$ \\
        $\delta\GKap^*$ & $\num[round-mode=places,round-precision=0]{2951}$ & $\num[round-mode=places,round-precision=0]{628}$ & $\num[round-mode=places,round-precision=2]{1.0683}$ & $\num[round-mode=places,round-precision=2]{1.6829}$ & $\num[round-mode=places,round-precision=1]{111.6543}$ \\
        \midrule
        \multicolumn{6}{l}{Bayesian inference} \\
        \midrule
        $\GKap^*$ & $\num[round-mode=places,round-precision=0]{152306}$ & $\num[round-mode=places,round-precision=0]{79392}$ & $\num[round-mode=places,round-precision=1]{282.97}$ & $\num[round-mode=places,round-precision=2]{40.31}$ & $\num[round-mode=places,round-precision=1]{3453.7}$ \\
        $\delta\GKap^*$ & $\num[round-mode=places,round-precision=0]{11803}$ & $\num[round-mode=places,round-precision=0]{3526}$ & $\num[round-mode=places,round-precision=2]{1.627}$ & $\num[round-mode=places,round-precision=2]{2.59}$ & $\num[round-mode=places,round-precision=1]{162.65}$ \\
        \bottomrule
    \end{tabular}
%    }
\end{table}
The determined parameters $\GKap^* = \lbrace \GKap\INDe^{*T}, \GKap\INDp^{*T} \rbrace^T$ of both methods are in good agreement. Both methods yield reasonable parameter uncertainties. However, the Bayesian approach yields larger uncertainties compared to the NLS method for the plasticity parameters. 
This can be attributed to the fact that the LS uncertainties are based on linearization and Gaussian assumptions, and they only hold exactly in the asymptotic regime of very large data sets. The posterior histograms for the plasticity parameters are also clearly non-Gaussian. Although the sampling-based Bayesian approach can handle these types of uncertainties, the associated sampling cost can be very high. Instead of the hierarchical Bayesian approach, outlined in Section~\ref{sec:twoStep_bayes}, one could also determine the full posterior $p(\GKap| \GV{d})$ directly. The hierarchical setting was chosen here, since it closely resembles the frequentist two-step procedure.

It should further be mentioned that -- in contrast to \citep{dileephartmann2022,troegerhartmann2022}, for example, where the Gaussian error propagation is employed -- a different approach is followed here to compute the uncertainties of subsequently identified parameters. The main advantage of the procedure, which is explained in Section~\ref{sec:twoStep_frequentist} and in more detail in Appendix~\ref{sec:two-step-appendix}, is that the majority of the required quantities, such as the Jacobians, are already known from the model calibration. Thus, the computation of the partial derivatives required for the Gaussian error propagation, see Eq.~\eqref{eq:uncertainty}, is circumvented, and the uncertainty estimation can be easily performed in a post-processing step.

%*************************************
\section{Conclusions}
While increasingly large amounts of data can be obtained from experimentation and measurement techniques, these data are usually sparse, i.e., they provide a reduced amount of information compared to a full-field simulation, and it tends to be noisy. A question that often arises in this context is how such data can be related to physical models and high-fidelity 3D fields, that can be computed from such models. Thus, apart from model calibration using real experiments, there is an increasing need for solutions to interpret the experimental data. As a consequence, parameter identification, which in the past played a role mainly in constitutive modeling combined with experimental mechanics, is increasingly gaining attention across disciplines.

The present manuscript provides a coherent overview of well-established experiments and measurement possibilities as well as known and newly developed computational approaches for parameter identification in the context of solid mechanics. By following a classification into reduced and all-at-once approaches
from the inverse problems community, most of the methods available in the literature can be classified. The all-at-once formulation combines the model residual and the difference between simulated and observed data into a single objective function, and the reduced formulation is obtained as the limit when the discrete model equation is enforced as a strong constraint. Interestingly, the virtual fields method can be recovered at the other end of the spectrum when the data-related objective is enforced strictly. This holistic view, which is one of the main contributions of this work, highlights the connections between the different methods -- which then allows for a better understanding of identifiability and the need for regularization, which is often crucial in parameter estimation problems. Moreover, this structured framework allows to easily formulate new approaches to parameter identification, two of which are put forth in this work. Since data are usually noisy, an uncertainty quantification perspective is included in the considerations as well. Bayesian uncertainty quantification, in particular, provides a parallel derivation for well-established regularization approaches. Our main contribution in this regard is the formulation of a parameter estimation and uncertainty quantification method for material models, proceeding in two subsequent steps, whereby we compare frequentist and Bayesian approaches. 

The review is accompanied by two numerical examples. The first is a linear elastic problem with artificially generated data, for which the different reduced and all-at-once approaches as well as virtual fields methods are applied and compared. The second example focuses on two-step inference in a frequentist setting and hierachical Bayesian inference for an inelastic material using real-world experimental data within the classical finite element least-squares approach.

From an application point of view, not all methods have the same level of maturity. Whereas reduced approaches are already widely used in practice, the all-at-once paradigm still needs further developments to be applicable to large-scale problems. The main difficulty is the large parameter space consisting of both material parameters and state vector. Additionally, the sensitivity to noise, which we observed in our numerical examples, needs to be reduced. Still, the all-at-once approach is of interest not only because of the unified perspective but also because efficient iterative solutions seem possible. While the paper mainly compares and discusses the first-order necessary conditions of different approaches, black box and mildly intrusive adaptations are possible in many cases, which is often desirable in practice. As a minimum implementation requirement, access to the parametric model residuals and the state quantities is needed to compute the objective function value.

We hope that the present study might help researchers and engineers working on different aspects of parameter identification to gain a better understanding of the potentials and pitfalls in parameter identification. In this sense, the unified framework may serve as an anchor point for the development of new methods by leveraging knowledge from statistics, optimization, machine learning, and finite elements.

\section{Acknowledgement}
The support of the German Research Foundation is gratefully acknowledged in the following projects:
\begin{itemize}
	\item DFG GRK2075-2 (Ulrich Römer and Henning Wessels): \textit{Modelling the constitutional evolution of building materials and structures with respect to aging}.
	\item DFG 501798687 (Henning Wessels): \textit{Monitoring data-driven life cycle management with AR based on adaptive, AI-supported corrosion prediction for reinforced concrete structures under combined impacts}. Subproject of SPP 2388: \textit{Hundred plus - Extending the Lifetime of Complex Engineering Structures through Intelligent Digitalization}.
\end{itemize}
Moritz Flaschel and Laura De Lorenzis acknowledge support from the Swiss National Science Foundation (SNF), project number 200021\_204316.

%*************************************
\section*{Code and data availability statement}
All the codes and experimental data will be released as open-source upon publication in the peer-reviewed literature.

%*************************************
\appendix

\section{Processing of experimental and numerical full-field data}
\label{sec:processingData}

The identification of material parameters from full-field data requires particular attention to the extraction of the data from both experiment as well as numerical model. Therefore, the procedure is explained in detail with a focus on the non-linear least-squares method using finite elements, see Sec.~\ref{sec:nlsfemdic}.

For full-field measurement data, one obtains from experiment $\hat{E}$, $\hat{E}=1,\ldots,\numexp$, a data vector $\GV{d}^{\,(\hat{E})}\elm{\numexpE}$. The data are displacements, in-plane stretches/strains, or in-plane strain components at spatial positions on the surface of the specimen at particular times. In addition, it can be extended by discrete force information from a load cell. In other words, in the case of $\numNe$ (temporal) load, i.e.,\ time steps, each time step consists of $\numdE$ entries (discrete spatially distributed displacements, in-plane principal stretches, in-plane strains, and forces concerned, \ldots), i.e.,\ $\numexpE = \numNe \numdE$. In experiment $\hat{E}$, there is a sampling rate that provides the time points $t_m$, $m=1,\ldots,\numNe$, at which we need to evaluate. The data of each experiment is compiled in the vector $\GV{d}^{\,(\hat{E})T} = \{\GV{d}_0^{\,(\hat{E})T}, \ldots, \GV{d}_\numNe^{\,(\hat{E})T}\}$, where $\GV{d}_m^{\,(\hat{E})} \elm{\numdE}$ symbolizes the data from one experiment at time $t_m$. If all data, from all considered experiments, is assembled, the entire data vector $\GVT{d} = \{\GV{d}^{\,(1)T}, \ldots, \GV{d}^{\,(\numexp)T}\}$, $\GV{d} \elm \numD$ with $\numD = \sum_{\hat{E}=1}^{\numexp} \numexpE$ is obtained. This includes either experiments that are repeated several times, or completely different loading paths, boundary conditions, or geometries.

On the other hand, the finite element model provides for each re-computed experiment $\hat{E}$, displacements ($\GV{u}\INDn^{\,(\hat{E})}$ and $\GVq{u}\INDn^{\,(\hat{E})}$) -- or strains/stretches, \ldots --, and reaction forces $\GV{p}\INDn^{\,(\hat{E})}$ for the time points $\tn^{\,(\hat{E})}$, $n=1,\ldots,N^{\,(\hat{E})}$. Obviously, the temporal and the spatial points of the experiment and the finite element model are different. Consequently, the finite element results are temporally and spatially interpolated to match the time points and the measurement points of the experiment. In the time domain,  a linear interpolation is chosen to align the experimental data with the model time data because sensitivities (particular derivatives), which should not be interpolated, are required later on. Accordingly, the dimension of the data vectors $\GV{d}$ is as large as it is given by the finite element simulations.

Some remarks concerning the digital image correlation data need to be added. Initially, DIC-data can only provide information about the deformation of the surface of a sub-area of a specimen or component. These displacements partly contain rigid body motions (based on the insufficient machine stiffness or the test equipment), which are difficult to consider in a finite element program. Some commercial DIC suppliers provide a rigid-body compensation (unfortunately, without specifying the mathematical background and procedures). Here, it thus makes more sense not to use the surface displacements, but the surface strains. This is done on the level of displacements in \cite{rosemenzel2020} or on the level of strains/stretches in \cite{hartmanngilbert2021}. 

Commercial finite element programs have a similar problem: Only a few programs are available for the comparison with DIC-data of the surface strains, where it is possible to evaluate the strains or measures, which can be determined from the strains (principal strains, shear angles, \ldots) at arbitrary points of the surface. Here, it could help to employ evaluation tools that are based either on triangulation, \cite{hsuschwabrigamontihumphrey1994,orteu2009,hartmannrodriguez2018}, or, for example, global strain determination tools, \cite{hartmannmuellerlohsetroeger2021}. These concepts can be applied to both the DIC-data as well as to the finite element nodal displacement data. This has the additional advantage that the same interpolation scheme and strain evaluation is applied to both systems, i.e.,\ DIC and FEM. Furthermore, the sensitivity analysis, i.e.,\ the derivatives, can be computed analytically.

Usually, the displacements -- or resulting quantities such as the principal strains or stretches -- are only compared with the assigned quantities of the experiment in a sub-area of the finite element (surface) domain, i.e.,\ only the subset  $\GVt{u}_n^{(\hat{E})} = \GMt{M}^{(\hat{E})}{\U}_n^{(\hat{E})}$, $\GVt{u}_n^{(\hat{E})} \elm{\numutE}$ of the finite element nodal displacements $\U_n^{(\hat{E})}\elm{\numuE}$ are included. For this purpose, the assignment matrix $\GMt{M}^{(\hat{E})} \elmm{\numutE}{\numuE}$ is introduced. In addition, single forces should be included, which are determined from the reaction forces $\GV{p}\INDn$, $\GV{F}_{\text{FEM}n}^{(\hat{E})} = -\GMq{M}^{(\hat{E})} \GV{p}\INDn$, $\GV{F}_{\text{FEM}n}^{(\hat{E})} \elm{\numFE}$. In a tensile experiment $\numFE=1$, whereas in a biaxial tensile test $\numFE=2$. For tension-torsion problems $\numFE=2$, namely the axial force and the torque. The matrix $\GMq{M}^{(\hat{E})}$ is exploited to extract the reaction force(s) / moment(s) from the Lagrange multipliers. The consideration of reaction forces in the NLS-FEM-DIC approach can be found in \cite{pottiertoussaintvacher2011,hartmanngilbert2021} for inelastic materials.

\section{System matrices for linear expressions in the parameters}
\label{ap:system_matrices}

In the case of linear elasticity, see the finite element matrix representation in Eq.~\eqref{eq:linelastsystem}, the term $\GM{K}(\GKap) \U + \GMq{K}(\GKap) \Uq$, which is linear in the material parameters $\GKap$, can be reformulated. For this purpose, the elasticity matrix is reformulated
\begin{equation}
    \label{eq:elasticitymatrix}
    \LM{C}\INDel
    = \sum_{k=1}^6 \sum_{l=1}^6 c_{kl} \LV{e}_k \LVT{e}_l
    = [\LV{c}_1 \; \ldots \; \LV{c}_6]
\end{equation}
with the unit vectors $\LV{e}_k \elm{6}$ with zero entries, and only a $1$ on the $k$-th entry. $\LV{c}_k \elm{6}$ represents the $k$-th column vector of the matrix $\LM{C}\INDel$. This expression is inserted into Eq.~\eqref{eq:system_hartmann}, and with Eq.~\eqref{eq:elementstiff} as well as ~\eqref{eq:Kdefinitions}, we obtain
\begin{equation}
    \GM{K}(\GKap) \U + \GMq{K}(\GKap) \Uq =
    \sum_{e=1}^\nel {\GM{Z}^{\,e}}^T
    \Bigg\{
        \sum_{j=1}^\numgu w^{e,j} {\LM{B}^{e,j}}^T
        \left[\sum_{k=1}^6 \sum_{l=1}^6 c_{kl} \LV{e}_k \LVT{e}_l\right] \LV{E}^{e,j} \det \LM{J}^{e,j}
    \Bigg\}. 
\end{equation}
The expression $\LM{C}\INDel \LV{E}^{e,j}$ with $\LV{E}^{e,j} =
\LM{B}^{e,j} \{\GM{Z}^{\, e} \U + \GMq{Z}^{\; e} \Uh\}$ is reformulated
\begin{equation}
    \left[\sum_{k=1}^6 \sum_{l=1}^6 c_{kl} \LV{e}_k \LVT{e}_l\right] \LV{E}^{e,j}
    =
    \sum_{k=1}^6 \sum_{l=1}^6 ({\LV{E}^{e,j}}^T c_{kl} \LV{e}_l) \LV{e}_k 
    =
    \sum_{k=1}^6 ({\LV{E}^{e,j}}^T \LV{c}_{k}) \LV{e}_k
    =
    \underbrace{
    \begin{bmatrix}
        {\LV{E}^{e,j}}^T & &  \\
        & \ddots & \\
        & & {\LV{E}^{e,j}}^T   
    \end{bmatrix}}_{\LM{E}\INDS^{e,j} \elmm{6}{\numkappa}}
    \GKap
\end{equation}
with the vector of material parameters
\begin{equation}
    \GKap = \{\LVT{c}_1 \; \ldots \; \LVT{c}_6\}^T. 
\end{equation}
This leads to the final linear expression
\begin{equation}
    \label{eq:linAkappa}
    \GM{K}(\GKap) \U + \GMq{K}(\GKap) \Uq 
    =
    \GM{A}\INDS(\U,\Uq) \GKap
    =
    \GVq{p}
\end{equation}
with
\begin{equation}
    \label{eq:Asdef}
    \GM{A}\INDS
    =
    \sum_{e=1}^\nel {\GM{Z}^{\,e}}^T
    \LM{a}\INDS^{(e)},
    \quad
    \GM{A}\INDS \elmm{\numu}{\numkappa}
\end{equation}
and the element contribution
\begin{equation}
    \LM{a}\INDS^{(e)} 
    = 
    \sum_{j=1}^\numgu w^{e,j} {\LM{B}^{e,j}}^T
    \LM{E}\INDS^{e,j} \det \LM{J}^{e,j}.
\end{equation}
Of course, by exploiting symmetry conditions (less entries in $\GKap$) or the specific structure of the elasticity matrix $\LM{C}$, the matrix $\LM{a}\INDS^{(e)}$ has to be adapted. This holds for the consideration of omitting zero-multiplications in the matrix-matrix product as well. 

Analogously, 
\begin{equation}
    \label{eq:resultingp}
    \GV{p} = \GMq{K}^T(\GKap) \U + \GMqq{K}(\GKap) \Uq = \GMq{A}\INDS(\U,\Uq) \GKap
\end{equation}
can be obtained with
\begin{equation}
    \label{eq:Asqdef}
    \GMq{A}\INDS
    =
    \sum_{e=1}^\nel {\GMq{Z}^{\,e}}^T
    \LM{a}\INDS^{(e)},
    \quad
    \GMq{A}\INDS \elmm{\nump}{\numkappa}.
\end{equation}

With respect to the investigation here, the special consideration of equivalent nodal forces which are non-zero is of special interest. Furthermore, a scaling of the non-zero elements is performed, and the influence of the reaction force on the inverse problem is treated. For this purpose, the equivalent nodal vector $\GVq{p}$ is reassembled into zero and non-zero values,
\begin{equation}
    \mzweiv{\GV{0}}{\GVqc{p}}
    = \GM{M} \GVq{p}
\end{equation}
where $\GM{M}$ symbolizes a reordering matrix. Thus, Eq.~\eqref{eq:linAkappa} reads
\begin{equation}
    \label{eq:Askappq}
    [\GM{M} \GM{A}\INDS] \GKap = 
    \begin{bmatrix}
        \GM{A}\INDS\EXPf \\
        \GM{A}\INDS\EXPp
    \end{bmatrix}
    \GKap
    =
    \mzweiv{\GV{0}}{\GVqc{p}}.
\end{equation}
Second, a resulting force $\check{p}$ is required for comparison purposes to experimental data (of course, the investigation can be extended to further resulting forces as it is necessary in, for example, biaxial tensile tests. However, this is omitted for the sake of brevity). This is done by summing up particular entries of $\GV{p}$ in Eq.~\eqref{eq:resultingp} 
\begin{equation}
    \label{eq:pcheck}
    \check{p} 
    = \GVT{m} \GV{p} 
    = \underbrace{\big[\GVT{m} \GMq{A}\INDS\big]}_{\GMq{A}\INDS\EXPr} \GKap
    = \GMq{A}\INDS\EXPr \GKap.
\end{equation}
$\LV{m}\elm{\nump}$ might be a vector containing ones or zeros to sum up the entries in $\GV{p}\elm{\nump}$. This relationship is weighted to obtain similar magnitudes of the entries for the system of equations that will be derived later,
\begin{equation}
    \label{eqmf:system_virtual_fields_method}
    \GM{A} \GKap = \GVc{p}
\end{equation}
with
\begin{equation}
    \GM{A} 
    = 
    \begin{bmatrix}
        \GM{A}\INDS\EXPf \\
        \sqrt{\sigma_r} \; \GMq{A}\INDS\EXPr
    \end{bmatrix}, \qquad
    \GVc{p} 
    = 
    \begin{bmatrix}
    \GV{0} \\
    \sqrt{\sigma_r} \check{p}
    \end{bmatrix}.
\end{equation}
$\sigma_r$ is chosen for scaling purposes.
A similar treatment is carried out for the expressions using the stiffness matrix representation of Eqns.~\eqref{eq:linAkappa} and \eqref{eq:resultingp}.
Eq.~\eqref{eq:linAkappa} is decomposed into
\begin{equation*}
    \GVqq{p} := \GM{M} \GVq{p}
    = \GM{M} \{\GM{K} \U + \GMq{K} \Uq \}
    = \GM{M} \GM{K} \U + \GM{M} \GMq{K} \Uq
\end{equation*}
such that
\begin{equation}
\GVqq{p}
=
\mzweiv{\GV{0}}{\GVqc{p}}
=
\begin{bmatrix}
    \GM{K}\EXPf & \GMq{K}\EXPf \\
    \GM{K}\EXPp & \GMq{K}\EXPp
\end{bmatrix}
\mzweiv{\GV{u}}{\GVq{u}},
\end{equation}
i.e.,\ nodes where no equivalent nodal forces act are explicitly set to zero.

Next, the resulting force \eqref{eq:pcheck} is reformulated
\begin{equation}
\label{eq:Kfr}
\check{p}
=
\GVT{m} \GV{p} 
=
\underbrace{\GVT{m} \GMq{K}^T}_{\GMq{K}\EXPr} \GV{u} + \underbrace{\GVT{m} \GMqq{K}}_{\GMqq{K}\EXPr} \GVq{u}.
\end{equation}
Once again compiling the terms above into a large system
\begin{equation}
    \underbrace{
    \begin{bmatrix}
        \GM{K}\EXPf \\
        \sqrt{\sigma_r} \; \GMq{K}\EXPr
    \end{bmatrix}
    }_{\GM{K}\EXPfr}
    \GV{u}
    +
    \underbrace{
    \begin{bmatrix}
        \GMq{K}\EXPf \\
        \sqrt{\sigma_r} \; \GMqq{K}\EXPr
    \end{bmatrix}
    }_{\GMq{K}\EXPfr}
    \GVq{u}
    =
    \underbrace{\mzweiv{\GV{0}}{\sqrt{\sigma_r} \check{p}}}_{\GVc{p}},
\end{equation}
we finally obtain the system
\begin{equation}
    \label{eqmf:system_AAO_FEM}
    \GM{K}\EXPfr \GV{u} + \GMq{K}\EXPfr \GVq{u} = \GVc{p}.
\end{equation}
A node-wise implementation for plane problems can be found in \cite{flaschel_unsupervised_2021}. 

\section{Artificial neural networks}
\label{sec:ann}

An \acf{ANN} is a global, smooth function approximator $\mathcal{U}$ defining a mapping from an input space $\mathbb{R}^{N}$ to an output space $\mathbb{R}^{M}$, such that
\begin{equation}\label{eq:ann_mapping}
    \mathcal{U}: \mathbb{R}^{N} \rightarrow \mathbb{R}^{M}.
\end{equation}
\acp{ANN} typically consist of a high number of computational units called neurons, which are arranged in an input, an output, and any number of so-called hidden layers. A fully connected \ac{FFNN} with a total of $L+1$ layers has $L-1$ hidden layers, where layer $0$ is the input layer and layer $L$ the output layer. The neurons of each two successive layers are connected, and a weight is associated to the connection. The weight of the connection between neuron $k$ in layer $l-1$ and neuron $j$ in layer $l$ for $l \in [1, L]$ is denoted by $w^{l}_{jk}$. All weights between layer $l-1$ and $l$ can then be combined in the weight matrix $\GM{W}^{l}$ with entries $w^{l}_{jk}$. In addition, the neurons in the hidden layers and the output layer are parameterized by a constant that is added to the weighted input, namely the bias parameter. 
The bias of neuron $j$ in layer $l$ is denoted by $b^{l}_{j}$, and the biases of all neurons in layer $l$ can be combined in vector $\bm{b}^{l}$ with entries $b^{l}_{j}$. The output of the neurons in the hidden layers and the output layer are computed from the sum of their weighted inputs and their bias as an argument of an activation function. 

According to the above notation, the mapping from an input to an output vector by an \ac{FFNN} can be formulated as follows: The weighted input $z^{l}_{j}$ of neuron $j$ in the hidden layer or output layer $l$ with an upstream layer consisting of $N_{l-1}$ neurons is defined by
\begin{equation}\label{eq:ann_weighted_input}
    z^{l}_{j}=\sum_{k=1}^{N_{l-1}}w^{l}_{jk}y^{l-1}_{k}+b^{l}_{j}, \; \forall \, l \in [1, L],
\end{equation}
where $y^{l-1}_{k}$ is the output of neuron $k$ in the upstream layer $l-1$, given by
\begin{equation}\label{eq:ann_output}
    y^{l-1}_{k}=\sigma^{l-1}(z^{l-1}_{k}).
\end{equation}
Here, $\sigma^{l-1}$ is the activation function of the neurons in layer $l-1$. Inserting (\ref{eq:ann_output}) in (\ref{eq:ann_weighted_input}), we obtain in symbolic notation
\begin{equation}\label{eq:ann_weighted_input_symbolic}
    \bm{z}^{l}=\bm{W}^{l}\sigma^{l-1}(\bm{z}^{l-1})+\bm{b}^{l}, \; \forall \, l \in [1, L],
\end{equation}
where $\bm{z}^{l-1}$ contains the weighted inputs of all neurons in layer $l-1$, and $\sigma^{l-1}$ is applied element-wise. In the hidden layers, non-linear functions are usually used as activation functions, such as hyperbolic tangent. The activation of the output neurons is typically computed by the identity. For the input layer $l=0$,
\begin{equation}\label{eq:ann_input}
    \boldsymbol{\sigma}^{0}(\boldsymbol{z}^{0})=\boldsymbol{x}
\end{equation}
applies, where $\bm{x}$ is the input vector to the \ac{FFNN}. Given (\ref{eq:ann_weighted_input})-(\ref{eq:ann_input}), the output vector $\bm{y}^{L}$ of the \ac{FFNN} as a function of $\bm{x}$ can be  recursively defined as follows:
\begin{equation}\label{eq:ann_definition}
    \begin{split}
    &\boldsymbol{y}^{L}=\boldsymbol{\sigma}^{L}(\boldsymbol{z}^{L}) \\
    &\boldsymbol{z}^{L}=\boldsymbol{W}^{L}\boldsymbol{\sigma}^{L-1}(\boldsymbol{z}^{L-1})+\boldsymbol{b}^{L} \\
    &\boldsymbol{z}^{L-1}=\boldsymbol{W}^{L-1}\boldsymbol{\sigma}^{L-2}(\boldsymbol{z}^{L-2})+\boldsymbol{b}^{L-1}\\
    &\vdots \\
    &\boldsymbol{z}^{2}=\boldsymbol{W}^{2}\boldsymbol{\sigma}^{1}(\boldsymbol{z}^{1})+\boldsymbol{b}^{2} \\
    &\boldsymbol{z}^{1}=\boldsymbol{W}^{1}\boldsymbol{x}+\boldsymbol{b}^{1}.
    \end{split}
\end{equation}
The definition in Eq.~\eqref{eq:ann_definition} demonstrates that \acp{FFNN} are highly parameterized, nonlinearly composed functions.

During the so-called training process, the parameters of the \ac{FFNN} $\mathcal{U}$ are adjusted to approximate the mapping between the inputs $\bm{x}_{i}$ and outputs $\bm{d}_{i}$, represented by the $N_{i}$ training points in the training data set $\GV{T}^{\text{train}}=\{\bm{x}_{i}, \bm{d}_{i}\}_{i=1}^{N_{i}}$ as closely as possible. The objective of this optimization problem is the loss function $\phi(\bm{\theta}; \GV{T}^{\text{train}})$  depending on the trainable \ac{FFNN} parameters $\bm{\theta}=\{\bm{W}^{l},\bm{b}^{l}\}_{1\leq l \leq L}$ and the training data $\GV{T}^{\text{train}}$. The problem of finding the optimal \ac{FFNN} parameters $\bm{\theta^{*}}$ can be formulated as
\begin{equation}
    \label{eq:ann_optimization_problem_forward}
    \bm{\theta^{*}}=\argmin_{\bm{\theta}}\phi(\bm{\theta};\GV{T}^{\text{train}}).
\end{equation}
The loss function is usually defined in terms of the mean squared error, such that
\begin{equation}
    \label{eq:ann_loss_function_mse}
    \phi(\bm{\theta};\GV{T}^{\text{train}}) = \frac{1}{N_{i}}\sum_{i}\norm{\mathcal{U}(\bm{\theta}; \bm{x}_{i}) - \bm{d}_{i}}^{2}.
\end{equation}
It is generally possible to choose other metrics, depending on the problem.

For the optimization of parameters, it is common to use gradient-based optimization algorithms. The gradient of the loss function with respect to the \ac{FFNN} parameters $\bm{\theta}$ can be calculated using automatic differentiation \cite{baydin_AutomaticDifferentiationMachine_2018}.

It is well known that \ac{FFNN} are universal function approximators \cite{cybenkoApproximationSuperpositionsSigmoidal1989, hornik_MultilayerFeedforwardNetworks_1989, li_SimultaneousApproximationsMultivariate_1996}. Provided that an \ac{FFNN} has a sufficient number of parameters, the \ac{FFNN}, according to the universal approximation theorem, can theoretically approximate any continuous function and its derivatives to an arbitrarily small error. It should be noted, however, that the question of optimal training of \acp{FFNN} -- to reach their full potential -- has not yet been solved. For a more in-depth introduction to deep learning, we refer to standard textbooks, for example,  \cite{goodfellow_DeepLearning_2016}.

\section{Sensitivities in least-squares method using finite elements}
\label{sec:sensnlsfemdic}

Since the simulation response $\GV{s}(\GKap)$ in Eq.~\eqref{eq:sensitivity} is composed of each experiment, and each re-simulation of the experiment leads to a system of non-linear equations \eqref{eq:BEforceII} depending on the material parameters
\begin{equation}
  \label{eq:BEforceIIofkappa}
  \begin{aligned}
    \GV{g}\big(\tnp,\Unp(\GKap),\Qnp(\GKap),\GKap\big) &= \GV{0}, \\
    \GV{l}\big(\tnp,\Unp(\GKap),\Qnp(\GKap),\Qn(\GKap),\GKap\big) &= \GV{0},  \\
    \GV{p}\INDNP(\GKap) 
    &= \GVq{g}\big(\tnp,\Unp(\GKap),\Qnp(\GKap),\GKap\big),
  \end{aligned}
\end{equation}
the sensitivity \eqref{eq:sensitivity} of each temporal point $\tn$ is required. The sensitivities can be computed applying the chain-rule to Eq.~\eqref{eq:BEforceIIofkappa}$_{1,2}$ yielding -- after computing the resulting linear system with several right-hand sides
\begin{equation}
\label{eq:sensiU}
    \Bigg[
    \dif{\GV{g}}{\Unp} - \dif{\GV{g}}{\Qnp} \Bigg[\dif{\GV{l}}{\Qnp}\Bigg]^{-1} \dif{\GV{l}}{\Unp}
    \Bigg]
    \difn{\Unp}{\GKap}
    =-\dif{\GV{g}}{\GKap} + \dif{\GV{g}}{\Qnp} \Bigg[\dif{\GV{l}}{\Qnp}\Bigg]^{-1} 
    \left[
        \dif{\GV{l}}{\GKap} + \dif{\GV{l}}{\Qn} \difn{\Qn}{\GKap}
    \right],
\end{equation}
i.e.\ $\DS \mathsf{d} \Unp/\mathsf{d} \GKap$, see, for details, \cite{hartmann2017,hartmanngilbert2021}. Here, the index $(E)$ of the experiment is omitted for brevity. The leading coefficient matrix represents the tangential stiffness matrix, which can be approximated by the last one compiled in the iterative Newton-step of the Multilevel Newton algorithm or the Newton-Raphson scheme. Since we only use a subset of the nodal displacements with respect to the least-squares method, we obtain
\begin{equation}
    \label{eq:sensU}
    \difn{\GVt{u}\INDNP}{\GKap}
  = \GMt{M} \difn{\Unp}{\GKap}.
\end{equation}
If the experimental surface data are mapped to the finite element mesh, the derivative of the interpolation scheme of the projection to the displacements in between an element has to be performed as well. Thus, it is recommended to project the nodal displacement information to the DIC-data in order to circumvent this step. 

Next, the sensitivity of the reaction force has to be calculated by the Lagrange-multiplier
\begin{equation}
    \label{eq:senslambda}
    \difn{\GV{p}\INDNP}{\GKap}
    = 
    \Bigg[\dif{\GVq{g}}{\Unp} - \dif{\GVq{g}}{\Qnp} \Bigg[\dif{\GV{l}}{\Qnp}\Bigg]^{-1} \dif{\GV{l}}{\Unp} \Bigg] \difn{\Unp}{\GKap}
    -
    \dif{\GVq{g}}{\Qnp} \Bigg[\dif{\GV{l}}{\Qnp}\Bigg]^{-1} \dif{\GV{l}}{\GKap} + \dif{\GVq{g}}{\GKap},
\end{equation}
i.e.,\ the sensitivity of a single reaction force, as an example, reads
\begin{equation}
  \label{eq:sensF}
  \difn{F_{\text{FEM}n+1}}{\GKap}
  =
  \GVq{M}^{T} \difn{\GV{p}\INDNP}{\GKap}.
\end{equation}
In \cite{hartmanngilbert2021}, it is also explained which quantities can be calculated on Gauss-point level, where either automatic differentiation schemes, \cite{rothehartmann2014,seidlgranzow2022}, or the code generation approach of \cite{Korelc2002a,Korelc1997,Korelc1999,Korelc2009} can be applied.

If principal strains are compared within the NLS-method, the evaluation tool of the interpolation scheme to calculate the strains has to be considered as well, but this once again requires the derivative \eqref{eq:sensU}, see \cite{hartmanngilbert2021}.

The applied approach to determine the sensitivities is called ``internal numerical differentiation'', see \cite{schittkowskibook2002}. The alternative approach is labeled as ``external numerical differentiation''. In that case, the derivatives are obtained by numerical differentiation
\begin{equation}
  \label{eq:ENDsensitivity}
  \difn{\GVt{u}\INDn}{\GKap}
  \approx
%  \sum_{i=1}^\numd
  \sum_{j=1}^\numkappa
  \frac{\GVt{u}\INDNP(\GKap + \Delta \kappa_j \GVq{e}_j) - \GVt{u}\INDNP(\GKap)}
       {\Delta \kappa_j} % \GV{e}_i
       \GVq{e}^T_j,
\end{equation}
with $\GVq{e}_j \elm{\numkappa}$ (all entries are zero except
for one entry which has a $1$ in the $j$-th row), i.e.,\ the finite element program has to be called $\numkappa+1$ times. $\Delta \kappa_j$ is a small quantity depending on the precision of the implementation, see \cite{pressteukolskyvetterlingflannery92}, where we apply double precision arithmetic. This has advantages for black box finite element programs. To obtain the quantities $\GVt{u}\INDn(\GKap + \Delta \kappa_j \GVq{e}_j)$ and $\GVt{u}\INDn(\GKap)$ the computations
\begin{equation}
    \label{eq:callsFEM}
    \begin{split}
        \GV{g}(\tnp,\Unp,\Qnp;\GKap) &= \GV{0} \\
        \GV{l}(\tnp,\Unp,\Qnp;\GKap) &= \GV{0}
    \end{split}
    \quad \leadsto \quad
    \left\{
    \begin{matrix}
      \Unp(\GKap) \\ \Qnp(\GKap) \\ \Glam\INDNP(\GKap)
    \end{matrix}
    \right.
\end{equation}
and 
\begin{equation}
    \label{eq:callsFEMdelta}
    \begin{matrix}
        \GV{g}(\tnp,\Unp,\Qnp;\GKap + \Delta \kappa_j \GVq{e}_j) = \GV{0} \\
        \GV{l}(\tnp,\Unp,\Qnp;\GKap + \Delta \kappa_j \GVq{e}_j) = \GV{0}
    \end{matrix}
    \quad \leadsto \quad
    \left\{
    \begin{matrix}
        \Unp(\GKap + \Delta \kappa_j \GVq{e}_j) \\ 
        \Qnp(\GKap + \Delta \kappa_j \GVq{e}_j) \\
        \Glam\INDNP(\GKap + \Delta \kappa_j \GVq{e}_j)
    \end{matrix}
    \right.
\end{equation}
are required (here, the post-processing step \eqref{eq:BEforceII}$_2$ is omitted for brevity). In this sense, the same time steps are required for both calculations. Otherwise, additional interpolations are necessary.
An application using the commercial finite element program Abaqus within a gradient-based optimization scheme is given by \cite{martinsandradecamposthuillier2020} (among many others), and a comparison of the required computational times with respect to internal numerical differentiation is provided in \cite{hartmanngilbert2021}.

\section{Derivation of finite element all-at-once approach}
\label{sec:derivAAOFEM}

In view of Eq.~\eqref{eq:resultDAEAAO}, in the special form \eqref{eq:DAEreduced} with $\GV{g}\INDa$ defined by Eq.~\eqref{eq:alleqlm}, we have 
\begin{equation}
    \GV{F}(t,\GVP{y},\GV{y},\GKap)
    =
    \left\{
    \begin{matrix}
        \GV{g}(t,\U,\GV{q},\GKap) - \GVq{p}(t)\\
        \GVq{g}(t,\U,\GV{q},\GKap) - \GV{p}\\
        \GVP{q}(t) - \GV{r}(t,\U,\GV{q},\GKap)
    \end{matrix}
    \right\}
\end{equation}
with
\begin{equation}
    \GV{y} =
    \left\{
        \begin{matrix}
            \U \\ \GV{p} \\ \GV{q}
        \end{matrix}
    \right\}
    \quad \text{and} \quad
    \GV{O}(\GV{y}) =
    \left\{
        \begin{matrix}
            \GV{O}\INDu(\U) \\ \GV{O}\INDp(\GV{p}) \\ \GV{0}
        \end{matrix}
    \right\}.
\end{equation}
This implies the derivatives
\begin{equation}
    \dif{\GV{F}}{\GVP{y}}
    =
    \begin{bmatrix}
        \GM{0} \\ \GV{0} \\ \GM{I}
    \end{bmatrix},
    \quad 
    \dif{\GV{F}}{\GV{y}}
    =
    \begin{bmatrix}
        \dif{\GV{g}}{\U} & \GM{0} & \dif{\GV{g}}{\GV{q}} \\[1ex] 
        \dif{\GVq{g}}{\U} & -\GM{I} & \dif{\GVq{g}}{\GV{q}} \\[1ex] 
        \dif{\GV{r}}{\U} & \GM{0} & -\dif{\GV{r}}{\GV{q}}
    \end{bmatrix},
    \quad
    \dif{\GV{F}}{\GKap}
    =
    \begin{bmatrix}
        \dif{\GV{g}}{\GKap} \\[1ex] \dif{\GVq{g}}{\GKap} \\[1ex] -\dif{\GV{r}}{\GKap}
    \end{bmatrix},
\end{equation}
as well as
\begin{equation}
    \difn{\GV{O}}{\GV{y}} 
    =
    \begin{bmatrix}
        \difn{\GV{O}\INDu}{\U} & \GM{0} & \GM{0} \\[1ex]
        \GM{0} & \difn{\GV{O}\INDp}{\GV{p}} & \GM{0} \\[1ex]
        \GM{0} & \GM{0} & \GM{0}
    \end{bmatrix}.
\end{equation}
Evaluating property \eqref{eq:qpminr}, then Eqns.~\eqref{eq:resultDAEAAO}$_{1,2}$ can be concretized to Eqns.~\eqref{eq:concreteFEAAO} using the aforementioned functional matrices.

\section{Iterative methods}
\label{app:Landweber}
Iterative methods can be used to solve the optimization problems for both reduced and all-at-once approaches. Once again following \cite{kaltenbacher2016regularization}, we only consider the Landweber iteration as a special case of gradient descent applied to the least-squares functional. Interestingly, one can avoid explicit regularization in this case, i.e., we can set $\gamma_{\text{S}}=\gamma_{\text{P}}=0$. The reduced approach iterative Landweber step reads
\begin{equation}
    \label{eq:Landweber_reduced}    
    \GKap_{k+1} = \GKap_{k} - \mu_k \GVT{J}(\GKap) \{\GV{s}(\GKap) - \GV{d}\},
\end{equation}
whereas the all-at-once approach leads to the following iterative update
\begin{equation}
    \label{eq:Landweber_all-at-once}    
    \GV{\beta}_{k+1} = \GV{\beta}_k - \mu_k \GVtT{J}(\GV{\beta})\{\GVt{s}(\GV{\beta}) - \GVt{d}\}. 
\end{equation}
The step sizes $\mu_k$ are chosen to ensure convergence. Note that, here, regularization is applied implicitly during iteration as discussed in \cite{kaltenbacher2016regularization}.

Working out the individual cases in more detail, one iteration in the reduced setting results in 
\begin{equation}
    \label{eq:Landweber-reduced}
    \begin{aligned}
        \GV{F}(\GV{y}_k,\GKap_k) &= \GV{0}, \\
        \left[\dif{\GV{F}(\GV{y}_k,\GKap_k)}{\GV{y}}\right]^T \GV{w}_k &= - \left[\difn{\GV{O}(\GV{y}_k)}{\GV{y}}\right]^\top \left\{ \GV{O}(\GV{y}_k) - \GV{d}  \right\}, \\
        \GKap_{k+1} &= \GKap_{k} - \mu_k \left[\dif{\GV{F}(\GV{y}_k,\GKap_k)}{\GKap}\right]^T \GV{w}_k,
        \end{aligned}
\end{equation}
which involves solving a non-linear mechanics problem together with its linearized adjoint. The corresponding all-at-once formulation reads
\begin{equation}
    \label{eq:Landweber-all-at-once}
    \begin{aligned}
        \GKap_{k+1} &= \GKap_{k} - \mu_k \left[\dif{\GV{F}(\GV{y}_k,\GKap_k)}{\GKap}\right]^T \GV{F}(\GV{y}_k,\GKap_k), \\
        \GV{y}_{k+1} &= \GV{y}_{k} - \mu_k \Bigg\{\left[\dif{\GV{F}(\GV{y}_k,\GKap_k)}{\GV{y}}\right]^\top \GV{F}(\GV{y}_k,\GKap_k) + \left[\difn{\GV{O}(\GV{y}_k)}{\GV{y}}\right]^\top \left\{ \GV{O}(\GV{y}_k) - \GV{d}  \right\}\Bigg\},
        \end{aligned}
\end{equation}
which does not involve any solution of a linear or non-linear model.

\section{Details on the two-step inference procedure}
\label{sec:two-step-appendix}

The derivation is an adaption of the general concepts presented in \cite{wooldridge2010econometric} to the considered two-step elastic-plastic setting. We denote with $m\INDe$ and $m\INDp$ the number of loading steps, representing the number of data vectors in the elastic and plastic regime, respectively. Resuming from Section \ref{sec:twoStep_frequentist}, the starting point is 
\begin{equation}
\GKap\INDe^* \sim \mathcal{N}\left(\GKap\INDen,\left[\GV{J}\INDe(\GKap\INDe^*)^\top \GV{\Sigma}_{\GV{r}\INDe}\GV{J}\INDe(\GKap\INDe^*) \right]^{-1} \right).
\end{equation}
This uncertainty in $\GKap\INDe^*$ now needs to be considered for the estimation of the plasticity parameters. The matrix $\GV{Q}$ can be approximated asymptotically as $\GV{Q}^* = \GV{Q}(\GKap\INDe^*,\GKap\INDp^*) = 2/(m\INDp) [\GV{J}\INDp^*]^\top \GV{J}\INDp^*$, where $\GV{J}\INDp^* = \GV{J}\INDp(\GKap\INDe^*,\GKap\INDp^*) \in \mathbb{R}^{m\INDp n \INDD \times n\INDp}$, for brevity, and plugged into
\begin{equation}
\sqrt{m\INDp} \{\GKap\INDp^* - \GKap\INDpn\}  = - \left[ \GV{Q}^* \right]^{-1} \sqrt{m\INDp} \nabla_{\GKap\INDp} \phi\INDp(\GKap\INDe^*,\GKap\INDpn).
\end{equation}
While inspecting the distribution of the right-hand-side, two independent sources of randomness appear: the uncertainty in $\GV{d}\INDp\in\mathbb{R}^{m\INDp n\INDD}$ and the uncertainty in $\GKap^*\INDe$. If $\GKap\INDe^*$ is replaced with $\GKap\INDen$ in the gradient of $\phi\INDp$, because of 
\begin{equation}
\nabla_{\GKap\INDp} \phi\INDp(\GKap\INDe^*,\GKap\INDpn)  = \nabla_{\GKap\INDp} \phi\INDp(\GKap\INDen,\GKap\INDpn)
+ \nabla_{\GKap\INDe} \nabla_{\GKap\INDp} \phi\INDp(\GKap\INDe^+,\GKap\INDpn)\{\GKap\INDe^* - \GKap\INDen\},
\end{equation}
which holds due to the mean value theorem, an additional term appears as 

\begin{equation}
\sqrt{m\INDp} \{\GKap\INDp^* - \GKap\INDpn\}  = -\left[\GV{Q}^* \right]^{-1} \sqrt{m\INDp}\left\{ \underbrace{\nabla_{\GKap\INDp} \phi\INDp(\GKap\INDen,\GKap\INDpn)}_{\GV{v}_1} + \underbrace{\nabla_{\GKap\INDe} \nabla_{\GKap\INDp} \phi\INDp(\GKap\INDe^+,\GKap\INDpn)\{\GKap\INDe^* - \GKap\INDen\}}_{\GV{v}_2} \right \}.
\end{equation}
With the definition
\begin{equation}
    \GV{J}\INDpe = \nabla_{\GKap\INDe} \left\{ \GV{s}\INDp(\GKap\INDen,\GKap\INDpn) - \GV{d}\INDp \right\},
\end{equation}
the vectors $\GV{v}_1$ and $\GV{v}_2$ are asymptotically equivalent to
\begin{align}
    \GV{v}_1 &= [\GV{J}\INDp^0]^\top \left\{ \GV{s}\INDp(\GKap\INDen,\GKap\INDpn) - \GV{d}\INDp \right\} \\
    \GV{v}_2 &= \left[[\GV{J}\INDp^0]^\top \GV{J}\INDpe^0 + [\nabla_{\GV{\kappa\INDe}}(\GV{J}\INDp^0)^\top] \left\{ \GV{s}\INDp(\GKap\INDen,\GKap\INDpn) - \GV{d}\INDp \right\}\right]\{\GKap\INDe^* - \GKap\INDen\}, 
\end{align}
where we abbreviate $\GV{J}\INDp^0 := \GV{J}\INDp(\GKap\INDen,\GKap\INDpn)$ and $\GV{J}\INDpe^0 := \GV{J}\INDpe(\GKap\INDen,\GKap\INDpn)$. Note that 
\begin{align}
\GV{c} &:= \underbrace{[\nabla_{\GV{\kappa\INDe}}[\GV{J}\INDp^0]^\top]}_{\GV{F}\INDpe^0} \left\{ \GV{s}\INDp(\GKap\INDen,\GKap\INDpn) - \GV{d}\INDp \right\}\{\GKap\INDe^* - \GKap\INDen\} \\
&= \sum_{k=1}^{n\INDe} \sum_{i=1}^{m\INDp}  [\GV{F}\INDpe^0]_{i,j,k} \left\{ \GV{s}\INDp(\GKap\INDen,\GKap\INDpn) - \GV{d}\INDp \right\}_j \{\GKap\INDe^* - \GKap\INDen\}_k \\
&= \sum_{k=1}^{n\INDe} [\GV{B}\INDpe^0]_{ik} \{\GKap\INDe^* - \GKap\INDen\}_k,
\end{align}
where tensor notation is omitted for simplicity. 

What is left is the computation of
\begin{align}
\label{eq:Zmatrix_1}
\GV{Z}(\GKap\INDen,\GKap\INDpn) &= \mathrm{Cov}_{\GV{d}\INDp,\GKap\INDe^*}\left\{\frac{2}{\sqrt{m\INDp}} \{\GV{v}_1 + \GV{v}_2\} \right\} \\
\label{eq:Zmatrix_2}
&= \frac{4}{m\INDp} \mathrm{Cov}_{\GV{d}\INDp}\left\{ \GV{v}_1 \right\} + \frac{4}{m\INDp} \mathrm{Cov}_{\GV{d}\INDp,\GKap\INDe^*}\left\{ \GV{v}_2 \right\} + \frac{8}{m\INDp} \mathrm{Cov}_{\GV{d}\INDp,\GKap\INDe^*}\left\{\GV{v}_1,\GV{v}_2\right\}.
\end{align}
Computing each term we obtain 
\begin{align}
    \mathrm{Cov}_{\GV{d}\INDp}\left\{ \GV{v}_1 \right\} &=   [\GV{J}\INDp^0]^\top \GV{\Sigma}_{\GV{r}\INDp} \GV{J}\INDp^0,\\
    \mathrm{Cov}_{\GV{d}\INDp,\GKap\INDe^*}\left\{ \GV{v}_{2a} \right\} &= \mathbb{E}_{\GV{d}\INDp,\GKap\INDe^*}[ [\GV{J}\INDp^0]^\top \GV{J}\INDpe^0 \{\GKap\INDe^* - \GKap\INDen\} \left \{[\GV{J}\INDp^0]^\top \GV{J}\INDpe^0 \{\GKap\INDe^* - \GKap\INDen\} \right \}^\top ] \\
    &= [\GV{J}\INDp^0]^\top \GV{J}\INDpe^0  \mathrm{Cov}\{\GKap\INDe^*\} [\GV{J}\INDpe^0]^\top \GV{J}\INDp^0
\end{align}
and moreover, 
\begin{align}
    \mathrm{Cov}_{\GV{d}\INDp,\GKap\INDe^*}\left\{ \GV{v}_{2b} \right\} &= \mathbb{E}_{\GV{d}\INDp,\GKap\INDe^*}
    \left[ \GV{c} \GV{c}^\top
    \right] \\
    &= \mathbb{E}_{\GV{d}\INDp,\GKap\INDe^*} \left[ \GV{B}\INDpe^0\{\GKap\INDe^* - \GKap\INDen\} \{\GKap\INDe^* - \GKap\INDen\}^\top  [\GV{B}\INDpe^0]^\top \right] \\
    &= \mathbb{E}_{\GV{d}\INDp} [
    \GV{F}\INDpe^0 \left\{ \GV{s}\INDp(\GKap\INDen,\GKap\INDpn) - \GV{d}\INDp \right\}    \mathrm{Cov}[\GKap\INDe^*] \left [\GV{F}\INDpe^0 \left\{ \GV{s}\INDp(\GKap\INDen,\GKap\INDpn) - \GV{d}\INDp \right\} \right ]^\top ] \\
    &= \GV{G}\INDpe^0 \mathrm{Cov}\{\GKap\INDe^*\}.
\end{align}
In the derivation, we assumed independence of the residuals, which resulted in 
\begin{align}
    &[ \mathbb{E}_{\GV{d}\INDp} [
     \GV{F}\INDpe^0 \left\{ \GV{s}\INDp(\GKap\INDen,\GKap\INDpn) - \GV{d}\INDp \right\} \mathrm{Cov}\{\GKap\INDe^*\} \left [\GV{F}\INDpe^0 \left\{ \GV{s}\INDp(\GKap\INDen,\GKap\INDpn) - \GV{d}\INDp \right\} \right ]^\top ] ]_{j,j'} \nonumber \\
    &= \mathbb{E}_{\GV{d}\INDp} ( 
    \sum_{i,i'=1}^{m} \sum_{k,k'=1}^{n\INDe}  [\GV{F}\INDpe^0]_{j,i,k} [\GV{F}\INDpe^0]_{j',i',k'} [\mathrm{Cov}\{\GKap\INDe^*\}]_{k,k'} \left \{ \GV{s}\INDp(\GKap\INDen,\GKap\INDpn) - \GV{d}\INDp \right\}_i \left \{ \GV{s}\INDp(\GKap\INDen,\GKap\INDpn) - \GV{d}\INDp \right\}_{i'}
    ) \\
    &= 
    \sum_{k,k'=1}^{n\INDe} \underbrace{\sum_{i=1}^{m} [\GV{F}\INDpe^0]_{j,i,k} [\GV{F}\INDpe^0]_{j',i,k'} [\mathrm{Cov}\{\GV{r}\INDp\}]_{i,i}}_{=: [\GV{G}\INDpe^0]_{j,k,j',k'}} [\mathrm{Cov}\{\GKap\INDe^*\}]_{k,k'}.
\end{align}
The last term in Eq.~\eqref{eq:Zmatrix_2} is obtained as
\begin{equation}
    \mathrm{Cov}_{\GV{d}\INDp,\GKap\INDe^*}\left\{\GV{v}_1,\GV{v}_2\right\} = \GV{0}.
\end{equation}
To see this, observe that 
\begin{align}
    \mathrm{Cov}_{\GV{d}\INDp,\GKap\INDe^*}\left\{\GV{v}_1,\GV{v}_2\right\}
    &=  \mathbb{E}_{\GV{d}\INDp,\GKap\INDe^*}[ \GV{v}_1 \GV{v}_2^\top] - \mathbb{E}_{\GV{d}\INDp,\GKap\INDe^*}\{ \GV{v}_1\} \mathbb{E}_{\GV{d}\INDp,\GKap\INDe^*}\{\GV{v}_2\}^\top \\
    &= \mathbb{E}_{\GV{d}\INDp}[ \GV{v}_1 \mathbb{E}_{\GKap\INDe^*}\{\GV{v}_2\}^\top] - \mathbb{E}_{\GV{d}\INDp,\GKap\INDe^*}\{ \GV{v}_1\} \mathbb{E}_{\GV{d}\INDp}\{ \mathbb{E}_{\GKap\INDe^*}\{\GV{v}_2\}\}^\top
\end{align}
and that $\mathbb{E}_{\GKap\INDe^*}\{\GV{v}_2\}=\GV{0}$ holds.

In summary, the covariance of the estimated parameters reads
\begin{equation}
\GV{Z}(\GKap\INDen,\GKap\INDpn) = \frac{4}{m\INDp} [ [\GV{J}\INDp^0]^\top \GV{\Sigma}_{\GV{r},p} \GV{J}\INDp^0 + [\GV{J}\INDp^0]^\top \GV{J}\INDpe^0  \mathrm{Cov}\{\GKap\INDe^*\} [\GV{J}\INDpe^0]^\top \GV{J}\INDp^0 + \sigma\INDp^2 \GV{G}\INDpe^0 \mathrm{Cov}\{\GKap\INDe^*\} ].
\end{equation}
Assuming uncorrelated plastic residuals $\GV{r}\INDp$ with common variance $\sigma\INDp^2$ results in Eq.~\eqref{eq:zMatrix}.

\section{Applied viscoplasticity model}
\label{ap:constModels}

Here, the constitutive model of a small strain Perzyna-type viscoplasticity model with von Mises yield function is summarized. The constitutive model fits into the structure of Eq.~\eqref{eq:inelasticlarge}. The stress algorithm developed in \cite{HartmannHaupt1993,HartmannLuehrsHaupt1997} integrates the evolution equations with an elastic-predictor/inelastic-corrector  method. Further, non-linear kinematic hardening of Armstrong and Frederick-type is considered, \cite{ArmstrongFrederick1966}, while no isotropic hardening is taken into account here, for the sake of brevity. $\T{X}$ defined the kinematic hardening variable, whereas $\T{E}_{\text{v}}$ symbolize the viscous strains. $\dot{s}$ represent the rate of viscous arc length, and $\lambda_{\text{v}}$ symbolizes the ``plastic'' multiplier, which is given by a constitutive model for the case of viscoplasticity. Utilizing the limit $\eta \rightarrow 0$ yields the rate-independent small strain elasto-plasticity material, which is applied in Section~\ref{sec:smallStrainPlasticityExp}. The constitutive model is compiled in Tab.~\ref{tab:plasticityModel}.
\begin{table}[ht]
    \centering
    \caption{Summary of constitutive equations for a model of viscoplastic (von Mises-type viscoplasticity with non-linear kinematic hardening)}
    \label{tab:plasticityModel}
    \renewcommand{\arraystretch}{1.2}
    \newcommand*{\SetEqNum}{%
      \refstepcounter{equation}%
      \thetag\theequation
    }
    {
    \resizebox{0.475\textwidth}{!}{
    \begin{tabular}{>{\centering\arraybackslash}m{1.5cm}|c|c r}
        \rowcolor{HgrauTU}
        &
        \text{Elasticity}
        &
        \multicolumn{2}{c}{\text{Viscoplasticity}} \\
        \hline
        \cellcolor{HgrauTU}
        Yield function
        &
        \multicolumn{2}{l}{
        \rule[-2ex]{0ex}{6ex}
        $\begin{aligned}
          \displaystyle f(\T{\sigma},\T{X})
          =
          \frac12 (\T{\sigma}-\T{X})^{\text{D}} \cdot (\T{\sigma}-\T{X})^{\text{D}} - \frac13 k^2
        \end{aligned}$
        \rule[-3ex]{0ex}{6ex}
        }
        &
        \SetEqNum\label{eq:yieldFun} \\
        \hline
        \cellcolor{HgrauTU}
        Loading condition
        &
        $f < 0$ %or
        &
        \multicolumn{2}{c}{$f \geq 0$} \\
        \hline
        \cellcolor{HgrauTU}
        Elasticity relation
        &
        \multicolumn{2}{l}{
        \rule[-3ex]{0ex}{7ex}
        $\begin{aligned}
          \T{\sigma} =
          K(\tr{\T{E}})\, \T{I} + 2 G(\T{E} - \T{E}_{\text{v}})^{\text{D}}
        \end{aligned}$}
        &
        \SetEqNum\label{eq:elasticityRelation} \\
        \hline
        \cellcolor{HgrauTU}
        \text{Flow rule}
        &
        \multicolumn{1}{l}
        {$\dot{\T{E}}_{\text{v}} = \T{0}$}
        &
        \multicolumn{1}{|l}{
        \rule[-2.5ex]{0ex}{7ex}
        $\begin{aligned}
          \dot{\T{E}}_{\text{v}} &=
          \lambda_\text{v}
          \frac{(\T{\sigma}-\T{X})^{\text{D}}}{\vert\vert(\T{\sigma}-\T{X})^{\text{D}}\vert\vert}
        \end{aligned}$
        \rule[-3.5ex]{0ex}{7ex}
        }
        &
        \SetEqNum\label{eq:flowRule1} \\
        \hline
        \cellcolor{HgrauTU}
        Kinematic hardening
        &
        \multicolumn{1}{l}
        {$\dot{\T{X}} = \T{0}$}
        &
        \multicolumn{1}{|l}{
        $\begin{aligned}
          \dot{\T{X}} & =
          c \dot{\T{E}}_{\text{v}} - b \dot{s} \T{X}
        \end{aligned}$
        }
        &
        \SetEqNum\label{eq:kinematicHardening} \\
        \hline
        \cellcolor{HgrauTU}
        Abbrevia-\linebreak tions
        &
        \multicolumn{2}{l}{
        \rule[-2.5ex]{0ex}{7ex}
        $\begin{aligned}
          \dot{s}
          =
          \lambda_\text{v} \sqrt{\frac23}, \qquad \lambda_\text{v} =
          \frac{1}{\eta}
          \Big\langle
          \frac%
          {f(\T{\sigma},\T{X})}
          {\sigma_0}
          \Big\rangle^{r}
        \end{aligned}$
        \rule[-3.5ex]{0ex}{7ex}
        }
        &
        \SetEqNum\label{eq:plasticMultiplier} \\
        \hline
      \cellcolor{HgrauTU}
      Material parameters
      &
        \multicolumn{2}{l}{
        \rule[-3ex]{0ex}{7ex}
        $\begin{aligned}
          \boldsymbol{\kappa} = \{K,\,G,\,k,\,b,\,c,\,\eta,\,r\}\elm{7}
        \end{aligned}$}
        &
        \SetEqNum\label{eq:matparviscoplasticity} \\
        \hline
    \end{tabular}
    }
    }
\end{table}
The case distinction in elastic and viscoplastic deformations is enabled with Macauley brackets, $\langle x \rangle = x$ for $x > 0$ and  $\langle x \rangle = 0$ for $x \le 0$. The parameter $\sigma_0 = \SI{1}{\N\per\mm\squared}$ is assumed for obtaining a dimensionless quantity in the Macauley brackets. Thus, $\sigma_0$ is not seen as a material parameter. The material parameters as shown in Tab.~\ref{tab:plasticityModel} are: bulk modulus $K$, shear modulus $G$, yield stress $k$, the parameters $b$ and $c$ for describing the non-linear kinematic hardening, the viscosity $\eta$, and the exponent $r$ of the viscosity function. 

In the context of the general representation \eqref{eq:smallproblem}, the viscous strains and the back stress tensor define the internal variables, $\LV{q}^T = \{\LV{E}_\text{v}^T, \LV{X}^T\}$. The evolution equations $\LV{r}(\LV{E},\LV{q},\GKap)$ are given by Eqns.~\eqref{eq:flowRule1} and  \eqref{eq:kinematicHardening}, which contain a case distinction.

\bibliographystyle{ieeetr}
\bibliography{      
    main,
    literature_flaschel
}
%\bibliography{      
%    ./literature/references,
%    ./literature/literatur,
%    ./literature/myownbooks,
%    ./literature/hartmann-databasis,
%    ./literature/dissertationen,
%    ./literature/misc,
%    ./literature/literature_anton,
%    ./literature/literature_flaschel,
%    ./literature/literature_wessels
%}  %%% Uncomment this line and comment out the ``thebibliography'' section below to use the external .bib file (using bibtex) .

% \end{thebibliography}

\end{document}